\documentclass[longauth]{aa} 
\usepackage{txfonts}
\usepackage{natbib}
\usepackage[rightcaption]{sidecap}
\usepackage{graphicx}   
\newcommand{\GE}{\emph{Gamow}}
\newcommand{\eg}{\textit{e.g.}}
\newcommand{\Log}{\text{Log}}

\begin{document} 

   \title{Fires in the deep: The luminosity distribution of early-time gamma-ray-burst afterglows in light of the {\it Gamow Explorer} sensitivity requirements\thanks{This study is partially based on data collected under ESO programmes 089.A-0067(B) (PI: Fynbo), 091.A-0703(A) (PI: Kr\"uhler), 091.A-0703(B) (PI: Kr\"uhler), 091.C-0934(B) (PI: Kaper).} \thanks{Table B.1 is only available in electronic form at the CDS via anonymous ftp to cdsarc.u-strasbg.fr (130.79.128.5) or via http://cdsweb.u-strasbg.fr/cgi-bin/qcat?J/A+A/.
}}

\author{D. A. Kann\thanks{Deceased} \inst{1,2} 
          \and
           N. E. White\thanks{E-mail: newhite@gwu.edu} \inst{3}
          \and
            G. Ghirlanda \inst{4}
          \and
            S. R. Oates \inst{5}
          \and
            A. Melandri \inst{4}
          \and
            M. Jel\'inek \inst{6}
          \and
            A. de Ugarte Postigo \inst{7}
          \and
            A. J. Levan \inst{8,9}
          \and
            A. Martin-Carrillo \inst{10}
          \and
            G. S.-H. Paek \inst{11}
          \and
            L. Izzo \inst{12,13}
          \and
            M. Blazek \inst{14}
          \and
            C. C. Th\"one \inst{6}
          \and
            J. F. Ag\"u\'i Fern\'andez \inst{14}
          \and
            R. Salvaterra \inst{15}
          \and
            N. R. Tanvir \inst{16}
          \and
            T.-C. Chang \inst{17}
          \and
            P. O'Brien \inst{16}
          \and
            A. Rossi \inst{18}
          \and
            D. A. Perley \inst{19}
          \and
            M. Im \inst{11}
          \and
            D. B. Malesani \inst{8}
          \and
            A. Antonelli \inst{20}
          \and
            S. Covino \inst{4}
          \and
            C. Choi \inst{11,21}
          \and
            P. D'Avanzo \inst{4}
          \and
            V. D'Elia \inst{20,22}
          \and
            S. Dichiara \inst{23,24,25}
          \and
            H.M. Fausey \inst{3}
          \and
            D. Fugazza \inst{4}
          \and
            A. Gomboc \inst{26}
          \and
            K. M. Gorski \inst{17,27}
          \and
            J. Granot \inst{28,29,3}
          \and
            C. Guidorzi \inst{18,30,31}
          \and
          L. Hanlon \inst{10}
          \and
            D. H. Hartmann \inst{32}
          \and
            R. Hudec \inst{6,33,34}
          \and
            H. D. Jun \inst{11,35}
          \and
            J. Kim \inst{11,36}
          \and
            Y. Kim \inst{21}
          \and
            S. Klose \inst{37}
          \and
            W. Klu{\'z}niak \inst{38}
          \and
            S. Kobayashi \inst{19}
          \and
            C. Kouveliotou \inst{3}
          \and
            A. Lidz \inst{39}
          \and
            M. Marongiu \inst{40}
          \and
            R. Martone \inst{2,30}
            \and
            P. Meintjes \inst{41}
          \and
            C. G. Mundell \inst{42,43}
            \and
            D. Murphy \inst{10}
          \and
            K. Nalewajko \inst{38}
          \and
            W.-K. Park \inst{21}
          \and
            D. Sz\'ecsi \inst{44}
          \and
            R. J. Smith \inst{19}
          \and
            B. Stecklum \inst{37}
          \and
            I. A. Steele \inst{19}
          \and
            J. \v Strobl \inst{6}
          \and
            H.-I- Sung \inst{21}
          \and
            A. Updike \inst{45}
          \and
            Y. Urata \inst{46}
          \and
            A.J. van der Horst \inst{3}
          }

   \institute{Hessian Research Cluster ELEMENTS, Giersch Science Center, Max-von-Laue-Stra$\beta$e 12, Goethe University Frankfurt, Campus Riedberg, D-60438 Frankfurt am Main, Germany 
           \and
    Instituto de Astrof\'isica de Andaluc\'ia (IAA-CSIC), Glorieta de la Astronom\'ia s/n, 18008 Granada, Spain 
         \and
            Department of Physics, George Washington University, Corcoran Hall, 725 21st Street NW, Washington DC 20052, USA 
         \and
            INAF - Osservatorio Astronomico di Brera, Via E. Bianchi 46, I-23807, Merate (LC), Italy 
         \and
            Department of Physics, Lancaster University, Lancaster, LA1 4YB, UK 
         \and
            Astronomical Institute of the Czech Academy of Sciences (ASU-CAS), Fri\v cova 298, 251 65 Ond\v rejov, Czech Republic 
         \and 
         Aix Marseille Univ, CNRS, LAM Marseille, France 
         \and 
            Department of Astrophysics/IMAPP, Radboud University, Nijmegen, Netherlands 
        \and 
            Department of Physics, University of Warwick, Coventry, UK 
         \and
            School of Physics and Centre for Space Research, University College Dublin, Dublin 4, Ireland 
         \and
             SNU Astronomy Research Center, Dept. of Physics \& Astronomy, Seoul National University, 1 Gwanak-ro, Gwanak-gu, Seoul 08826, Republic of Korea 
         \and 
            DARK, Niels Bohr Institute, University of Copenhagen, Jagtvej 128, 2200 Copenhagen, Denmark 
            \and
            Osservatorio Astronomico di Capodimonte, Istituto Nazionale di Astrofisica (INAF), Salita Moiariello, 16, I-80131 Naples, Italy 
         \and
            Centro Astronómico Hispano-Alemán, Observatorio de Calar Alto, Sierra de los Filabres, 04550 Gérgal, Spain 
         \and 
            INAF IASF-Milano, Via Alfonso Corti 12, I-20133 Milano, Italy 
         \and 
            School of Physics and Astronomy, University of Leicester, University Rd, Leicester, LE1 7RH, UK 
         \and 
            Jet Propulsion Lab, 4800 Oak Grove Dr, Pasadena, CA 91109, USA 
         \and 
            INAF - Osservatorio di Astrofisica e Scienza dello Spazio, Via Piero Gobetti 93/3, 40129 Bologna, Italy 
         \and
            Astrophysics Research Institute, Liverpool John Moores University, IC2, Liverpool Science Park, 146 Brownlow Hill, Liverpool L3 5RF, UK 
         \and
            INAF - Osservatorio Astronomico di Roma, Via Frascati 33, I-00040 Monte Porzio Catone (RM), Italy 
         \and
            Korea Astronomy and Space Science Institute, 776 Daedukdae-ro, Yuseong-Gu, Daejeon 34055, Republic of Korea 
        \and
            Space Science Data Center (SSDC) - Agenzia Spaziale Italiana (ASI), Via del Politecnico, I-00133 Roma, Italy 
         \and
            Department of Astronomy, University of Maryland, College Park, MD 20742–4111, USA 
         \and
            Astrophysics Science Division, NASA Goddard Space Flight Center, 8800 Greenbelt Rd, Greenbelt, MD 20771, USA 
         \and
            Department of Astronomy and Astrophysics, The Pennsylvania State University, 525 Davey Lab, University Park, PA 16802, USA 
         \and
            Center for Astrophysics and Cosmology, University of Nova Gorica, Vipavska 13, 5000 Nova Gorica, Slovenia 
         \and
            Nicolaus Copernicus Superior School, ul. Nowogrodzka 47A, 00-695, Warsaw, Poland 
        \and
            Astrophysics Research Center of the Open university (ARCO), The Open University of Israel, P.O Box 808, Ra'anana 43537, Israel 
        \and
            Department of Natural Sciences, The Open University of Israel, P.O Box 808, Ra'anana 43537, Israel 
        \and
            Department of Physics and Earth Science, University of Ferrara, Via Saragat 1, 44122 Ferrara, Italy 
         \and
            INFN – Sezione di Ferrara, Via Saragat 1, 44122 Ferrara, Italy 
         \and
            Clemson University, Department of Physics \& Astronomy, Clemson, SC 29634 
        \and
            Czech Technical University in Prague, Faculty of Electrical Engineering, Prague, Czech Republic 
        \and
            V. P. Engelgardt Astronomical Observatory, Kazan Federal University, Kazan, Republic of Tatarstan 
         \and
            Department of Physics, Northwestern College, Orange City, IA 51041, U.S.A. 
       \and
            Daegu National Science Museum, 20, Techno-daero 6-gil, Yuga-myeon, Dalseong-gun, Daegu 43023, Republic of Korea 
         \and
            Th\"uringer Landessternwarte Tautenburg, Sternwarte 5, 07778 Tautenburg, Germany 
         \and
            Nicolaus Copernicus Astronomical Center, Polish Academy of Sciences, Bartycka 18, 00-716 Warsaw, Poland 
         \and
            Department of Physics and Astronomy, University of Pennsylvania, Philadelphia, PA 19104, USA 
        \and
            INAF -- Osservatorio Astronomico di Cagliari - via della Scienza 5 - I-09047 Selargius, Italy 
        \and
            Department of Physics, University of the Free State, PO Box 339, Bloemfontein 9300, South Africa 
         \and
            European Space Agency (ESA), European Space Astronomy Centre, E-28692 Villanueva de la Canñada, Madrid, Spain 
        \and
            Department of Physics, University of Bath, Claverton Down, Bath, BA2 7AY 
         \and
            Institute of Astronomy, Faculty of Physics, Astronomy and Informatics, Nicolaus Copernicus University, Grudzi\k{a}dzka 5, 87-100 Toru\'n, Poland 
        \and
            Roger Williams University, Bristol, RI, USA 
        \and
            Institute of Astronomy, National Central University, Chung-Li 32054, Taiwan 
             }

   \date{Received 5 October 2023 / Accepted 15 February 2024}

\authorrunning{Kann et al.}
\titlerunning{High-redshift GRB Afterglow Predictions}
  
  \abstract
   {Gamma-ray bursts (GRBs) are ideal probes of the Universe at high redshift ($z$), pinpointing the locations of the earliest star-forming galaxies and providing bright backlights with simple featureless power-law spectra that can be used to spectrally fingerprint the intergalactic medium and host galaxy during the period of reionization. Future missions such as \emph{Gamow Explorer} (hereafter \GE{})\ are being proposed to unlock this potential by increasing the rate of identification of high-$z$ ($z>5$) GRBs in order to rapidly trigger observations from $6-10$ m ground telescopes, the James Webb Space Telescope (JWST), and the upcoming  Extremely Large Telescopes (ELTs).}
   {\GE{} was proposed to the NASA 2021 Medium-Class Explorer (MIDEX) program as a fast-slewing satellite featuring a wide-field lobster-eye X-ray telescope (LEXT) to detect and localize GRBs with arcminute accuracy, and a narrow-field multi-channel photo-$z$ infrared telescope (PIRT) to measure their photometric redshifts for $>80\%$ of the LEXT detections using the Lyman-$\alpha$ dropout technique. We use a large sample of observed GRB afterglows to derive the PIRT sensitivity requirement.}
   {We compiled a complete sample of GRB optical--near-infrared (optical--NIR) afterglows from 2008 to 2021, adding a total of 66 new afterglows to our earlier sample, including all known high-$z$ GRB afterglows. This sample is expanded with over 2837 unpublished data points for 40 of these GRBs. We performed full light-curve and spectral-energy-distribution analyses of these afterglows to derive their true luminosity at very early times. We compared the high-$z$ sample to the comparison sample at lower redshifts. For all the light curves, where possible, we determined the brightness at the time of the initial finding chart of \GE{}, at different high redshifts and in different NIR bands. This was validated using a theoretical approach to predicting the afterglow brightness. We then followed the evolution of the luminosity to predict requirements for ground- and space-based follow-up. Finally, we discuss the potential biases between known GRB afterglow samples and those to be detected by \GE{}.}
   {We find that the luminosity distribution of high-$z$ GRB afterglows is comparable to those at lower redshift, and we therefore are able to use the afterglows of lower-$z$ GRBs as proxies for those at high $z$. We find that a PIRT sensitivity of $15\,\mu$Jy (21 mag AB) in a 500~s exposure simultaneously in five NIR bands within 1000s of the GRB trigger will meet the \GE{} mission requirements. Depending on the $z$ and NIR band, we find that between 75\% and 85\% of all afterglows at $z>5$ will be recovered by \GE{} at $5\sigma$ detection significance, allowing the determination of a robust photo-$z$. As a check for possible observational biases and selection effects, we compared the results with those obtained through population-synthesis models, and find them to be consistent.}
   {\GE{} and other high-$z$ GRB missions will be capable of using a relatively modest 0.3m  onboard NIR photo-$z$ telescope to rapidly identify and report high-$z$ GRBs for further follow-up by larger facilities, opening a new window onto the era of reionization and the high-redshift Universe.}


   \maketitle
%

\section{Introduction}


At their peak, gamma-ray bursts (GRBs) are the most luminous events in the Universe and occur when either massive stars reach their final stage in a supernova explosion or when binary compact objects  ---one of which is a neutron star (for a review see \citealp{zhang_2018})--- merge. In both cases, a relativistic jet emerges powered by accretion onto a newly formed black hole, which results in a bright panchromatic afterglow. In the first few hours to days following a GRB, the afterglow is brighter by several orders of magnitude than a conventional supernova and can be detected to very high redshift ($z\sim 10$ and in principle to $z\sim 20$ {\it viz.} \citealp{Akerlof1999Nature,Boer2006ApJ,Kann2007AJ,Racusin2008Nature,Bloom2009ApJ,Perley2011AJ,Cucchiara2011ApJ,zhi-ping,burns_2023}). As the class of ``long/soft'' GRBs \citep{Mazets1981ApSS,Kouveliotou1993ApJ} ---also known as ``type II'' GRBs in a more physically motivated classification scheme \citep{Zhang2009ApJ}--- is linked to the explosions of massive stars \citep[for reviews, see \eg,][]{Woosley2006ARAA,Hjorth2012Book,Cano2017AdAst}, the detection and localization of a long GRB points to a young region forming massive stars within a galaxy.

The intrinsic spectra of GRB afterglows are nonthermal synchrotron radiation (e.g., \citealt{Rossi2011AA,Zheng2012ApJ}, see \eg, \citealt{Pe'er2015AdAst,Kumar2015PhR} for reviews of GRB emission physics). Therefore, the intrinsic spectrum within an observing band can be described by a simple power law, or a smoothly broken power law \citep{Sari1998ApJ,GranotSari2002}. This not only means that they are interesting objects to study in their own right, but also implies that they are ideal backlight sources with which to probe the intra- and intergalactic medium, especially with rapid high-signal-to-noise-ratio (S/N) and high-resolution spectroscopy \citep{Vreeswijk2007AA,Vreeswijk2011AA,Prochaska2009ApJ,Sheffer2009ApJ,D'Elia2009ApJ,D'Elia2010MNRAS,deUgartePostigo2012AA,Kruhler2013AA,Thoene2013MNRAS,Heintz2019AA}.

GRBs are especially interesting tools with which to study the high-redshift ($z>5$) Universe \citep{Salvaterra2015,Yuan2016SSRv}. Not only do they pinpoint extremely distant and very faint star forming galaxies \citep[\eg,][]{Tanvir2012ApJ,Basa2012AA,Salvaterra2013,McGuire2016ApJ}, but, given sufficient S/N, their afterglow spectra also enable the study of the era of reionization \citep{Totani2006PASJ,Gallerani2008MNRAS,Vangioni2015MNRAS,Lidz2021}, the phenomenon of UV leakage \citep{Tanvir2019MNRAS,Vielfaure2020AA}, and the transparency of the Gunn-Peterson trough \citep{Chornock2013ApJ,Chornock2014arXiv,Hartoog2015AA}. GRBs furthermore allow the study of cosmic chemical enrichment \citep{Sparre2014,Saccardi2023}, the evolution of dust at high $z$ \citep{Perley2010MNRAS,Jang2011ApJ,Zafar2011ApJ,Zafar2011AA,Bolmer2018AA}, and, with sufficiently large samples, trace the star-formation history of the Universe \citep{Lloyd-Ronning2002ApJ,Kistler2008ApJ,Li2008MNRAS,Wang2009MNRAS,Wang2011ApJ,Virgili2011MNRAS,Ishida2011MNRAS,Robertson2012ApJ,Wang2013AA,Hao2020ApJS,Palmerio2020arXiv,Ghirlanda2022}. 
Moreover, GRBs at high redshifts can constrain the possible evolution of the initial mass function with redshift ( \citealt{Fryer2022}) and the existence of Pop-III stars \citep{Burlon2008ApJ,Ma2015,Ma2017}.

However, the detectability of GRBs at high redshifts represents a problem \citep[\eg,][]{Qin2010MNRAS,Ghirlanda2015MNRAS}. Before the launch of the \emph{Neil Gehrels Swift Observatory} satellite (henceforth \emph{Swift}, \citealt{Gehrels2004apJ}), the most distant known GRB was GRB 000131 at $z=4.50$ \citep{Andersen2000AA}. \emph{Swift}'s coded-mask imager, the Burst Alert Telescope \citep[BAT,][]{Barthelmy2005SSRv} along with novel image-based triggering schemes \citep{Lien2014ApJ}, the rapid repointing capability, and the arcsecond localization capabilities of the onboard X-ray Telescope \citep[XRT,][]{Burrows2005SSRv} promised a general increase  not only 
in the detection but also in the precise localization rate, and especially an increase in the detection rate of high-$z$ GRBs. In a certain sense, \emph{Swift} has fulfilled this promise with the first detection of a GRB at $z>6$, GRB 050904 \citep{Cusumano2007AA,Haislip2006Nature,Kawai2006Nature}, and the subsequent extension to the spectroscopic (GRB 090423, $z=8.26$, \citealt{Tanvir2009Nature,Salvaterra2009Nature}) and photometric (GRB 090429B, $z\approx9.4$, \citealt{Cucchiara2011ApJ}) redshift record holders.

Over the past 18 years, Swift has discovered 17 GRBs at $z > 5$, representing $\sim$3\% of the total for which redshifts have been obtained. As the bright GRB afterglow fades quickly within a day or two, it is critical to obtain early high-S/N, medium-resolution (R$\gtrsim2500$) near-infrared (NIR) spectroscopy for these events.  To date, only four have medium-resolution NIR spectra, and all at redshift $z < 6.3$, when reionization was largely complete.
A bottleneck is that large ground-based telescopes are currently required to identify  high-redshift candidates (\eg,   as optical dropout sources),
which adds unacceptable delays (see Appendix \ref{redshift delays}).
Even GRBs in the redshift range of $z\approx4.5-6$ have been rare among detections (GRB 160327A, \citealt{deUgartePostigo2016GCN19245}; GRB 201014A, \citealt{deUgartePostigo2020GCN28650}; GRB 201221A, \citealt{Malesani2020GCN29100}). 

Analyses of near redshift-complete samples suggest that only 1--2\% of \emph{Swift} bursts originate at $z>6$ \citep{Perley2016shoals1}. If GRBs are to fulfil their promise as cosmological probes in the era of the James Webb Space Telescope (JWST, \citealt{Gardner2006SSRv}) and the upcoming 30-40 m class ground-based optical/NIR Extremely Large Telescopes (ELTs), future GRB missions must not only mirror \emph{Swift}'s capabilities of rapidly and precisely localizing GRBs, but also yield an increased total rate of GRBs, with the sensitivity optimized to increase the fraction from high-$z$. With increased rates approaching 1 or 2 GRBs per day, it will be important for future GRB missions to provide confirmed redshifts in order to avoid triggering large telescopes to follow up every GRB. 

There are proposed missions that include a NIR telescope to directly identify high-$z$ events using the photo-$z$ technique, so as to immediately flag the interesting bursts for follow-up by JWST and large ($>6$m) ground-based telescopes of these rare high-$z$ GRB events. These include \GE{}  \citep{White2021SPIE}, the High-$z$ Gamma-ray bursts for Unraveling the Dark Ages Mission (\emph{HiZ-GUNDAM}) \citep{Yonetoku2014SPIE,Kinugawa2019ApJ}, and the Transient High-Energy Sky and Early Universe Surveyor (\emph{THESEUS}) \citep{Amati2018AdSpR,Amati2021THESEUS,Tanvir2021THESEUS,Ghirlanda2021THESEUS}. The science goals and concepts of these missions are similar to those previously and unsuccessfully proposed a decade ago to NASA:  \emph{JANUS}  \citep{Burrows2010SPIE,Roming2012MSAIS} and \emph{Lobster} \citep{Gehrels_Lobster}. This paper considers the history of GRB afterglow measurements in order to assess the requirements for future GRB missions and follow-up by $>6$m observatories to study a large sample GRBs from the high-redshift Universe. We focus on using the predicted GRB afterglow NIR brightness from high redshift to specify the required  sensitivity of the \GE{} NIR telescope. This work is applicable to all the mission concepts mentioned. 

The paper is organized as follows. In Sect. \ref{Gamow} we give a brief overview of \GE{}. In Sect. \ref{sample} we extend the work of \citep[K6, K10, K11 henceforth]{Kann2006ApJ,Kann2010ApJ,Kann2011ApJ} and introduce our new, expanded sample of optical--NIR afterglows to provide light curves shifted to various high-$z$ redshifts, which can then be used to set the sensitivity requirements. To check for observational biases and/or selection effects in Sect. \ref{simulations}, we approach the problem from the standpoint of GRB and afterglow predictions from population synthesis models of long GRB afterglows (\eg,\citealt{Ghirlanda2015MNRAS}). We discuss our results in Sect. \ref{results}, including limitations and outcomes from the current capabilities and requirements of \GE{}, before concluding in Sect. 
\ref{summary and conclusions}. We present more details on the expanded afterglow sample  in Appendix \ref{sampledetails}, both at high-$z$ and at low-$z,$ and the challenge of rapidly obtaining redshifts using ground-based facilities  in Appendix \ref{redshift delays}.

In our calculations, we assume a flat Universe with a matter density $\Omega_M=0.27$, a cosmological constant $\Omega_\Lambda=0.73$, and a Hubble constant $H_0=71$ km s$^{-1}$ Mpc$^{-1}$ \citep{Spergel2003ApJS}, to remain in agreement with our older sample papers\footnote{Use of Planck cosmological parameters \citep{Planck2020AA} does not change any conclusions of this work.}. Errors are given at the $1\sigma$ level, and upper limits at the $3\sigma$ level for a parameter of interest.

\section{The \GE{} explorer mission}
\label{Gamow}

\GE{} was proposed to the NASA 2021 Medium-Class Explorer (MIDEX) opportunity \citep{White2021SPIE}. It features a wide-field X-ray detector (LEXT, a Lobster Eye X-ray Telescope covering 0.3 to 5 keV) \citep{Feldman2021SPIE} to detect the GRBs across a wide field of 0.41 sr with a localization accuracy of $<3$\arcmin. A rapidly slewing spacecraft points a narrow field of view multichannel NIR telescope (PIRT, Photo-$z$ InfraRed Telescope \citep{Seiffert_2021}) to detect the afterglow in the $0.5-2.4\;\mu$m band, with 5 channels.  \GE{} will be orbiting around L2 so as to minimize viewing constraints (\eg, Earth limb) that limit low Earth orbiting missions such as \emph{Swift}. \GE{} will observe within the JWST's field of regard (sun angle 85$\deg$ to 135$\deg$), to ensure high$-z$ GRBs are available for follow-up. 

It is expected PIRT will start taking data $\approx100$ s after the trigger, and the first exposure of 500 s duration will be used to detect the GRB afterglow, measure its position to arc second precision and its brightness simultaneously in five different filters to determine a photometric redshift estimate via the Lyman dropout technique \citep{Fausey2023, Steidel1992, Haislip2006Nature,Kruhler2011AA}. This redshift determination will identify high-$z$ ($z\gtrsim5$) candidates and alert observers to allow selective observations with $6-10$ m class ground- and space-based telescopes. While the \GE{} 2021 MIDEX proposal was not successful, it is expected that a variation on the \GE{} concept, \emph{THESEUS}, or \emph{HiZ GUNDAM} will be necessary to survey and employ high$-z$ GRBs for cosmology.

A $z>5$ sample of $>20$ GRBs with high signal to noise follow-up spectra with R $\sim$ 2500 is required to determine the profile of reionization with sufficient precision to distinguish between slow and fast reionization models \citep{Lidz2021}. To accomplish this within a typical $2-3$ yr prime mission lifetime requires a high-$z$ detection rate $\gtrsim10$ times that of \emph{Swift}. For the case we are considering of \GE{} LEXT must be sensitive enough to detect GRBs that are potentially faint because of high luminosity distance, stretched out in duration by cosmological time dilation, and spectrally soft due to redshifting. The PIRT must be able to significantly detect the associated afterglows \emph{and} measure their photo-$z$ to a precision reliable enough to trigger observations on the most advanced astronomical facilities. To obtain the required sample and minimize the cost of the LEXT (which scales with field of view) it is essential to not miss high$-z$ GRBs (\emph{approx.} one every month). This requires a PIRT sensitivity sufficient to retrieve the redshift for at least $\simeq 80\%$ of GRB afterglows. This is to be compared to the \emph{Swift} redshift retrieval rate of 30\% \citep{Perley2016ApJ} using primarily ground-based telescopes.

\section{Sample selection and analysis}
\label{sample}

Over the past two decades, we have been building a database of GRB afterglow and host galaxy photometry\footnote{This database is now being prepared for online access \citep{Blazek2020SPIE}.} \citep[][K6, K10, K11]{Zeh2004Apj,Zeh2006ApJ}. Using this database as a foundation, we here describe how we expand it to create an extended sample with which we will be able to study the early time luminosity distribution of GRB afterglows.

\subsection{The high-redshift sample}

\begin{figure*}
\begin{center}
        \centering
            \includegraphics[width=\columnwidth]{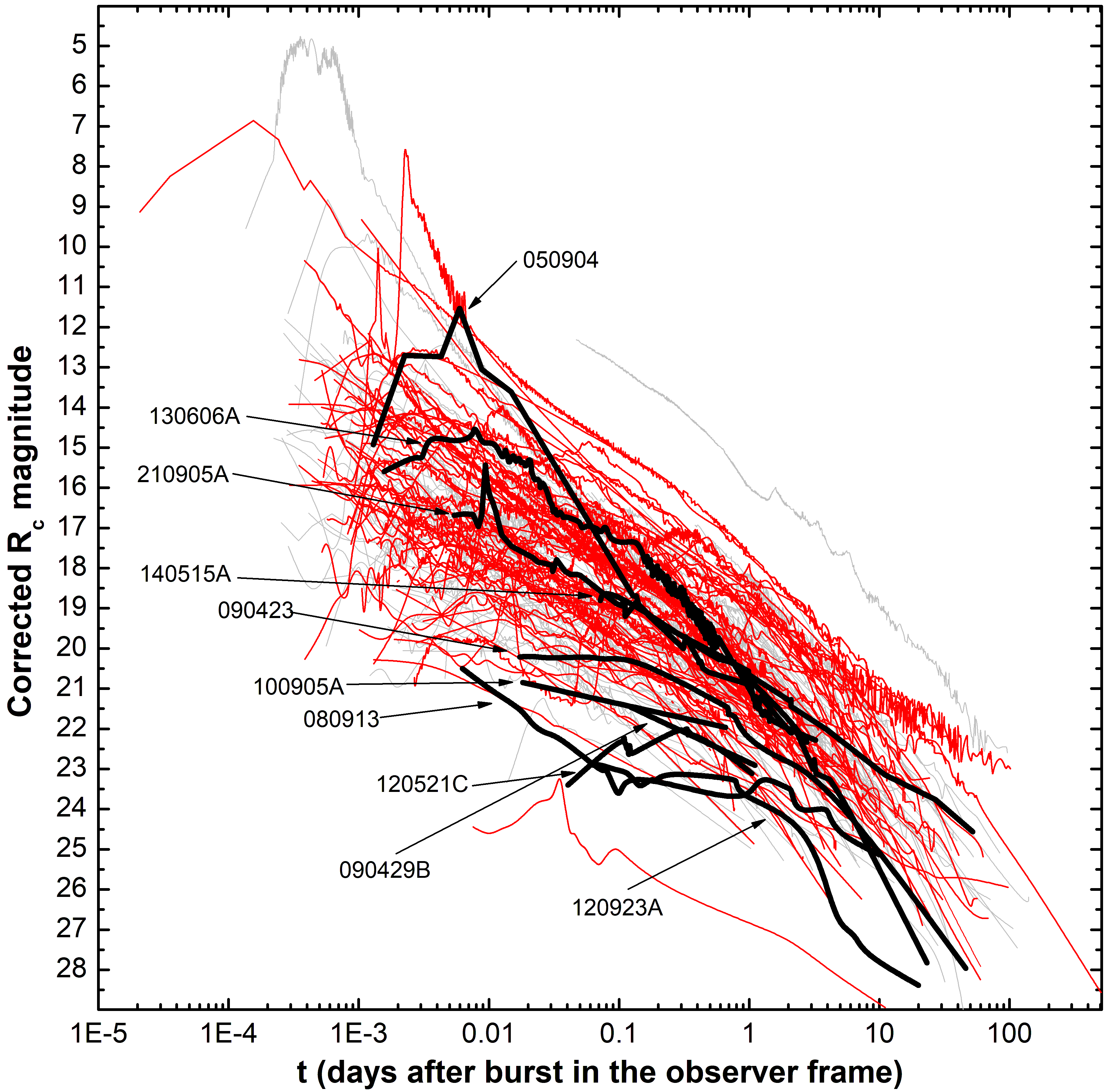} 
            \includegraphics[width=\columnwidth]{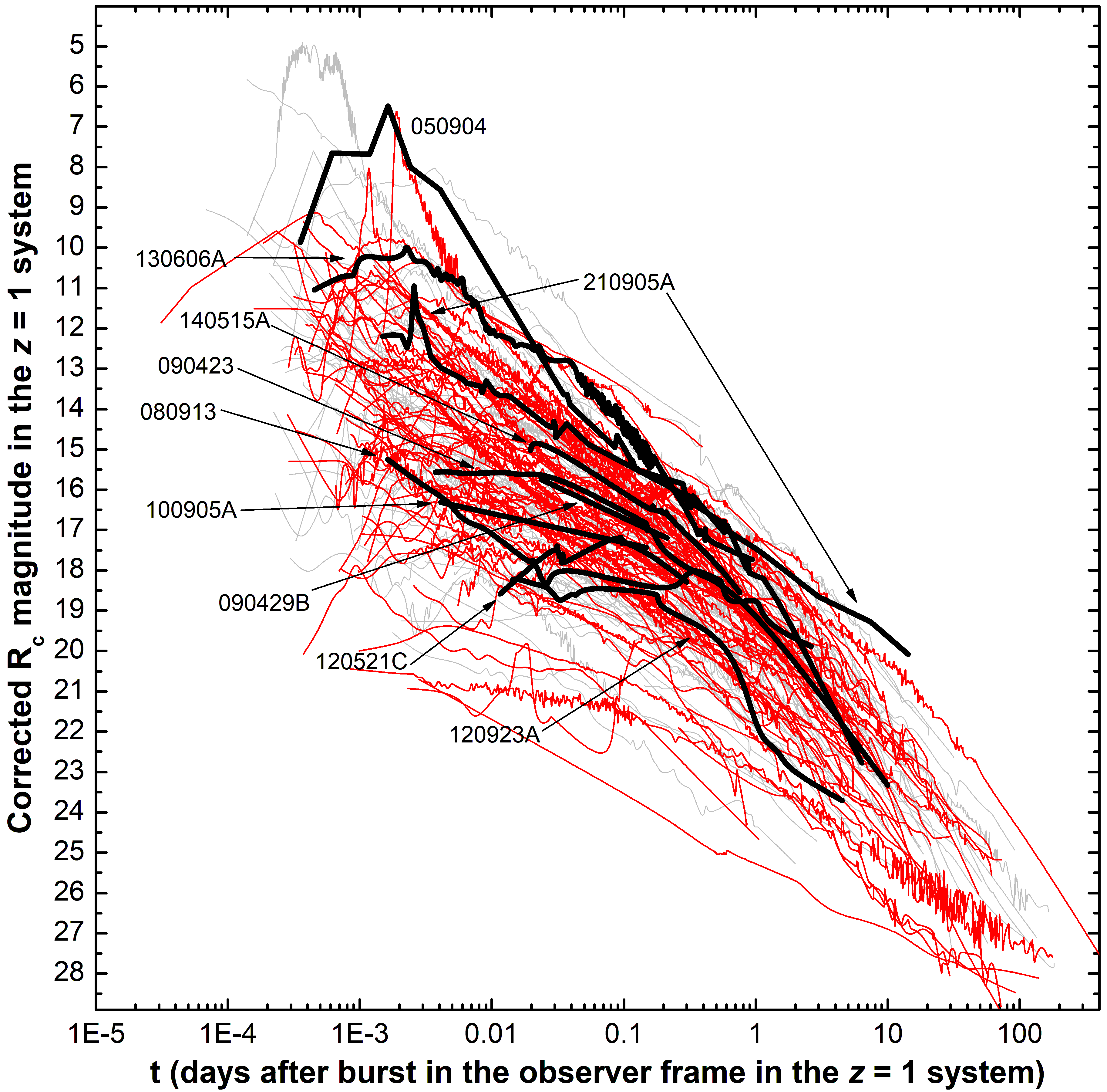}
                  \caption{GRB afterglow sample from 2008 to 2021. Left panel: The observer frame light curves. These have been corrected for Galactic foreground extinction as well as, where applicable, the host-galaxy contribution and the supernova contribution. Light-gray curves represent the pre-\emph{Swift} and \emph{Swift} era samples of K6, K10, K11 (Type II GRB afterglows only). Red curves represent new GRB afterglow light curves presented in this work. Thick black curves represent the $z\gtrsim6$ high-$z$ sample. These light curves have been constructed by using the intrinsic spectral slopes to extrapolate the magnitude from redder bands to the $R_C$ band, essentially assuming the Universe is completely transparent. For these, the GRBs are labeled. Right panel: GRB afterglow light curves when shifted to all originate from $z=1$. See text for more details.}
            \label{KPobs}
\end{center}
\end{figure*}


To study the luminosity distribution of GRB afterglows at high redshift, one must first address the question of whether GRBs at low and high redshifts are the same \citep{Littlejohns2013MNRAS} or if there might be some optical luminosity evolution \citep{Coward2013MNRAS}.

So far, \emph{Swift} has enabled the discovery of a total of ten GRBs at $z\gtrsim6$, with a wide disparity in follow-up quality, mainly linked to their redshift. The GRBs at $z\approx5.9-6.3$, such as GRBs 050904, 130606A, 140515A, and 210905A, were still able to be observed with almost no Lyman-$\alpha$ damping in the $z^\prime$ band, allowing significantly better light curve coverage, in addition, the afterglows of GRBs 050904, 130606A, and 210905A were very luminous (see Appendix \ref{hizgrbsdetails} for detailed descriptions of our analysis for all these GRB afterglows). Also, for these three events, high quality spectroscopy was obtained. On the other hand, the most distant known event, GRB 090429B \citep{Cucchiara2011ApJ}, has only four detections, no spectroscopy\footnote{In fact infrared spectroscopy was attempted at Gemini-North, but had to be curtailed due to high winds.}, and assumptions on the underlying spectrum need to be made to analyze the SED (in the terminology of K10, it is only a ``Silver Sample'' GRB). However, we include all of them to allow the largest possible sample.

The analysis follows the outline described in K10. After correction for Galactic foreground extinction \citep{Schlafly2011ApJ}, the afterglow light curves are modeled with either a simple power-law (SPL) or a smoothly broken power-law (BPL) function, depending on what is needed. Hereby, if possible, the host galaxy magnitude in each band is taken into account separately. The fit uses, if possible, all available afterglow data. An achromatic evolution is assumed (and then checked via the fit)and therefore the fit parameters (prebreak decay index $\alpha_1$, post-break decay index $\alpha_2$, break time $t_b$ in days, and break smoothness $n$) are shared parameters ($n$ often has to be fixed, usually to a value of 10, see \citealt{Zeh2006ApJ}, but see also \citealt{Lamb2021MNRAS}). The derived normalization of the fit (derived at 1 d for a SPL fit, and at break time assuming $n=\infty$ for a BPL fit) then represent a spectral energy distribution (SED) of the GRB afterglow, not just derived as a ``slice of time'' but using the entire data set (and the assumption of achromaticity). If the afterglow shows variability beyond what can be fit with a BPL, then this data is either excluded (\eg, short term flares) or separate fits are undertaken (double peaked light curves, steep shallow steep evolution). In such cases, we obtain several SEDs that can be checked for color evolution, and if none is found, they are jointly fit in a similar process to that used for the afterglow light curves in different bands. The details for each GRB are given in Appendix \ref{hizgrbsdetails}.

The SEDs are fit with four models, namely an SPL (no dust), and
extinction curves derived from local galaxies \citep[Milky Way (MW), Large (LMC), and Small Magellanic Clouds (SMC),][]{Pei1992ApJ}. While it may seem counter intuitive to use local Universe dust models to describe dust at high-$z$, it has been shown that most low extinction cases are fit well by SMC dust \citep[\eg, K6, K10, ][]{Starling2007ApJ,Schady2007MNRAS,Schady2010MNRAS}, though some clear cases of deviating dust models have been found \citep[\eg,][]{Perley2010MNRAS,Jang2011ApJ}. Furthermore, with the ``large'' sample, we confirm what had already been pointed out by \eg, K6, K11, namely that dust extinction at high-$z$ is generally lower than at low-$z$, though certain biases apply. In this sample, only GRBs 090429B \citep{Cucchiara2011ApJ} and 120521C \citep{Laskar2014ApJ} show evidence for small amounts ($A_V\approx0.10-0.14$ mag) of dust along the line of sight (but see citations on the discussion on dust signatures in the afterglow of GRB 050904, Appendix \ref{GRB050904}).

We should note that our sample of ten high-$z$ GRBs represents only those with data sufficiently good to derive at least a photo-$z$ and allow their classification as high-$z$ events. There are more examples of high-$z$ GRB candidates known. For example, 
GRB 060116 was suggested to lie at $z\approx6.60$ based on a photometric analysis \citep{Kocevski2006GCN4528,Kocevski2006GCN4540,D'Avanzo2006GCN4532,Malesani2006GCN4541,Piranomonte2006GCN4583,Grazian2006GCN4545}, however, a lower-$z$ solution with a significantly larger host galaxy dust extinction was also possible. Furthermore, the burst lay behind a complex region of Galactic dust, making the foreground extinction for this event high but also poorly determined \citep{Tanvir2006GCN4602}. Then, \cite{Chrimes2019MNRAS} report on observations of GRB 100205A, which had an afterglow that was yet again significantly fainter than that of GRB 090429B, and no reasonably precise photo-$z$ could be determined, but it could lie at up to $z\approx8$.

We show the high-$z$ sample in comparison to the joint sample of K6, K10, K11 in the left panel of Fig. \ref{KPobs}. At 1 d after the GRB (in the observer frame) we find that the afterglows of high-$z$ GRBs are, in general, in the fainter half of the long GRB afterglow brightness distribution. As we have shown that dust along the line of sight plays a minor role, at best this is most likely a pure distance effect.

As we have derived the intrinsic spectral slope as well as any dust extinction, we can use the method of K6 to correct the afterglows to a common redshift of $z=1$ (additionally taking the empirical correction for Lyman damping into account, see above). This sample is shown in the right panel of Fig. \ref{KPobs}. It is immediately clear that the afterglows of high-$z$ GRBs are distributed similarly in brightness/luminosity as the comparison sample. 

We derive the magnitudes at one day for the high-$z$ sample, which is possible by interpolation for five GRB afterglows (of GRBs 050904, 080913, 090423, 120923A, and 210905A). It needs only a very short extrapolation in the case of GRB 130606B, and needs longer extrapolations in the cases of GRBs 090429B, 100905A, 120521C, 140515A, though never more than 0.7 dex. We then use knowledge of the intrinsic spectral slope to transform the derived magnitudes at $z=1$ to absolute magnitudes $M_R$, and finally to $M_B$ to compare directly to the results derived in K10\footnote{Originally, K6 had used $M_B$ to allow a direct comparison to quasar samples, which are usually given in $M_B$.}. We find that the mean of the luminosity distribution is $\overline{M_B}=-23.92\pm0.42$ mag (FWHM 1.25 mag). K10 divided their ``Golden Sample'' (comprised of \emph{Swift}-era GRB afterglows from that work and pre-\emph{Swift} GRB afterglows from K6) into two populations at $z<1.4$ and $z\geq1.4$, with the 43 higher-$z$ afterglows yielding $\overline{M_B}=-23.78\pm0.23$ mag (FWHM 1.51 mag), fully overlapping our result for our $z\gtrsim6$ sample. We therefore conclude there is no evidence for significant afterglow luminosity evolution between $z\gtrsim6$ to $1.4<z\lesssim6$.

Now, the problem that arises is that most $z\gtrsim6$ GRB afterglows have only been observed, and have only really been observable, by large ($\diameter\gtrsim2$ m) telescopes with NIR capabilities, such as the 2.2 m MPG/GROND, the 3.8 m United Kingdom InfraRed Telescope (UKIRT)/Wide Field Infrared Camera (WFCAM; \citealt{CasaliM2007}), and ``big glass'' such as the 8.2 m VLT/(ISAAC or HAWKI), 8.1 m Gemini/NIRI, and 10.0 m Keck/MOSFIRE \citep{Chrimes2019MNRAS}. This implies that few high-$z$ GRB afterglows (specifically the aforementioned, very luminous GRBs 050904, 130606A, as well as, with some extrapolation, GRBs 080913, 210905A) have actually been detected during the time after trigger that \GE{} will be attaining its first finding chart; and this sample is very clearly biased toward the most luminous high redshift events as only those have been detectable by rapidly slewing telescope robots of small aperture. The issue is exacerbated as, above $z\approx6$, the Lyman-$\alpha$ break increasingly cuts off flux in the SDSS $z^\prime$ band, presenting a further obstacle to optical detection.
Hence we lack a good census of the early afterglow behavior of typical high-$z$ bursts.

\subsection{Extended early time sample}
\label{ExtSample}

As we have shown that the luminosity of GRB afterglows at the targeted high redshifts are directly comparable to the lower-$z$ sample of K6, K10, K11, we can use those afterglows as proxies, shifting them to high-$z$ with our knowledge of the intrinsic spectral slope. We can then use them in lieu of actual high-$z$ GRB afterglows, assuming they are detected at an early enough time, hereby greatly increasing our sample. We not only shift them in terms of distance and time, but also shift them to NIR bands, which, at the target redshift, are not affected by Lyman damping. These are: $J@z=5$, $J@z=6$, $H@z=8$, $K@z=10$, and $K@z=15$. However, almost all pre-\emph{Swift} afterglows are useless for this exercise as their initial detections are too late. Only GRB 990123 and its extremely early prompt flash \citep{Akerlof1999Nature} can be used directly, for GRBs 021004 and 021211, we are able to back extrapolate the data in a secure manner and add them to the sample as well. We note that such a back extrapolation may yield a result that is not only insecure by up to 0.5 mag, but may be even significantly incorrect, \eg, in the case where a well determined, smooth decay is back extrapolated, but in reality it experienced a steep rise and turnover shortly before the first actual detection. Of course, in such cases, there is no way to know this for a fact. Afterglows whose earliest behavior is a rise to peak are more secure, however, even in such cases, prompt emission linked variability may be superimposed, such as in the cases of GRBs 080603A \citep{Guidorzi2011MNRAS} and 161023A \citep{deUgartePostigo2018AA}. However, we will come to see that most GRB afterglows are more luminous than the assumed detection threshold by margins so large that even a result dimmer by several magnitudes will not make a difference to the simple binary classification of ``brighter than the threshold'' or not.

The position of the injection frequency, the frequency at which the lowest energy electrons are radiating, could affect our analyses \citep{mangano2007}. In general, even at relatively early times, the injection frequency is expected for typical afterglow parameters to be at lower frequencies than optical/NIR. However, in a few cases, afterglow spectra have been modeled including the crossing of the optical/NIR bands by the injection frequency. While a detailed modeling could be needed in those cases, we do not expect that our statistical conclusions could be affected by the injection frequency position in any relevant way.

The Type II GRB samples of K10, K11 extend to late 2009, however, many well-observed GRBs from 2008 and 2009 are missing as large swathes of their photometric data were still unpublished at the time. These samples yield a total of 31 GRB afterglows that can be used in our sample, but this is still a small number. We therefore undertook the task of mining the literature for events from the years 2008 to 2021 to add to this sample. These events need to fulfil the K6, K10, K11 criteria namely:

\begin{itemize}
\item A well-measured redshift, to allow us to correctly model and subtract host-galaxy dust extinction, and determine the shift in magnitude $dRc$ to $z=1$ (and the further magnitude shifts to the high-$z$ targets given above).
\item Multicolor detections to allow the creation of a usable SED for extinction determination (``Golden'' or ``Silver Sample'' following K10).
\end{itemize}

A second set of criteria establishes this as the \emph{Early Light Curve Sample}:

\begin{itemize}
\item Detection of the afterglow at early times. This criterion was {\it ad hoc}, ``by eye'', leading to some cases with a full analysis which, in the end, could not be included in the sample as the first detection was too late, and no secure back-extrapolation could be achieved. We still list the analyses of these afterglows in Appendix \ref{ELCsampledetails} but point out they were not included in the rest of the study.
\item The existence of a publication in the literature featuring a well calibrated data set, and not just data from GCN Circulars\footnote{\url{https://gcn.gsfc.nasa.gov/gcn3_archive.html}}. The reason for this is twofold. For one, it has kept the literature mining within a reasonable scope, as such publications usually feature figures that allow a rapid evaluation of the quality of the light curve data, whereas GRBs with only GCN values only yield this information \emph{post facto}, after collecting and plotting all available data. Secondly, such data from actual publications (usually from refereed journals, but some are from conference proceedings, which may not be refereed) generally yield a trustworthy ``backbone'', which is often multicolor, against which other data can be compared. We make an exception for GRB 180720B, for which our team obtained well-calibrated multi-epoch, multicolor data, which we present here for the first time (see Sect. \ref{adddata}).
\end{itemize}

These selection criteria yield a total of 45 new afterglows in the sample, furthermore four GRB afterglows already published in K10/K11 were reanalyzed with additional data, and we add a further 14 GRB afterglows from a separate sample dealing with the SNe associated with GRBs \citep{Kann2019AA,Kann2021AA}. These latter events are only studied in terms of their early afterglow luminosity here, we defer further analysis to a future publication.

The data are gathered, plotted, cleaned of outliers, and fit according to the methods detailed further above. Hereby, potential multicolor host galaxy and supernova data are taken into account, and the host galaxy and supernova contributions are then subtracted to yield pure afterglow light curves. These are then combined (individually for each GRB) into composite $R_C$ light curves using the derived normalization, fully analog to what has been described in K6, K10. These composite light curves are shown in red in Fig. \ref{KPobs} (left panel: observer frame and right panel: shifted to $z = 1$). It can be seen that compared to the samples presented in K10, K11, several even less luminous GRB afterglows have been added, here we especially highlight GRB 120714B associated with SN 2012eb, which is, over a large time-span, the least luminous well-observed afterglow so far \citep{Klose2019AA}. Generally, however, the new sample is in excellent agreement with the distribution of the K6, K10, K11 samples, albeit with an increase in very early detections, which was the aim all along.

Any detailed discussion of salient light curve features is deferred to future publications. Here we concentrate on the early luminosity distribution. As it is expected that \GE{} will typically be on target within 100 s and observe for 500 s, we derive the logarithmic meantime of 245 s post-trigger as the point in time where we derive the flux density. This is an observer-frame time point, and the higher the redshift of the GRB is, the earlier we are probing into the rest-frame evolution of the light curve, a circumstance which will make \GE{} a powerful tool for probing the early relationship between prompt emission and rest-frame UV/optical emission. As most of our light curves exhibit regular evolution over this time span, deviations from this time point stemming from an asymmetric flux distribution over the exposure are expected to be low. Furthermore, this allows us to derive the flux density with a simple linear interpolation between the two measurements closest in time. 

\begin{figure*}
         \centering 
         \includegraphics[width=\linewidth]{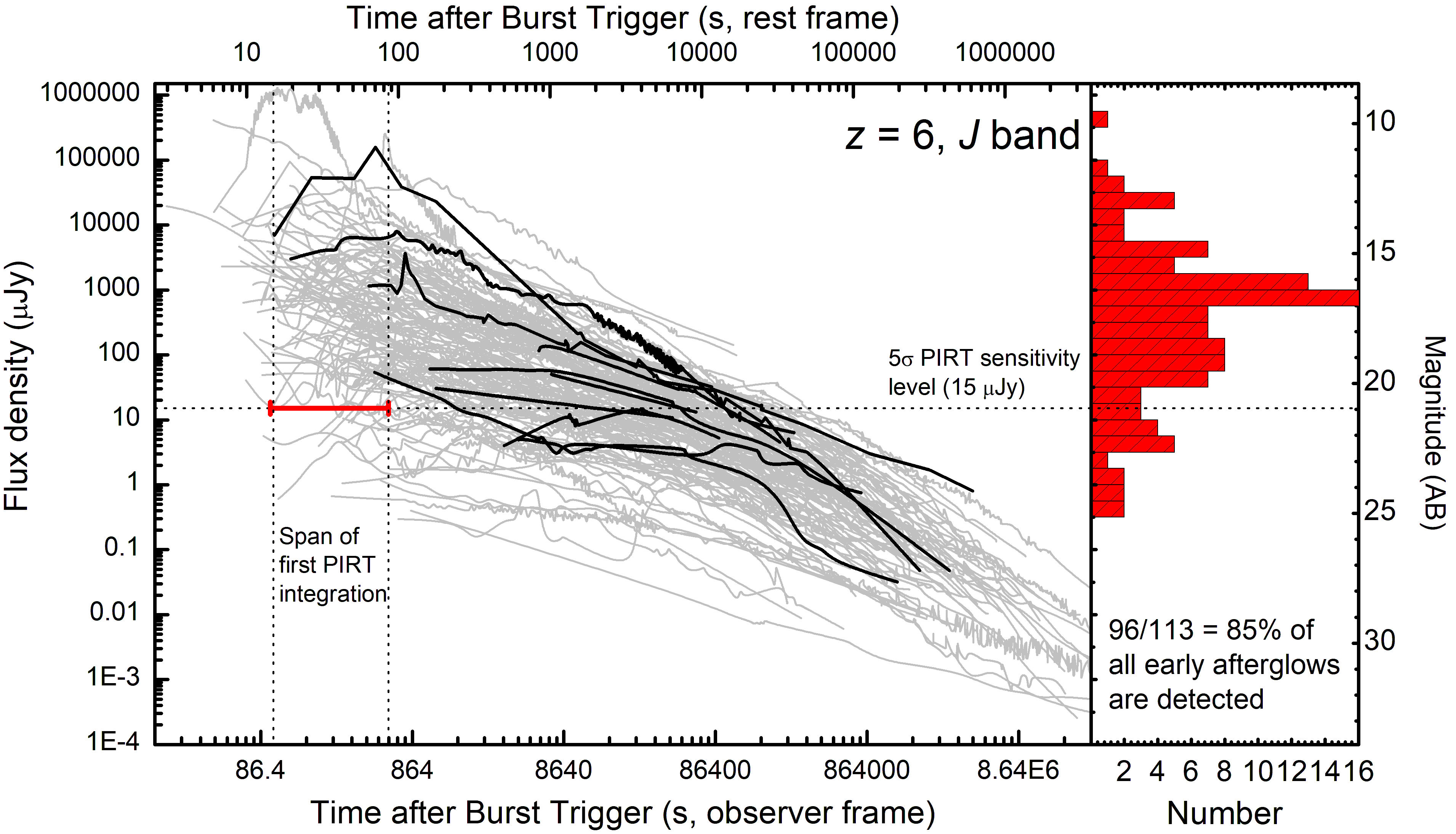} 
         \caption{Light curves of GRB afterglows and the sensitivity of the initial \GE{} image. Afterglow light curves have been corrected for Galactic extinction, host extinction, and, where necessary, for the host-galaxy and supernova components. All afterglows are shifted to $z=6$ using knowledge of the intrinsic spectral slope, and flux densities have been converted to the $J$ band. Thick black lines are the high-$z$ sample of GRB afterglows lying at $z\gtrsim6$. It can be seen that they agree with the distribution of afterglow luminosities at lower redshifts. The red horizontal bar with arrows represents the time span of the first image \GE{} will take after slewing to a GRB position, a 500 s integration which we assume starts at 100 s after the trigger and reaches a limit of 15 $\mu$Jy (21 mag in the AB system). The situation for the $J$ band at $z=5$, the $H$ band at $z=8$, and the $K$ band at $z=10$, $z=15$ are similar, the time dilation and distance-induced luminosity decrease mostly cancel each other out for the higher redshifts. The distribution of afterglow brightness during the first 1000s after the GRB trigger is shown in the histogram in the right panel.}
         \label{J6KP}
\end{figure*}

Finally, for the $J@z=5$, $J@z=6$, $H@z=8$, $K@z=10$, and $K@z=15$ light curve plots, we derive 116, 115, 95, 82, and 53 detections, respectively. We show one example in Fig. \ref{J6KP}. We do not differentiate between the K6, K10, K11 samples and the new Early Light Curve sample anymore, but still highlight the high-$z$ GRB afterglows up to 2021. The red two-headed arrow represents the time span during which \GE{} will take its initial finding chart, and it is placed at the expected $5\sigma$ detection limit of 15 $\mu$Jy. It is clearly visible that almost all GRB afterglows which actually have follow-up at such early times are detectable by \GE{}. Plots for the other filter-redshift combinations look very similar. As redshift increases, afterglows become fainter (distance increases) and stretch out (time dilation), but they also become more luminous as we move to redder filters. These effects cancel each other out for the most part, but the fainter/more stretched out effect dominates, implying it becomes harder and harder to detect enough afterglows. From a purely practical standpoint, the sample also decreases with higher redshift as even very rapid observations now correspond to times which are after the 500 s initial finding chart. We also point out that a significant fraction of early afterglows show a rising behavior, implying they will be significantly fainter during the finding chart than purely by the effect of increasing distance. Therefore, it is an informed strategy for further follow-up with \GE{} to obtain further and deeper finding charts, as the afterglow may not ``pop up'' until many hundred to perhaps several thousand seconds after the initial GRB, something that is discussed more in Sect. \ref{Followup}.

\subsection{Additional data}
\label{adddata}

To improve the results of our sample (\eg, additional filters), and especially to extend the number of early detections, here we analyze and present so-far unpublished data on a total of 40 GRBs. Here, we highlight data sets for GRB 080413B (126 data points), GRB 080605 (104 data points), GRB 080721A (102 data points), GRB 081203A (144 data points), GRB 091018 (139 data points), GRB 091020 (111 data points), GRB 091208B (161 data points), GRB 100906A (230 data points), GRB 110213A (132 data points), GRB 131030A (261 data points),  GRB 141221A (112 data points), GRB 150910A (109 data points), and GRB 180720B (132 data points), the last GRB not having a well-calibrated data set in the literature prior to this publication yet.

Observations have been obtained from these telescopes:
\begin{itemize}
\item The \emph{Swift} 30cm modified Ritchey-Chretien UltraViolet/Optical Telescope UVOT \citep{Roming2005SSRv}, which yields data in three UV lenticular filters $uvw2$, $uvm2$, and $uvw1$, three optical filters $u$, $b$, and $v$, and unfiltered ($white$). Early data are taken in ``event mode'' and can be split up into a finer time resolution.
\item The 0.6m Rapid Eye Mount REM \citep{Zerbi2001AN}, equipped with the optical/NIR camera ROSS (Johnson-Cousins optical filters, 2MASS $JHK_S$ filters), and later the optical camera ROSS2 (SDSS filters) and the NIR camera REMIR ($Z$ and 2MASS $JHK$ filters), situated at ESO La Silla Observatory, Chile.
\item The 1.34m lens/2m mirror \emph{Tautenberg} classical Schmidt telescope of the Th\"uringer Landessternwarte Tautenberg, Thuringia, Germany, equipped with a $2k\times2k$ CCD detector and Johnson-Cousins $BVR_CI_C$ filters as well as a $Z$ filter.
\item The 2.0m Liverpool telescope (LT), equipped with RATCam and SDSS filters, located at Observatorio Roque de los Muchachos (ORM), La Palma, Canary Islands, Spain \citep{Steele2004SPIE}, as well as its copies, the Faulkes Telescopes North (FTN) and South (FTS) located at Haleakal\={a} Observatory, Maui, Hawaii, USA; and Siding Springs Observatory, New South Wales, Australia, respectively, as well as the 1.0m telescope at McDonald Observatory, Texas, USA, which are now part of the Las Cumbr\'es Observatory Global Telescope (LCOGT) Network \citep{Brown2013PASP}.
\item The 0.5m D50, and the 0.25m Burst Alert Robotic Telescope \citep[BART, with Near-Field and Wide-Field detectors, NF, WF respectively;][]{Jelinek05} at the Astronomick\'y \'ustav Akademie v\v ed \v Cesk\'e republiky (As\'U), Ond\v rejov, Czech Republic, yielding $R_C$ or unfiltered $CR$ (calibrated to $R_C$) observations.
\item The 0.4m Watcher telescope, equipped with an $R_C$ filter or unfiltered, at Boyden observatory, near Bloemfontein, South Africa \citep{Ferrero2010}.
\item The 0.9m T90 and 1.5 m T150 telescopes at the Observatorio Sierra Nevada (OSN), Granada, Spain, equipped with Johnson-Cousins filters.
\item The 0.8m Javalambre Auxiliary Survey Telescope JAST/T80, equipped with a $2\deg^2$ wide-field imager and SDSS filters \citep{Cenarro2019}.
\item Telescopes used by the team of the Seoul National University (SNU): The 1.0m LOAO telescope located at Mt. Lemmon Optical Astronomy Observatory, Tucson, Arizona, USA  \citep{HanW2005, LeeI2010}. The 1.8m Bohyunsan telescope at the  Bohyunsan observatory, Korea, equipped with the Korea Astronomy and Space Science Institute Near-Infrared Camera System (Bohyunsan/KASINICS; \citealt{MoonB2008}). The 2.1m Otto Struve Telescope, located at McDonald Observatory,Texas, USA, equipped with the Camera for QUasars in EArly uNiverse (CQUEAN; \citealt{ParkWK2012}). The 1.5m AZT-22 telescope of Maidanak Observatory, Uzbekistan, equipped with the Seoul National University 4kx4k Camera (SNUCAM; \citealt{ImM2010}). The UKIRT/WFCAM on Mauna Kea, Hawaii, USA \citep{CasaliM2007}. 
\item A small number of data points have been acquired by the 8.2m Very Large Telescope equipped with the FOcal Reducer and low dispersion Spectrograph 2 (FORS2) and the X-shooter acquisition camera, Cerro Paranal, Chile, the F/Photometric Robotic Atmospheric Monitor (FRAM) at the Pierre Auger Observatory (PAO) in Malargue, Argentina, and the 1.0m Anna L. Nickel telescope, Lick Observatory, California, USA. The 10.4m Gran Telescopio Canarias (GTC) High PERformance CAMera (HiPERCAM) located at the Roque de los Muchachos Observatory on the island of La Palma, in the Canary Islands, Spain \citep{Dhillon2021}.
\end{itemize}

UVOT usually begins observing the fields of GRBs within the first minutes after the \emph{Swift}/BAT trigger. Observations are typically taken in both event and image modes. In Table \ref{obslog}, we only give data for certain filters for each GRB if there is at least one detection in that filter, in that case, all data, including upper limits, are presented. Before extracting count rates from the event lists, the astrometry was refined following the methodology of \cite{Oates09MNRAS}. The source counts were extracted initially using a source region of 5\arcsec{} radius. When the count rate dropped to below 0.5 counts/s, we used a source region of 3\arcsec{} radius. In order to be consistent with the UVOT calibration, these count rates were then corrected to 5\arcsec{} using the curve of growth contained in the calibration files. Depending on the GRB, background counts were extracted using one or more circular regions located in source-free regions. The count rates were obtained from the event and image lists using the \emph{Swift} tools \texttt{uvotevtlc} and \texttt{uvotsource}, respectively. They were converted to magnitudes using the UVOT photometric zero-points \citep{Poole2008MNRAS,Breeveld2011AIPC}. To improve the S/N ratio, the count rates in each filter were binned using $\Delta t/t=0.1$ or $\Delta t/t=0.2$, depending on circumstances, leading to longer but deeper exposures at later times. The early event-mode $white$ and $u$ finding charts were usually bright enough to be split into multiple exposures.

Ground-based photometry was generally analyzed using standard procedures: Bias-subtraction, flat-fielding, and stacking, where necessary. Field calibration was obtained against on-chip comparison stars from the Sloan Digital Sky Survey \citep{Alam2015ApJS} or the Pan-STARRS survey \citep{Chambers2016a}. These values were either used directly for observations obtained in SDSS filters, or converted to Johnson-Cousins filters using the transformations of Lupton\footnote{\url{http://classic.sdss.org/dr7/algorithms/sdssUBVRITransform.html}}. Zero point 1-sigma errors of the calibration are usually $0.02-0.07$ mag, rarely higher. This systematic error is added in quadrature to the statistical measurement errors. NIR observations are calibrated against the 2 Micron All-Sky Survey 2MASS \citep{Skrutskie2006AJ}.

The \emph{Tautenberg} data of GRB 080605 had to be specially analyzed. This bright afterglow occurred in a crowded field, with a total of three stars nearby \citep{Kann2008GCN7864}. These stars were separated from the afterglow in VLT and LT/FTN images, but the large pixel scale of the \emph{Tautenberg} camera (1\farcs12) led to them being blended together. About 84 to 86 d after the GRB trigger, we revisited the field and acquired deep template imaging under good conditions in $VR_CIC$. These template images were subtracted from fully reduced images containing the afterglow using the High Order Transform of PSF ANd Template Subtraction (\texttt{HOTPANTS}) software \citep{Becker2015ascl}. For S/N reasons, the three obtained $V$ images as well as the four final $I_C$ images of the first epoch were stacked. All images in the second ($R_CI_C$) and third ($I_C$ only) epochs were stacked. The afterglow was not detected in the third epoch. In a few cases, the input images had a smaller PSF than the template images (in $I_C$ only). In these cases, as well as in the case of one further $I_C$ image where \texttt{HOTPANTS} did not converge, we modeled the PSF of stars in the image and used that model to subtract the two nearby bright stars. This leaves the afterglow as well as an even closer, faint star. We applied the PSF subtraction to the reference images, and determined the contribution of the faint third star, which was then subtracted in flux space. The derived \emph{Tautenberg} magnitudes agree excellently with the contemporaneous LT measurements.

Reduction and analysis of data from the telescopes at As\'U, namely the Ond\v rejov 0.5m D50, the 0.25m BART/NF \citep{bart},WF, as well as the 0.3m FRAM/PAO, are described in \cite{Jelinek2019AN}, including the weighted image co-addition technique.

\section{Theoretical approach via GRB simulations}
\label{simulations}

The afterglow sample studied in this work is mostly composed of \emph{Swift}/BAT bursts (with a few INTEGRAL GRBs, such as GRBs 161023A and 210312B, as this satellite also delivers positions with arcmin precision within seconds). Given its detection threshold, \emph{Swift}/BAT introduces a redshift-dependent bias on the minimum luminosity of detectable GRBs. 
Any future mission design with improved sensitivity, like \GE{}, will access less luminous GRBs than \emph{Swift}/BAT. It is worth exploring how this systematic difference reflects on the afterglow luminosity in comparison to the current afterglow sample of GRBs detected by \emph{Swift}/BAT. 

The long GRB population is simulated following the prescription of \cite{Ghirlanda2015MNRAS}. We assume that the long GRB rate is $\propto (1+z)^{\delta}\psi_{\star}(z)$, where $\psi_{\star}(z)$ is the cosmic star formation rate \citep{Li2008MNRAS} and $\delta=1.7$ \citep{Salvaterra2012ApJ}. GRBs are assigned a luminosity according to a broken power-law probability density function with low (high) end power-law slopes -1.3 (-2.5) and break $L_{\star}=10^{52}$ erg s$^{-1}$. This function is defined within the limits $[10^{46},10^{56}]$ erg s$^{-1}$. In order to compute the flux of each simulated burst we assume a Band function \citep{band1993} for the prompt emission spectrum with low (high) energy spectral index assigned randomly from uniform distributions centered at -1 (-2.5) and standard deviation 0.1. We link the GRB luminosity to its rest frame peak energy via the Yonetoku  correlation $\Log(L_{\rm iso})=-27.02+0.57\Log(E_{\rm peak})$ \citep{yonetoku2004,nava2012} where we account also for a 0.30 dex the scatter around this correlation. Similarly the isotropic equivalent energy is assigned following the Amati correlation \citep{Amati2002AA,nava2012} $\Log(E_{\rm iso})=-34.46+0.7\Log(E_{\rm peak})$ with a scatter 0.25 dex. 

\begin{figure*}
    \centering
    \includegraphics[scale=1.11]{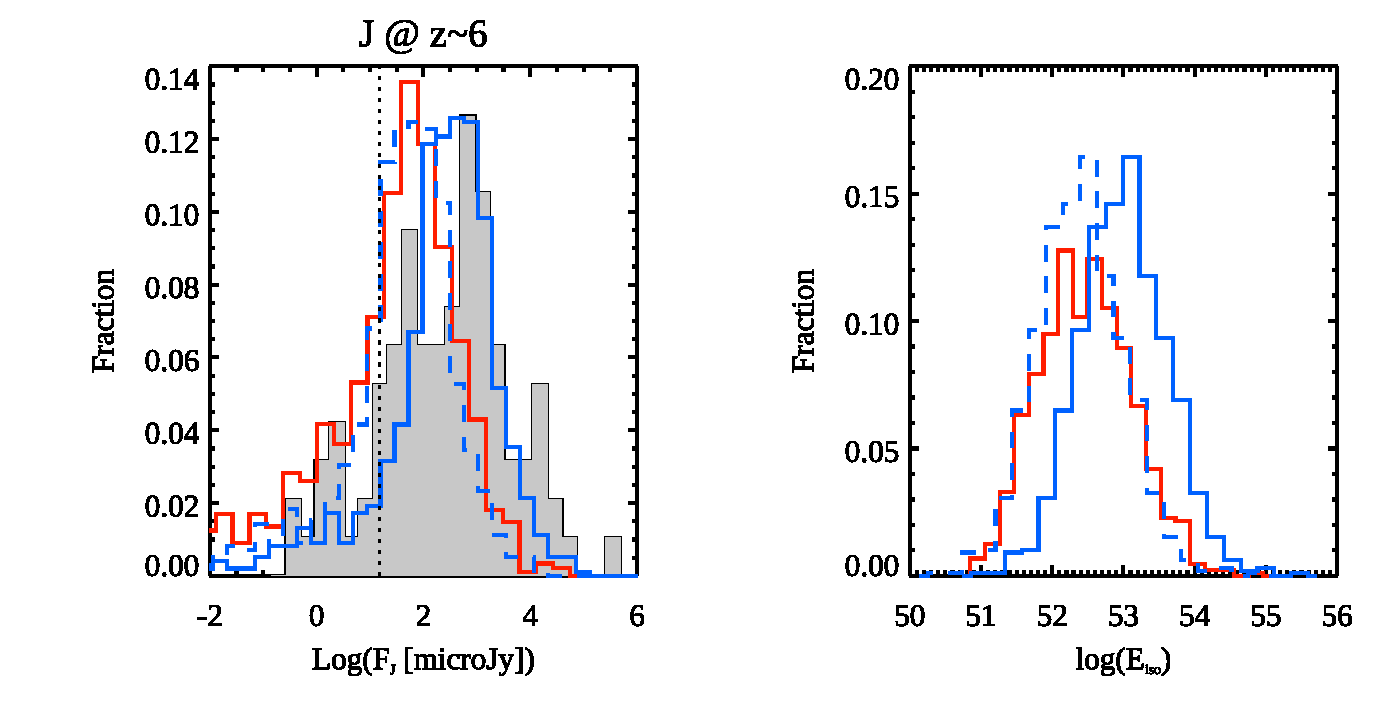}
    \caption{Simulated GRB samples detectable by \emph{Swift} and \GE{} at z$\sim 6$. {\it Right panel:} Distributions of the isotropic equivalent energy of simulated \emph{Swift} (blue) and \GE{} LEXT (red) GRBs. The dashed histograms in the right-hand panel corresponds to a rescaling of the \emph{Swift} GRB distribution by a factor $\sim$4 to reproduce the \GE{} (red) distribution. {\it Left panel:} Afterglow flux density distributions in the $J$ band. This corresponds to the central frequency $1.84\times10^{14}$ Hz of observed \emph{Swift} GRBs (solid filled gray histogram from Fig. \ref{J6KP}) compared with the distribution of $J$ flux densities at 560 sec (observer frame) of simulated \emph{Swift} bursts (blue histogram) and simulated LEXT GRBs (red histogram). The dashed blue histogram shows the afterglow flux density rescaled by the energy factor ratio (see text). Dotted vertical line shows the flux density threshold of 15$\mu$Jy. }
    \label{population}
\end{figure*}

The afterglow emission is computed through the public code afterglowpy \citep{Ryan2020ApJ} assuming distributions of the microphysical parameters as reported in Table 1 of \cite{Campana2022}. 
By considering the \emph{Swift}/BAT flux limit and \GE{} LEXT instrumental design \citep{Feldman2021SPIE}, we compare in the right panel of Fig. \ref{population} the distributions of the prompt-emission isotropic-equivalent energy $E_{\rm iso}$ of GRBs detectable by \emph{Swift}/BAT (blue histogram) and LEXT (red histogram) at $z\sim 6$. Owing to its softer energy range ($0.3-5$ keV) and better sensitivity, LEXT detects less energetic GRBs from all redshifts: there is a systematic difference of $\sim4$ between the two distributions (the dashed blue line histogram shows the \emph{Swift} distribution shifted by this factor).  

Within the standard fireball model (\eg, \citealt{Meszaros1997ApJ}), the afterglow luminosity depends on the outflow isotropic equivalent kinetic energy  $E_{\rm k,iso}=\frac{1-\eta}{\eta}E_{\rm iso}$
 which can be related to the prompt emission isotropic equivalent energy $E_{\rm iso}$ through the $\gamma$--ray emission efficiency parameter $\eta$. 
 Moreover, afterglow emission depends on the external shock efficiency of accelerating relativistic particles and amplifying seed magnetic fields. Following the approach explained in \citealt{Campana2022}, we adopt the afterglow model of \cite{Ryan2020ApJ} to simulate afterglow emission 
 produced in a constant density external medium\footnote{While a wind environment is expected to be produced by the mass loss during the pre-explosive phases of the massive star progenitor of long GRBs, current multiwavelength afterglow observations seem to slightly prefer an ISM solution \citep{Aksulu2022,Schulze2011}.} by a uniform jet. The free model parameters regulating the shock micro-physics and the external  medium density are calibrated by reproducing the afterglow flux density distributions of \emph{Swift} GRBs in the X-rays (at 11 hours) and optical ($R_C$) bands at 600 s, 1, 11, 24 h of the BAT6 complete sample \citep{D'Avanzo2014MNRAS,Melandri2014AABAT6}. 

The left panel of Fig. \ref{population} compares the simulated afterglow flux density of GRBs at $z\sim6$ detectable by \emph{Swift} (blue) and the observed sample of GRB afterglows collected in this work and redshifted to $z=6$ (solid filled gray histogram from Fig. \ref{J6KP}). These two distributions have similar central values. The afterglow flux density distribution of GRBs detected by the \GE{} at $z\sim6$ instead is on average dimmer by a factor $\sim6$. This is mainly determined by the smaller kinetic energy (of which $E_{\rm iso}$ is a proxy) of bursts detected by LEXT. Indeed, by assuming a standard afterglow model, the flux in the optical-NIR band, sampling the frequency range between the maximum frequency and the cooling one \citep{Starling2007ApJ,Greiner2003ApJ}, should scale as $E_{k}^{(p+3)/4}$ \citep{Panaitescu2000ApJ} in the constant density external medium case, assuming all the other parameters are fixed.
For a typical value of the shock-accelerated electron energy distribution $p=2.3$ the difference between the \emph{Swift} and \GE{} isotropic energies (Fig. \ref{population} right panel) accounts for the factor of 6 in the flux density distributions (left panel). The corresponding rescaled flux density distribution is shown by the dashed blue histogram in the left panel of Fig.\ref{population}.

\section{Results}
\label{results}
\subsection{The sensitivity of \GE{} and the Early Light Curve sample}

\begin{figure*}
\begin{center}
         \centering 
         \includegraphics[width=0.32\textwidth]{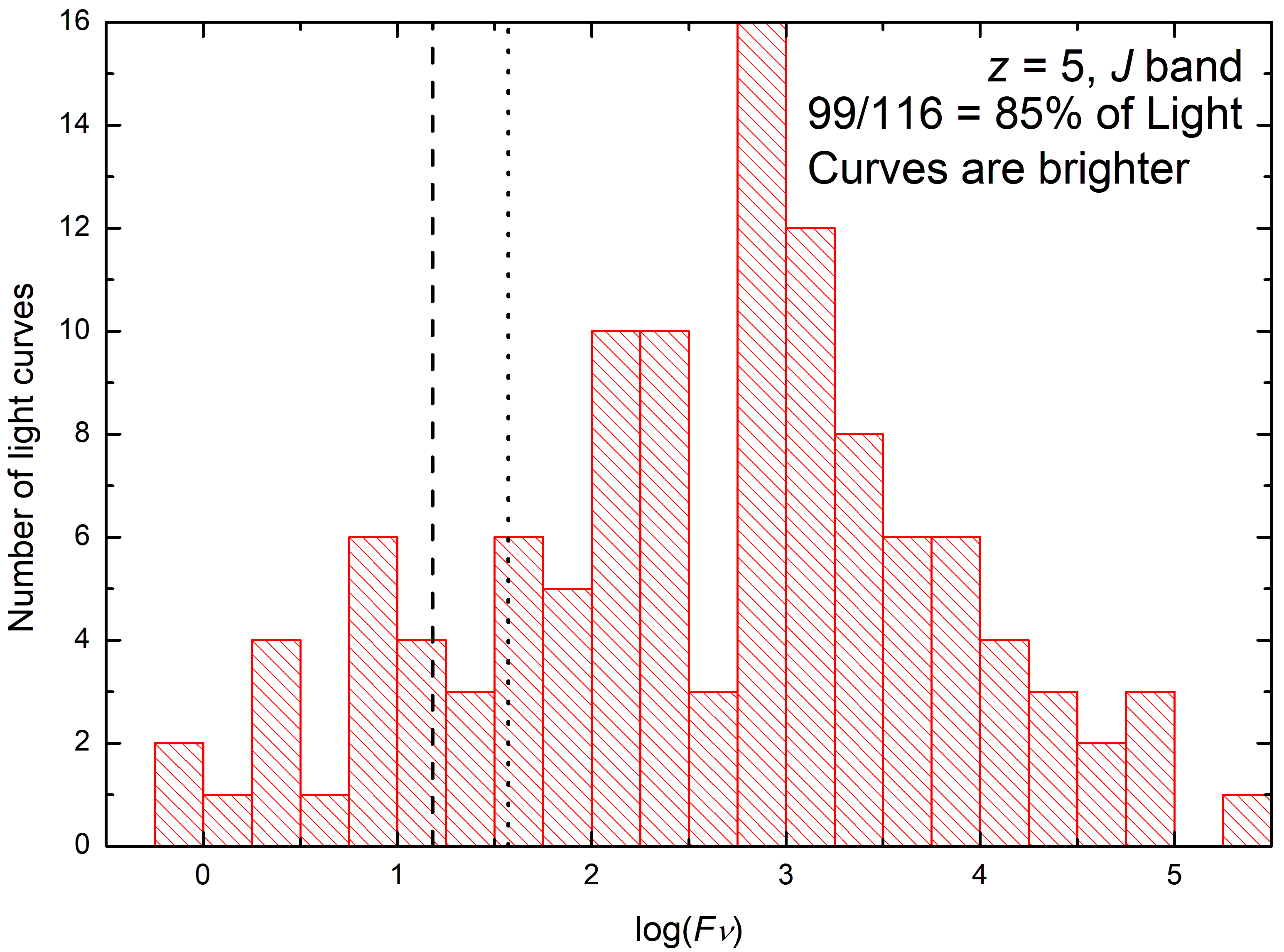} 
         \includegraphics[width=0.32\textwidth]{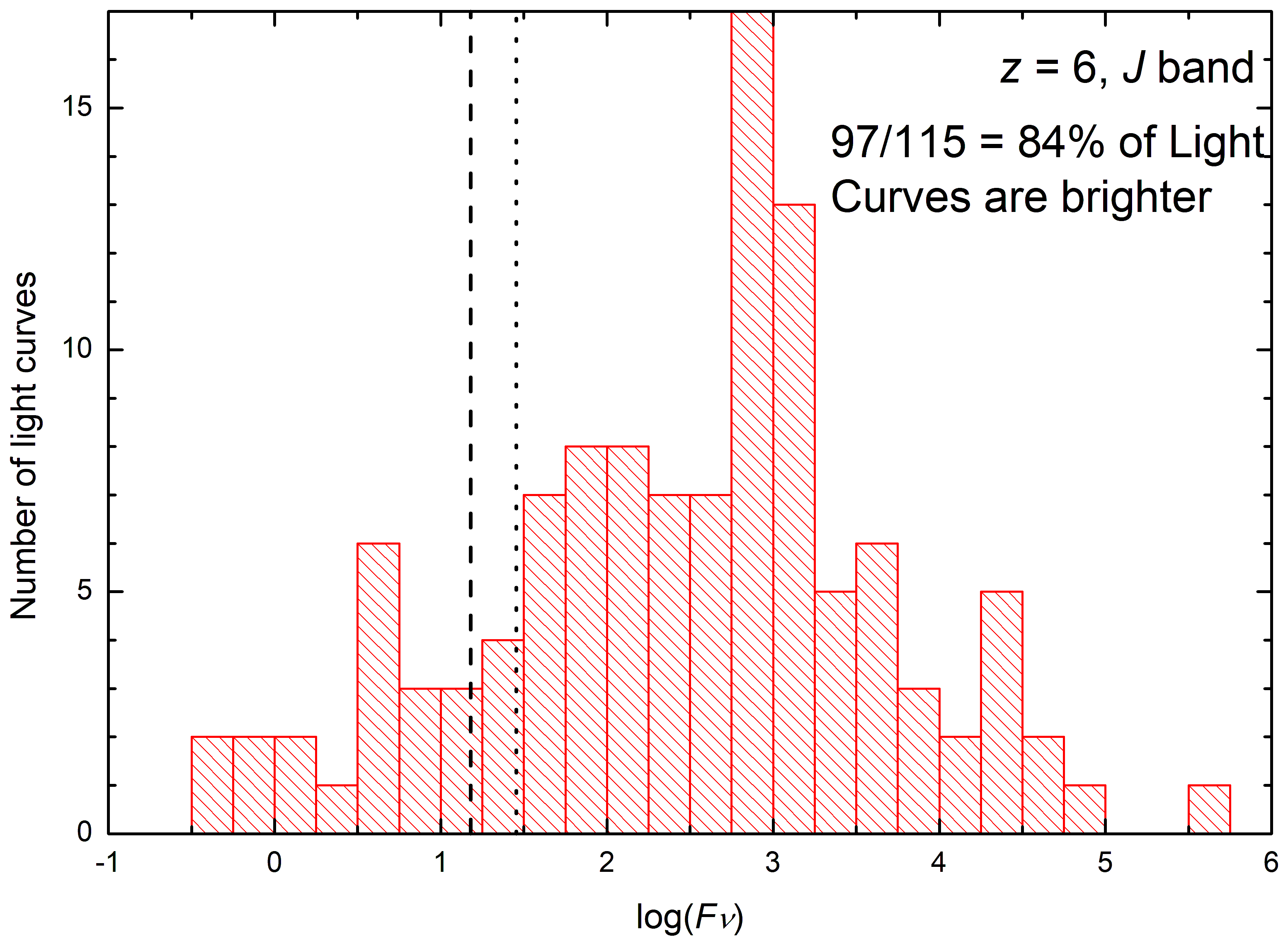} 
         \includegraphics[width=0.32\textwidth]{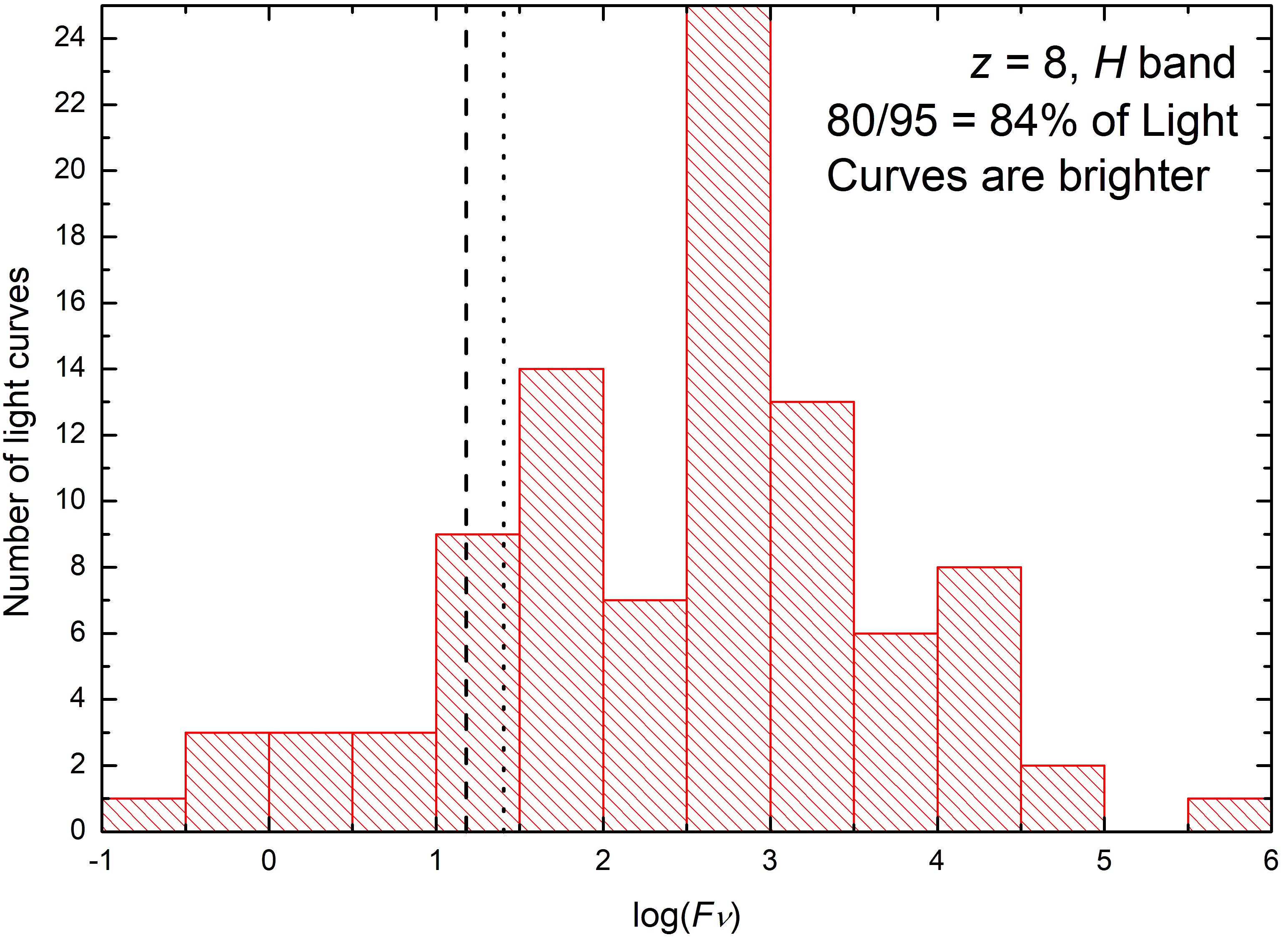} 
         \includegraphics[width=0.32\textwidth]{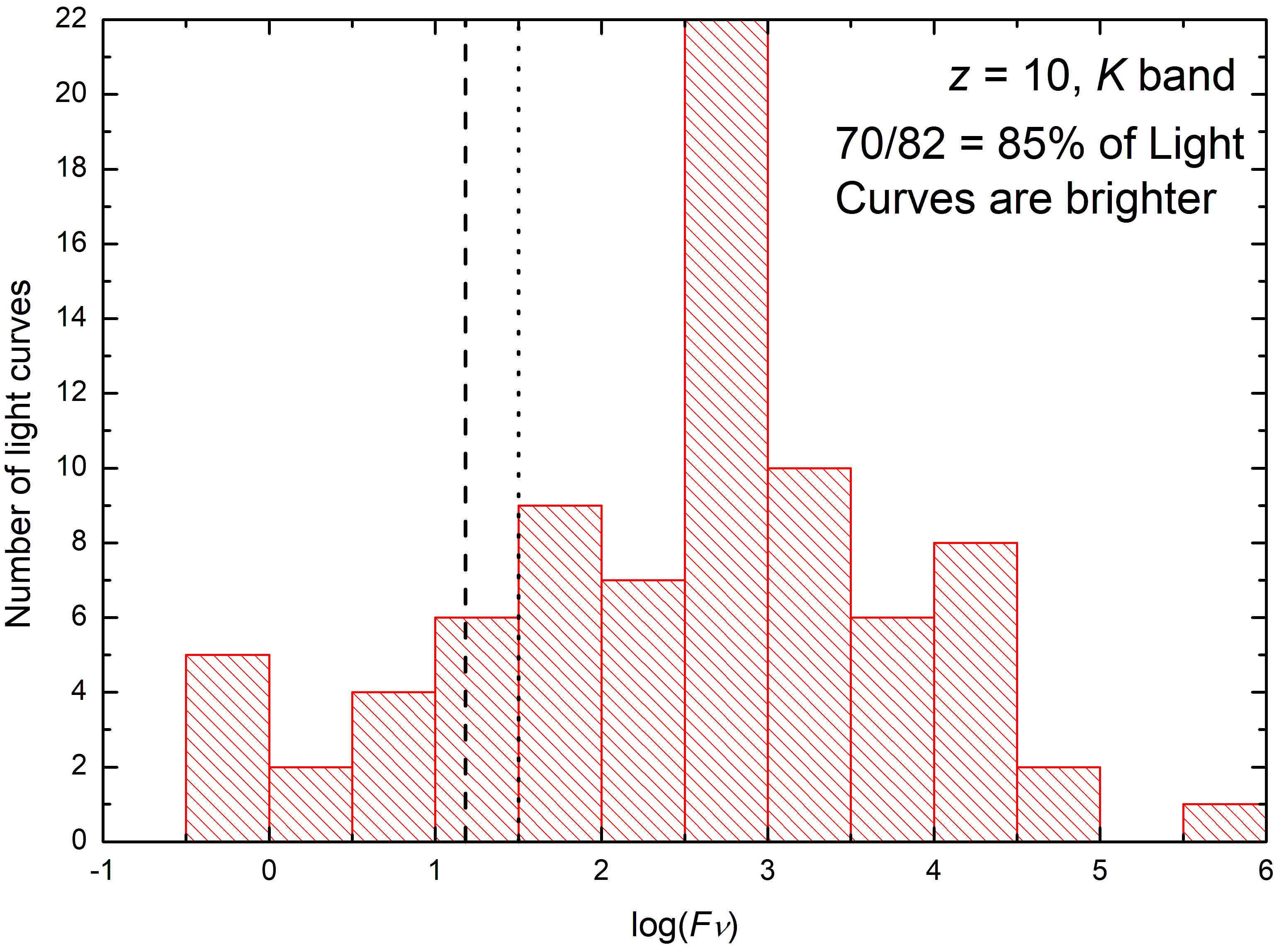} 
         \includegraphics[width=0.32\textwidth]{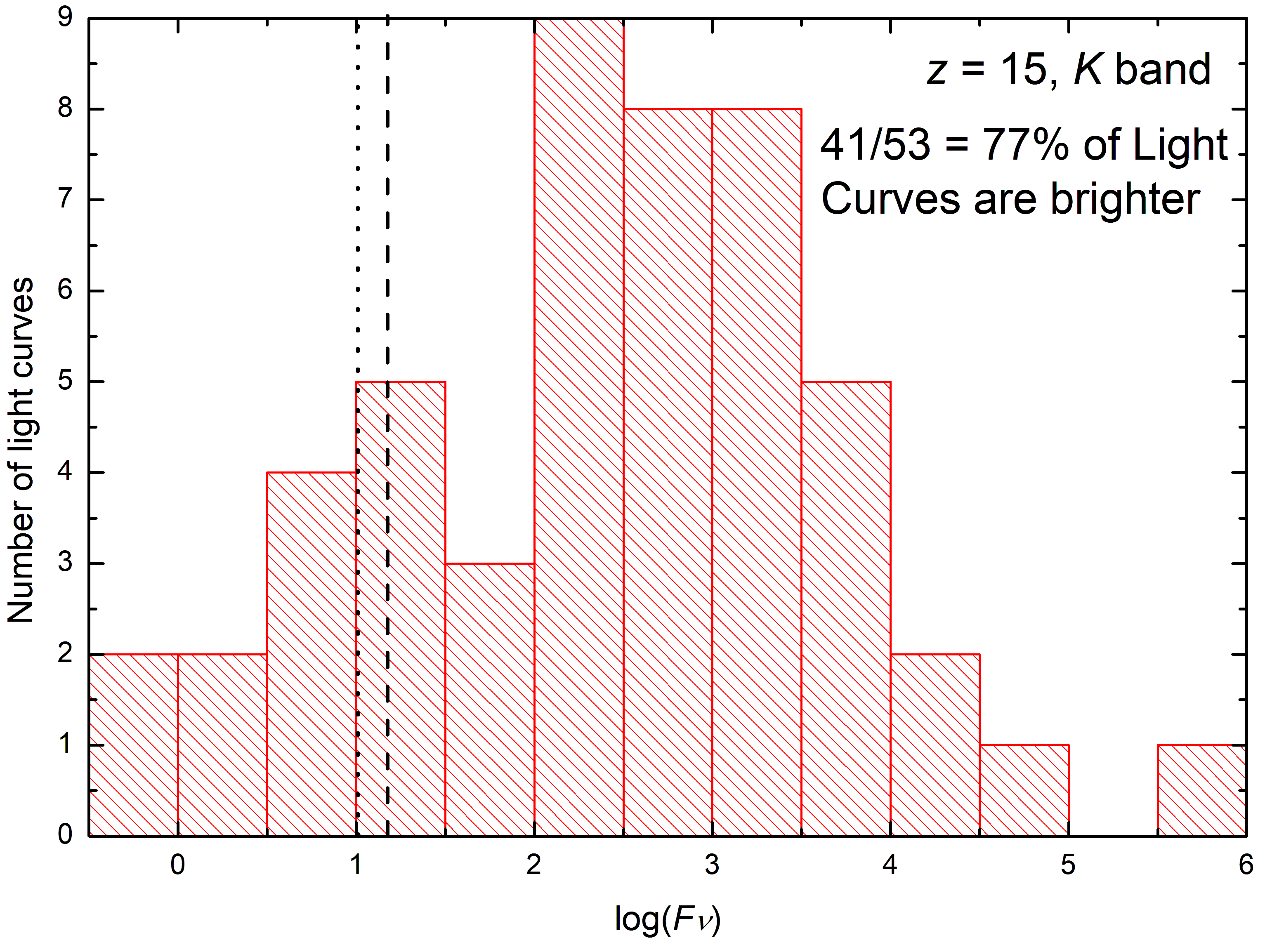} 
         \caption{Distribution of early GRB afterglow luminosities. These are measured at 245 s, the logarithmic center of an observation stretching from 100 s to 600 s after the trigger in the observer frame. The legends indicate to which redshift and which band the afterglows have been shifted, as well as the number of afterglows that are brighter than 15 $\mu$Jy. The vertical dashed lines show the 15 $\mu$Jy sensitivity requirement. The dotted vertical dashed line shows the flux threshold required to recover 80\% of redshifts, which exceeds the requirement for all but $z \simeq 15$.}
         \label{Histo}
\end{center}
\end{figure*}

Using the results on the flux density at 245 s after trigger (observer-frame) that we derived for the Early Light Curve sample, we are now able to check how many early afterglows would be detected at at least $5\sigma$ significance by \GE{}, depending on the band and the assumed redshift. Histograms for all five combinations studied here are shown in Fig. \ref{Histo}.

With the exception of the extreme case of $K@z=15$ (so all the way up to $K@z=10$), we find that \GE{} will recover $84\%-86\%$ of all early afterglows.
We can also ask the inverse question: If we set the goal of 80\% early afterglow recovery, what would the limiting flux densities PIRT needs to achieve? To do so, we simply check the flux densities of the GRB afterglow detections spanning the 80\% demarcation, and do a linear interpolation. For example, for the $J@z=5$ sample, we have 114 detections, so the brightest 91 detections contribute to the recovered sample, with the exact demarcation being 91.2. The 91st brightest afterglow is detected at 46.3 $\mu$Jy, whereas the 92nd brightest is 39.4 $\mu$Jy, a difference of 6.9 $\mu$Jy , 20\% of that is 1.4 $\mu$Jy and thus the threshold is at 44.9 $\mu$Jy. For the other filter-redshift combinations, we find 33.8 $\mu$Jy, 35.2 $\mu$Jy, and 30.3 $\mu$Jy, respectively, and for $K@z=15$ we find 11.0 $\mu$Jy, as for this sample PIRT with a sensitivity of 15 $\mu$Jy is capable of recovering only 77\% of all afterglows. In conclusion, overall, an 80\% recovery rate with PIRT could be reached even if it is only half as sensitive, with a $5\sigma$ limit of 30 $\mu$Jy in 500 s. However, as the true luminosity distribution of very high redshift GRB afterglows at early times is unknown (combined with the different sample in terms of prompt-emission properties LEXT will detect), and may indeed be fainter than our redshifted sample, as our simulations show, this should not be a place for cutting corners, and the 15 $\mu$Jy limit provides a necessary safety margin.

The improved sensitivity and softer energy range extension of \GE\ make it more efficient in detecting high redshift slightly fainter than \emph{Swift} bursts. Based on the results shown in Fig.\ref{population}, the theoretical approach leads to 70\%-75\% of recovery of all early afterglows by PIRT. However, this fraction may increase if there is a mild evolution of the characteristic luminosity of the GRB progenitors with redshift as recently found  \citep{Ghirlanda2022}.

\section{Discussion}
\label{discussion}
\subsection{Meeting the Gamow requirements}
Our results using past observations of GRB optical--NIR afterglows transformed to high redshift combined with the photo-$z$ simulations of \cite{Fausey2023} demonstrate that a space-based NIR multiband telescope with a sensitivity of 15 $\mu$Jy in 500s can provide photo-$z$ measurements with sufficient precision to determine whether or not a GRB is from high redshift $z > 5$. This can be achieved with a modest 30cm diameter telescope, cooled to below 210 K to eliminate its thermal radiation for sensitivity out to 2.4 microns \citep{Seiffert_2021}. 

This result is based on past observations, which will have selection biases. \GE{} with a lower energy response for the LEXT compared to traditional GRB detectors will be sensitive to lower isotropic luminosity GRBs, especially at high redshift. K11 and K10 show that there is a mild correlation between prompt isotropic bolometric luminosity and the R$_c$ band afterglow at 1 day, with an order of magnitude lower over three orders of magnitude of GRB luminosity. As noted by K11 there is almost a factor of 200 (over 5 mag) scatter in optical magnitude for a given GRB luminosity (see figure 13 in K11), most likely caused by differences in the jet viewing angle and circumstellar material. As a further validation our population modelling combined with a simple theoretical model for the jet interaction with a uniform circumstellar material shows that at most there is a overall factor of 4 lower average isotropic bolometric luminosity, and overall a factor of $\sim$ 6 flux reduction, which is also consistent with the large scatter in the K11 figure 13. Even in this worst case scenario, the 15 $\mu$Jy sensitivity threshold recovers $\approx 75\%$ of the $z>5$ GRBs.

Another concern is that while the comparison sample is corrected for extinction, there maybe dust in the high redshift host galaxies which may impact the predictions. In the rest-frame (F)UV even small amounts of LOS extinction can have a large impact. Potentially GRB which would be only ``somewhat extinguished'' at low-$z$ become  invisible at high-$z$. This \emph{complete sample} analysis suggest there cannot be a large number of dusty high$-z$ GRB host galaxies -- which maybe consistent with the expectation of less dust at high redshift. JWST observations will likely provide more constraints on dust formation at high redshift. 

\subsection{Follow-up strategies with Big Glass and JWST}
\label{Followup}

\begin{figure}
      \centering 
         \includegraphics[width=\columnwidth]{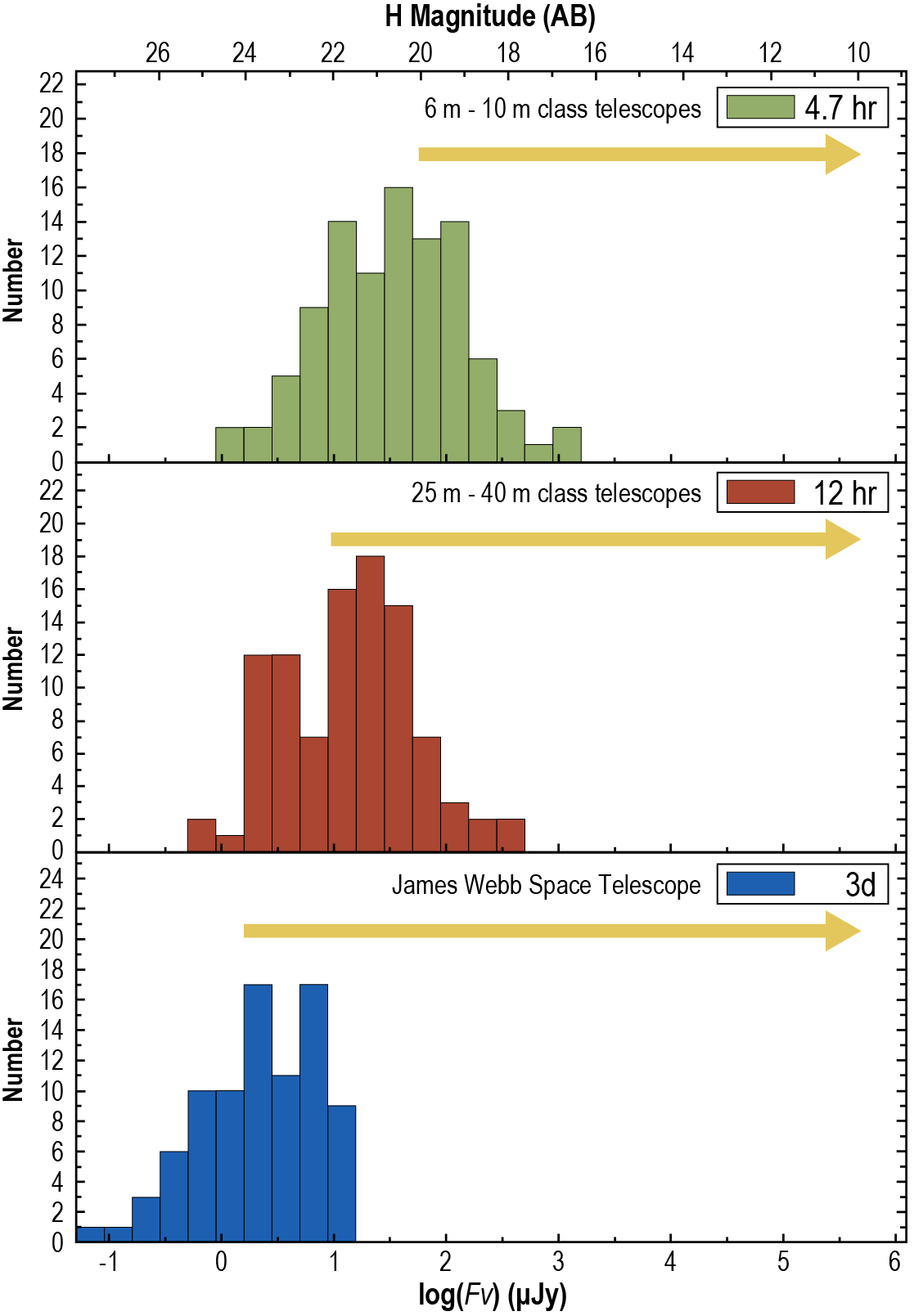} 
         \caption{Flux-density distributions at different observer-frame time points of the $z=8$ $H$ band sample. These represent a 4.7 hr ground-based reaction with current 6 m to 10 m class telescopes (top), a 12 hr reaction with future 25 m to 40 m telescopes (middle), and a 3 day disruptive TOO reaction by JWST (bottom). With the arrows, we indicate the magnitude span over which each of the indicated telescope classes can acquire the required NIR spectrum ($\sim$ 2500 resolution) with a few hour integration time. }
         \label{H8histo}
\end{figure}

Figure~\ref{H8histo} shows the distribution of flux-density at 4.7 hr, 12 hr and 3 day (observer-frame) after the GRB trigger of the $z=8$ $H$ band sample. 4.7 hr was used based on concept of operations simulations which show that this is the average time for one of the three major observatory sites to become available for \emph{Gamow} detected GRBs. Also shown are approximate sensitivities with a 6m-10m NIR spectroscopy with $\sim$ 2500 resolution and a signal to noise ration of $\sim 20$ per spectral resolution element in a few hours exposure. From this, it is clear that for the current largest telescopes it is essential to begin spectroscopy within a few hours. For the coming 25m to 40 m telescopes this can be relaxed to $\sim$ 12 hr. Follow-up by JWST using a 3 day disruptive TOO is feasible for a large fraction of high$-z$ GRBs. A cascading strategy would be to first attempt spectroscopy from the ground, then trigger JWST if the former does not succeed because of weather, badly placed atmospheric telluric lines or being simply too high a redshift (where JWST excels). 

It is important to note that cosmological dimming is offset almost completely by the time-stretching involved in increased redshift. As an example, we studied the luminosity sample for the ``detected'' sample at two days after redshift notification (2.011 d after trigger), and find for $J@z=6$ a mean and standard deviation: $\overline{J}=22.53\pm1.34$ mag. For $H@z=8$, it is: $\overline{H}=22.64\pm1.29$ mag. And for $K@z=10$, it is:  
$\overline{K}=22.64\pm1.37$ mag. We also derived mean magnitudes of the sample (in this case the representative $H@z=8$ ``detect'' sample) for even later times to check the feasibility of JWST follow-up at 5, 10, and 14 d, the latter case representing the time delay when a ToO request to JWST does not fall into the disruptive ToO category (which are limited in number). We find $\overline{H}=23.68\pm1.43$ mag,
$\overline{H}=24.55\pm1.50$ mag, and
$\overline{H}=24.99\pm1.50$ mag, respectively. Using again $H<24$ mag as a threshold for useful spectroscopy from JWST, we find that 69\% (50/72), 41\% (24/59), and 32\% (16/50) are still bright enough, respectively. Hence, while a not insignificant portion of GRB afterglows remain viable spectroscopic targets for JWST even two weeks after the GRB, the sample becomes increasingly biased toward the brightest afterglows. Additionally, as the afterglow evolution cannot be predicted, even initially bright afterglows should, for best results, be observed within the first week after the GRB.

%
%
%
\section{Summary and conclusions}
\label{summary and conclusions}

We took a large sample of GRB optical and NIR afterglows from 2008 to 2021 and used them to predict the expected afterglow brightness verses time for GRBs from high-$z$ ($z>5$). This sample is used to set the requirements for future GRB missions optimized to find high-$z$ GRBs and use them for cosmological studies of early star formation, galaxy evolution, metal enrichment, and reionization. We compared our findings to predictions from population studies combined with theoretical predictions and find good agreement. Our conclusion is that for future missions designed to survey and then use high-$z$ GRBs for cosmology, an onboard NIR telescope with multiband filters such as PIRT is essential in order to carry out the following functions:
\begin{enumerate}
\item Determine for every GRB whether there is a sufficiently bright IR afterglow that can be spectroscopically followed up with large-telescope $(>6$ m)\ observatories.
\item Determine the redshift of the IR afterglow to flag which of many hundreds of GRBs are from high redshift in order to identify the few high-$z$ objects worthy of follow-up observation.
\item Determine the position of the IR transient with sufficient accuracy that a large ground-based telescope can directly follow up and place the afterglow in the spectrometer slit.
\item Do all of this within 1000 s, so that spectroscopy of the afterglow can start when the afterglow is bright; and if possible within 1 hour.
\end{enumerate}
We conclude by noting that these capabilities will continue the legacy of \emph{Swift} \citep{GEHRELS20152} and will address many astrophysics questions in tandem with new ground-based facilities such as the Vera Ruben Observatory, and future gravitational wave observatories. 

\begin{acknowledgements}

Based in part on observations made at the Sierra Nevada Observatory, operated by the Instituto de Astrof\'isica de Andaluc\'ia (IAA-CSIC).
DAK acknowledges support from Spanish National Research Project RTI2018-098104-J-I00 (GRBPhot). DAK is indebted to U. Laux, C. H\"ogner, and F. Ludwig for long years of observational support at the Th\"uringer Landessternwarte Tautenburg, and S. Ertel, M. R\"oder, H. Meusinger, as well as A. Nicuesa Guelbenzu for observation time and further support.
AdUP and CCT acknowledge support from Ram\'on y Cajal fellowships RyC-2012-09975 and RyC-2012-09984 and the Spanish Ministry of Economy and Competitiveness through projects AYA2014-58381-P and AYA2017-89384-P, AdUP furthermore from the BBVA foundation.


We thank T. Sakamoto, M. Jang, Y. Jeon, M. Karouzos, E. Kang, P. Choi, and S. Pak for either assisting us obtaining the UKIRT, LOAO, Maidanak, and McDonald data for their role in securing the observational resources.
MI and GSHP acknowledge the support from the National Research Foundation (NRF) grants, No. 2020R1A2C3011091, and No. 2021M3F7A1084525, and the KASI R \& D program (Program No. 2020-1-600-05) supervized by the Ministry of Science and ICT (MSIT) of Korea.
HDJ was supported by the NRF grants No. 2022R1C1C2013543 and No. 2021M3F7AA1084525, and YK was supported by the NRF grant No. 2021R1C1C2091550, both funded by MSIT. Y.K. was supported by the National Research Foundation of Korea (NRF) grant funded by the Korean Government (MSIT) (No. 2021R1C1C2091550 and 2022 R1A4A3031306).

MM acknowledges financial support from the Italian Ministry of University and Research - Project Proposal CIR01\_00010.

This research was funded in part by the National Science Center (NCN), Poland under grant number OPUS 2021/41/B/ST9/00757

This work made use of data supplied by the UK \emph{Swift} Science Data Centre at the University of Leicester.

The Pan-STARRS1 Surveys (PS1) and the PS1 public science archive have been made possible through contributions by the Institute for Astronomy, the University of Hawaii, the Pan-STARRS Project Office, the Max-Planck Society and its participating institutes, the Max Planck Institute for Astronomy, Heidelberg and the Max Planck Institute for Extraterrestrial Physics, Garching, The Johns Hopkins University, Durham University, the University of Edinburgh, the Queen's University Belfast, the Harvard-Smithsonian Center for Astrophysics, the Las Cumbres Observatory Global Telescope Network Incorporated, the National Central University of Taiwan, the Space Telescope Science Institute, the National Aeronautics and Space Administration under Grant No. NNX08AR22G issued through the Planetary Science Division of the NASA Science Mission Directorate, the National Science Foundation Grant No. AST-1238877, the University of Maryland, Eotvos Lorand University (ELTE), the Los Alamos National Laboratory, and the Gordon and Betty Moore Foundation.

Funding for SDSS-III has been provided by the Alfred P. Sloan Foundation, the Participating Institutions, the National Science Foundation, and the U.S. Department of Energy Office of Science. The SDSS-III web site is \url{http://www.sdss3.org/}. SDSS-III is managed by the Astrophysical Research Consortium for the Participating Institutions of the SDSS-III Collaboration including the University of Arizona, the Brazilian Participation Group, Brookhaven National Laboratory, Carnegie Mellon University, University of Florida, \'ithe French Participation Group, the German Participation Group, Harvard University, the Instituto de Astrof\'isica de Canarias, the Michigan State/Notre Dame/JINA Participation Group, Johns Hopkins University, Lawrence Berkeley National Laboratory, Max Planck Institute for Astrophysics, Max Planck Institute for Extraterrestrial Physics, New Mexico State University, New York University, Ohio State University, Pennsylvania State University, University of Portsmouth, Princeton University, the Spanish Participation Group, University of Tokyo, University of Utah, Vanderbilt University, University of Virginia, University of Washington, and Yale University.

This work includes the data taken by the Mt. Lemmon Astronomy Observatory 1.0 m and the Bohyunsan Optical Astronomy Observatory 1.8 m telescope of the Korea Astronomy \& Space Science Institute (KASI), the 2.1 m Otto-Struve telescope of the McDonald Observatory of The University of Texas at Austin, and UKIRT which was supported by NASA and operated under an agreement between the University of Hawaii, the University of Arizona, and Lockheed Martin Advanced Technology Center; UKIRT operations were enabled through the cooperation of the Joint Astronomy Centre of the Science and Technology 

\end{acknowledgements}

\bibliographystyle{aa} 
\bibliography{mybib,morebib,morebib_im} 

\begin{thebibliography}{789}
\expandafter\ifx\csname natexlab\endcsname\relax\def\natexlab#1{#1}\fi

\bibitem[{{Abdalla} {et~al.}(2019){Abdalla}, {Adam}, {Aharonian}, {Ait
  Benkhali}, {Ang{\"u}ner}, {Arakawa}, {Arcaro}, {Armand}, {Ashkar}, {Backes},
  {Barbosa Martins}, {Barnard}, {Becherini}, {Berge}, {Bernl{\"o}hr},
  {Bissaldi}, {Blackwell}, {B{\"o}ttcher}, {Boisson}, {Bolmont}, {Bonnefoy},
  {Bregeon}, {Breuhaus}, {Brun}, {Brun}, {Bryan}, {B{\"u}chele}, {Bulik},
  {Bylund}, {Capasso}, {Caroff}, {Carosi}, {Casanova}, {Cerruti}, {Chand},
  {Chandra}, {Chen}, {Colafrancesco}, {Cury{\l}o}, {Davids}, {Deil}, {Devin},
  {deWilt}, {Dirson}, {Djannati-Ata{\"\i}}, {Dmytriiev}, {Donath},
  {Doroshenko}, {Dyks}, {Egberts}, {Emery}, {Ernenwein}, {Eschbach}, {Feijen},
  {Fegan}, {Fiasson}, {Fontaine}, {Funk}, {F{\"u}{\ss}ling}, {Gabici},
  {Gallant}, {Gat{\'e}}, {Giavitto}, {Giunti}, {Glawion}, {Glicenstein},
  {Gottschall}, {Grondin}, {Hahn}, {Haupt}, {Heinzelmann}, {Henri}, {Hermann},
  {Hinton}, {Hofmann}, {Hoischen}, {Holch}, {Holler}, {Horns}, {Huber},
  {Iwasaki}, {Jamrozy}, {Jankowsky}, {Jankowsky}, {Jardin-Blicq},
  {Jung-Richardt}, {Kastendieck}, {Katarzy{\'n}ski}, {Katsuragawa}, {Katz},
  {Khangulyan}, {Kh{\'e}lifi}, {King}, {Klepser}, {Klu{\'z}niak}, {Komin},
  {Kosack}, {Kostunin}, {Kreter}, {Lamanna}, {Lemi{\`e}re}, {Lemoine-Goumard},
  {Lenain}, {Leser}, {Levy}, {Lohse}, {Lypova}, {Mackey}, {Majumdar},
  {Malyshev}, {Marandon}, {Marcowith}, {Mares}, {Mariaud}, {Mart{\'\i}-Devesa},
  {Marx}, {Maurin}, {Meintjes}, {Mitchell}, {Moderski}, {Mohamed}, {Mohrmann},
  {Moore}, {Moulin}, {Muller}, {Murach}, {Nakashima}, {de Naurois},
  {Ndiyavala}, {Niederwanger}, {Niemiec}, {Oakes}, {O'Brien}, {Odaka}, {Ohm},
  {de Ona Wilhelmi}, {Ostrowski}, {Oya}, {Panter}, {Parsons}, {Perennes},
  {Petrucci}, {Peyaud}, {Piel}, {Pita}, {Poireau}, {Priyana Noel}, {Prokhorov},
  {Prokoph}, {P{\"u}hlhofer}, {Punch}, {Quirrenbach}, {Raab}, {Rauth},
  {Reimer}, {Reimer}, {Remy}, {Renaud}, {Rieger}, {Rinchiuso}, {Romoli},
  {Rowell}, {Rudak}, {Ruiz-Velasco}, {Sahakian}, {Sailer}, {Saito}, {Sanchez},
  {Santangelo}, {Sasaki}, {Schlickeiser}, {Sch{\"u}ssler}, {Schulz}, {Schutte},
  {Schwanke}, {Schwemmer}, {Seglar-Arroyo}, {Senniappan}, {Seyffert}, {Shafi},
  {Shiningayamwe}, {Simoni}, {Sinha}, {Sol}, {Specovius}, {Spir-Jacob},
  {Stawarz}, {Steenkamp}, {Stegmann}, {Steppa}, {Takahashi}, {Tavernier},
  {Taylor}, {Terrier}, {Tiziani}, {Tluczykont}, {Trichard}, {Tsirou}, {Tsuji},
  {Tuffs}, {Uchiyama}, {van der Walt}, {van Eldik}, {van Rensburg}, {van
  Soelen}, {Vasileiadis}, {Veh}, {Venter}, {Vincent}, {Vink}, {V{\"o}lk},
  {Vuillaume}, {Wadiasingh}, {Wagner}, {White}, {Wierzcholska}, {Yang},
  {Yoneda}, {Zacharias}, {Zanin}, {Zdziarski}, {Zech}, {Ziegler}, {Zorn},
  {{\.Z}ywucka}, {de Palma}, {Axelsson}, \& {Roberts}}]{Abdalla2019Nature}
{Abdalla}, H., {Adam}, R., {Aharonian}, F., {et~al.} 2019, \nat, 575, 464

\bibitem[{{Ackermann} {et~al.}(2014){Ackermann}, {Ajello}, {Asano}, {Atwood},
  {Axelsson}, {Baldini}, {Ballet}, {Barbiellini}, {Baring}, {Bastieri},
  {Bechtol}, {Bellazzini}, {Bissaldi}, {Bonamente}, {Bregeon}, {Brigida},
  {Bruel}, {Buehler}, {Burgess}, {Buson}, {Caliandro}, {Cameron}, {Caraveo},
  {Cecchi}, {Chaplin}, {Charles}, {Chekhtman}, {Cheung}, {Chiang}, {Chiaro},
  {Ciprini}, {Claus}, {Cleveland}, {Cohen-Tanugi}, {Collazzi}, {Cominsky},
  {Connaughton}, {Conrad}, {Cutini}, {D'Ammando}, {de Angelis}, {DeKlotz}, {de
  Palma}, {Dermer}, {Desiante}, {Diekmann}, {Di Venere}, {Drell},
  {Drlica-Wagner}, {Favuzzi}, {Fegan}, {Ferrara}, {Finke}, {Fitzpatrick},
  {Focke}, {Franckowiak}, {Fukazawa}, {Funk}, {Fusco}, {Gargano}, {Gehrels},
  {Germani}, {Gibby}, {Giglietto}, {Giles}, {Giordano}, {Giroletti}, {Godfrey},
  {Granot}, {Grenier}, {Grove}, {Gruber}, {Guiriec}, {Hadasch}, {Hanabata},
  {Harding}, {Hayashida}, {Hays}, {Horan}, {Hughes}, {Inoue}, {Jogler},
  {J{\'o}hannesson}, {Johnson}, {Kawano}, {Kn{\"o}dlseder}, {Kocevski}, {Kuss},
  {Lande}, {Larsson}, {Latronico}, {Longo}, {Loparco}, {Lovellette}, {Lubrano},
  {Mayer}, {Mazziotta}, {McEnery}, {Michelson}, {Mizuno}, {Moiseev}, {Monzani},
  {Moretti}, {Morselli}, {Moskalenko}, {Murgia}, {Nemmen}, {Nuss}, {Ohno},
  {Ohsugi}, {Okumura}, {Omodei}, {Orienti}, {Paneque}, {Pelassa}, {Perkins},
  {Pesce-Rollins}, {Petrosian}, {Piron}, {Pivato}, {Porter}, {Racusin},
  {Rain{\`o}}, {Rando}, {Razzano}, {Razzaque}, {Reimer}, {Reimer}, {Ritz},
  {Roth}, {Ryde}, {Sartori}, {Parkinson}, {Scargle}, {Schulz}, {Sgr{\`o}},
  {Siskind}, {Sonbas}, {Spandre}, {Spinelli}, {Tajima}, {Takahashi}, {Thayer},
  {Thayer}, {Thompson}, {Tibaldo}, {Tinivella}, {Torres}, {Tosti}, {Troja},
  {Usher}, {Vandenbroucke}, {Vasileiou}, {Vianello}, {Vitale}, {Winer}, {Wood},
  {Yamazaki}, {Younes}, {Yu}, {Zhu}, {Bhat}, {Briggs}, {Byrne}, {Foley},
  {Goldstein}, {Jenke}, {Kippen}, {Kouveliotou}, {McBreen}, {Meegan},
  {Paciesas}, {Preece}, {Rau}, {Tierney}, {van der Horst}, {von Kienlin},
  {Wilson-Hodge}, {Xiong}, {Cusumano}, {La Parola}, \&
  {Cummings}}]{Ackermann2014Science}
{Ackermann}, M., {Ajello}, M., {Asano}, K., {et~al.} 2014, Science, 343, 42

\bibitem[{{Ackermann} {et~al.}(2013){Ackermann}, {Ajello}, {Asano}, {Baldini},
  {Barbiellini}, {Baring}, {Bastieri}, {Bellazzini}, {Blandford}, {Bonamente},
  {Borgland}, {Bottacini}, {Bregeon}, {Brigida}, {Bruel}, {Buehler}, {Buson},
  {Caliandro}, {Cameron}, {Caraveo}, {Cecchi}, {Charles}, {Chaves},
  {Chekhtman}, {Chiang}, {Ciprini}, {Claus}, {Cohen-Tanugi}, {Conrad},
  {Cutini}, {D'Ammando}, {de Angelis}, {de Palma}, {Dermer}, {Silva}, {Drell},
  {Drlica-Wagner}, {Favuzzi}, {Fegan}, {Focke}, {Franckowiak}, {Fukazawa},
  {Fusco}, {Gargano}, {Gasparrini}, {Gehrels}, {Giglietto}, {Giordano},
  {Giroletti}, {Glanzman}, {Godfrey}, {Granot}, {Greiner}, {Grenier}, {Grove},
  {Guiriec}, {Hadasch}, {Hanabata}, {Hayashida}, {Hays}, {Hughes}, {Jackson},
  {Jogler}, {J{\'o}hannesson}, {Johnson}, {Kn{\"o}dlseder}, {Kocevski}, {Kuss},
  {Lande}, {Larsson}, {Latronico}, {Longo}, {Loparco}, {Lovellette}, {Lubrano},
  {Mazziotta}, {McEnery}, {Mehault}, {M{\'e}sz{\'a}ros}, {Michelson},
  {Mitthumsiri}, {Mizuno}, {Monte}, {Monzani}, {Moretti}, {Morselli},
  {Moskalenko}, {Murgia}, {Naumann-Godo}, {Norris}, {Nuss}, {Nymark}, {Ohno},
  {Ohsugi}, {Omodei}, {Orienti}, {Orlando}, {Paneque}, {Perkins},
  {Pesce-Rollins}, {Piron}, {Pivato}, {Racusin}, {Rain{\`o}}, {Rando},
  {Razzano}, {Razzaque}, {Reimer}, {Reimer}, {Romoli}, {Roth}, {Ryde},
  {Sanchez}, {Sgr{\`o}}, {Siskind}, {Sonbas}, {Spinelli}, {Stamatikos},
  {Takahashi}, {Tanaka}, {Thayer}, {Thayer}, {Tibaldo}, {Tinivella}, {Tosti},
  {Troja}, {Usher}, {Vandenbroucke}, {Vasileiou}, {Vianello}, {Vitale},
  {Waite}, {Winer}, {Wood}, {Yang}, {Gruber}, {Bhat}, {Bissaldi}, {Briggs},
  {Burgess}, {Connaughton}, {Foley}, {Kippen}, {Kouveliotou}, {McBreen},
  {McGlynn}, {Paciesas}, {Pelassa}, {Preece}, {Rau}, {van der Horst}, {von
  Kienlin}, {Kann}, {Filgas}, {Klose}, {Kr{\"u}hler}, {Fukui}, {Sako},
  {Tristram}, {Oates}, {Ukwatta}, \& {Littlejohns}}]{Ackermann2013ApJ}
{Ackermann}, M., {Ajello}, M., {Asano}, K., {et~al.} 2013, \apj, 763, 71

\bibitem[{{Ajello} {et~al.}(2020){Ajello}, {Arimoto}, {Axelsson}, {Baldini},
  {Barbiellini}, {Bastieri}, {Bellazzini}, {Berretta}, {Bissaldi}, {Blandford},
  {Bonino}, {Bottacini}, {Bregeon}, {Bruel}, {Buehler}, {Burns}, {Buson},
  {Cameron}, {Caputo}, {Caraveo}, {Cavazzuti}, {Chen}, {Chiaro}, {Ciprini},
  {Cohen-Tanugi}, {Costantin}, {Cutini}, {D'Ammando}, {DeKlotz}, {de la Torre
  Luque}, {de Palma}, {Desai}, {Di Lalla}, {Di Venere}, {Fana Dirirsa},
  {Fegan}, {Franckowiak}, {Fukazawa}, {Funk}, {Fusco}, {Gargano}, {Gasparrini},
  {Giglietto}, {Gill}, {Giordano}, {Giroletti}, {Granot}, {Green}, {Grenier},
  {Grondin}, {Guiriec}, {Hays}, {Horan}, {J{\'o}hannesson}, {Kocevski},
  {Kovac'evic'}, {Kuss}, {Larsson}, {Latronico}, {Lemoine-Goumard}, {Li},
  {Liodakis}, {Longo}, {Loparco}, {Lovellette}, {Lubrano}, {Maldera},
  {Malyshev}, {Manfreda}, {Mart{\'\i}-Devesa}, {Mazziotta}, {McEnery}, {Mereu},
  {Meyer}, {Michelson}, {Mitthumsiri}, {Mizuno}, {Monzani}, {Moretti},
  {Morselli}, {Moskalenko}, {Negro}, {Nuss}, {Omodei}, {Orienti}, {Orlando},
  {Palatiello}, {Paliya}, {Paneque}, {Pei}, {Persic}, {Pesce-Rollins},
  {Petrosian}, {Piron}, {Poon}, {Porter}, {Principe}, {Racusin}, {Rain{\`o}},
  {Rando}, {Rani}, {Razzano}, {Razzaque}, {Reimer}, {Reimer}, {Ryde}, {Saz
  Parkinson}, {Serini}, {Sgr{\`o}}, {Siskind}, {Spandre}, {Spinelli}, {Tajima},
  {Takagi}, {Takahashi}, {Tak}, {Thayer}, {Thompson}, {Torres}, {Troja},
  {Valverde}, {Van Klaveren}, {Wood}, {Yassine}, {Zaharijas}, {Mailyan},
  {Bhat}, {Briggs}, {Cleveland}, {Giles}, {Goldstein}, {Hui}, {Malacaria},
  {Preece}, {Roberts}, {Veres}, {Wilson-Hodge}, {Kienlin}, {Cenko}, {O'Brien},
  {Beardmore}, {Lien}, {Osborne}, {Tohuvavohu}, {D'Elia}, {D'A{\`\i}}, {Perri},
  {Gropp}, {Klingler}, {Capalbi}, {Tagliaferri}, {Stamatikos}, \& {De
  Pasquale}}]{Ajello2020ApJ}
{Ajello}, M., {Arimoto}, M., {Axelsson}, M., {et~al.} 2020, \apj, 890, 9

\bibitem[{{Akerlof} {et~al.}(1999){Akerlof}, {Balsano}, {Barthelmy}, {Bloch},
  {Butterworth}, {Casperson}, {Cline}, {Fletcher}, {Frontera}, {Gisler},
  {Heise}, {Hills}, {Kehoe}, {Lee}, {Marshall}, {McKay}, {Miller}, {Piro},
  {Priedhorsky}, {Szymanski}, \& {Wren}}]{Akerlof1999Nature}
{Akerlof}, C., {Balsano}, R., {Barthelmy}, S., {et~al.} 1999, \nat, 398, 400

\bibitem[{{Akerlof} {et~al.}(2011){Akerlof}, {Zheng}, {Pandey}, \&
  {McKay}}]{Akerlof2011ApJ}
{Akerlof}, C.~W., {Zheng}, W., {Pandey}, S.~B., \& {McKay}, T.~A. 2011, \apj,
  726, 22

\bibitem[{{Akitaya} {et~al.}(2014){Akitaya}, {Moritani}, {Ui}, {Kanda}, \&
  {Yoshida}}]{Akitaya2014GCN16163}
{Akitaya}, H., {Moritani}, Y., {Ui}, T., {Kanda}, Y., \& {Yoshida}, M. 2014,
  GRB Coordinates Network, 16163

\bibitem[{{Aksulu} {et~al.}(2022){Aksulu}, {Wijers}, {van Eerten}, \& {van der
  Horst}}]{Aksulu2022}
{Aksulu}, M.~D., {Wijers}, R.~A.~M.~J., {van Eerten}, H.~J., \& {van der
  Horst}, A.~J. 2022, \mnras, 511, 2848

\bibitem[{{Alam} {et~al.}(2015){Alam}, {Albareti}, {Allende Prieto}, {Anders},
  {Anderson}, {Anderton}, {Andrews}, {Armengaud}, {Aubourg}, {Bailey}, {Basu},
  {Bautista}, {Beaton}, {Beers}, {Bender}, {Berlind}, {Beutler}, {Bhardwaj},
  {Bird}, {Bizyaev}, {Blake}, {Blanton}, {Blomqvist}, {Bochanski}, {Bolton},
  {Bovy}, {Shelden Bradley}, {Brandt}, {Brauer}, {Brinkmann}, {Brown},
  {Brownstein}, {Burden}, {Burtin}, {Busca}, {Cai}, {Capozzi}, {Carnero
  Rosell}, {Carr}, {Carrera}, {Chambers}, {Chaplin}, {Chen}, {Chiappini},
  {Chojnowski}, {Chuang}, {Clerc}, {Comparat}, {Covey}, {Croft}, {Cuesta},
  {Cunha}, {da Costa}, {Da Rio}, {Davenport}, {Dawson}, {De Lee}, {Delubac},
  {Deshpande}, {Dhital}, {Dutra-Ferreira}, {Dwelly}, {Ealet}, {Ebelke},
  {Edmondson}, {Eisenstein}, {Ellsworth}, {Elsworth}, {Epstein}, {Eracleous},
  {Escoffier}, {Esposito}, {Evans}, {Fan}, {Fern{\'a}ndez-Alvar}, {Feuillet},
  {Filiz Ak}, {Finley}, {Finoguenov}, {Flaherty}, {Fleming}, {Font-Ribera},
  {Foster}, {Frinchaboy}, {Galbraith-Frew}, {Garc{\'\i}a},
  {Garc{\'\i}a-Hern{\'a}ndez}, {Garc{\'\i}a P{\'e}rez}, {Gaulme}, {Ge},
  {G{\'e}nova-Santos}, {Georgakakis}, {Ghezzi}, {Gillespie}, {Girardi},
  {Goddard}, {Gontcho}, {Gonz{\'a}lez Hern{\'a}ndez}, {Grebel}, {Green},
  {Grieb}, {Grieves}, {Gunn}, {Guo}, {Harding}, {Hasselquist}, {Hawley},
  {Hayden}, {Hearty}, {Hekker}, {Ho}, {Hogg}, {Holley-Bockelmann}, {Holtzman},
  {Honscheid}, {Huber}, {Huehnerhoff}, {Ivans}, {Jiang}, {Johnson},
  {Kinemuchi}, {Kirkby}, {Kitaura}, {Klaene}, {Knapp}, {Kneib}, {Koenig},
  {Lam}, {Lan}, {Lang}, {Laurent}, {Le Goff}, {Leauthaud}, {Lee}, {Lee},
  {Licquia}, {Liu}, {Long}, {L{\'o}pez-Corredoira}, {Lorenzo-Oliveira},
  {Lucatello}, {Lundgren}, {Lupton}, {Mack}, {Mahadevan}, {Maia}, {Majewski},
  {Malanushenko}, {Malanushenko}, {Manchado}, {Manera}, {Mao}, {Maraston},
  {Marchwinski}, {Margala}, {Martell}, {Martig}, {Masters}, {Mathur},
  {McBride}, {McGehee}, {McGreer}, {McMahon}, {M{\'e}nard}, {Menzel},
  {Merloni}, {M{\'e}sz{\'a}ros}, {Miller}, {Miralda-Escud{\'e}}, {Miyatake},
  {Montero-Dorta}, {More}, {Morganson}, {Morice-Atkinson}, {Morrison},
  {Mosser}, {Muna}, {Myers}, {Nandra}, {Newman}, {Neyrinck}, {Nguyen},
  {Nichol}, {Nidever}, {Noterdaeme}, {Nuza}, {O'Connell}, {O'Connell},
  {O'Connell}, {Ogando}, {Olmstead}, {Oravetz}, {Oravetz}, {Osumi}, {Owen},
  {Padgett}, {Padmanabhan}, {Paegert}, {Palanque-Delabrouille}, {Pan},
  {Parejko}, {P{\^a}ris}, {Park}, {Pattarakijwanich}, {Pellejero-Ibanez},
  {Pepper}, {Percival}, {P{\'e}rez-Fournon}, {á¹rez-Ra`fols}, {Petitjean},
  {Pieri}, {Pinsonneault}, {Porto de Mello}, {Prada}, {Prakash},
  {Price-Whelan}, {Protopapas}, {Raddick}, {Rahman}, {Reid}, {Rich}, {Rix},
  {Robin}, {Rockosi}, {Rodrigues}, {Rodr{\'\i}guez-Torres}, {Roe}, {Ross},
  {Ross}, {Rossi}, {Ruan}, {Rubi{\~n}o-Mart{\'\i}n}, {Rykoff},
  {Salazar-Albornoz}, {Salvato}, {Samushia}, {S{\'a}nchez}, {Santiago},
  {Sayres}, {Schiavon}, {Schlegel}, {Schmidt}, {Schneider}, {Schultheis},
  {Schwope}, {Sc{\'o}ccola}, {Scott}, {Sellgren}, {Seo}, {Serenelli}, {Shane},
  {Shen}, {Shetrone}, {Shu}, {Silva Aguirre}, {Sivarani}, {Skrutskie},
  {Slosar}, {Smith}, {Sobreira}, {Souto}, {Stassun}, {Steinmetz}, {Stello},
  {Strauss}, {Streblyanska}, {Suzuki}, {Swanson}, {Tan}, {Tayar}, {Terrien},
  {Thakar}, {Thomas}, {Thomas}, {Thompson}, {Tinker}, {Tojeiro}, {Troup},
  {Vargas-Maga{\~n}a}, {Vazquez}, {Verde}, {Viel}, {Vogt}, {Wake}, {Wang},
  {Weaver}, {Weinberg}, {Weiner}, {White}, {Wilson}, {Wisniewski},
  {Wood-Vasey}, {Ye`che}, {York}, {Zakamska}, {Zamora}, {Zasowski}, {Zehavi},
  {Zhao}, {Zheng}, {Zhou}, {Zhou}, {Zou}, \& {Zhu}}]{Alam2015ApJS}
{Alam}, S., {Albareti}, F.~D., {Allende Prieto}, C., {et~al.} 2015, \apjs, 219,
  12

\bibitem[{{Amati} {et~al.}(2002){Amati}, {Frontera}, {Tavani}, {in't Zand},
  {Antonelli}, {Costa}, {Feroci}, {Guidorzi}, {Heise}, {Masetti}, {Montanari},
  {Nicastro}, {Palazzi}, {Pian}, {Piro}, \& {Soffitta}}]{Amati2002AA}
{Amati}, L., {Frontera}, F., {Tavani}, M., {et~al.} 2002, \aap, 390, 81

\bibitem[{{Amati} {et~al.}(2018){Amati}, {O'Brien}, {G{\"o}tz}, {Bozzo},
  {Tenzer}, {Frontera}, {Ghirlanda}, {Labanti}, {Osborne}, {Stratta}, {Tanvir},
  {Willingale}, {Attina}, {Campana}, {Castro-Tirado}, {Contini}, {Fuschino},
  {Gomboc}, {Hudec}, {Orleanski}, {Renotte}, {Rodic}, {Bagoly}, {Blain},
  {Callanan}, {Covino}, {Ferrara}, {Le Floch}, {Marisaldi}, {Mereghetti},
  {Rosati}, {Vacchi}, {D'Avanzo}, {Giommi}, {Piranomonte}, {Piro}, {Reglero},
  {Rossi}, {Santangelo}, {Salvaterra}, {Tagliaferri}, {Vergani}, {Vinciguerra},
  {Briggs}, {Campolongo}, {Ciolfi}, {Connaughton}, {Cordier}, {Morelli},
  {Orlandini}, {Adami}, {Argan}, {Atteia}, {Auricchio}, {Balazs}, {Baldazzi},
  {Basa}, {Basak}, {Bellutti}, {Bernardini}, {Bertuccio}, {Braga}, {Branchesi},
  {Brandt}, {Brocato}, {Budtz-Jorgensen}, {Bulgarelli}, {Burderi}, {Camp},
  {Capozziello}, {Caruana}, {Casella}, {Cenko}, {Chardonnet}, {Ciardi},
  {Colafrancesco}, {Dainotti}, {D'Elia}, {De Martino}, {De Pasquale}, {Del
  Monte}, {Della Valle}, {Drago}, {Evangelista}, {Feroci}, {Finelli},
  {Fiorini}, {Fynbo}, {Gal-Yam}, {Gendre}, {Ghisellini}, {Grado}, {Guidorzi},
  {Hafizi}, {Hanlon}, {Hjorth}, {Izzo}, {Kiss}, {Kumar}, {Kuvvetli}, {Lavagna},
  {Li}, {Longo}, {Lyutikov}, {Maio}, {Maiorano}, {Malcovati}, {Malesani},
  {Margutti}, {Martin-Carrillo}, {Masetti}, {McBreen}, {Mignani}, {Morgante},
  {Mundell}, {Nargaard-Nielsen}, {Nicastro}, {Palazzi}, {Paltani}, {Panessa},
  {Pareschi}, {Pe'er}, {Penacchioni}, {Pian}, {Piedipalumbo}, {Piran}, {Rauw},
  {Razzano}, {Read}, {Rezzolla}, {Romano}, {Ruffini}, {Savaglio}, {Sguera},
  {Schady}, {Skidmore}, {Song}, {Stanway}, {Starling}, {Topinka}, {Troja}, {van
  Putten}, {Vanzella}, {Vercellone}, {Wilson-Hodge}, {Yonetoku}, {Zampa},
  {Zampa}, {Zhang}, {Zhang}, {Zhang}, {Zhang}, {Antonelli}, {Bianco}, {Boci},
  {Boer}, {Botticella}, {Boulade}, {Butler}, {Campana}, {Capitanio}, {Celotti},
  {Chen}, {Colpi}, {Comastri}, {Cuby}, {Dadina}, {De Luca}, {Dong}, {Ettori},
  {Gandhi}, {Geza}, {Greiner}, {Guiriec}, {Harms}, {Hernanz}, {Hornstrup},
  {Hutchinson}, {Israel}, {Jonker}, {Kaneko}, {Kawai}, {Wiersema}, {Korpela},
  {Lebrun}, {Lu}, {MacFadyen}, {Malaguti}, {Maraschi}, {Melandri}, {Modjaz},
  {Morris}, {Omodei}, {Paizis}, {P{\'a}ta}, {Petrosian}, {Rachevski}, {Rhoads},
  {Ryde}, {Sabau-Graziati}, {Shigehiro}, {Sims}, {Soomin}, {Sz{\'e}csi},
  {Urata}, {Uslenghi}, {Valenziano}, {Vianello}, {Vojtech}, {Watson}, \&
  {Zicha}}]{Amati2018AdSpR}
{Amati}, L., {O'Brien}, P., {G{\"o}tz}, D., {et~al.} 2018, Advances in Space
  Research, 62, 191

\bibitem[{{Amati} {et~al.}(2021){Amati}, {O'Brien}, {G{\"o}tz}, {Bozzo},
  {Santangelo}, {Tanvir}, {Frontera}, {Mereghetti}, {Osborne}, {Blain}, {Basa},
  {Branchesi}, {Burderi}, {Caballero-Garc{\'\i}a}, {Castro-Tirado},
  {Christensen}, {Ciolfi}, {De Rosa}, {Doroshenko}, {Ferrara}, {Ghirlanda},
  {Hanlon}, {Heddermann}, {Hutchinson}, {Labanti}, {Le Floch}, {Lerman},
  {Paltani}, {Reglero}, {Rezzolla}, {Rosati}, {Salvaterra}, {Stratta},
  {Tenzer}, \& {Theseus Consortium}}]{Amati2021THESEUS}
{Amati}, L., {O'Brien}, P.~T., {G{\"o}tz}, D., {et~al.} 2021, Experimental
  Astronomy, 52, 183

\bibitem[{{Andersen} {et~al.}(2000){Andersen}, {Hjorth}, {Pedersen}, {Jensen},
  {Hunt}, {Gorosabel}, {M{\o}ller}, {Fynbo}, {Kippen}, {Thomsen}, {Olsen},
  {Christensen}, {Vestergaard}, {Masetti}, {Palazzi}, {Hurley}, {Cline},
  {Kaper}, \& {Jaunsen}}]{Andersen2000AA}
{Andersen}, M.~I., {Hjorth}, J., {Pedersen}, H., {et~al.} 2000, \aap, 364, L54

\bibitem[{{Andreev} {et~al.}(2015){Andreev}, {Mazaeva}, {Sergeev}, {Volnova},
  \& {Pozanenko}}]{Andreev2015GCN18306}
{Andreev}, M., {Mazaeva}, E., {Sergeev}, A., {Volnova}, A., \& {Pozanenko}, A.
  2015, GRB Coordinates Network, 18306

\bibitem[{{Andreev} {et~al.}(2008{\natexlab{a}}){Andreev}, {Sergeev}, {Babina},
  \& {Pozanenko}}]{Andreev2008GCN8596}
{Andreev}, M., {Sergeev}, A., {Babina}, J., \& {Pozanenko}, A.
  2008{\natexlab{a}}, GRB Coordinates Network, 8596

\bibitem[{{Andreev} {et~al.}(2008{\natexlab{b}}){Andreev}, {Sergeev}, {Babina},
  \& {Pozanenko}}]{Andreev2008GCN8615}
{Andreev}, M., {Sergeev}, A., {Babina}, J., \& {Pozanenko}, A.
  2008{\natexlab{b}}, GRB Coordinates Network, 8615

\bibitem[{{Andreev} {et~al.}(2009{\natexlab{a}}){Andreev}, {Sergeev},
  {Parakhin}, {Karpov}, {Kuznietsova}, {Petkov}, \&
  {Pozanenko}}]{Andreev2009GCN10273}
{Andreev}, M., {Sergeev}, A., {Parakhin}, N., {et~al.} 2009{\natexlab{a}}, GRB
  Coordinates Network, 10273

\bibitem[{{Andreev} {et~al.}(2009{\natexlab{b}}){Andreev}, {Sergeev}, \&
  {Pozanenko}}]{Andreev2009GCN10207}
{Andreev}, M., {Sergeev}, A., \& {Pozanenko}, A. 2009{\natexlab{b}}, GCN
  Circulars, 10207

\bibitem[{{Andreev} {et~al.}(2010{\natexlab{a}}){Andreev}, {Sergeev}, \&
  {Pozanenko}}]{Andreev2010GCN11191}
{Andreev}, M., {Sergeev}, A., \& {Pozanenko}, A. 2010{\natexlab{a}}, GRB
  Coordinates Network, 11191

\bibitem[{{Andreev} {et~al.}(2010{\natexlab{b}}){Andreev}, {Sergeev},
  {Pozanenko}, {Parakhin}, {Velichko}, {Borachok}, \&
  {Petkov}}]{Andreev2010GCN11200}
{Andreev}, M., {Sergeev}, A., {Pozanenko}, A., {et~al.} 2010{\natexlab{b}}, GRB
  Coordinates Network, 11200

\bibitem[{{Andreev} {et~al.}(2010{\natexlab{c}}){Andreev}, {Sergeev},
  {Volnova}, \& {Pozanenko}}]{Andreev2010GCN10694}
{Andreev}, M., {Sergeev}, A., {Volnova}, A., \& {Pozanenko}, A.
  2010{\natexlab{c}}, GRB Coordinates Network, 10694

\bibitem[{{Antonelli} {et~al.}(2008){Antonelli}, {D'Avanzo}, {Malesani},
  {Covino}, {Fugazza}, {Calzoletti}, {Campana}, {Chincarini}, {Conciatore},
  {Cutini}, {D'Elia}, {D'Alessio}, {Fiore}, {Goldoni}, {Guetta}, {Guidorzi},
  {Israel}, {Maiorano}, {Masetti}, {Melandri}, {Meurs}, {Nicastro}, {Palazzi},
  {Pian}, {Piranomonte}, {Piranomonte}, {Stella}, {Stratta}, {Tagliaferri},
  {Tosti}, {Testa}, {Vergani}, \& {Vitali}}]{Antonelli2008GCN7597}
{Antonelli}, L.~A., {D'Avanzo}, P., {Malesani}, D., {et~al.} 2008, GRB
  Coordinates Network, 7597

\bibitem[{{Antonelli} {et~al.}(2009){Antonelli}, {D'Avanzo}, {Perna}, {Amati},
  {Covino}, {Cutini}, {D'Elia}, {Gallozzi}, {Grazian}, {Palazzi},
  {Piranomonte}, {Rossi}, {Spiro}, {Stella}, {Testa}, {Chincarini}, {di Paola},
  {Fiore}, {Fugazza}, {Giallongo}, {Maiorano}, {Masetti}, {Pedichini},
  {Salvaterra}, {Tagliaferri}, \& {Vergani}}]{Antonelli2009AA}
{Antonelli}, L.~A., {D'Avanzo}, P., {Perna}, R., {et~al.} 2009, \aap, 507, L45

\bibitem[{{Ashall} {et~al.}(2019){Ashall}, {Mazzali}, {Pian}, {Woosley},
  {Palazzi}, {Prentice}, {Kobayashi}, {Holmbo}, {Levan}, {Perley},
  {Stritzinger}, {Bufano}, {Filippenko}, {Melandri}, {Oates}, {Rossi},
  {Selsing}, {Zheng}, {Castro-Tirado}, {Chincarini}, {D'Avanzo}, {De Pasquale},
  {Emery}, {Fruchter}, {Hurley}, {Moller}, {Nomoto}, {Tanaka}, \&
  {Valeev}}]{Ashall2017Nature}
{Ashall}, C., {Mazzali}, P.~A., {Pian}, E., {et~al.} 2019, \mnras, 487, 5824

\bibitem[{{Band} {et~al.}(1993){Band}, {Matteson}, {Ford}, {Schaefer},
  {Palmer}, {Teegarden}, {Cline}, {Briggs}, {Paciesas}, {Pendleton}, {Fishman},
  {Kouveliotou}, {Meegan}, {Wilson}, \& {Lestrade}}]{band1993}
{Band}, D., {Matteson}, J., {Ford}, L., {et~al.} 1993, \apj, 413, 281

\bibitem[{{Bardho} {et~al.}(2016){Bardho}, {Gendre}, {Rossi}, {Amati},
  {Haislip}, {Klotz}, {Palazzi}, {Reichart}, {Trotter}, \&
  {Bo{\"e}r}}]{Bardho2016MNRAS}
{Bardho}, O., {Gendre}, B., {Rossi}, A., {et~al.} 2016, \mnras, 459, 508

\bibitem[{{Barthelmy} {et~al.}(2005){Barthelmy}, {Barbier}, {Cummings},
  {Fenimore}, {Gehrels}, {Hullinger}, {Krimm}, {Markwardt}, {Palmer},
  {Parsons}, {Sato}, {Suzuki}, {Takahashi}, {Tashiro}, \&
  {Tueller}}]{Barthelmy2005SSRv}
{Barthelmy}, S.~D., {Barbier}, L.~M., {Cummings}, J.~R., {et~al.} 2005, \ssr,
  120, 143

\bibitem[{{Basa} {et~al.}(2012){Basa}, {Cuby}, {Savaglio}, {Boissier},
  {Cl{\'e}ment}, {Flores}, {Le Borgne}, \& {Mazure}}]{Basa2012AA}
{Basa}, S., {Cuby}, J.~G., {Savaglio}, S., {et~al.} 2012, \aap, 542, A103

\bibitem[{{Becerra} {et~al.}(2021){Becerra}, {De Colle}, {Cant{\'o}}, {Lizano},
  {Gonz{\'a}lez}, {Granot}, {Klotz}, {Watson}, {Fraija}, {Araudo}, {Troja},
  {Atteia}, {Lee}, {Turpin}, {Bloom}, {Boer}, {Butler}, {Gonz{\'a}lez},
  {Kutyrev}, {Prochaska}, {Ramirez-Ruiz}, {Richer}, \&
  {Rom{\'a}n-Z{\'u}{\~n}iga}}]{Becerra2021ApJ}
{Becerra}, R.~L., {De Colle}, F., {Cant{\'o}}, J., {et~al.} 2021, \apj, 908, 39

\bibitem[{{Becerra} {et~al.}(2017{\natexlab{a}}){Becerra}, {Watson}, {Lee},
  {Fraija}, {Butler}, {Bloom}, {Capone}, {Cucchiara}, {de Diego}, {Fox},
  {Gehrels}, {Georgiev}, {Gonz{\'a}lez}, {Kutyrev}, {Littlejohns}, {Prochaska},
  {Ramirez-Ruiz}, {Richer}, {Rom{\'a}n-Z{\'u}{\~n}iga}, {Toy}, \&
  {Troja}}]{Becerra2017ApJ}
{Becerra}, R.~L., {Watson}, A.~M., {Lee}, W.~H., {et~al.} 2017{\natexlab{a}},
  \apj, 837, 116

\bibitem[{{Becerra} {et~al.}(2017{\natexlab{b}}){Becerra}, {Watson}, {Lee},
  {Fraija}, {Butler}, {Bloom}, {Capone}, {Cucchiara}, {de Diego}, {Fox},
  {Gehrels}, {Georgiev}, {Gonz{\'a}lez}, {Kutyrev}, {Littlejohns}, {Prochaska},
  {Ramirez-Ruiz}, {Richer}, {Rom{\'a}n-Z{\'u}{\~n}iga}, {Toy}, \&
  {Troja}}]{Becerra2017ApJErr}
{Becerra}, R.~L., {Watson}, A.~M., {Lee}, W.~H., {et~al.} 2017{\natexlab{b}},
  \apj, 842, 68

\bibitem[{{Becker}(2015)}]{Becker2015ascl}
{Becker}, A. 2015, {HOTPANTS: High Order Transform of PSF ANd Template
  Subtraction}

\bibitem[{{Belkin} {et~al.}(2020){Belkin}, {Pozanenko}, {Mazaeva}, {Volnova},
  {Minaev}, {Tominaga}, {Grebenev}, {Chelovekov}, {Buckley}, {Blinnikov},
  {Volvach}, {Volvach}, {Inasaridze}, {Klunko}, {Molotov}, {Reva},
  {Rumyantsev}, \& {Chestnov}}]{Belkin2020AstL}
{Belkin}, S.~O., {Pozanenko}, A.~S., {Mazaeva}, E.~D., {et~al.} 2020, Astronomy
  Letters, 46, 783

\bibitem[{{Berger} {et~al.}(2007){Berger}, {Chary}, {Cowie}, {Price},
  {Schmidt}, {Fox}, {Cenko}, {Djorgovski}, {Soderberg}, {Kulkarni}, {McCarthy},
  {Gladders}, {Peterson}, \& {Barger}}]{Berger2007ApJ}
{Berger}, E., {Chary}, R., {Cowie}, L.~L., {et~al.} 2007, \apj, 665, 102

\bibitem[{{Berger} {et~al.}(2011){Berger}, {Chornock}, {Holmes}, {Foley},
  {Cucchiara}, {Wolf}, {Podsiadlowski}, {Fox}, \& {Roth}}]{Berger2011ApJ}
{Berger}, E., {Chornock}, R., {Holmes}, T.~R., {et~al.} 2011, \apj, 743, 204

\bibitem[{{Berger} {et~al.}(2008){Berger}, {Fox}, {Cucchiara}, \&
  {Cenko}}]{Berger2008GCN8335}
{Berger}, E., {Fox}, D.~B., {Cucchiara}, A., \& {Cenko}, S.~B. 2008, GRB
  Coordinates Network, 8335

\bibitem[{{Bersier}(2011)}]{Bersier2011GCN12216}
{Bersier}, D. 2011, GRB Coordinates Network, 12216

\bibitem[{{Beskin} {et~al.}(2010){Beskin}, {Karpov}, {Bondar}, {Greco},
  {Guarnieri}, {Bartolini}, \& {Piccioni}}]{Beskin2010ApJL}
{Beskin}, G., {Karpov}, S., {Bondar}, S., {et~al.} 2010, \apjl, 719, L10

\bibitem[{{Bikmaev} {et~al.}(2014{\natexlab{a}}){Bikmaev}, {Khamitov},
  {Irtuganov}, {Sakhibullin}, {Burenin}, {Pavlinsky}, {Sunyaev}, {Kirbiyik},
  {Kiziloglu}, \& {Gogus}}]{Bikmaev2014GCN16482}
{Bikmaev}, I., {Khamitov}, I., {Irtuganov}, E., {et~al.} 2014{\natexlab{a}},
  GRB Coordinates Network, 16482

\bibitem[{{Bikmaev} {et~al.}(2014{\natexlab{b}}){Bikmaev}, {Khamitov},
  {Melnikov}, {Sakhibullin}, {Burenin}, {Pavlinsky}, {Sunyaev}, {Kirbiyik},
  {Kiziloglu}, \& {Gogus}}]{Bikmaev2014GCN16185}
{Bikmaev}, I., {Khamitov}, I., {Melnikov}, S., {et~al.} 2014{\natexlab{b}}, GRB
  Coordinates Network, 16185

\bibitem[{{Blanchard} {et~al.}(2016){Blanchard}, {Berger}, \&
  {Fong}}]{Blanchard2016ApJ}
{Blanchard}, P.~K., {Berger}, E., \& {Fong}, W.-f. 2016, \apj, 817, 144

\bibitem[{{Bla{\v{z}}ek} {et~al.}(2020){Bla{\v{z}}ek}, {de Ugarte Postigo},
  {Kann}, {Th{\"o}ne}, {Ag{\"u}{\'\i} Fern{\'a}ndez}, \&
  {Izzo}}]{Blazek2020SPIE}
{Bla{\v{z}}ek}, M., {de Ugarte Postigo}, A., {Kann}, D.~A., {et~al.} 2020, in
  Society of Photo-Optical Instrumentation Engineers (SPIE) Conference Series,
  Vol. 11452, Society of Photo-Optical Instrumentation Engineers (SPIE)
  Conference Series, 1145218

\bibitem[{{Bloom} \& {Nugent}(2010)}]{Bloom2010GCN10433}
{Bloom}, J.~S. \& {Nugent}, P.~E. 2010, GRB Coordinates Network, 10433

\bibitem[{{Bloom} {et~al.}(2009){Bloom}, {Perley}, {Li}, {Butler}, {Miller},
  {Kocevski}, {Kann}, {Foley}, {Chen}, {Filippenko}, {Starr}, {Macomber},
  {Prochaska}, {Chornock}, {Poznanski}, {Klose}, {Skrutskie}, {Lopez}, {Hall},
  {Glazebrook}, \& {Blake}}]{Bloom2009ApJ}
{Bloom}, J.~S., {Perley}, D.~A., {Li}, W., {et~al.} 2009, \apj, 691, 723

\bibitem[{{Bo{\"e}r} {et~al.}(2006){Bo{\"e}r}, {Atteia}, {Damerdji}, {Gendre},
  {Klotz}, \& {Stratta}}]{Boer2006ApJ}
{Bo{\"e}r}, M., {Atteia}, J.~L., {Damerdji}, Y., {et~al.} 2006, \apjl, 638, L71

\bibitem[{{Bolmer} {et~al.}(2018){Bolmer}, {Greiner}, {Kr{\"u}hler}, {Schady},
  {Ledoux}, {Tanvir}, \& {Levan}}]{Bolmer2018AA}
{Bolmer}, J., {Greiner}, J., {Kr{\"u}hler}, T., {et~al.} 2018, \aap, 609, A62

\bibitem[{{Breeveld} {et~al.}(2011){Breeveld}, {Landsman}, {Holland}, {Roming},
  {Kuin}, \& {Page}}]{Breeveld2011AIPC}
{Breeveld}, A.~A., {Landsman}, W., {Holland}, S.~T., {et~al.} 2011, in American
  Institute of Physics Conference Series, Vol. 1358, American Institute of
  Physics Conference Series, ed. J.~E. {McEnery}, J.~L. {Racusin}, \&
  N.~{Gehrels}, 373--376

\bibitem[{{Breeveld} \& {Mangano}(2011)}]{Breeveld2011GCN11969}
{Breeveld}, A.~A. \& {Mangano}, V. 2011, GRB Coordinates Network, 11969

\bibitem[{{Breeveld} \& {Page}(2012)}]{Breeveld2012GCN13448}
{Breeveld}, A.~A. \& {Page}, M.~J. 2012, GRB Coordinates Network, 13448

\bibitem[{{Breeveld} \& {Siegel}(2014)}]{Breeveld2014GCN16198}
{Breeveld}, A.~A. \& {Siegel}, M.~H. 2014, GRB Coordinates Network, 16198

\bibitem[{{Breeveld} \& {Stratta}(2012)}]{Breeveld2012GCN13226}
{Breeveld}, A.~A. \& {Stratta}, G. 2012, GRB Coordinates Network, 13226, 1

\bibitem[{{Breeveld} \& {Troja}(2013)}]{Breeveld2013GCN15414}
{Breeveld}, A.~A. \& {Troja}, E. 2013, GRB Coordinates Network, 15414

\bibitem[{{Brennan} {et~al.}(2008){Brennan}, {Reichart}, {Nysewander},
  {Lacluyze}, {Ivarsen}, {Crain}, {Schubel}, {Foster}, {Haislip}, {Styblova},
  \& {Trotter}}]{Brennan2008GCN7629}
{Brennan}, T., {Reichart}, D., {Nysewander}, M., {et~al.} 2008, GRB Coordinates
  Network, 7629

\bibitem[{{Brivio} {et~al.}(2022){Brivio}, {Covino}, {D'Avanzo}, {Wiersema},
  {Maund}, {Bernardini}, {Campana}, \& {Melandri}}]{Brivio2022AA}
{Brivio}, R., {Covino}, S., {D'Avanzo}, P., {et~al.} 2022, \aap, 666, A179

\bibitem[{{Brown} {et~al.}(2013){Brown}, {Baliber}, {Bianco}, {Bowman},
  {Burleson}, {Conway}, {Crellin}, {Depagne}, {De Vera}, {Dilday}, {Dragomir},
  {Dubberley}, {Eastman}, {Elphick}, {Falarski}, {Foale}, {Ford}, {Fulton},
  {Garza}, {Gomez}, {Graham}, {Greene}, {Haldeman}, {Hawkins}, {Haworth},
  {Haynes}, {Hidas}, {Hjelstrom}, {Howell}, {Hygelund}, {Lister}, {Lobdill},
  {Martinez}, {Mullins}, {Norbury}, {Parrent}, {Paulson}, {Petry}, {Pickles},
  {Posner}, {Rosing}, {Ross}, {Sand}, {Saunders}, {Shobbrook}, {Shporer},
  {Street}, {Thomas}, {Tsapras}, {Tufts}, {Valenti}, {Vander Horst}, {Walker},
  {White}, \& {Willis}}]{Brown2013PASP}
{Brown}, T.~M., {Baliber}, N., {Bianco}, F.~B., {et~al.} 2013, \pasp, 125, 1031

\bibitem[{{Buckley} {et~al.}(2016{\natexlab{a}}){Buckley}, {Potter}, {Kniazev},
  {Kotze}, {Lipunov}, {Gorbovskoy}, {Kornilov}, {Kuvshinov}, {Tyurina},
  {Balanutsa}, {Kuznetsov}, {Chazov}, {Tlatov}, {Parhomenko}, {Dormidontov},
  {Senik}, {Yurkov}, {Gabovich}, {Sergienko}, {Ivanov}, {Yazev}, {Budnev},
  {Gres}, {Chuvalaev}, {Poleshchuk}, {Podesta}, {Mallamaci}, {Lopez},
  {Podesta}, {Levato}, {Saffe}, {Rebolo}, {Serra}, {Lodieu}, {Israelian}, \&
  {Suarez-Andres}}]{Buckley2016GCN20330}
{Buckley}, D., {Potter}, S., {Kniazev}, A., {et~al.} 2016{\natexlab{a}}, GRB
  Coordinates Network, 20330

\bibitem[{{Buckley} {et~al.}(2015){Buckley}, {Potter}, {Kniazev}, {Kotze},
  {Rebolo}, {Serra}, {Lodieu}, {Israelian}, {Suarez-Andres}, {Gorbovskoy},
  {Lipunov}, {Tyurina}, {Kornilov}, {Balanutsa}, {Kuznetsov}, {Kuvshinov},
  {Vlasenko}, {Popova}, {Ivanov}, {Yazev}, {Budnev}, {Gres}, {Chuvalaev},
  {Poleshchuk}, {Tlatov}, {Parhomenko}, {Dormidontov}, {Sennik}, {Yurkov},
  {Sergienko}, {Varda}, {Sinyakov}, {Krushinski}, {Zalozhnih}, {Levato},
  {Saffe}, {Mallamaci}, {Lopez}, \& {Podest}}]{Buckley2015GCN18511}
{Buckley}, D., {Potter}, S., {Kniazev}, A., {et~al.} 2015, GRB Coordinates
  Network, 18511

\bibitem[{{Buckley} {et~al.}(2016{\natexlab{b}}){Buckley}, {Hamanowicz},
  {Martin-Carrillo}, {Razzaque}, {Garrigoux}, {Hanlon}, {Kotze}, \&
  {Kuhn}}]{Buckley2016GCN20322}
{Buckley}, D.~A.~H., {Hamanowicz}, A., {Martin-Carrillo}, A., {et~al.}
  2016{\natexlab{b}}, GRB Coordinates Network, 20322

\bibitem[{{Burlon} {et~al.}(2008){Burlon}, {Ghirlanda}, {Ghisellini},
  {Lazzati}, {Nava}, {Nardini}, \& {Celotti}}]{Burlon2008ApJ}
{Burlon}, D., {Ghirlanda}, G., {Ghisellini}, G., {et~al.} 2008, \apjl, 685, L19

\bibitem[{Burns {et~al.}(2023)Burns, Svinkin, Fenimore, Kann, Fernández,
  Frederiks, Hamburg, Lesage, Temiraev, Tsvetkova, Bissaldi, Briggs, Dalessi,
  Dunwoody, Fletcher, Goldstein, Hui, Hristov, Kocevski, Lysenko, Mailyan,
  Mangan, McBreen, Racusin, Ridnaia, Roberts, Ulanov, Veres, Wilson-Hodge, \&
  Wood}]{burns_2023}
Burns, E., Svinkin, D., Fenimore, E., {et~al.} 2023, The Astrophysical Journal
  Letters, 946, L31

\bibitem[{{Burrows} {et~al.}(2005){Burrows}, {Hill}, {Nousek}, {Kennea},
  {Wells}, {Osborne}, {Abbey}, {Beardmore}, {Mukerjee}, {Short}, {Chincarini},
  {Campana}, {Citterio}, {Moretti}, {Pagani}, {Tagliaferri}, {Giommi},
  {Capalbi}, {Tamburelli}, {Angelini}, {Cusumano}, {Br{\"a}uninger}, {Burkert},
  \& {Hartner}}]{Burrows2005SSRv}
{Burrows}, D.~N., {Hill}, J.~E., {Nousek}, J.~A., {et~al.} 2005, \ssr, 120, 165

\bibitem[{{Burrows} {et~al.}(2010){Burrows}, {Roming}, {Fox}, {Herter},
  {Falcone}, {Bil{\'e}n}, {Nousek}, \& {Kennea}}]{Burrows2010SPIE}
{Burrows}, D.~N., {Roming}, P.~W.~A., {Fox}, D.~B., {et~al.} 2010, in Society
  of Photo-Optical Instrumentation Engineers (SPIE) Conference Series, Vol.
  7732, Space Telescopes and Instrumentation 2010: Ultraviolet to Gamma Ray,
  ed. M.~{Arnaud}, S.~S. {Murray}, \& T.~{Takahashi}, 77321U

\bibitem[{{Butler} {et~al.}(2015{\natexlab{a}}){Butler}, {Watson}, {Kutyrev},
  {Lee}, {Richer}, {Fox}, {Prochaska}, {Bloom}, {Cucchiara}, {Troja},
  {Littlejohns}, {Ramirez-Ruiz}, {de Diego}, {Georgiev}, {Gonzalez},
  {Roman-Zuniga}, {Gehrels}, {Moseley}, {Capone}, {Golkhou}, \&
  {Toy}}]{Butler2015GCN18302}
{Butler}, N., {Watson}, A.~M., {Kutyrev}, A., {et~al.} 2015{\natexlab{a}}, GRB
  Coordinates Network, 18302

\bibitem[{{Butler} {et~al.}(2015{\natexlab{b}}){Butler}, {Watson}, {Kutyrev},
  {Lee}, {Richer}, {Fox}, {Prochaska}, {Bloom}, {Cucchiara}, {Troja},
  {Littlejohns}, {Ramirez-Ruiz}, {de Diego}, {Georgiev}, {Gonzalez},
  {Roman-Zuniga}, {Gehrels}, {Moseley}, {Capone}, {Golkhou}, \&
  {Toy}}]{Butler2015GCN18312}
{Butler}, N., {Watson}, A.~M., {Kutyrev}, A., {et~al.} 2015{\natexlab{b}}, GRB
  Coordinates Network, 18312

\bibitem[{{Campana} {et~al.}(2011){Campana}, {D'Avanzo}, {Lazzati}, {Covino},
  {Tagliaferri}, \& {Panagia}}]{Campana2011MNRAS}
{Campana}, S., {D'Avanzo}, P., {Lazzati}, D., {et~al.} 2011, \mnras, 418, 1511

\bibitem[{{Campana} {et~al.}(2022){Campana}, {Ghirlanda}, {Salvaterra},
  {Gonzalez}, {Landoni}, {Pariani}, {Riva}, {Riva}, {Smartt}, {Tanvir}, \&
  {Vergani}}]{Campana2022}
{Campana}, S., {Ghirlanda}, G., {Salvaterra}, R., {et~al.} 2022, Nature
  Astronomy, 6, 1101

\bibitem[{{Campana} {et~al.}(2021){Campana}, {Lazzati}, {Perna}, {Grazia
  Bernardini}, \& {Nava}}]{Campana2021AA}
{Campana}, S., {Lazzati}, D., {Perna}, R., {Grazia Bernardini}, M., \& {Nava},
  L. 2021, \aap, 649, A135

\bibitem[{{Campana} {et~al.}(2007){Campana}, {Lazzati}, {Ripamonti}, {Perna},
  {Covino}, {Tagliaferri}, {Moretti}, {Romano}, {Cusumano}, \&
  {Chincarini}}]{Campana2007ApJ}
{Campana}, S., {Lazzati}, D., {Ripamonti}, E., {et~al.} 2007, \apjl, 654, L17

\bibitem[{{Cano} {et~al.}(2011){Cano}, {Bersier}, {Guidorzi}, {Margutti},
  {Svensson}, {Kobayashi}, {Melandri}, {Wiersema}, {Pozanenko}, {van der
  Horst}, {Pooley}, {Fernandez-Soto}, {Castro-Tirado}, {Postigo}, {Im},
  {Kamble}, {Sahu}, {Alonso-Lorite}, {Anupama}, {Bibby}, {Burgdorf}, {Clay},
  {Curran}, {Fatkhullin}, {Fruchter}, {Garnavich}, {Gomboc}, {Gorosabel},
  {Graham}, {Gurugubelli}, {Haislip}, {Huang}, {Huxor}, {Ibrahimov}, {Jeon},
  {Jeon}, {Ivarsen}, {Kasen}, {Klunko}, {Kouveliotou}, {Lacluyze}, {Levan},
  {Loznikov}, {Mazzali}, {Moskvitin}, {Mottram}, {Mundell}, {Nugent},
  {Nysewander}, {O'Brien}, {Park}, {Peris}, {Pian}, {Reichart}, {Rhoads},
  {Rol}, {Rumyantsev}, {Scowcroft}, {Shakhovskoy}, {Small}, {Smith}, {Sokolov},
  {Starling}, {Steele}, {Strom}, {Tanvir}, {Tsapras}, {Urata}, {Vaduvescu},
  {Volnova}, {Volvach}, {Wijers}, {Woosley}, \& {Young}}]{Cano2011MNRAS}
{Cano}, Z., {Bersier}, D., {Guidorzi}, C., {et~al.} 2011, \mnras, 413, 669

\bibitem[{{Cano} {et~al.}(2014{\natexlab{a}}){Cano}, {de Ugarte Postigo},
  {Pozanenko}, {Butler}, {Th{\"o}ne}, {Guidorzi}, {Kr{\"u}hler}, {Gorosabel},
  {Jakobsson}, {Leloudas}, {Malesani}, {Hjorth}, {Melandri}, {Mundell},
  {Wiersema}, {D'Avanzo}, {Schulze}, {Gomboc}, {Johansson}, {Zheng}, {Kann},
  {Knust}, {Varela}, {Akerlof}, {Bloom}, {Burkhonov}, {Cooke}, {de Diego},
  {Dhungana}, {Farina}, {Ferrante}, {Flewelling}, {Fox}, {Fynbo}, {Gehrels},
  {Georgiev}, {Gonz{\'a}lez}, {Greiner}, {G{\"u}ver}, {Hartoog}, {Hatch},
  {Jelinek}, {Kehoe}, {Klose}, {Klunko}, {Kopa{\v c}}, {Kutyrev}, {Krugly},
  {Lee}, {Levan}, {Linkov}, {Matkin}, {Minikulov}, {Molotov}, {Prochaska},
  {Richer}, {Rom{\'a}n-Z{\'u}{\~n}iga}, {Rumyantsev},
  {S{\'a}nchez-Ram{\'{\i}}rez}, {Steele}, {Tanvir}, {Volnova}, {Watson}, {Xu},
  \& {Yuan}}]{Cano2014AA}
{Cano}, Z., {de Ugarte Postigo}, A., {Pozanenko}, A., {et~al.}
  2014{\natexlab{a}}, \aap, 568, A19

\bibitem[{{Cano} {et~al.}(2017{\natexlab{a}}){Cano}, {Izzo}, {de Ugarte
  Postigo}, {Th{\"o}ne}, {Kr{\"u}hler}, {Heintz}, {Malesani}, {Geier},
  {Fuentes}, {Chen}, {Covino}, {D'Elia}, {Fynbo}, {Goldoni}, {Gomboc},
  {Hjorth}, {Jakobsson}, {Kann}, {Milvang-Jensen}, {Pugliese},
  {S{\'a}nchez-Ram{\'{\i}}rez}, {Schulze}, {Sollerman}, {Tanvir}, \&
  {Wiersema}}]{Cano2017AA}
{Cano}, Z., {Izzo}, L., {de Ugarte Postigo}, A., {et~al.} 2017{\natexlab{a}},
  \aap, 605, A107

\bibitem[{{Cano} {et~al.}(2015){Cano}, {Malesani}, {de Ugarte Postigo}, {Xu},
  {Saario}, \& {Jakobsson}}]{Cano2015GCN18279}
{Cano}, Z., {Malesani}, D., {de Ugarte Postigo}, A., {et~al.} 2015, GRB
  Coordinates Network, 18279

\bibitem[{{Cano} {et~al.}(2014{\natexlab{b}}){Cano}, {Malesani}, {Geier},
  {Jensen}, {Taddia}, \& {Fremling}}]{Cano2014GCN16169}
{Cano}, Z., {Malesani}, D., {Geier}, S., {et~al.} 2014{\natexlab{b}}, GRB
  Coordinates Network, 16169

\bibitem[{{Cano} {et~al.}(2009){Cano}, {Melandri}, {Mundell}, {Gomboc},
  {Bersier}, {Clay}, {Kobayashi}, {Mottram}, {Smith}, {Steele}, \&
  {Guidorzi}}]{Cano2009GCN10262}
{Cano}, Z., {Melandri}, A., {Mundell}, C.~G., {et~al.} 2009, GRB Coordinates
  Network, 10262

\bibitem[{{Cano} {et~al.}(2017{\natexlab{b}}){Cano}, {Wang}, {Dai}, \&
  {Wu}}]{Cano2017AdAst}
{Cano}, Z., {Wang}, S.-Q., {Dai}, Z.-G., \& {Wu}, X.-F. 2017{\natexlab{b}},
  Advances in Astronomy, 2017, 8929054

\bibitem[{Cappellazzo {et~al.}(2022)Cappellazzo, Zafar, Corcho-Caballero, Kann,
  López-Sánchez, \& Ahmad}]{Capellazzo2022}
Cappellazzo, E., Zafar, T., Corcho-Caballero, P., {et~al.} 2022, Monthly
  Notices of the Royal Astronomical Society, 517, 6022

\bibitem[{{Casali} {et~al.}(2007){Casali}, {Adamson}, {Alves de Oliveira},
  {Almaini}, {Burch}, {Chuter}, {Elliot}, {Folger}, {Foucaud}, {Hambly},
  {Hastie}, {Henry}, {Hirst}, {Irwin}, {Ives}, {Lawrence}, {Laidlaw}, {Lee},
  {Lewis}, {Lunney}, {McLay}, {Montgomery}, {Pickup}, {Read}, {Rees}, {Robson},
  {Sekiguchi}, {Vick}, {Warren}, \& {Woodward}}]{CasaliM2007}
{Casali}, M., {Adamson}, A., {Alves de Oliveira}, C., {et~al.} 2007, \aap, 467,
  777

\bibitem[{{Castro-Tirado} {et~al.}(2013){Castro-Tirado},
  {S{\'a}nchez-Ram{\'\i}rez}, {Ellison}, {Jel{\'\i}nek},
  {Mart{\'\i}n-Carrillo}, {Bromm}, {Gorosabel}, {Bremer}, {Winters}, {Hanlon},
  {Meegan}, {Topinka}, {Pandey}, {Guziy}, {Jeong}, {Sonbas}, {Pozanenko},
  {Cunniffe}, {Fern{\'a}ndez-Mu{\~n}oz}, {Ferrero}, {Gehrels}, {Hudec},
  {Kub{\'a}nek}, {Lara-Gil}, {Mu{\~n}oz-Mart{\'\i}nez},
  {P{\'e}rez-Ram{\'\i}rez}, {{\v{S}}trobl}, {{\'A}lvarez-Iglesias},
  {Inasaridze}, {Rumyantsev}, {Volnova}, {Hellmich}, {Mottola}, {Castro
  Cer{\'o}n}, {Cepa}, {G{\"o}{\u{g}}{\"u}{\c{s}}}, {G{\"u}ver}, {{\"O}nal
  Ta{\c{s}}}, {Park}, {Sabau-Graziati}, \& {Tejero}}]{Castro-Tirado2013arXiv}
{Castro-Tirado}, A.~J., {S{\'a}nchez-Ram{\'\i}rez}, R., {Ellison}, S.~L.,
  {et~al.} 2013, arXiv e-prints, arXiv:1312.5631

\bibitem[{{Cenarro, A. J.} {et~al.}(2019){Cenarro, A. J.}, {Moles, M.},
  {Cristóbal-Hornillos, D.}, {Marín-Franch, A.}, {Ederoclite, A.}, {Varela,
  J.}, {López-Sanjuan, C.}, {Hernández-Monteagudo, C.}, {Angulo, R. E.},
  {Vázquez Ramió, H.}, {Viironen, K.}, {Bonoli, S.}, {Orsi, A. A.}, {Hurier,
  G.}, {San Roman, I.}, {Greisel, N.}, {Vilella-Rojo, G.}, {Díaz-García, L.
  A.}, {Logroño-García, R.}, {Gurung-López, S.}, {Spinoso, D.},
  {Izquierdo-Villalba, D.}, {Aguerri, J. A. L.}, {Allende Prieto, C.},
  {Bonatto, C.}, {Carvano, J. M.}, {Chies-Santos, A. L.}, {Daflon, S.}, {Dupke,
  R. A.}, {Falcón-Barroso, J.}, {Gonçalves, D. R.}, {Jiménez-Teja, Y.},
  {Molino, A.}, {Placco, V. M.}, {Solano, E.}, {Whitten, D. D.}, {Abril, J.},
  {Antón, J. L.}, {Bello, R.}, {Bielsa de Toledo, S.}, {Castillo-Ramírez,
  J.}, {Chueca, S.}, {Civera, T.}, {Díaz-Martín, M. C.},
  {Domínguez-Martínez, M.}, {Garzarán-Calderaro, J.}, {Hernández-Fuertes,
  J.}, {Iglesias-Marzoa, R.}, {Iñiguez, C.}, {Jiménez Ruiz, J. M.}, {Kruuse,
  K.}, {Lamadrid, J. L.}, {Lasso-Cabrera, N.}, {López-Alegre, G.},
  {López-Sainz, A.}, {Maícas, N.}, {Moreno-Signes, A.}, {Muniesa, D. J.},
  {Rodríguez-Llano, S.}, {Rueda-Teruel, F.}, {Rueda-Teruel, S.},
  {Soriano-Laguía, I.}, {Tilve, V.}, {Valdivielso, L.}, {Yanes-Díaz, A.},
  {Alcaniz, J. S.}, {Mendes de Oliveira, C.}, {Sodré, L.}, {Coelho, P.},
  {Lopes de Oliveira, R.}, {Tamm, A.}, {Xavier, H. S.}, {Abramo, L. R.},
  {Akras, S.}, {Alfaro, E. J.}, {Alvarez-Candal, A.}, {Ascaso, B.}, {Beasley,
  M. A.}, {Beers, T. C.}, {Borges Fernandes, M.}, {Bruzual, G. R.}, {Buzzo, M.
  L.}, {Carrasco, J. M.}, {Cepa, J.}, {Cortesi, A.}, {Costa-Duarte, M. V.}, {De
  Prá, M.}, {Favole, G.}, {Galarza, A.}, {Galbany, L.}, {Garcia, K.},
  {González Delgado, R. M.}, {González-Serrano, J. I.}, {Gutiérrez-Soto, L.
  A.}, {Hernandez-Jimenez, J. A.}, {Kanaan, A.}, {Kuncarayakti, H.}, {Landim,
  R. C. G.}, {Laur, J.}, {Licandro, J.}, {Lima Neto, G. B.}, {Lyman, J. D.},
  {Maíz Apellániz, J.}, {Miralda-Escudé, J.}, {Morate, D.},
  {Nogueira-Cavalcante, J. P.}, {Novais, P. M.}, {Oncins, M.}, {Oteo, I.},
  {Overzier, R. A.}, {Pereira, C. B.}, {Rebassa-Mansergas, A.}, {Reis, R. R.
  R.}, {Roig, F.}, {Sako, M.}, {Salvador-Rusiñol, N.}, {Sampedro, L.},
  {Sánchez-Blázquez, P.}, {Santos, W. A.}, {Schmidtobreick, L.}, {Siffert, B.
  B.}, {Telles, E.}, \& {Vilchez, J. M.}}]{Cenarro2019}
{Cenarro, A. J.}, {Moles, M.}, {Cristóbal-Hornillos, D.}, {et~al.} 2019, A\&A,
  622, A176

\bibitem[{{Cenko}(2011)}]{Cenko2011GCN12771}
{Cenko}, S.~B. 2011, GRB Coordinates Network, 12771

\bibitem[{{Cenko} {et~al.}(2006){Cenko}, {Berger}, {Djorgovski}, {Mahabal}, \&
  {Fox}}]{Cenko2006GCN5155}
{Cenko}, S.~B., {Berger}, E., {Djorgovski}, S.~G., {Mahabal}, A.~A., \& {Fox},
  D.~B. 2006, GRB Coordinates Network, 5155

\bibitem[{{Cenko} {et~al.}(2010){Cenko}, {Bloom}, {Perley}, \&
  {Cobb}}]{Cenko2010GCN10443}
{Cenko}, S.~B., {Bloom}, J.~S., {Perley}, D.~A., \& {Cobb}, B.~E. 2010, GRB
  Coordinates Network, 10443

\bibitem[{{Cenko} \& {Perley}(2014{\natexlab{a}})}]{Cenko2014GCN16129}
{Cenko}, S.~B. \& {Perley}, D.~A. 2014{\natexlab{a}}, GRB Coordinates Network,
  16129

\bibitem[{{Cenko} \& {Perley}(2014{\natexlab{b}})}]{Cenko2014GCN16153}
{Cenko}, S.~B. \& {Perley}, D.~A. 2014{\natexlab{b}}, GRB Coordinates Network,
  16153

\bibitem[{{Cenko} {et~al.}(2009){Cenko}, {Perley}, {Junkkarinen}, {Burbidge},
  {Diego}, \& {Miller}}]{Cenko2009GCN9518}
{Cenko}, S.~B., {Perley}, D.~A., {Junkkarinen}, V., {et~al.} 2009, GRB
  Coordinates Network, 9518

\bibitem[{{Chambers} {et~al.}(2016){Chambers}, {Magnier}, {Metcalfe},
  {Flewelling}, {Huber}, {Waters}, {Denneau}, {Draper}, {Farrow}, {Finkbeiner},
  {Holmberg}, {Koppenhoefer}, {Price}, {Saglia}, {Schlafly}, {Smartt},
  {Sweeney}, {Wainscoat}, {Burgett}, {Grav}, {Heasley}, {Hodapp}, {Jedicke},
  {Kaiser}, {Kudritzki}, {Luppino}, {Lupton}, {Monet}, {Morgan}, {Onaka},
  {Stubbs}, {Tonry}, {Banados}, {Bell}, {Bender}, {Bernard}, {Botticella},
  {Casertano}, {Chastel}, {Chen}, {Chen}, {Cole}, {Deacon}, {Frenk},
  {Fitzsimmons}, {Gezari}, {Goessl}, {Goggia}, {Goldman}, {Grebel}, {Hambly},
  {Hasinger}, {Heavens}, {Heckman}, {Henderson}, {Henning}, {Holman}, {Hopp},
  {Ip}, {Isani}, {Keyes}, {Koekemoer}, {Kotak}, {Long}, {Lucey}, {Liu},
  {Martin}, {McLean}, {Morganson}, {Murphy}, {Nieto-Santisteban}, {Norberg},
  {Peacock}, {Pier}, {Postman}, {Primak}, {Rae}, {Rest}, {Riess}, {Riffeser},
  {Rix}, {Roser}, {Schilbach}, {Schultz}, {Scolnic}, {Szalay}, {Seitz},
  {Shiao}, {Small}, {Smith}, {Soderblom}, {Taylor}, {Thakar}, {Thiel},
  {Thilker}, {Urata}, {Valenti}, {Walter}, {Watters}, {Werner}, {White},
  {Wood-Vasey}, \& {Wyse}}]{Chambers2016a}
{Chambers}, K.~C., {Magnier}, E.~A., {Metcalfe}, N., {et~al.} 2016,
  arXiv:1612.05560

\bibitem[{{Chen} {et~al.}(2021){Chen}, {Peng}, {Du}, {Yin}, \&
  {Wu}}]{Chen2021ApJ}
{Chen}, J.-M., {Peng}, Z.-Y., {Du}, T.-T., {Yin}, Y., \& {Wu}, H. 2021, \apj,
  920, 53

\bibitem[{{Chen} {et~al.}(2008){Chen}, {Huang}, {Chen}, {Huang}, {Urata}, \&
  {Marshall}}]{Chen2008GCN7990}
{Chen}, T.~W., {Huang}, L.~C., {Chen}, Y.~T., {et~al.} 2008, GRB Coordinates
  Network, 7990

\bibitem[{{Chester} \& {Beardmore}(2012)}]{Chester2012GCN12880}
{Chester}, M.~M. \& {Beardmore}, A.~P. 2012, GRB Coordinates Network, 12880

\bibitem[{{Chester} \& {Lien}(2015)}]{Chester2015GCN17531}
{Chester}, M.~M. \& {Lien}, A.~Y. 2015, GRB Coordinates Network, 17531

\bibitem[{{Chester} {et~al.}(2014){Chester}, {Sonbas}, {Page}, {Malesani}, \&
  {Oates}}]{Chester2014GCN16147}
{Chester}, M.~M., {Sonbas}, E., {Page}, K.~L., {Malesani}, D., \& {Oates},
  S.~R. 2014, GRB Coordinates Network, 16147

\bibitem[{{Choi} {et~al.}(2014){Choi}, {Im}, {Sung}, \&
  {Urata}}]{Choi2014GCN16149}
{Choi}, C., {Im}, M., {Sung}, H.~I., \& {Urata}, Y. 2014, GRB Coordinates
  Network, 16149

\bibitem[{{Chornock} {et~al.}(2014){Chornock}, {Berger}, {Fox}, {Fong},
  {Laskar}, \& {Roth}}]{Chornock2014arXiv}
{Chornock}, R., {Berger}, E., {Fox}, D.~B., {et~al.} 2014, arXiv e-prints,
  arXiv:1405.7400

\bibitem[{{Chornock} {et~al.}(2013){Chornock}, {Berger}, {Fox}, {Lunnan},
  {Drout}, {Fong}, {Laskar}, \& {Roth}}]{Chornock2013ApJ}
{Chornock}, R., {Berger}, E., {Fox}, D.~B., {et~al.} 2013, \apj, 774, 26

\bibitem[{{Chrimes} {et~al.}(2019){Chrimes}, {Levan}, {Stanway}, {Berger},
  {Bloom}, {Cenko}, {Cobb}, {Cucchiara}, {Fruchter}, {Gompertz}, {Hjorth},
  {Jakobsson}, {Lyman}, {O'Brien}, {Perley}, {Tanvir}, {Wheatley}, \&
  {Wiersema}}]{Chrimes2019MNRAS}
{Chrimes}, A.~A., {Levan}, A.~J., {Stanway}, E.~R., {et~al.} 2019, \mnras, 488,
  902

\bibitem[{{Christie} {et~al.}(2009){Christie}, {de Ugarte Postigo}, \&
  {Natusch}}]{Christie2009GCN9396}
{Christie}, G.~W., {de Ugarte Postigo}, A., \& {Natusch}, T. 2009, GRB
  Coordinates Network, 9396

\bibitem[{{Clemens} {et~al.}(2008){Clemens}, {Kupcu Yoldas}, {Greiner},
  {Kruehler}, {Yoldas}, \& {Szokoly}}]{Clemens2008GCN7851}
{Clemens}, C., {Kupcu Yoldas}, A., {Greiner}, J., {et~al.} 2008, GRB
  Coordinates Network, 7851

\bibitem[{{Cobb}(2009{\natexlab{a}})}]{Cobb2009GCN10110}
{Cobb}, B.~E. 2009{\natexlab{a}}, GRB Coordinates Network, 10110

\bibitem[{{Cobb}(2009{\natexlab{b}})}]{Cobb2009GCN10111}
{Cobb}, B.~E. 2009{\natexlab{b}}, GRB Coordinates Network, 10111

\bibitem[{{Cobb}(2012)}]{Cobb2012GCN13928}
{Cobb}, B.~E. 2012, GRB Coordinates Network, 13928

\bibitem[{{Cobb} {et~al.}(2010){Cobb}, {Bloom}, {Perley}, {Morgan}, {Cenko}, \&
  {Filippenko}}]{Cobb2010ApJ}
{Cobb}, B.~E., {Bloom}, J.~S., {Perley}, D.~A., {et~al.} 2010, \apjl, 718, L150

\bibitem[{{Covino}(2014)}]{Covino2014GCN17208}
{Covino}, S. 2014, GRB Coordinates Network, 17208

\bibitem[{{Covino} \& {Fugazza}(2018)}]{Covino2018GCN23021}
{Covino}, S. \& {Fugazza}, D. 2018, GRB Coordinates Network, 23021

\bibitem[{{Coward} {et~al.}(2017){Coward}, {Gendre}, {Tanga}, {Turpin},
  {Zadko}, {Dodson}, {Devog{\'e}le}, {Howell}, {Kennewell}, {Bo{\"e}r},
  {Klotz}, {Dornic}, {Moore}, \& {Heary}}]{Coward2017PASA}
{Coward}, D.~M., {Gendre}, B., {Tanga}, P., {et~al.} 2017, \pasa, 34, e005

\bibitem[{{Coward} {et~al.}(2013){Coward}, {Howell}, {Branchesi}, {Stratta},
  {Guetta}, {Gendre}, \& {Macpherson}}]{Coward2013MNRAS}
{Coward}, D.~M., {Howell}, E.~J., {Branchesi}, M., {et~al.} 2013, \mnras, 432,
  2141

\bibitem[{{Coward} {et~al.}(2010){Coward}, {Todd}, {Vaalsta}, {Laas-Bourez},
  {Klotz}, {Imerito}, {Yan}, {Luckas}, {Fletcher}, {Zadnik}, {Burman}, {Blair},
  {Zadko}, {Bo{\"e}r}, {Thierry}, {Howell}, {Gordon}, {Ahmat}, {Moore}, \&
  {Frost}}]{Coward2010PASA}
{Coward}, D.~M., {Todd}, M., {Vaalsta}, T.~P., {et~al.} 2010, \pasa, 27, 331

\bibitem[{{Crouzet} \& {Malesani}(2018)}]{Crouzet2018GCN22988}
{Crouzet}, N. \& {Malesani}, D.~B. 2018, GRB Coordinates Network, 22988

\bibitem[{{Cucchiara} {et~al.}(2011{\natexlab{a}}){Cucchiara}, {Bloom}, \&
  {Cenko}}]{Cucchiara2011GCN12202}
{Cucchiara}, A., {Bloom}, J.~S., \& {Cenko}, S.~B. 2011{\natexlab{a}}, GRB
  Coordinates Network, 12202

\bibitem[{{Cucchiara} {et~al.}(2011{\natexlab{b}}){Cucchiara}, {Cenko},
  {Bloom}, {Melandri}, {Morgan}, {Kobayashi}, {Smith}, {Perley}, {Li}, {Hora},
  {da Silva}, {Prochaska}, {Milne}, {Butler}, {Cobb}, {Worseck}, {Mundell},
  {Steele}, {Filippenko}, {Fumagalli}, {Klein}, {Stephens}, {Bluck}, \&
  {Mason}}]{Cucchiara2011ApJ2}
{Cucchiara}, A., {Cenko}, S.~B., {Bloom}, J.~S., {et~al.} 2011{\natexlab{b}},
  \apj, 743, 154

\bibitem[{{Cucchiara} {et~al.}(2011{\natexlab{c}}){Cucchiara}, {Levan}, {Fox},
  {Tanvir}, {Ukwatta}, {Berger}, {Kr{\"u}hler}, {K{\"u}pc{\"u} Yolda{\c s}},
  {Wu}, {Toma}, {Greiner}, {Olivares}, {Rowlinson}, {Amati}, {Sakamoto},
  {Roth}, {Stephens}, {Fritz}, {Fynbo}, {Hjorth}, {Malesani}, {Jakobsson},
  {Wiersema}, {O'Brien}, {Soderberg}, {Foley}, {Fruchter}, {Rhoads},
  {Rutledge}, {Schmidt}, {Dopita}, {Podsiadlowski}, {Willingale}, {Wolf},
  {Kulkarni}, \& {D'Avanzo}}]{Cucchiara2011ApJ}
{Cucchiara}, A., {Levan}, A.~J., {Fox}, D.~B., {et~al.} 2011{\natexlab{c}},
  \apj, 736, 7

\bibitem[{{Cucchiara} \& {Perley}(2013)}]{Cucchiara2013GCN15144}
{Cucchiara}, A. \& {Perley}, D. 2013, GRB Coordinates Network, 15144

\bibitem[{{Cucchiara} {et~al.}(2015){Cucchiara}, {Veres}, {Corsi}, {Cenko},
  {Perley}, {Lien}, {Marshall}, {Pagani}, {Toy}, {Capone}, {Frail}, {Horesh},
  {Modjaz}, {Butler}, {Littlejohns}, {Watson}, {Kutyrev}, {Lee}, {Richer},
  {Klein}, {Fox}, {Prochaska}, {Bloom}, {Troja}, {Ramirez-Ruiz}, {de Diego},
  {Georgiev}, {Gonz{\'a}lez}, {Rom{\'a}n-Z{\'u}{\~n}iga}, {Gehrels}, \&
  {Moseley}}]{Cucchiara2015ApJ}
{Cucchiara}, A., {Veres}, P., {Corsi}, A., {et~al.} 2015, \apj, 812, 122

\bibitem[{{Cusumano} {et~al.}(2006){Cusumano}, {Mangano}, {Chincarini},
  {Panaitescu}, {Burrows}, {La Parola}, {Sakamoto}, {Campana}, {Mineo},
  {Tagliaferri}, {Angelini}, {Barthelemy}, {Beardmore}, {Boyd}, {Cominsky},
  {Gronwall}, {Fenimore}, {Gehrels}, {Giommi}, {Goad}, {Hurley}, {Kennea},
  {Mason}, {Marshall}, {M{\'e}sz{\'a}ros}, {Nousek}, {Osborne}, {Palmer},
  {Roming}, {Wells}, {White}, \& {Zhang}}]{Cusumano2006Nature}
{Cusumano}, G., {Mangano}, V., {Chincarini}, G., {et~al.} 2006, \nat, 440, 164

\bibitem[{{Cusumano} {et~al.}(2007){Cusumano}, {Mangano}, {Chincarini},
  {Panaitescu}, {Burrows}, {La Parola}, {Sakamoto}, {Campana}, {Mineo},
  {Tagliaferri}, {Angelini}, {Barthelmy}, {Beardmore}, {Boyd}, {Cominsky},
  {Gronwall}, {Fenimore}, {Gehrels}, {Giommi}, {Goad}, {Hurley}, {Immler},
  {Kennea}, {Mason}, {Marshal}, {M{\'e}sz{\'a}ros}, {Nousek}, {Osborne},
  {Palmer}, {Roming}, {Wells}, {White}, \& {Zhang}}]{Cusumano2007AA}
{Cusumano}, G., {Mangano}, V., {Chincarini}, G., {et~al.} 2007, \aap, 462, 73

\bibitem[{{D'Avanzo} {et~al.}(2014{\natexlab{a}}){D'Avanzo}, {Covino},
  {Malesani}, {Rossi}, \& {Tagliaferri}}]{D'Avanzo2014GCN16166}
{D'Avanzo}, P., {Covino}, S., {Malesani}, D., {Rossi}, A., \& {Tagliaferri}, G.
  2014{\natexlab{a}}, GRB Coordinates Network, 16166

\bibitem[{{D'Avanzo} {et~al.}(2006){D'Avanzo}, {Malesani}, \&
  {Antonelli}}]{D'Avanzo2006GCN4532}
{D'Avanzo}, P., {Malesani}, D., \& {Antonelli}, L.~A. 2006, GRB Coordinates
  Network, 4532

\bibitem[{{D'Avanzo} {et~al.}(2014{\natexlab{b}}){D'Avanzo}, {Malesani},
  {D'Elia}, {Antonelli}, {Tagliaferri}, {Vergani}, {Fiorenzano}, \&
  {Mainella}}]{D'Avanzo2014GCN16493}
{D'Avanzo}, P., {Malesani}, D., {D'Elia}, V., {et~al.} 2014{\natexlab{b}}, GRB
  Coordinates Network, 16493

\bibitem[{{D'Avanzo} {et~al.}(2018){D'Avanzo}, {Melandri}, {Covino}, \&
  {Fugazza}}]{DAvanzo2018GCN22536}
{D'Avanzo}, P., {Melandri}, A., {Covino}, S., \& {Fugazza}, D. 2018, GRB
  Coordinates Network, 22536

\bibitem[{{D'Avanzo} {et~al.}(2014{\natexlab{c}}){D'Avanzo}, {Melandri},
  {Malesani}, {Pursimo}, {Baglio}, \& {Andreoni}}]{D'Avanzo2014GCN15953}
{D'Avanzo}, P., {Melandri}, A., {Malesani}, D., {et~al.} 2014{\natexlab{c}},
  GRB Coordinates Network, 15953

\bibitem[{{D'Avanzo} {et~al.}(2014{\natexlab{d}}){D'Avanzo}, {Salvaterra},
  {Bernardini}, {Nava}, {Campana}, {Covino}, {D'Elia}, {Ghirlanda},
  {Ghisellini}, {Melandri}, {Sbarufatti}, {Vergani}, \&
  {Tagliaferri}}]{D'Avanzo2014MNRAS}
{D'Avanzo}, P., {Salvaterra}, R., {Bernardini}, M.~G., {et~al.}
  2014{\natexlab{d}}, \mnras, 442, 2342

\bibitem[{{De Cia} {et~al.}(2010){De Cia}, {Vreeswijk}, \&
  {Jakobsson}}]{DeCia2010GCN11170}
{De Cia}, A., {Vreeswijk}, P.~M., \& {Jakobsson}, P. 2010, GRB Coordinates
  Network, 11170

\bibitem[{{De Pasquale} {et~al.}(2015){De Pasquale}, {Kuin}, {Oates},
  {Schulze}, {Cano}, {Guidorzi}, {Beardmore}, {Evans}, {Uhm}, {Zhang}, {Page},
  {Kobayashi}, {Castro-Tirado}, {Gorosabel}, {Sakamoto}, {Fatkhullin},
  {Pandey}, {Im}, {Chandra}, {Frail}, {Gao}, {Kopa{\v{c}}}, {Jeon}, {Akerlof},
  {Huang}, {Pak}, {Park}, {Gomboc}, {Melandri}, {Zane}, {Mundell}, {Saxton},
  {Holland}, {Virgili}, {Urata}, {Steele}, {Bersier}, {Tanvir}, {Sokolov}, \&
  {Moskvitin}}]{DePasquale2015MNRAS}
{De Pasquale}, M., {Kuin}, N.~P.~M., {Oates}, S., {et~al.} 2015, \mnras, 449,
  1024

\bibitem[{{De Pasquale} {et~al.}(2016{\natexlab{a}}){De Pasquale}, {Oates},
  {Racusin}, {Kann}, {Zhang}, {Pozanenko}, {Volnova}, {Trotter}, {Frank},
  {Cucchiara}, {Troja}, {Sbarufatti}, {Butler}, {Schulze}, {Cano}, {Page},
  {Castro-Tirado}, {Gorosabel}, {Lien}, {Fox}, {Littlejohns}, {Bloom},
  {Prochaska}, {de Diego}, {Gonzalez}, {Richer}, {Rom{\'a}n-Z{\'u}{\~n}iga},
  {Watson}, {Gehrels}, {Moseley}, {Kutyrev}, {Zane}, {Hoette}, {Russell},
  {Rumyantsev}, {Klunko}, {Burkhonov}, {Breeveld}, {Reichart}, \&
  {Haislip}}]{DePasquale2016MNRAS}
{De Pasquale}, M., {Oates}, S.~R., {Racusin}, J.~L., {et~al.}
  2016{\natexlab{a}}, \mnras, 455, 1027

\bibitem[{{De Pasquale} \& {Pagani}(2009)}]{dePasquale2009GCN10267}
{De Pasquale}, M. \& {Pagani}, C. 2009, GRB Coordinates Network, 10267

\bibitem[{{De Pasquale} {et~al.}(2016{\natexlab{b}}){De Pasquale}, {Page},
  {Kann}, {Oates}, {Schulze}, {Zhang}, {Cano}, {Gendre}, {Malesani}, {Rossi},
  {Troja}, {Piro}, {Bo{\"e}r}, {Stratta}, \& {Gehrels}}]{DePasquale2016MNRAS2}
{De Pasquale}, M., {Page}, M.~J., {Kann}, D.~A., {et~al.} 2016{\natexlab{b}},
  \mnras, 462, 1111

\bibitem[{{De Pasquale} \& {Parsons}(2008)}]{DePasquale2008GCN8603}
{De Pasquale}, M. \& {Parsons}, A. 2008, GRB Coordinates Network, 8603

\bibitem[{{de Ugarte Postigo} {et~al.}(2011){de Ugarte Postigo},
  {Castro-Tirado}, \& {Gorosabel}}]{deUgartePostigo2011GCN11978}
{de Ugarte Postigo}, A., {Castro-Tirado}, A.~J., \& {Gorosabel}, J. 2011, GRB
  Coordinates Network, 11978

\bibitem[{{de Ugarte Postigo} {et~al.}(2009{\natexlab{a}}){de Ugarte Postigo},
  {Castro-Tirado}, {Gorosabel}, {Jelinek}, {Kubanek}, {Uv}, {Cunniffe},
  {Guziy}, {Yock}, {Allen}, {Bond}, \&
  {Christie}}]{deUgartePostigo2009GCN10255}
{de Ugarte Postigo}, A., {Castro-Tirado}, A.~J., {Gorosabel}, J., {et~al.}
  2009{\natexlab{a}}, GRB Coordinates Network, 10255

\bibitem[{{de Ugarte Postigo} {et~al.}(2005){de Ugarte Postigo},
  {Castro-Tirado}, {Gorosabel}, {J{\'o}hannesson}, {Bj{\"o}rnsson},
  {Gudmundsson}, {Bremer}, {Pak}, {Tanvir}, {Castro Cer{\'o}n}, {Guzyi},
  {Jel{\'{\i}}nek}, {Klose}, {P{\'e}rez-Ram{\'{\i}}rez}, {Aceituno}, {Campo
  Bagat{\'{\i}}n}, {Covino}, {Cardiel}, {Fathkullin}, {Henden}, {Huferath},
  {Kurata}, {Malesani}, {Mannucci}, {Ruiz-Lapuente}, {Sokolov}, {Thiele},
  {Wisotzki}, {Antonelli}, {Bartolini}, {Boattini}, {Guarnieri}, {Piccioni},
  {Pizzichini}, {del Principe}, {di Paola}, {Fugazza}, {Ghisellini}, {Hunt},
  {Konstantinova}, {Masetti}, {Palazzi}, {Pian}, {Stefanon}, {Testa}, \&
  {Tristram}}]{deUgarte2005AA}
{de Ugarte Postigo}, A., {Castro-Tirado}, A.~J., {Gorosabel}, J., {et~al.}
  2005, \aap, 443, 841

\bibitem[{{de Ugarte Postigo} {et~al.}(2012){de Ugarte Postigo}, {Fynbo},
  {Th{\"o}ne}, {Christensen}, {Gorosabel}, {Milvang-Jensen}, {Schulze},
  {Jakobsson}, {Wiersema}, {S{\'a}nchez-Ram{\'\i}rez}, {Leloudas}, {Zafar},
  {Malesani}, \& {Hjorth}}]{deUgartePostigo2012AA}
{de Ugarte Postigo}, A., {Fynbo}, J.~P.~U., {Th{\"o}ne}, C.~C., {et~al.} 2012,
  \aap, 548, A11

\bibitem[{{de Ugarte Postigo} {et~al.}(2009{\natexlab{b}}){de Ugarte Postigo},
  {Gorosabel}, {Malesani}, {Fynbo}, \& {Levan}}]{deUgartePostigo2009GCN9381}
{de Ugarte Postigo}, A., {Gorosabel}, J., {Malesani}, D., {Fynbo}, J.~P.~U., \&
  {Levan}, A.~J. 2009{\natexlab{b}}, GRB Coordinates Network, 9381

\bibitem[{{de Ugarte Postigo} {et~al.}(2020{\natexlab{a}}){de Ugarte Postigo},
  {Kann}, {Blazek}, {Agui Fernandez}, {Thoene}, {Gomez Velarde}, \& {Perez
  Romero}}]{deUgartePostigo2020GCN28650}
{de Ugarte Postigo}, A., {Kann}, D.~A., {Blazek}, M., {et~al.}
  2020{\natexlab{a}}, GRB Coordinates Network, 28650

\bibitem[{{de Ugarte Postigo} {et~al.}(2019){de Ugarte Postigo}, {Kann},
  {Thoene}, \& {Izzo}}]{deUgartePostigo2019GCN23692}
{de Ugarte Postigo}, A., {Kann}, D.~A., {Thoene}, C.~C., \& {Izzo}, L. 2019,
  GRB Coordinates Network, 23692

\bibitem[{{de Ugarte Postigo} {et~al.}(2015{\natexlab{a}}){de Ugarte Postigo},
  {Kruehler}, {Flores}, \& {Fynbo}}]{deUgartePostigo2015GCN17523}
{de Ugarte Postigo}, A., {Kruehler}, T., {Flores}, H., \& {Fynbo}, J.~P.~U.
  2015{\natexlab{a}}, GRB Coordinates Network, 17523

\bibitem[{{de Ugarte Postigo} {et~al.}(2009{\natexlab{c}}){de Ugarte Postigo},
  {Kubanek}, {Jelinek}, {Castro-Tirado}, {Gorosabel}, {Cunniffe}, {Guziy},
  {Allen}, \& {Yock}}]{deUgartePostigo2009GCN10043}
{de Ugarte Postigo}, A., {Kubanek}, P., {Jelinek}, M., {et~al.}
  2009{\natexlab{c}}, GRB Coordinates Network, 1043

\bibitem[{{de Ugarte Postigo} {et~al.}(2016){de Ugarte Postigo}, {Tanvir},
  {Cano}, {Izzo}, {Fynbo}, {Sanchez-Ramirez}, {Thoene}, \&
  {Pesev}}]{deUgartePostigo2016GCN19245}
{de Ugarte Postigo}, A., {Tanvir}, N.~R., {Cano}, Z., {et~al.} 2016, GRB
  Coordinates Network, 19245

\bibitem[{{de Ugarte Postigo} {et~al.}(2015{\natexlab{b}}){de Ugarte Postigo},
  {Thoene}, {Lombardi}, \& {Perez}}]{deUgartePostigo2015GCN18274}
{de Ugarte Postigo}, A., {Thoene}, C., {Lombardi}, G., \& {Perez}, A.
  2015{\natexlab{b}}, GRB Coordinates Network, 18274

\bibitem[{{de Ugarte Postigo} {et~al.}(2018{\natexlab{a}}){de Ugarte Postigo},
  {Th{\"o}ne}, {Bensch}, {van der Horst}, {Kann}, {Cano}, {Izzo}, {Goldoni},
  {Mart{\'\i}n}, {Filgas}, {Schady}, {Gorosabel}, {Bikmaev}, {Bremer},
  {Burenin}, {Castro-Tirado}, {Covino}, {Fynbo}, {Garcia-Appadoo}, {de
  Gregorio-Monsalvo}, {Jel{\'\i}nek}, {Khamitov}, {Kamble}, {Kouveliotou},
  {Kr{\"u}hler}, {Leloudas}, {Melnikov}, {Nardini}, {Perley}, {Petitpas},
  {Pooley}, {Rau}, {Rol}, {S{\'a}nchez-Ram{\'\i}rez}, {Starling}, {Tanvir},
  {Wiersema}, {Wijers}, \& {Zafar}}]{deUgartePostigo2018AA2}
{de Ugarte Postigo}, A., {Th{\"o}ne}, C.~C., {Bensch}, K., {et~al.}
  2018{\natexlab{a}}, \aap, 620, A190

\bibitem[{{de Ugarte Postigo} {et~al.}(2018{\natexlab{b}}){de Ugarte Postigo},
  {Th{\"o}ne}, {Bolmer}, {Schulze}, {Mart{\'\i}n}, {Kann}, {D'Elia}, {Selsing},
  {Martin-Carrillo}, {Perley}, {Kim}, {Izzo}, {S{\'a}nchez-Ram{\'\i}rez},
  {Guidorzi}, {Klotz}, {Wiersema}, {Bauer}, {Bensch}, {Campana}, {Cano},
  {Covino}, {Coward}, {De Cia}, {de Gregorio-Monsalvo}, {De Pasquale}, {Fynbo},
  {Greiner}, {Gomboc}, {Hanlon}, {Hansen}, {Hartmann}, {Heintz}, {Jakobsson},
  {Kobayashi}, {Malesani}, {Martone}, {Meintjes}, {Micha{\l}owski}, {Mundell},
  {Murphy}, {Oates}, {Salmon}, {van Soelen}, {Tanvir}, {Turpin}, {Xu}, \&
  {Zafar}}]{deUgartePostigo2018AA}
{de Ugarte Postigo}, A., {Th{\"o}ne}, C.~C., {Bolmer}, J., {et~al.}
  2018{\natexlab{b}}, \aap, 620, A119

\bibitem[{{de Ugarte Postigo} {et~al.}(2020{\natexlab{b}}){de Ugarte Postigo},
  {Th{\"o}ne}, {Mart{\'\i}n}, {Japelj}, {Levan}, {Micha{\l}owski}, {Selsing},
  {Kann}, {Schulze}, {Palmerio}, {Vergani}, {Tanvir}, {Bensch}, {Covino},
  {D'Elia}, {De Pasquale}, {Fruchter}, {Fynbo}, {Hartmann}, {Heintz}, {van der
  Horst}, {Izzo}, {Jakobsson}, {Ng}, {Perley}, {Rossi}, {Sbarufatti},
  {Salvaterra}, {S{\'a}nchez-Ram{\'\i}rez}, {Watson}, \&
  {Xu}}]{deUgartePostigo2020AA}
{de Ugarte Postigo}, A., {Th{\"o}ne}, C.~C., {Mart{\'\i}n}, S., {et~al.}
  2020{\natexlab{b}}, \aap, 633, A68

\bibitem[{{de Ugarte Postigo} {et~al.}(2013){de Ugarte Postigo}, {Xu},
  {Leloudas}, {Kr\"uhler}, {Malesani}, {Gorosabel}, {Thoene},
  {Sanchez-Ramirez}, {Schulze}, {Fynbo}, {Hjorth}, {Jakobsson}, \&
  {Cabrera-Lavers}}]{deUgartePostigo2013GCN14646}
{de Ugarte Postigo}, A., {Xu}, D., {Leloudas}, G., {et~al.} 2013, GCN
  Circulars, 14646

\bibitem[{{D'Elia} {et~al.}(2010){D'Elia}, {Fiore}, {Goldoni}, {D'Odorico},
  {Campana}, {Covino}, {D'Avanzo}, {Meurs}, {Norci}, \&
  {Tagliaferri}}]{D'Elia2010MNRAS}
{D'Elia}, V., {Fiore}, F., {Goldoni}, P., {et~al.} 2010, \mnras, 401, 385

\bibitem[{{D'Elia} {et~al.}(2009){D'Elia}, {Fiore}, {Perna}, {Krongold},
  {Covino}, {Fugazza}, {Lazzati}, {Nicastro}, {Antonelli}, {Campana},
  {Chincarini}, {D'Avanzo}, {Della Valle}, {Goldoni}, {Guetta}, {Guidorzi},
  {Meurs}, {Mirabel}, {Molinari}, {Norci}, {Piranomonte}, {Stella}, {Stratta},
  {Tagliaferri}, \& {Ward}}]{D'Elia2009ApJ}
{D'Elia}, V., {Fiore}, F., {Perna}, R., {et~al.} 2009, \apj, 694, 332

\bibitem[{{D'Elia} {et~al.}(2014){D'Elia}, {Fynbo}, {Goldoni}, {Covino}, {de
  Ugarte Postigo}, {Ledoux}, {Calura}, {Gorosabel}, {Malesani}, {Matteucci},
  {S{\'a}nchez-Ram{\'\i}rez}, {Savaglio}, {Castro-Tirado}, {Hartoog}, {Kaper},
  {Mu{\~n}oz-Darias}, {Pian}, {Piranomonte}, {Tagliaferri}, {Tanvir},
  {Vergani}, {Watson}, \& {Xu}}]{DElia2014AA}
{D'Elia}, V., {Fynbo}, J.~P.~U., {Goldoni}, P., {et~al.} 2014, \aap, 564, A38

\bibitem[{Dhillon {et~al.}(2021)Dhillon, Bezawada, Black, Dixon, Gamble, Gao,
  Henry, Kerry, Littlefair, Lunney, Marsh, Miller, Parsons, Ashley, Breedt,
  Brown, Dyer, Green, Pelisoli, Sahman, Wild, Ives, Mehrgan, Stegmeier,
  Dubbeldam, Morris, Osborn, Wilson, Casares, Muñoz-Darias, Pallé,
  Rodríguez-Gil, Shahbaz, Torres, de Ugarte Postigo, Cabrera-Lavers,
  Corradi, Domínguez, \& García-Alvarez}]{Dhillon2021}
Dhillon, V.~S., Bezawada, N., Black, M., {et~al.} 2021, Monthly Notices of the
  Royal Astronomical Society, 507, 350

\bibitem[{{Dichiara} {et~al.}(2015{\natexlab{a}}){Dichiara}, {Guidorzi},
  {Kobayashi}, {Gomboc}, \& {Mundell}}]{Dichiara2015GCN18266}
{Dichiara}, S., {Guidorzi}, C., {Kobayashi}, S., {Gomboc}, A., \& {Mundell}, C.
  2015{\natexlab{a}}, GRB Coordinates Network, 18266

\bibitem[{{Dichiara} {et~al.}(2015{\natexlab{b}}){Dichiara}, {Kopac},
  {Guidorzi}, {Kobayashi}, \& {Gomboc}}]{Dichiara2015GCN18520}
{Dichiara}, S., {Kopac}, D., {Guidorzi}, C., {Kobayashi}, S., \& {Gomboc}, A.
  2015{\natexlab{b}}, GRB Coordinates Network, 18520

\bibitem[{{Duan} \& {Wang}(2019)}]{Duan2019ApJ}
{Duan}, M.-Y. \& {Wang}, X.-G. 2019, \apj, 884, 61

\bibitem[{{Elenin} {et~al.}(2010{\natexlab{a}}){Elenin}, {Molotov}, \&
  {Pozanenko}}]{Elenin2010GCN11129}
{Elenin}, L., {Molotov}, I., \& {Pozanenko}, A. 2010{\natexlab{a}}, GRB
  Coordinates Network, 11129

\bibitem[{{Elenin} {et~al.}(2010{\natexlab{b}}){Elenin}, {Molotov}, {Volnova},
  \& {Pozanenko}}]{Elenin2010GCN11133}
{Elenin}, L., {Molotov}, I., {Volnova}, A., \& {Pozanenko}, A.
  2010{\natexlab{b}}, GRB Coordinates Network, 11133

\bibitem[{{Elenin} {et~al.}(2012){Elenin}, {Volnova}, \&
  {Pozanenko}}]{Elenin2012GCN12871}
{Elenin}, L., {Volnova}, A., \& {Pozanenko}, A. 2012, GRB Coordinates Network,
  12871

\bibitem[{{Ershova} {et~al.}(2020){Ershova}, {Lipunov}, {Gorbovskoy},
  {Tyurina}, {Kornilov}, {Zimnukhov}, {Gabovich}, {Gress}, {Budnev}, {Yurkov},
  {Vladimirov}, {Kuznetsov}, {Balanutsa}, {Rebolo}, {Serra-Ricart}, {Buckley},
  {Podesta}, {Levato}, {Lopez}, {Podesta}, {Francile}, {Mallamaci}, {Yazev},
  {Vlasenko}, {Tlatov}, {Senik}, {Grinshpun}, {Chasovnikov}, {Topolev},
  {Pozdnyakov}, {Zhirkov}, {Kuvshinov}, \& {Balakin}}]{Ershova2020ARep}
{Ershova}, O.~A., {Lipunov}, V.~M., {Gorbovskoy}, E.~S., {et~al.} 2020,
  Astronomy Reports, 64, 126

\bibitem[{{Fatkhullin} {et~al.}(2008){Fatkhullin}, {Moskvitin}, {Posanenko},
  {Sonbas}, {Volnova}, {Rumyantsev}, {Kouprianov}, {Sokolov}, \&
  {Castro-Tirado}}]{Fatkhullin2008GCN8695}
{Fatkhullin}, T., {Moskvitin}, A., {Posanenko}, A., {et~al.} 2008, GRB
  Coordinates Network, 8695

\bibitem[{Fausey {et~al.}(2023)Fausey, van der Horst, White, Seiffert,
  Willems, Young, Kann, Ghirlanda, Salvaterra, Tanvir, Levan, Moss, Chang,
  Fruchter, Guiriec, Hartmann, Kouveliotou, Granot, \& Lidz}]{Fausey2023}
Fausey, H.~M., van der Horst, A.~J., White, N.~E., {et~al.} 2023, Monthly
  Notices of the Royal Astronomical Society, 526, 4599

\bibitem[{{Feldman} {et~al.}(2021){Feldman}, {O'Brien}, {White}, {Baumgartner},
  {Thomas}, {Lodge}, {Bautz}, \& {Hinrichsen}}]{Feldman2021SPIE}
{Feldman}, C., {O'Brien}, P., {White}, N., {et~al.} 2021, in Society of
  Photo-Optical Instrumentation Engineers (SPIE) Conference Series, Vol. 11822,
  Optics for EUV, X-Ray, and Gamma-Ray Astronomy X, ed. S.~L. {O'Dell}, J.~A.
  {Gaskin}, \& G.~{Pareschi}, 118221D

\bibitem[{{Ferrante} {et~al.}(2014){Ferrante}, {Guver}, {Flewelling}, {Kehoe},
  \& {Dhungana}}]{Ferrante2014GCN16145}
{Ferrante}, F.~V., {Guver}, T., {Flewelling}, H., {Kehoe}, R., \& {Dhungana},
  G. 2014, GRB Coordinates Network, 16145

\bibitem[{Ferrero {et~al.}(2010)Ferrero, Hanlon, Felletti, French, Melady,
  Mcbreen, Kubánek, Jelínek, Mcbreen, Meintjes, Calitz, \&
  Hoffman}]{Ferrero2010}
Ferrero, A., Hanlon, L., Felletti, R., {et~al.} 2010, Advances in Astronomy,
  2010

\bibitem[{{Filgas} {et~al.}(2012){Filgas}, {Greiner}, {Schady}, {de Ugarte
  Postigo}, {Oates}, {Nardini}, {Kr{\"u}hler}, {Panaitescu}, {Kann}, {Klose},
  {Afonso}, {Allen}, {Castro-Tirado}, {Christie}, {Dong}, {Elliott}, {Natusch},
  {Nicuesa Guelbenzu}, {Olivares E.}, {Rau}, {Rossi}, {Sudilovsky}, \&
  {Yock}}]{Filgas2012AA}
{Filgas}, R., {Greiner}, J., {Schady}, P., {et~al.} 2012, \aap, 546, A101

\bibitem[{{Filgas} {et~al.}(2011{\natexlab{a}}){Filgas}, {Greiner}, {Schady},
  {Kr{\"u}hler}, {Updike}, {Klose}, {Nardini}, {Kann}, {Rossi}, {Sudilovsky},
  {Afonso}, {Clemens}, {Elliott}, {Nicuesa Guelbenzu}, {Olivares E.}, \&
  {Rau}}]{Filgas2011AA}
{Filgas}, R., {Greiner}, J., {Schady}, P., {et~al.} 2011{\natexlab{a}}, \aap,
  535, A57

\bibitem[{{Filgas} {et~al.}(2011{\natexlab{b}}){Filgas}, {Kr{\"u}hler},
  {Greiner}, {Rau}, {Palazzi}, {Klose}, {Schady}, {Rossi}, {Afonso},
  {Antonelli}, {Clemens}, {Covino}, {D'Avanzo}, {K{\"u}pc{\"u} Yolda{\c s}},
  {Nardini}, {Nicuesa Guelbenzu}, {Olivares}, {Updike}, \& {Yolda{\c
  s}}}]{Filgas2011AA2}
{Filgas}, R., {Kr{\"u}hler}, T., {Greiner}, J., {et~al.} 2011{\natexlab{b}},
  \aap, 526, A113

\bibitem[{{Fitzpatrick} \& {Massa}(1986)}]{Fitzpatrick1986ApJ}
{Fitzpatrick}, E.~L. \& {Massa}, D. 1986, \apj, 307, 286

\bibitem[{{Fitzpatrick} \& {Massa}(2007)}]{Fitzpatrick2007ApJ}
{Fitzpatrick}, E.~L. \& {Massa}, D. 2007, \apj, 663, 320

\bibitem[{{Flewelling} {et~al.}(2013){Flewelling}, {Schultz}, {Primak},
  {Chambers}, {Magnier}, {Sweeney}, {Waters}, {Chastel}, {Huber}, \&
  {Smith}}]{Flewelling2013GCN14538}
{Flewelling}, H., {Schultz}, A., {Primak}, N., {et~al.} 2013, GCN Circulars,
  14538

\bibitem[{{Fong} {et~al.}(2014){Fong}, {Chornock}, {Fox}, \&
  {Berger}}]{Fong2014GCN16274}
{Fong}, W., {Chornock}, R., {Fox}, D., \& {Berger}, E. 2014, GRB Coordinates
  Network, 16274

\bibitem[{{Fox} {et~al.}(2006){Fox}, {Cummings}, {Gehrels}, {Holland},
  {Hunsberger}, {Kennea}, {Krimm}, {Markwardt}, {Marshall}, {Morris}, {Palmer},
  \& {Stamatikos}}]{Fox2006GCN5150}
{Fox}, D.~B., {Cummings}, J.~R., {Gehrels}, N., {et~al.} 2006, GRB Coordinates
  Network, 5150

\bibitem[{{Fraija} {et~al.}(2019){Fraija}, {Dichiara}, {Pedreira},
  {Galvan-Gamez}, {Becerra}, {Montalvo}, {Montero}, {Betancourt Kamenetskaia},
  \& {Zhang}}]{Fraija2019ApJ}
{Fraija}, N., {Dichiara}, S., {Pedreira}, A.~C. C. d. E.~S., {et~al.} 2019,
  \apj, 885, 29

\bibitem[{{Frail} {et~al.}(2006){Frail}, {Cameron}, {Kasliwal}, {Nakar},
  {Price}, {Berger}, {Gal-Yam}, {Kulkarni}, {Fox}, {Soderberg}, {Schmidt},
  {Ofek}, \& {Cenko}}]{Frail2006ApJ}
{Frail}, D.~A., {Cameron}, P.~B., {Kasliwal}, M., {et~al.} 2006, \apjl, 646,
  L99

\bibitem[{{Friis} {et~al.}(2015){Friis}, {De Cia}, {Kr{\"u}hler}, {Fynbo},
  {Ledoux}, {Vreeswijk}, {Watson}, {Malesani}, {Gorosabel}, {Starling},
  {Jakobsson}, {Varela}, {Wiersema}, {Drachmann}, {Trotter}, {Th{\"o}ne}, {de
  Ugarte Postigo}, {D'Elia}, {Elliott}, {Maturi}, {Goldoni}, {Greiner},
  {Haislip}, {Kaper}, {Knust}, {LaCluyze}, {Milvang-Jensen}, {Reichart},
  {Schulze}, {Sudilovsky}, {Tanvir}, \& {Vergani}}]{Friis2015MNRAS}
{Friis}, M., {De Cia}, A., {Kr{\"u}hler}, T., {et~al.} 2015, \mnras, 451, 167

\bibitem[{Fryer {et~al.}(2022)Fryer, Lien, Fruchter, Ghirlanda, Hartmann,
  Salvaterra, Sanderbeck, \& Johnson}]{Fryer2022}
Fryer, C.~L., Lien, A.~Y., Fruchter, A., {et~al.} 2022, The Astrophysical
  Journal, 929, 111

\bibitem[{{Fugazza} {et~al.}(2012){Fugazza}, {D'Avanzo}, {Melandri}, \&
  {Antonelli}}]{Fugazza2012GCN12882}
{Fugazza}, D., {D'Avanzo}, P., {Melandri}, A., \& {Antonelli}, L.~A. 2012, GRB
  Coordinates Network, 12882

\bibitem[{{Fujiwara} {et~al.}(2016){Fujiwara}, {Saito}, {Tachibana}, {Yoshii},
  {Ono}, {Harita}, {Muraki}, {Morita}, {Ozawa}, {Saisho}, {Yatsu}, \&
  {Kawai}}]{Fujiwara2016GCN20314}
{Fujiwara}, T., {Saito}, Y., {Tachibana}, Y., {et~al.} 2016, GRB Coordinates
  Network, 20314

\bibitem[{{Fujiwara} {et~al.}(2014){Fujiwara}, {Yoshii}, {Saito}, {Tachibana},
  {Ohuchi}, {Kurita}, {Ono}, {Yatsu}, \& {Kawai}}]{Fujiwara2014GCN16173}
{Fujiwara}, T., {Yoshii}, T., {Saito}, Y., {et~al.} 2014, GRB Coordinates
  Network, 16173

\bibitem[{{Fukui} {et~al.}(2008){Fukui}, {Itow}, {Sumi}, \&
  {Tristram}}]{Fukui2008GCN7622}
{Fukui}, A., {Itow}, Y., {Sumi}, T., \& {Tristram}, P. 2008, GRB Coordinates
  Network, 7622

\bibitem[{{Fynbo} {et~al.}(2009){Fynbo}, {Jakobsson}, {Prochaska}, {Malesani},
  {Ledoux}, {de Ugarte Postigo}, {Nardini}, {Vreeswijk}, {Wiersema}, {Hjorth},
  {Sollerman}, {Chen}, {Th{\"o}ne}, {Bj{\"o}rnsson}, {Bloom}, {Castro-Tirado},
  {Christensen}, {De Cia}, {Fruchter}, {Gorosabel}, {Graham}, {Jaunsen},
  {Jensen}, {Kann}, {Kouveliotou}, {Levan}, {Maund}, {Masetti},
  {Milvang-Jensen}, {Palazzi}, {Perley}, {Pian}, {Rol}, {Schady}, {Starling},
  {Tanvir}, {Watson}, {Xu}, {Augusteijn}, {Grundahl}, {Telting}, \&
  {Quirion}}]{Fynbo2009ApJS}
{Fynbo}, J.~P.~U., {Jakobsson}, P., {Prochaska}, J.~X., {et~al.} 2009, \apjs,
  185, 526

\bibitem[{{Fynbo} {et~al.}(2014){Fynbo}, {Kr{\"u}hler}, {Leighly}, {Ledoux},
  {Vreeswijk}, {Schulze}, {Noterdaeme}, {Watson}, {Wijers}, {Bolmer}, {Cano},
  {Christensen}, {Covino}, {D'Elia}, {Flores}, {Friis}, {Goldoni}, {Greiner},
  {Hammer}, {Hjorth}, {Jakobsson}, {Japelj}, {Kaper}, {Klose}, {Knust},
  {Leloudas}, {Levan}, {Malesani}, {Milvang-Jensen}, {M{\o}ller}, {Nicuesa
  Guelbenzu}, {Oates}, {Pian}, {Schady}, {Sparre}, {Tagliaferri}, {Tanvir},
  {Th{\"o}ne}, {de Ugarte Postigo}, {Vergani}, {Wiersema}, {Xu}, \&
  {Zafar}}]{Fynbo2014AA}
{Fynbo}, J.~P.~U., {Kr{\"u}hler}, T., {Leighly}, K., {et~al.} 2014, \aap, 572,
  A12

\bibitem[{{Galeev} {et~al.}(2009){Galeev}, {Bikmaev}, {Sakhibullin}, {Burenin},
  {Pavlinsky}, {Sunyaev}, {Khamitov}, {Eker}, {Kiziloglu}, \&
  {Gogus}}]{Galeev2009GCN9548}
{Galeev}, A., {Bikmaev}, I., {Sakhibullin}, N., {et~al.} 2009, GCN Circulars,
  9548

\bibitem[{{Gallerani} {et~al.}(2008){Gallerani}, {Salvaterra}, {Ferrara}, \&
  {Choudhury}}]{Gallerani2008MNRAS}
{Gallerani}, S., {Salvaterra}, R., {Ferrara}, A., \& {Choudhury}, T.~R. 2008,
  \mnras, 388, L84

\bibitem[{{Gardner} {et~al.}(2006){Gardner}, {Mather}, {Clampin}, {Doyon},
  {Greenhouse}, {Hammel}, {Hutchings}, {Jakobsen}, {Lilly}, {Long}, {Lunine},
  {McCaughrean}, {Mountain}, {Nella}, {Rieke}, {Rieke}, {Rix}, {Smith},
  {Sonneborn}, {Stiavelli}, {Stockman}, {Windhorst}, \&
  {Wright}}]{Gardner2006SSRv}
{Gardner}, J.~P., {Mather}, J.~C., {Clampin}, M., {et~al.} 2006, \ssr, 123, 485

\bibitem[{{Garnavich} \& {Rose}(2014)}]{Garnavich2014GCN16492}
{Garnavich}, P. \& {Rose}, B. 2014, GRB Coordinates Network, 16492

\bibitem[{{Gehrels} {et~al.}(2012){Gehrels}, {Barthelmy}, \&
  {Cannizzo}}]{Gehrels_Lobster}
{Gehrels}, N., {Barthelmy}, S.~D., \& {Cannizzo}, J.~K. 2012, in New Horizons
  in Time Domain Astronomy, ed. E.~{Griffin}, R.~{Hanisch}, \& R.~{Seaman},
  Vol. 285, 41--46

\bibitem[{Gehrels \& Cannizzo(2015)}]{GEHRELS20152}
Gehrels, N. \& Cannizzo, J. 2015, Journal of High Energy Astrophysics, 7, 2,
  swift 10 Years of Discovery, a novel approach to Time Domain Astronomy

\bibitem[{{Gehrels} {et~al.}(2004){Gehrels}, {Chincarini}, {Giommi}, {Mason},
  {Nousek}, {Wells}, {White}, {Barthelmy}, {Burrows}, {Cominsky}, {Hurley},
  {Marshall}, {M{\'e}sz{\'a}ros}, {Roming}, {Angelini}, {Barbier}, {Belloni},
  {Campana}, {Caraveo}, {Chester}, {Citterio}, {Cline}, {Cropper}, {Cummings},
  {Dean}, {Feigelson}, {Fenimore}, {Frail}, {Fruchter}, {Garmire}, {Gendreau},
  {Ghisellini}, {Greiner}, {Hill}, {Hunsberger}, {Krimm}, {Kulkarni}, {Kumar},
  {Lebrun}, {Lloyd-Ronning}, {Markwardt}, {Mattson}, {Mushotzky}, {Norris},
  {Osborne}, {Paczynski}, {Palmer}, {Park}, {Parsons}, {Paul}, {Rees},
  {Reynolds}, {Rhoads}, {Sasseen}, {Schaefer}, {Short}, {Smale}, {Smith},
  {Stella}, {Tagliaferri}, {Takahashi}, {Tashiro}, {Townsley}, {Tueller},
  {Turner}, {Vietri}, {Voges}, {Ward}, {Willingale}, {Zerbi}, \&
  {Zhang}}]{Gehrels2004apJ}
{Gehrels}, N., {Chincarini}, G., {Giommi}, P., {et~al.} 2004, \apj, 611, 1005

\bibitem[{{Gendre} {et~al.}(2012){Gendre}, {Atteia}, {Bo{\"e}r}, {Colas},
  {Klotz}, {Kugel}, {Laas-Bourez}, {Rinner}, {Strajnic}, {Stratta}, \&
  {Vachier}}]{Gendre2012ApJ}
{Gendre}, B., {Atteia}, J.~L., {Bo{\"e}r}, M., {et~al.} 2012, \apj, 748, 59

\bibitem[{{Gendre} {et~al.}(2007){Gendre}, {Galli}, {Corsi}, {Klotz}, {Piro},
  {Stratta}, {Bo{\"e}r}, \& {Damerdji}}]{Gendre2007AA}
{Gendre}, B., {Galli}, A., {Corsi}, A., {et~al.} 2007, \aap, 462, 565

\bibitem[{{Gendre} {et~al.}(2013){Gendre}, {Stratta}, {Atteia}, {Basa},
  {Bo{\"e}r}, {Coward}, {Cutini}, {D'Elia}, {Howell}, {Klotz}, \&
  {Piro}}]{Gendre2013ApJ}
{Gendre}, B., {Stratta}, G., {Atteia}, J.~L., {et~al.} 2013, \apj, 766, 30

\bibitem[{{Ghirlanda} \& {Salvaterra}(2022)}]{Ghirlanda2022}
{Ghirlanda}, G. \& {Salvaterra}, R. 2022, \apj, 932, 10

\bibitem[{{Ghirlanda} {et~al.}(2015){Ghirlanda}, {Salvaterra}, {Ghisellini},
  {Mereghetti}, {Tagliaferri}, {Campana}, {Osborne}, {O'Brien}, {Tanvir},
  {Willingale}, {Amati}, {Basa}, {Bernardini}, {Burlon}, {Covino}, {D'Avanzo},
  {Frontera}, {G{\"o}tz}, {Melandri}, {Nava}, {Piro}, \&
  {Vergani}}]{Ghirlanda2015MNRAS}
{Ghirlanda}, G., {Salvaterra}, R., {Ghisellini}, G., {et~al.} 2015, \mnras,
  448, 2514

\bibitem[{{Ghirlanda} {et~al.}(2021){Ghirlanda}, {Salvaterra}, {Toffano},
  {Ronchini}, {Guidorzi}, {Oganesyan}, {Ascenzi}, {Bernardini}, {Camisasca},
  {Mereghetti}, {Nava}, {Ravasio}, {Branchesi}, {Castro-Tirado}, {Amati},
  {Blain}, {Bozzo}, {O'Brien}, {G{\"o}tz}, {Le Floch}, {Osborne}, {Rosati},
  {Stratta}, {Tanvir}, {Bogomazov}, {D'Avanzo}, {Hafizi}, {Mandhai},
  {Melandri}, {Peer}, {Topinka}, {Vergani}, \& {Zane}}]{Ghirlanda2021THESEUS}
{Ghirlanda}, G., {Salvaterra}, R., {Toffano}, M., {et~al.} 2021, Experimental
  Astronomy, 52, 277

\bibitem[{{Gomboc} {et~al.}(2008{\natexlab{a}}){Gomboc}, {Guidorzi},
  {Melandri}, {Steele}, {Mundell}, {Bersier}, {Bode}, {Burgdorf}, {Fraser},
  {Kobayashi}, {Mottram}, {Smith}, {O'Brien}, {Bannister}, \&
  {Tanvir}}]{Gomboc2008GCN7625}
{Gomboc}, A., {Guidorzi}, C., {Melandri}, A., {et~al.} 2008{\natexlab{a}}, GRB
  Coordinates Network, 7625

\bibitem[{{Gomboc} {et~al.}(2008{\natexlab{b}}){Gomboc}, {Guidorzi},
  {Melandri}, {Steele}, {Mundell}, {Bersier}, {Bode}, {Burgdorf}, {Fraser},
  {Kobayashi}, {Mottram}, {Smith}, {O'Brien}, {Bannister}, \&
  {Tanvir}}]{Gomboc2008GCN7626}
{Gomboc}, A., {Guidorzi}, C., {Melandri}, A., {et~al.} 2008{\natexlab{b}}, GRB
  Coordinates Network, 7626

\bibitem[{{Gomboc} {et~al.}(2008{\natexlab{c}}){Gomboc}, {Melandri}, {Smith},
  {Mundell}, {Steele}, {Bersier}, {Kobayashi}, {Carter}, {Burgdorf}, {Bode}, \&
  {Guidorzi}}]{Gomboc2008GCN7831}
{Gomboc}, A., {Melandri}, A., {Smith}, R.~J., {et~al.} 2008{\natexlab{c}}, GRB
  Coordinates Network, 7831

\bibitem[{{Gorbovskoy} {et~al.}(2009){Gorbovskoy}, {Kuvshinov}, {Lipunov},
  {Kornilov}, {Belinski}, {Krylov}, {Shatskiy}, {Tyurina}, {Balanutsa},
  {Chazov}, {Tlatov}, {Parkhomenko}, {Krushinski}, {Zalognikh}, {Kopytova},
  {Yazev}, {Ivanov}, {Budnev}, \& {Yurkov}}]{Gorbovskoy2009GCN10052}
{Gorbovskoy}, E., {Kuvshinov}, D., {Lipunov}, V., {et~al.} 2009, GRB
  Coordinates Network, 1052, 1

\bibitem[{{Gorbovskoy} {et~al.}(2015){Gorbovskoy}, {Lipunov}, {Tyurina},
  {Kornilov}, {Balanutsa}, {Kuznetsov}, {Kuvshinov}, {Ivanov}, {Yazev},
  {Budnev}, {Gres}, {Chuvalaev}, \& {Poleshchuk}}]{Gorbovskoy2015GCN18673}
{Gorbovskoy}, E., {Lipunov}, V., {Tyurina}, N., {et~al.} 2015, GRB Coordinates
  Network, 18673

\bibitem[{{Gorbovskoy} {et~al.}(2016){Gorbovskoy}, {Lipunov}, {Buckley},
  {Kornilov}, {Balanutsa}, {Tyurina}, {Kuznetsov}, {Kuvshinov}, {Gorbunov},
  {Vlasenko}, {Popova}, {Chazov}, {Potter}, {Kotze}, {Kniazev}, {Gress},
  {Budnev}, {Ivanov}, {Yazev}, {Tlatov}, {Senik}, {Dormidontov}, {Parhomenko},
  {Krushinski}, {Zalozhnich}, {Castro-Tirado}, {S{\'a}nchez-Ram{\'\i}rez},
  {Sergienko}, {Gabovich}, {Yurkov}, {Levato}, {Saffe}, {Mallamaci}, {Lopez},
  {Podest}, \& {Vladimirov}}]{Gorbovskoy2016MNRAS}
{Gorbovskoy}, E.~S., {Lipunov}, V.~M., {Buckley}, D.~A.~H., {et~al.} 2016,
  \mnras, 455, 3312

\bibitem[{{Gorbovskoy} {et~al.}(2013){Gorbovskoy}, {Lipunov}, {Kornilov},
  {Belinski}, {Kuvshinov}, {Tyurina}, {Sankovich}, {Krylov}, {Shatskiy},
  {Balanutsa}, {Chazov}, {Kuznetsov}, {Zimnukhov}, {Shumkov}, {Shurpakov},
  {Senik}, {Gareeva}, {Pruzhinskaya}, {Tlatov}, {Parkhomenko}, {Dormidontov},
  {Krushinsky}, {Punanova}, {Zalozhnyh}, {Popov}, {Burdanov}, {Yazev},
  {Budnev}, {Ivanov}, {Konstantinov}, {Gress}, {Chuvalaev}, {Yurkov},
  {Sergienko}, {Kudelina}, {Sinyakov}, {Karachentsev}, {Moiseev}, \&
  {Fatkhullin}}]{Gorbovskoy2013ARep}
{Gorbovskoy}, E.~S., {Lipunov}, V.~M., {Kornilov}, V.~G., {et~al.} 2013,
  Astronomy Reports, 57, 233

\bibitem[{{Gorbovskoy} {et~al.}(2012){Gorbovskoy}, {Lipunova}, {Lipunov},
  {Kornilov}, {Belinski}, {Shatskiy}, {Tyurina}, {Kuvshinov}, {Balanutsa},
  {Chazov}, {Kuznetsov}, {Zimnukhov}, {Kornilov}, {Sankovich}, {Krylov},
  {Ivanov}, {Chvalaev}, {Poleschuk}, {Konstantinov}, {Gress}, {Yazev},
  {Budnev}, {Krushinski}, {Zalozhnich}, {Popov}, {Tlatov}, {Parhomenko},
  {Dormidontov}, {Senik}, {Yurkov}, {Sergienko}, {Varda}, {Kudelina},
  {Castro-Tirado}, {Gorosabel}, {S{\'a}nchez-Ram{\'\i}rez}, {Jelinek}, \&
  {Tello}}]{Gorbovskoy2012MNRAS}
{Gorbovskoy}, E.~S., {Lipunova}, G.~V., {Lipunov}, V.~M., {et~al.} 2012,
  \mnras, 421, 1874

\bibitem[{{Gorosabel} {et~al.}(2009){Gorosabel}, {de Ugarte Postigo}, {Montes},
  {Klutsch}, \& {Castro-Tirado}}]{Gorosabel2009GCN9379}
{Gorosabel}, J., {de Ugarte Postigo}, A., {Montes}, D., {Klutsch}, A., \&
  {Castro-Tirado}, A.~J. 2009, GRB Coordinates Network, 9379

\bibitem[{{Gorosabel} {et~al.}(2011){Gorosabel}, {Jimenez}, {Castro-Tirado}, \&
  {de Ugarte Postigo}}]{Gorosabel2011GCN12206}
{Gorosabel}, J., {Jimenez}, D., {Castro-Tirado}, A.~J., \& {de Ugarte Postigo},
  A. 2011, GRB Coordinates Network, 12206

\bibitem[{{G{\"o}tz} {et~al.}(2009){G{\"o}tz}, {Paul}, {Basa}, {Wei}, {Zhang},
  {Atteia}, {Barret}, {Cordier}, {Claret}, {Deng}, {Fan}, {Hu}, {Huang},
  {Mandrou}, {Mereghetti}, {Qiu}, \& {Wu}}]{Gotz2009AIPC}
{G{\"o}tz}, D., {Paul}, J., {Basa}, S., {et~al.} 2009, in American Institute of
  Physics Conference Series, Vol. 1133, Gamma-ray Burst: Sixth Huntsville
  Symposium, ed. C.~{Meegan}, C.~{Kouveliotou}, \& N.~{Gehrels}, 25--30

\bibitem[{{Gou} {et~al.}(2007){Gou}, {Fox}, \& {M{\'e}sz{\'a}ros}}]{Gou2007ApJ}
{Gou}, L.~J., {Fox}, D.~B., \& {M{\'e}sz{\'a}ros}, P. 2007, \apj, 668, 1083

\bibitem[{Granot \& Sari(2002)}]{GranotSari2002}
Granot, J. \& Sari, R. 2002, Astrophys. J., 568, 820

\bibitem[{{Grazian} {et~al.}(2006){Grazian}, {Fernandez-Soto}, {Testa},
  {D'Avanzo}, {Antonelli}, {Malesani}, {Chincarini}, {Tagliaferri}, {Campana},
  {Covino}, {della Valle}, {Fiore}, {Piranomonte}, \&
  {Stella}}]{Grazian2006GCN4545}
{Grazian}, A., {Fernandez-Soto}, A., {Testa}, V., {et~al.} 2006, GRB
  Coordinates Network, 4545

\bibitem[{{Greiner} {et~al.}(2018){Greiner}, {Bolmer}, {Wieringa}, {van der
  Horst}, {Petry}, {Schulze}, {Knust}, {de Bruyn}, {Kr{\"u}hler}, {Wiseman},
  {Klose}, {Delvaux}, {Graham}, {Kann}, {Moin}, {Nicuesa-Guelbenzu}, {Schady},
  {Schmidl}, {Schweyer}, {Tanga}, {Tingay}, {van Eerten}, \&
  {Varela}}]{Greiner2018AA}
{Greiner}, J., {Bolmer}, J., {Wieringa}, M., {et~al.} 2018, \aap, 614, A29

\bibitem[{{Greiner} {et~al.}(2015{\natexlab{a}}){Greiner}, {Fox}, {Schady},
  {Kr{\"u}hler}, {Trenti}, {Cikota}, {Bolmer}, {Elliott}, {Delvaux}, {Perna},
  {Afonso}, {Kann}, {Klose}, {Savaglio}, {Schmidl}, {Schweyer}, {Tanga}, \&
  {Varela}}]{Greiner2015ApJ}
{Greiner}, J., {Fox}, D.~B., {Schady}, P., {et~al.} 2015{\natexlab{a}}, \apj,
  809, 76

\bibitem[{{Greiner} {et~al.}(2003){Greiner}, {Klose}, {Salvato}, {Zeh},
  {Schwarz}, {Hartmann}, {Masetti}, {Stecklum}, {Lamer}, {Lodieu}, {Scholz},
  {Sterken}, {Gorosabel}, {Burud}, {Rhoads}, {Mitrofanov}, {Litvak}, {Sanin},
  {Grinkov}, {Andersen}, {Castro Cer{\'o}n}, {Castro-Tirado}, {Fruchter},
  {Fynbo}, {Hjorth}, {Kaper}, {Kouveliotou}, {Palazzi}, {Pian}, {Rol},
  {Tanvir}, {Vreeswijk}, {Wijers}, \& {van den Heuvel}}]{Greiner2003ApJ}
{Greiner}, J., {Klose}, S., {Salvato}, M., {et~al.} 2003, \apj, 599, 1223

\bibitem[{{Greiner} {et~al.}(2009){Greiner}, {Kr{\"u}hler}, {Fynbo}, {Rossi},
  {Schwarz}, {Klose}, {Savaglio}, {Tanvir}, {McBreen}, {Totani}, {Zhang}, {Wu},
  {Watson}, {Barthelmy}, {Beardmore}, {Ferrero}, {Gehrels}, {Kann}, {Kawai},
  {Yolda{\c s}}, {M{\'e}sz{\'a}ros}, {Milvang-Jensen}, {Oates}, {Pierini},
  {Schady}, {Toma}, {Vreeswijk}, {Yolda{\c s}}, {Zhang}, {Afonso}, {Aoki},
  {Burrows}, {Clemens}, {Filgas}, {Haiman}, {Hartmann}, {Hasinger}, {Hjorth},
  {Jehin}, {Levan}, {Liang}, {Malesani}, {Pyo}, {Schulze}, {Szokoly}, {Terada},
  \& {Wiersema}}]{Greiner2009ApJ}
{Greiner}, J., {Kr{\"u}hler}, T., {Fynbo}, J.~P.~U., {et~al.} 2009, \apj, 693,
  1610

\bibitem[{{Greiner} {et~al.}(2011){Greiner}, {Kr{\"u}hler}, {Klose}, {Afonso},
  {Clemens}, {Filgas}, {Hartmann}, {K{\"u}pc{\"u} Yolda{\c{s}}}, {Nardini},
  {Olivares E.}, {Rau}, {Rossi}, {Schady}, \& {Updike}}]{Greiner2011AA}
{Greiner}, J., {Kr{\"u}hler}, T., {Klose}, S., {et~al.} 2011, \aap, 526, A30

\bibitem[{{Greiner} {et~al.}(2013){Greiner}, {Kr{\"u}hler}, {Nardini},
  {Filgas}, {Moin}, {de Breuck}, {Montenegro-Montes}, {Lundgren}, {Klose},
  {fonso}, {Bertoldi}, {Elliott}, {Kann}, {Knust}, {Menten}, {Nicuesa
  Guelbenzu}, {Olivares E.}, {Rau}, {Rossi}, {Schady}, {Schmidl}, {Siringo},
  {Spezzi}, {Sudilovsky}, {Tingay}, {Updike}, {Wang}, {Weiss}, {Wieringa}, \&
  {Wyrowski}}]{Greiner2013AA}
{Greiner}, J., {Kr{\"u}hler}, T., {Nardini}, M., {et~al.} 2013, \aap, 560, A70

\bibitem[{{Greiner} {et~al.}(2015{\natexlab{b}}){Greiner}, {Mazzali}, {Kann},
  {Kr{\"u}hler}, {Pian}, {Prentice}, {Olivares E.}, {Rossi}, {Klose},
  {Taubenberger}, {Knust}, {Afonso}, {Ashall}, {Bolmer}, {Delvaux}, {Diehl},
  {Elliott}, {Filgas}, {Fynbo}, {Graham}, {Guelbenzu}, {Kobayashi}, {Leloudas},
  {Savaglio}, {Schady}, {Schmidl}, {Schweyer}, {Sudilovsky}, {Tanga}, {Updike},
  {van Eerten}, \& {Varela}}]{Greiner2015Nat}
{Greiner}, J., {Mazzali}, P.~A., {Kann}, D.~A., {et~al.} 2015{\natexlab{b}},
  \nat, 523, 189

\bibitem[{{Gruber}(2012)}]{Gruber2012GCN12874}
{Gruber}, D. 2012, GRB Coordinates Network, 12874

\bibitem[{{Gruber} {et~al.}(2011){Gruber}, {Kr{\"u}hler}, {Foley}, {Nardini},
  {Burlon}, {Rau}, {Bissaldi}, {von Kienlin}, {McBreen}, {Greiner}, {Bhat},
  {Briggs}, {Burgess}, {Chaplin}, {Connaughton}, {Diehl}, {Fishman}, {Gibby},
  {Giles}, {Goldstein}, {Guiriec}, {van der Horst}, {Kippen}, {Kouveliotou},
  {Lin}, {Meegan}, {Paciesas}, {Preece}, {Tierney}, \&
  {Wilson-Hodge}}]{Gruber2011AA}
{Gruber}, D., {Kr{\"u}hler}, T., {Foley}, S., {et~al.} 2011, \aap, 528, A15

\bibitem[{{Guidorzi} {et~al.}(2014{\natexlab{a}}){Guidorzi}, {Dichiara},
  {Kopac}, \& {Gomboc}}]{Guidorzi2014GCN17209}
{Guidorzi}, C., {Dichiara}, S., {Kopac}, D., \& {Gomboc}, A.
  2014{\natexlab{a}}, GRB Coordinates Network, 17209

\bibitem[{{Guidorzi} {et~al.}(2015){Guidorzi}, {Japelj}, {Gomboc}, \&
  {Mundell}}]{Guidorzi2015GCN17530}
{Guidorzi}, C., {Japelj}, J., {Gomboc}, A., \& {Mundell}, C.~G. 2015, GRB
  Coordinates Network, 17530

\bibitem[{{Guidorzi} {et~al.}(2018){Guidorzi}, {Kobayashi}, {Mundell},
  {Gomboc}, \& {Steele}}]{Guidorzi2018GCN22534}
{Guidorzi}, C., {Kobayashi}, S., {Mundell}, C.~G., {Gomboc}, A., \& {Steele},
  I.~A. 2018, GRB Coordinates Network, 22534

\bibitem[{{Guidorzi} {et~al.}(2011){Guidorzi}, {Kobayashi}, {Perley},
  {Vianello}, {Bloom}, {Chandra}, {Kann}, {Li}, {Mundell}, {Pozanenko},
  {Prochaska}, {Antoniuk}, {Bersier}, {Filippenko}, {Frail}, {Gomboc},
  {Klunko}, {Melandri}, {Mereghetti}, {Morgan}, {O'Brien}, {Rumyantsev},
  {Smith}, {Steele}, {Tanvir}, \& {Volnova}}]{Guidorzi2011MNRAS}
{Guidorzi}, C., {Kobayashi}, S., {Perley}, D.~A., {et~al.} 2011, \mnras, 417,
  2124

\bibitem[{{Guidorzi} {et~al.}(2014{\natexlab{b}}){Guidorzi}, {Mundell},
  {Harrison}, {Margutti}, {Sudilovsky}, {Zauderer}, {Kobayashi}, {Cucchiara},
  {Melandri}, {Pandey}, {Berger}, {Bersier}, {D'Elia}, {Gomboc}, {Greiner},
  {Japelj}, {Kopa{\v{c}}}, {Kumar}, {Malesani}, {Mottram}, {O'Brien}, {Rau},
  {Smith}, {Steele}, {Tanvir}, \& {Virgili}}]{Guidorzi2014MNRAS}
{Guidorzi}, C., {Mundell}, C.~G., {Harrison}, R., {et~al.} 2014{\natexlab{b}},
  \mnras, 438, 752

\bibitem[{{Guidorzi} {et~al.}(2009){Guidorzi}, {Steele}, {Melandri}, {Bersier},
  {Mottram}, {Mundell}, {Smith}, {Gomboc}, {O'Brien}, {Bannister}, \&
  {Tanvir}}]{Guidorzi2009GCN9375}
{Guidorzi}, C., {Steele}, I.~A., {Melandri}, A., {et~al.} 2009, GRB Coordinates
  Network, 9375

\bibitem[{{Guiriec} {et~al.}(2016){Guiriec}, {Kouveliotou}, {Hartmann},
  {Granot}, {Asano}, {M{\'e}sz{\'a}ros}, {Gill}, {Gehrels}, \&
  {McEnery}}]{Guiriec2016ApJ}
{Guiriec}, S., {Kouveliotou}, C., {Hartmann}, D.~H., {et~al.} 2016, \apjl, 831,
  L8

\bibitem[{{Guver} {et~al.}(2014){Guver}, {Ferrante}, {Flewelling}, {Kehoe}, \&
  {Dhungana}}]{Guver2014GCN16120}
{Guver}, T., {Ferrante}, F.~V., {Flewelling}, H., {Kehoe}, R., \& {Dhungana},
  G. 2014, GRB Coordinates Network, 16120

\bibitem[{{Guver} {et~al.}(2011){Guver}, {Sonbas}, {Kaynar}, {Gogus}, \&
  {Eker}}]{Guver2011GCN12769}
{Guver}, T., {Sonbas}, E., {Kaynar}, S., {Gogus}, E., \& {Eker}, Z. 2011, GRB
  Coordinates Network, 12769

\bibitem[{{Hagen} \& {Troja}(2014)}]{Hagen2014GCN16662}
{Hagen}, L.~M.~Z. \& {Troja}, E. 2014, GRB Coordinates Network, 16662

\bibitem[{{Haislip} {et~al.}(2009{\natexlab{a}}){Haislip}, {Reichart},
  {Ivarsen}, {Lacluyze}, {Cominsky}, {McLin}, {Graves}, {Spear}, {Egger},
  {Foster}, {Moore}, {Oza}, {Schubel}, {Styblova}, {Trotter}, {Crain}, \&
  {Nysewander}}]{Haislip2009GCN10219}
{Haislip}, J., {Reichart}, D., {Ivarsen}, K., {et~al.} 2009{\natexlab{a}}, GCN
  Circulars, 10219

\bibitem[{{Haislip} {et~al.}(2009{\natexlab{b}}){Haislip}, {Reichart},
  {Ivarsen}, {Lacluyze}, {Egger}, {Foster}, {Moore}, {Oza}, {Schubel},
  {Styblova}, {Trotter}, {Crain}, \& {Nysewander}}]{Haislip2009GCN10230}
{Haislip}, J., {Reichart}, D., {Ivarsen}, K., {et~al.} 2009{\natexlab{b}}, GCN
  Circulars, 10230

\bibitem[{{Haislip} {et~al.}(2009{\natexlab{c}}){Haislip}, {Reichart},
  {Ivarsen}, {Lacluyze}, {Egger}, {Foster}, {Moore}, {Oza}, {Schubel},
  {Styblova}, {Trotter}, {Crain}, \& {Nysewander}}]{Haislip2009GCN10249}
{Haislip}, J., {Reichart}, D., {Ivarsen}, K., {et~al.} 2009{\natexlab{c}}, GCN
  Circulars, 10249

\bibitem[{{Haislip} {et~al.}(2006){Haislip}, {Nysewander}, {Reichart}, {Levan},
  {Tanvir}, {Cenko}, {Fox}, {Price}, {Castro-Tirado}, {Gorosabel}, {Evans},
  {Figueredo}, {MacLeod}, {Kirschbrown}, {Jelinek}, {Guziy}, {Postigo},
  {Cypriano}, {Lacluyze}, {Graham}, {Priddey}, {Chapman}, {Rhoads}, {Fruchter},
  {Lamb}, {Kouveliotou}, {Wijers}, {Bayliss}, {Schmidt}, {Soderberg},
  {Kulkarni}, {Harrison}, {Moon}, {Gal-Yam}, {Kasliwal}, {Hudec}, {Vitek},
  {Kubanek}, {Crain}, {Foster}, {Clemens}, {Bartelme}, {Canterna}, {Hartmann},
  {Henden}, {Klose}, {Park}, {Williams}, {Rol}, {O'Brien}, {Bersier}, {Prada},
  {Pizarro}, {Maturana}, {Ugarte}, {Alvarez}, {Fernandez}, {Jarvis}, {Moles},
  {Alfaro}, {Ivarsen}, {Kumar}, {Mack}, {Zdarowicz}, {Gehrels}, {Barthelmy}, \&
  {Burrows}}]{Haislip2006Nature}
{Haislip}, J.~B., {Nysewander}, M.~C., {Reichart}, D.~E., {et~al.} 2006, \nat,
  440, 181

\bibitem[{{Han} {et~al.}(2005){Han}, {Mack}, {Lee}, {Park}, {Jin}, {Kim},
  {Kim}, {Yuk}, {Lee}, \& {Bradstreet}}]{HanW2005}
{Han}, W., {Mack}, P., {Lee}, C.-U., {et~al.} 2005, \pasj, 57, 821

\bibitem[{{Hao} {et~al.}(2020){Hao}, {Cao}, {Lu}, {Chu}, {Fan}, {Yuan}, \&
  {Yuan}}]{Hao2020ApJS}
{Hao}, J.-M., {Cao}, L., {Lu}, Y.-J., {et~al.} 2020, \apjs, 248, 21

\bibitem[{{Harbeck} {et~al.}(2014{\natexlab{a}}){Harbeck}, {Kaur},
  {Delgado-Navarro}, {Orio}, \& {Hartmann}}]{Harbeck2014GCN16165}
{Harbeck}, D., {Kaur}, A., {Delgado-Navarro}, A., {Orio}, M., \& {Hartmann},
  D.~H. 2014{\natexlab{a}}, GRB Coordinates Network, 16165

\bibitem[{{Harbeck} {et~al.}(2014{\natexlab{b}}){Harbeck}, {Kaur},
  {Delgado-Navarro}, {Orio}, \& {Hartmann}}]{Harbeck2014GCN16175}
{Harbeck}, D., {Kaur}, A., {Delgado-Navarro}, A., {Orio}, M., \& {Hartmann},
  D.~H. 2014{\natexlab{b}}, GRB Coordinates Network, 16175

\bibitem[{{Hartoog} {et~al.}(2015){Hartoog}, {Malesani}, {Fynbo}, {Goto},
  {Kr{\"u}hler}, {Vreeswijk}, {De Cia}, {Xu}, {M{\o}ller}, {Covino}, {D'Elia},
  {Flores}, {Goldoni}, {Hjorth}, {Jakobsson}, {Krogager}, {Kaper}, {Ledoux},
  {Levan}, {Milvang-Jensen}, {Sollerman}, {Sparre}, {Tagliaferri}, {Tanvir},
  {de Ugarte Postigo}, {Vergani}, {Wiersema}, {Datson}, {Salinas}, {Mikkelsen},
  \& {Aghanim}}]{Hartoog2015AA}
{Hartoog}, O.~E., {Malesani}, D., {Fynbo}, J.~P.~U., {et~al.} 2015, \aap, 580,
  A139

\bibitem[{{Hartoog} {et~al.}(2013){Hartoog}, {Wiersema}, {Vreeswijk}, {Kaper},
  {Tanvir}, {Savaglio}, {Berger}, {Chornock}, {Covino}, {D'Elia}, {Flores},
  {Fynbo}, {Goldoni}, {Gomboc}, {Melandri}, {Pozanenko}, {Schaye}, {de Ugarte
  Postigo}, \& {Wijers}}]{Hartoog2013MNRAS}
{Hartoog}, O.~E., {Wiersema}, K., {Vreeswijk}, P.~M., {et~al.} 2013, \mnras,
  430, 2739

\bibitem[{{Heintz} {et~al.}(2017){Heintz}, {Fynbo}, {Jakobsson}, {Kr{\"u}hler},
  {Christensen}, {Watson}, {Ledoux}, {Noterdaeme}, {Perley}, {Rhodin},
  {Selsing}, {Schulze}, {Tanvir}, {M{\o}ller}, {Goldoni}, {Xu}, \&
  {Milvang-Jensen}}]{Heintz2017AA}
{Heintz}, K.~E., {Fynbo}, J.~P.~U., {Jakobsson}, P., {et~al.} 2017, \aap, 601,
  A83

\bibitem[{{Heintz} {et~al.}(2019){Heintz}, {Ledoux}, {Fynbo}, {Jakobsson},
  {Noterdaeme}, {Krogager}, {Bolmer}, {M{\o}ller}, {Vergani}, {Watson},
  {Zafar}, {De Cia}, {Tanvir}, {Malesani}, {Japelj}, {Covino}, \&
  {Kaper}}]{Heintz2019AA}
{Heintz}, K.~E., {Ledoux}, C., {Fynbo}, J.~P.~U., {et~al.} 2019, \aap, 621, A20

\bibitem[{{Heintz} {et~al.}(2018{\natexlab{a}}){Heintz}, {Malesani}, \&
  {Moran-Kelly}}]{Heintz2018GCN23478}
{Heintz}, K.~E., {Malesani}, D.~B., \& {Moran-Kelly}, S. 2018{\natexlab{a}},
  GRB Coordinates Network, 23478

\bibitem[{{Heintz} {et~al.}(2018{\natexlab{b}}){Heintz}, {Watson}, {Jakobsson},
  {Fynbo}, {Bolmer}, {Arabsalmani}, {Cano}, {Covino}, {D'Elia}, {Gomboc},
  {Japelj}, {Kaper}, {Krogager}, {Pugliese}, {S{\'a}nchez-Ram{\'\i}rez},
  {Selsing}, {Sparre}, {Tanvir}, {Th{\"o}ne}, {de Ugarte Postigo}, \&
  {Vergani}}]{Heintz2018MNRAS}
{Heintz}, K.~E., {Watson}, D., {Jakobsson}, P., {et~al.} 2018{\natexlab{b}},
  \mnras, 479, 3456

\bibitem[{{Hentunen} {et~al.}(2010{\natexlab{a}}){Hentunen}, {Nissinen}, \&
  {Salmi}}]{Hentunen2010GCN11173}
{Hentunen}, V.~P., {Nissinen}, M., \& {Salmi}, T. 2010{\natexlab{a}}, GRB
  Coordinates Network, 11173

\bibitem[{{Hentunen} {et~al.}(2010{\natexlab{b}}){Hentunen}, {Nissinen}, \&
  {Salmi}}]{Hentunen2010GCN11253}
{Hentunen}, V.~P., {Nissinen}, M., \& {Salmi}, T. 2010{\natexlab{b}}, GRB
  Coordinates Network, 11253

\bibitem[{{Hentunen} {et~al.}(2011{\natexlab{a}}){Hentunen}, {Nissinen}, \&
  {Salmi}}]{Hentunen2011GCN11966}
{Hentunen}, V.~P., {Nissinen}, M., \& {Salmi}, T. 2011{\natexlab{a}}, GRB
  Coordinates Network, 11966

\bibitem[{{Hentunen} {et~al.}(2011{\natexlab{b}}){Hentunen}, {Nissinen}, \&
  {Salmi}}]{Hentunen2011GCN12754}
{Hentunen}, V.~P., {Nissinen}, M., \& {Salmi}, T. 2011{\natexlab{b}}, GRB
  Coordinates Network, 12754

\bibitem[{{Hentunen} {et~al.}(2013){Hentunen}, {Nissinen}, \&
  {Salmi}}]{Hentunen2013GCN15418}
{Hentunen}, V.~P., {Nissinen}, M., \& {Salmi}, T. 2013, GRB Coordinates
  Network, 15418

\bibitem[{{Hermansson} {et~al.}(2013){Hermansson}, {Holmstrom}, \&
  {Johansson}}]{Hermansson2013GCN14596}
{Hermansson}, L., {Holmstrom}, P., \& {Johansson}, M. 2013, GCN Circulars,
  14596

\bibitem[{{Hjorth} \& {Bloom}(2012)}]{Hjorth2012Book}
{Hjorth}, J. \& {Bloom}, J.~S. 2012, {The Gamma-Ray Burst - Supernova
  Connection} (Cambridge University Press (Cambridge)), 169--190

\bibitem[{{Holland}(2006)}]{Holland2006GCN5158}
{Holland}, S.~T. 2006, GRB Coordinates Network, 5158

\bibitem[{{Holland} {et~al.}(2012){Holland}, {De Pasquale}, {Mao}, {Sakamoto},
  {Schady}, {Covino}, {Fan}, {Jin}, {D'Avanzo}, {Antonelli}, {D'Elia},
  {Chincarini}, {Fiore}, {Bhushan Pandey}, \& {Cobb}}]{Holland2012ApJ}
{Holland}, S.~T., {De Pasquale}, M., {Mao}, J., {et~al.} 2012, \apj, 745, 41

\bibitem[{{Holland} {et~al.}(2010){Holland}, {Kuin}, \&
  {Rowlinson}}]{Holland2010GCN10432}
{Holland}, S.~T., {Kuin}, N.~P.~M., \& {Rowlinson}, A. 2010, GRB Coordinates
  Network, 10432

\bibitem[{{Holland} \& {Pagani}(2012{\natexlab{a}})}]{Holland2012GCN13666}
{Holland}, S.~T. \& {Pagani}, C. 2012{\natexlab{a}}, GRB Coordinates Network,
  13666

\bibitem[{{Holland} \& {Pagani}(2012{\natexlab{b}})}]{Holland2012GCN13901}
{Holland}, S.~T. \& {Pagani}, C. 2012{\natexlab{b}}, GRB Coordinates Network,
  13901

\bibitem[{{Holland} \& {Rowlinson}(2010)}]{Holland2010GCN10436}
{Holland}, S.~T. \& {Rowlinson}, A. 2010, GRB Coordinates Network, 10436

\bibitem[{{Holland} \& {Sbarufatti}(2008)}]{Holland2008GCN7830}
{Holland}, S.~T. \& {Sbarufatti}, B. 2008, GRB Coordinates Network, 7830

\bibitem[{{Holland} {et~al.}(2008){Holland}, {Ward}, \&
  {Marshall}}]{Holland2008GCN7991}
{Holland}, S.~T., {Ward}, P.~A., \& {Marshall}, F.~E. 2008, GRB Coordinates
  Network, 7991

\bibitem[{{Honda} {et~al.}(2014){Honda}, {Takagi}, {Arai}, \&
  {Morihana}}]{Honda2014GCN16496}
{Honda}, S., {Takagi}, Y., {Arai}, A., \& {Morihana}, K. 2014, GRB Coordinates
  Network, 16496

\bibitem[{{Hu} {et~al.}(2019){Hu}, {Oates}, {Lipunov}, {Zhang},
  {Castro-Tirado}, {Jeong}, {S{\'a}nchez-Ram{\'\i}rez}, {Tello}, {Cunniffe},
  {Gorbovskoy}, {Caballero-Garc{\'\i}a}, {Pandey}, {Kornilov}, {Tyurina},
  {Kuznetsov}, {Balanutsa}, {Gress}, {Gorbunov}, {Vlasenko}, {Vladimirov},
  {Budnev}, {Balakin}, {Ershova}, {Krushinski}, {Gabovich}, {Yurkov},
  {Gorosabel}, {Moskvitin}, {Burenin}, {Sokolov}, {Delgado}, {Guziy},
  {Fernandez-Garc{\'\i}a}, \& {Park}}]{Hu2019AA}
{Hu}, Y.~D., {Oates}, S.~R., {Lipunov}, V.~M., {et~al.} 2019, \aap, 632, A100

\bibitem[{{Huang} {et~al.}(2017){Huang}, {Urata}, {Takahashi}, {Im}, {Yu},
  {Choi}, {Butler}, {Watson}, {Kutyrev}, {Lee}, {Klein}, {Fox}, {Littlejohns},
  {Cucchiara}, {Troja}, {Gonz{\'a}lez}, {Richer}, {Rom{\'a}n-Z{\'u}{\~n}iga},
  {Bloom}, {Prochaska}, {Gehrels}, {Moseley}, {Georgiev}, {de Diego}, \&
  {Ramirez-Ruiz}}]{Huang2017PASJ}
{Huang}, K., {Urata}, Y., {Takahashi}, S., {et~al.} 2017, \pasj, 69, 20

\bibitem[{{Huang} {et~al.}(2008){Huang}, {Chen}, {Chen}, {Huang}, \&
  {Urata}}]{Huang2008GCN7999}
{Huang}, L.~C., {Chen}, T.~W., {Chen}, Y.~T., {Huang}, K.~Y., \& {Urata}, Y.
  2008, GRB Coordinates Network, 7999

\bibitem[{{Huang} {et~al.}(2018){Huang}, {Wang}, {Zheng}, {Liang}, {Lin},
  {Zhong}, {Zhang}, {Huang}, {Filippenko}, \& {Zhang}}]{Huang2018ApJ}
{Huang}, L.-Y., {Wang}, X.-G., {Zheng}, W., {et~al.} 2018, \apj, 859, 163

\bibitem[{{Huang} {et~al.}(2020){Huang}, {Liang}, {Liu}, {Cheng}, \&
  {Wang}}]{Huang2020ApJ}
{Huang}, X.-L., {Liang}, E.-W., {Liu}, R.-Y., {Cheng}, J.-G., \& {Wang}, X.-Y.
  2020, \apjl, 903, L26

\bibitem[{{Ibrahimov} {et~al.}(2008){Ibrahimov}, {Karimov}, {Rumyantsev}, \&
  {Pozanenko}}]{Ibrahimov2008GCN7975}
{Ibrahimov}, M., {Karimov}, R., {Rumyantsev}, V., \& {Pozanenko}, A. 2008, GRB
  Coordinates Network, 7975

\bibitem[{{Im}(2013)}]{Im2013GCN14464}
{Im}, M. 2013, GCN Circulars, 14464

\bibitem[{{Im} \& {Choi}(2013)}]{ImnChoi2013}
{Im}, M. \& {Choi}, C. 2013, GRB Coordinates Network, 15432, 1

\bibitem[{{Im} {et~al.}(2010{\natexlab{a}}){Im}, {Choi}, {Jun}, {Kang},
  {Urata}, {Choi}, {Sakamoto}, \& {Gehrels}}]{Im2010GCN11222}
{Im}, M., {Choi}, C., {Jun}, H., {et~al.} 2010{\natexlab{a}}, GRB Coordinates
  Network, 11222

\bibitem[{{Im} {et~al.}(2010{\natexlab{b}}){Im}, {Choi}, {Jun}, {Kang},
  {Urata}, {Choi}, {Sakamoto}, \& {Gehrels}}]{Im2010GCN11232}
{Im}, M., {Choi}, C., {Jun}, H., {et~al.} 2010{\natexlab{b}}, GRB Coordinates
  Network, 11232

\bibitem[{{Im} {et~al.}(2010{\natexlab{c}}){Im}, {Choi}, {Jun}, {Kang},
  {Urata}, {Choi}, {Sakamoto}, {Tanvir}, \& {Levan}}]{Im2010GCN11208}
{Im}, M., {Choi}, C., {Jun}, H., {et~al.} 2010{\natexlab{c}}, GRB Coordinates
  Network, 11208

\bibitem[{{Im} {et~al.}(2019){Im}, {Paek}, \& {Choi}}]{Im2019GCN23757}
{Im}, M., {Paek}, G.~S.~H., \& {Choi}, C. 2019, GRB Coordinates Network, 23757

\bibitem[{{Im} {et~al.}(2012){Im}, {Sung}, \& {Urata}}]{Im2012GCN13823}
{Im}, M., {Sung}, H.~I., \& {Urata}, Y. 2012, GRB Coordinates Network, 13823

\bibitem[{{Im} {et~al.}(2013){Im}, {Sung}, \& {Urata}}]{Im2013GCN14800}
{Im}, M., {Sung}, H.~I., \& {Urata}, Y. 2013, GRB Coordinates Network, 14800

\bibitem[{{Im} {et~al.}(2010{\natexlab{d}}){Im}, {Ko}, {Cho}, {Choi}, {Jeon},
  {Lee}, \& {Ibrahimov}}]{ImM2010}
{Im}, M.-S., {Ko}, J.-W., {Cho}, Y.-S., {et~al.} 2010{\natexlab{d}}, Journal of
  Korean Astronomical Society, 43, 75

\bibitem[{{Ishida} {et~al.}(2011){Ishida}, {de Souza}, \&
  {Ferrara}}]{Ishida2011MNRAS}
{Ishida}, E.~E.~O., {de Souza}, R.~S., \& {Ferrara}, A. 2011, \mnras, 418, 500

\bibitem[{{Isogai} \& {Kawai}(2008)}]{Isogai2008GCN8629}
{Isogai}, M. \& {Kawai}, N. 2008, GRB Coordinates Network, 8629

\bibitem[{{Izzo} {et~al.}(2018{\natexlab{a}}){Izzo}, {de Ugarte Postigo},
  {Kann}, {Malesani}, {Heintz}, {Tanvir}, {D'Elia}, {Wiersema}, {Kouveliotou},
  \& {Levan}}]{Izzo2018GCN23488}
{Izzo}, L., {de Ugarte Postigo}, A., {Kann}, D.~A., {et~al.}
  2018{\natexlab{a}}, GRB Coordinates Network, 23488

\bibitem[{{Izzo} {et~al.}(2018{\natexlab{b}}){Izzo}, {Kann}, {de Ugarte
  Postigo}, {Thoene}, {Bensch}, {Blazek}, {Diaz-Martin}, \&
  {Rodriguez-Llano}}]{Izzo2018GCN23040}
{Izzo}, L., {Kann}, D.~A., {de Ugarte Postigo}, A., {et~al.}
  2018{\natexlab{b}}, GRB Coordinates Network, 23040

\bibitem[{{Jakobsson} {et~al.}(2004){Jakobsson}, {Hjorth}, {Fynbo}, {Watson},
  {Pedersen}, {Bj{\"o}rnsson}, \& {Gorosabel}}]{Jakobsson2004ApJ}
{Jakobsson}, P., {Hjorth}, J., {Fynbo}, J.~P.~U., {et~al.} 2004, \apjl, 617,
  L21

\bibitem[{{Jakobsson} {et~al.}(2010){Jakobsson}, {Malesani}, {Villforth},
  {Hjorth}, {Watson}, \& {Tanvir}}]{Jakobsson2010GCN10438}
{Jakobsson}, P., {Malesani}, D., {Villforth}, C., {et~al.} 2010, GRB
  Coordinates Network, 10438

\bibitem[{{Jakobsson} {et~al.}(2008){Jakobsson}, {Vreeswijk}, {Xu}, \&
  {Thoene}}]{Jakobsson2008GCN7832}
{Jakobsson}, P., {Vreeswijk}, P.~M., {Xu}, D., \& {Thoene}, C.~C. 2008, GRB
  Coordinates Network, 7832

\bibitem[{{Jang} {et~al.}(2011){Jang}, {Im}, {Lee}, {Urata}, {Huang},
  {Hirashita}, {Fan}, \& {Jiang}}]{Jang2011ApJ}
{Jang}, M., {Im}, M., {Lee}, I., {et~al.} 2011, \apjl, 741, L20

\bibitem[{{Jang} {et~al.}(2012{\natexlab{a}}){Jang}, {Im}, \&
  {Urata}}]{Jang2012GCN12898}
{Jang}, M., {Im}, M., \& {Urata}, Y. 2012{\natexlab{a}}, GRB Coordinates
  Network, 12898

\bibitem[{{Jang} {et~al.}(2012{\natexlab{b}}){Jang}, {Im}, \&
  {Urata}}]{Jang2012GCN13139}
{Jang}, M., {Im}, M., \& {Urata}, Y. 2012{\natexlab{b}}, GRB Coordinates
  Network, 13139

\bibitem[{{Janiuk} {et~al.}(2013){Janiuk}, {Charzy{\'n}ski}, \&
  {Bejger}}]{Janiuk2013AA}
{Janiuk}, A., {Charzy{\'n}ski}, S., \& {Bejger}, M. 2013, \aap, 560, A25

\bibitem[{{Japelj} {et~al.}(2015){Japelj}, {Covino}, {Gomboc}, {Vergani},
  {Goldoni}, {Selsing}, {Cano}, {D'Elia}, {Flores}, {Fynbo}, {Hammer},
  {Hjorth}, {Jakobsson}, {Kaper}, {Kopa{\v{c}}}, {Kr{\"u}hler}, {Melandri},
  {Piranomonte}, {S{\'a}nchez-Ram{\'\i}rez}, {Tagliaferri}, {Tanvir}, {de
  Ugarte Postigo}, {Watson}, \& {Wijers}}]{Japelj2015A&A}
{Japelj}, J., {Covino}, S., {Gomboc}, A., {et~al.} 2015, \aap, 579, A74

\bibitem[{{Japelj} {et~al.}(2016){Japelj}, {Vergani}, {Salvaterra}, {Hunt}, \&
  {Mannucci}}]{Japelj2016AA}
{Japelj}, J., {Vergani}, S.~D., {Salvaterra}, R., {Hunt}, L.~K., \& {Mannucci},
  F. 2016, \aap, 593, A115

\bibitem[{{Jel{\'\i}nek} {et~al.}(2016){Jel{\'\i}nek}, {Castro-Tirado},
  {Cunniffe}, {Gorosabel}, {V{\'\i}tek}, {Kub{\'a}nek}, {de Ugarte Postigo},
  {Guziy}, {Tello}, {P{\'a}ta}, {S{\'a}nchez-Ram{\'\i}rez}, {Oates}, {Jeong},
  {{\v{S}}trobl}, {Castillo-Carri{\'o}n}, {Mateo Sanguino}, {Rabaza},
  {P{\'e}rez-Ram{\'\i}rez}, {Fern{\'a}ndez-Mu{\~n}oz}, {de la Morena
  Carretero}, {Hudec}, {Reglero}, \& {Sabau-Graziati}}]{Jelinek2016AdAst}
{Jel{\'\i}nek}, M., {Castro-Tirado}, A.~J., {Cunniffe}, R., {et~al.} 2016,
  Advances in Astronomy, 2016, 192846

\bibitem[{{Jel{\'\i}nek} {et~al.}(2013){Jel{\'\i}nek}, {G{\'o}mez Gauna}, \&
  {Castro-Tirado}}]{Jelinek2013EAS}
{Jel{\'\i}nek}, M., {G{\'o}mez Gauna}, E., \& {Castro-Tirado}, A.~J. 2013, in
  EAS Publications Series, Vol.~61, EAS Publications Series, ed. A.~J.
  {Castro-Tirado}, J.~{Gorosabel}, \& I.~H. {Park}, 475--477

\bibitem[{{Jel{\'\i}nek} {et~al.}(2012){Jel{\'\i}nek}, {Gorosabel},
  {Castro-Tirado}, {de Ugarte Postigo}, {Guziy}, {Cunniffe}, {Kub{\'a}nek},
  {Prouza}, {V{\'\i}tek}, {Hudec}, {Reglero}, \&
  {Sabau-Graziati}}]{Jelinek2012AcPol}
{Jel{\'\i}nek}, M., {Gorosabel}, J., {Castro-Tirado}, A.~J., {et~al.} 2012,
  Acta Polytechnica, 52, 34

\bibitem[{{Jel{\'\i}nek} {et~al.}(2019){Jel{\'\i}nek}, {Kann}, {{\v{S}}trobl},
  \& {Hudec}}]{Jelinek2019AN}
{Jel{\'\i}nek}, M., {Kann}, D.~A., {{\v{S}}trobl}, J., \& {Hudec}, R. 2019,
  Astronomische Nachrichten, 340, 622

\bibitem[{{Jel{\'\i}nek} {et~al.}(2005){Jel{\'\i}nek}, {Kub{\'a}nek}, {Hudec},
  {Nekola}, {Topinka}, \& {{\v{S}}trobl}}]{Jelinek05}
{Jel{\'\i}nek}, M., {Kub{\'a}nek}, P., {Hudec}, R., {et~al.} 2005, in
  Astronomical Society of the Pacific Conference Series, Vol. 330, The
  Astrophysics of Cataclysmic Variables and Related Objects, ed. J.~M.
  {Hameury} \& J.~P. {Lasota}, 481

\bibitem[{{Jelinek} {et~al.}(2018){Jelinek}, {Strobl}, {Hudec}, \&
  {Polasek}}]{Jelinek2018GCN23024}
{Jelinek}, M., {Strobl}, J., {Hudec}, R., \& {Polasek}, C. 2018, GRB
  Coordinates Network, 23024

\bibitem[{{Jelinek} {et~al.}(2021){Jelinek}, {Strobl}, {Trcka}, {Hudec}, \&
  {Polasek}}]{Jelinek2021GCN}
{Jelinek}, M., {Strobl}, J., {Trcka}, S., {Hudec}, R., \& {Polasek}, C. 2021,
  GRB Coordinates Network, 29651

\bibitem[{{Jeon} {et~al.}(2011){Jeon}, {Im}, {Pak}, \&
  {Jeong}}]{Jeon2011GCN11967}
{Jeon}, Y., {Im}, M., {Pak}, S., \& {Jeong}, H. 2011, GRB Coordinates Network,
  11967

\bibitem[{{Jiang} {et~al.}(2021{\natexlab{a}}){Jiang}, {Kashikawa}, {Wang},
  {Walth}, {Ho}, {Cai}, {Egami}, {Fan}, {Ito}, {Liang}, {Schaerer}, \&
  {Stark}}]{Jiang2021NatAs1}
{Jiang}, L., {Kashikawa}, N., {Wang}, S., {et~al.} 2021{\natexlab{a}}, Nature
  Astronomy, 5, 256

\bibitem[{{Jiang} {et~al.}(2021{\natexlab{b}}){Jiang}, {Wang}, {Zhang},
  {Kashikawa}, {Ho}, {Cai}, {Egami}, {Walth}, {Yang}, {Zhang}, \&
  {Zhao}}]{Jiang2021NatAs2}
{Jiang}, L., {Wang}, S., {Zhang}, B., {et~al.} 2021{\natexlab{b}}, Nature
  Astronomy, 5, 262

\bibitem[{{Jiang} {et~al.}(2021{\natexlab{c}}){Jiang}, {Wang}, {Zhang},
  {Kashikawa}, {Ho}, {Cai}, {Egami}, {Walth}, {Yang}, {Zhang}, \&
  {Zhao}}]{Jiang2021arXiv}
{Jiang}, L., {Wang}, S., {Zhang}, B., {et~al.} 2021{\natexlab{c}}, Nature
  Astronomy, 5, 998

\bibitem[{{Jin} {et~al.}(2013){Jin}, {Covino}, {Della Valle}, {Ferrero},
  {Fugazza}, {Malesani}, {Melandri}, {Pian}, {Salvaterra}, {Bersier},
  {Campana}, {Cano}, {Castro-Tirado}, {D'Avanzo}, {Fynbo}, {Gomboc},
  {Gorosabel}, {Guidorzi}, {Haislip}, {Hjorth}, {Kobayashi}, {LaCluyze},
  {Marconi}, {Mazzali}, {Mundell}, {Piranomonte}, {Reichart},
  {S{\'a}nchez-Ram{\'{\i}}rez}, {Smith}, {Steele}, {Tagliaferri}, {Tanvir},
  {Valenti}, {Vergani}, {Vestrand}, {Walker}, \& {Wo{\'z}niak}}]{Jin2013ApJ}
{Jin}, Z.-P., {Covino}, S., {Della Valle}, M., {et~al.} 2013, \apj, 774, 114

\bibitem[{{Jin} {et~al.}(2023){Jin}, {Zhou}, {Wang}, {Geng}, {Covino}, {Wu},
  {Li}, {Fan}, {Wei}, \& {Wei}}]{zhi-ping}
{Jin}, Z.-P., {Zhou}, H., {Wang}, Y., {et~al.} 2023, Nature Astronomy

\bibitem[{{Jordana-Mitjans} {et~al.}(2020){Jordana-Mitjans}, {Mundell},
  {Kobayashi}, {Smith}, {Guidorzi}, {Steele}, {Shrestha}, {Gomboc}, {Marongiu},
  {Martone}, {Lipunov}, {Gorbovskoy}, {Buckley}, {Rebolo}, \&
  {Budnev}}]{Jordana-Mitjans2020ApJ}
{Jordana-Mitjans}, N., {Mundell}, C.~G., {Kobayashi}, S., {et~al.} 2020, \apj,
  892, 97

\bibitem[{{Kann} {et~al.}(2020){Kann}, {Blazek}, {de Ugarte Postigo}, \&
  {Th{\"o}ne}}]{Kann2020RNAAS}
{Kann}, D.~A., {Blazek}, M., {de Ugarte Postigo}, A., \& {Th{\"o}ne}, C.~C.
  2020, Research Notes of the American Astronomical Society, 4, 247

\bibitem[{{Kann} {et~al.}(2021{\natexlab{a}}){Kann}, {de Ugarte Postigo},
  {Thoene}, {Blazek}, {Agui Fernandez}, \& {Scarpa}}]{Kann2021GCN29653}
{Kann}, D.~A., {de Ugarte Postigo}, A., {Thoene}, C.~C., {et~al.}
  2021{\natexlab{a}}, GRB Coordinates Network, 29653

\bibitem[{{Kann} {et~al.}(2021{\natexlab{b}}){Kann}, {de Ugarte Postigo},
  {Thoene}, {Blazek}, {Agui Fernandez}, \& {Scarpa}}]{Kann2021GCN29655}
{Kann}, D.~A., {de Ugarte Postigo}, A., {Thoene}, C.~C., {et~al.}
  2021{\natexlab{b}}, GRB Coordinates Network, 29655

\bibitem[{{Kann} {et~al.}(2015){Kann}, {Delvaux}, \&
  {Greiner}}]{Kann2015GCN17522}
{Kann}, D.~A., {Delvaux}, C., \& {Greiner}, J. 2015, GRB Coordinates Network,
  17522

\bibitem[{{Kann} {et~al.}(2018{\natexlab{a}}){Kann}, {Izzo}, \&
  {Casanova}}]{Kann2018GCN22985}
{Kann}, D.~A., {Izzo}, L., \& {Casanova}, V. 2018{\natexlab{a}}, GRB
  Coordinates Network, 22985

\bibitem[{{Kann} {et~al.}(2010{\natexlab{a}}){Kann}, {Klose}, {Laux}, \&
  {Stecklum}}]{Kann2010GCN11187}
{Kann}, D.~A., {Klose}, S., {Laux}, U., \& {Stecklum}, B. 2010{\natexlab{a}},
  GRB Coordinates Network, 11187

\bibitem[{{Kann} {et~al.}(2006){Kann}, {Klose}, \& {Zeh}}]{Kann2006ApJ}
{Kann}, D.~A., {Klose}, S., \& {Zeh}, A. 2006, \apj, 641, 993

\bibitem[{{Kann} {et~al.}(2011){Kann}, {Klose}, {Zhang}, {Covino}, {Butler},
  {Malesani}, {Nakar}, {Wilson}, {Antonelli}, {Chincarini}, {Cobb}, {D'Avanzo},
  {D'Elia}, {Della Valle}, {Ferrero}, {Fugazza}, {Gorosabel}, {Israel},
  {Mannucci}, {Piranomonte}, {Schulze}, {Stella}, {Tagliaferri}, \&
  {Wiersema}}]{Kann2011ApJ}
{Kann}, D.~A., {Klose}, S., {Zhang}, B., {et~al.} 2011, \apj, 734, 96

\bibitem[{{Kann} {et~al.}(2010{\natexlab{b}}){Kann}, {Klose}, {Zhang},
  {Malesani}, {Nakar}, {Pozanenko}, {Wilson}, {Butler}, {Jakobsson}, {Schulze},
  {Andreev}, {Antonelli}, {Bikmaev}, {Biryukov}, {B{\"o}ttcher}, {Burenin},
  {Castro Cer{\'o}n}, {Castro-Tirado}, {Chincarini}, {Cobb}, {Covino},
  {D'Avanzo}, {D'Elia}, {Della Valle}, {de Ugarte Postigo}, {Efimov},
  {Ferrero}, {Fugazza}, {Fynbo}, {G{\aa}lfalk}, {Grundahl}, {Gorosabel},
  {Gupta}, {Guziy}, {Hafizov}, {Hjorth}, {Holhjem}, {Ibrahimov}, {Im},
  {Israel}, {Je{\'l}inek}, {Jensen}, {Karimov}, {Khamitov}, {Kizilo{\v g}lu},
  {Klunko}, {Kub{\'a}nek}, {Kutyrev}, {Laursen}, {Levan}, {Mannucci}, {Martin},
  {Mescheryakov}, {Mirabal}, {Norris}, {Ovaldsen}, {Paraficz}, {Pavlenko},
  {Piranomonte}, {Rossi}, {Rumyantsev}, {Salinas}, {Sergeev}, {Sharapov},
  {Sollerman}, {Stecklum}, {Stella}, {Tagliaferri}, {Tanvir}, {Telting},
  {Testa}, {Updike}, {Volnova}, {Watson}, {Wiersema}, \& {Xu}}]{Kann2010ApJ}
{Kann}, D.~A., {Klose}, S., {Zhang}, B., {et~al.} 2010{\natexlab{b}}, \apj,
  720, 1513

\bibitem[{{Kann} \& {Laux}(2009)}]{Kann2009GCN10077}
{Kann}, D.~A. \& {Laux}, U. 2009, GRB Coordinates Network, 10077

\bibitem[{{Kann} {et~al.}(2008{\natexlab{a}}){Kann}, {Laux}, \&
  {Ertel}}]{Kann2008GCN7865}
{Kann}, D.~A., {Laux}, U., \& {Ertel}, S. 2008{\natexlab{a}}, GRB Coordinates
  Network, 7865

\bibitem[{{Kann} {et~al.}(2008{\natexlab{b}}){Kann}, {Laux}, \&
  {Ertel}}]{Kann2008GCN7823}
{Kann}, D.~A., {Laux}, U., \& {Ertel}, S. 2008{\natexlab{b}}, GRB Coordinates
  Network, 7823

\bibitem[{{Kann} {et~al.}(2008{\natexlab{c}}){Kann}, {Laux}, \&
  {Ertel}}]{Kann2008GCN7864}
{Kann}, D.~A., {Laux}, U., \& {Ertel}, S. 2008{\natexlab{c}}, GRB Coordinates
  Network, 7864

\bibitem[{{Kann} {et~al.}(2008{\natexlab{d}}){Kann}, {Laux}, \&
  {Ertel}}]{Kann2008GCN7845}
{Kann}, D.~A., {Laux}, U., \& {Ertel}, S. 2008{\natexlab{d}}, GRB Coordinates
  Network, 7845

\bibitem[{{Kann} {et~al.}(2008{\natexlab{e}}){Kann}, {Laux}, {Klose}, {Ertel},
  \& {Greiner}}]{Kann2008GCN7829}
{Kann}, D.~A., {Laux}, U., {Klose}, S., {Ertel}, S., \& {Greiner}, J.
  2008{\natexlab{e}}, GRB Coordinates Network, 7829

\bibitem[{{Kann} {et~al.}(2009{\natexlab{a}}){Kann}, {Laux}, {Roeder}, \&
  {Meusinger}}]{Kann2009GCN10076}
{Kann}, D.~A., {Laux}, U., {Roeder}, M., \& {Meusinger}, H. 2009{\natexlab{a}},
  GRB Coordinates Network, 10076

\bibitem[{{Kann} {et~al.}(2009{\natexlab{b}}){Kann}, {Laux}, {Roeder}, \&
  {Meusinger}}]{Kann2009GCN10090}
{Kann}, D.~A., {Laux}, U., {Roeder}, M., \& {Meusinger}, H. 2009{\natexlab{b}},
  GRB Coordinates Network, 10090

\bibitem[{{Kann} {et~al.}(2010{\natexlab{c}}){Kann}, {Laux}, \&
  {Stecklum}}]{Kann2010GCN11236}
{Kann}, D.~A., {Laux}, U., \& {Stecklum}, B. 2010{\natexlab{c}}, GRB
  Coordinates Network, 11236

\bibitem[{{Kann} {et~al.}(2010{\natexlab{d}}){Kann}, {Ludwig}, \&
  {Stecklum}}]{Kann2010GCN11246}
{Kann}, D.~A., {Ludwig}, F., \& {Stecklum}, B. 2010{\natexlab{d}}, GRB
  Coordinates Network, 11246

\bibitem[{{Kann} {et~al.}(2007){Kann}, {Masetti}, \& {Klose}}]{Kann2007AJ}
{Kann}, D.~A., {Masetti}, N., \& {Klose}, S. 2007, \aj, 133, 1187

\bibitem[{{Kann} {et~al.}(2010{\natexlab{e}}){Kann}, {Nicuesa Guelbenzu},
  {Ludwig}, \& {Stecklum}}]{Kann2010GCN11247}
{Kann}, D.~A., {Nicuesa Guelbenzu}, A., {Ludwig}, F., \& {Stecklum}, B.
  2010{\natexlab{e}}, GRB Coordinates Network, 11247

\bibitem[{{Kann} {et~al.}(2010{\natexlab{f}}){Kann}, {Nicuesa Guelbenzu},
  {Ludwig}, \& {Stecklum}}]{Kann2010GCN11238}
{Kann}, D.~A., {Nicuesa Guelbenzu}, A., {Ludwig}, F., \& {Stecklum}, B.
  2010{\natexlab{f}}, GRB Coordinates Network, 11238

\bibitem[{{Kann} {et~al.}(2021{\natexlab{c}}){Kann}, {Rossi}, {Oates}, {Klose},
  {Blazek}, {Ag{\"u}{\'\i} Fern{\'a}ndez}, {de Ugarte Postigo}, \&
  {Th{\"o}ne}}]{Kann2021AA}
{Kann}, D.~A., {Rossi}, A., {Oates}, S.~R., {et~al.} 2021{\natexlab{c}}, arXiv
  e-prints, arXiv:2110.00110

\bibitem[{{Kann} {et~al.}(2018{\natexlab{b}}){Kann}, {Schady}, {Olivares},
  {Klose}, {Rossi}, {Perley}, {Zhang}, {Kr{\"u}hler}, {Greiner}, {Nicuesa
  Guelbenzu}, {Elliott}, {Knust}, {Cano}, {Filgas}, {Pian}, {Mazzali}, {Fynbo},
  {Leloudas}, {Afonso}, {Delvaux}, {Graham}, {Rau}, {Schmidl}, {Schulze},
  {Tanga}, {Updike}, \& {Varela}}]{Kann2018AA}
{Kann}, D.~A., {Schady}, P., {Olivares}, E.~F., {et~al.} 2018{\natexlab{b}},
  \aap, 617, A122

\bibitem[{{Kann} {et~al.}(2019{\natexlab{a}}){Kann}, {Schady}, {Olivares E.},
  {Klose}, {Rossi}, {Perley}, {Kr{\"u}hler}, {Greiner}, {Nicuesa Guelbenzu},
  {Elliott}, {Knust}, {Filgas}, {Pian}, {Mazzali}, {Fynbo}, {Leloudas},
  {Afonso}, {Delvaux}, {Graham}, {Rau}, {Schmidl}, {Schulze}, {Tanga},
  {Updike}, \& {Varela}}]{Kann2019AA}
{Kann}, D.~A., {Schady}, P., {Olivares E.}, F., {et~al.} 2019{\natexlab{a}},
  \aap, 624, A143

\bibitem[{{Kann} {et~al.}(2019{\natexlab{b}}){Kann}, {Schady}, {Olivares E.},
  {Klose}, {Rossi}, {Perley}, {Kr{\"u}hler}, {Greiner}, {Nicuesa Guelbenzu},
  {Elliott}, {Knust}, {Filgas}, {Pian}, {Mazzali}, {Fynbo}, {Leloudas},
  {Afonso}, {Delvaux}, {Graham}, {Rau}, {Schmidl}, {Schulze}, {Tanga},
  {Updike}, \& {Varela}}]{Kann2017AA_SN2011kl}
{Kann}, D.~A., {Schady}, P., {Olivares E.}, F., {et~al.} 2019{\natexlab{b}},
  \aap, 624, A143

\bibitem[{{Kann} {et~al.}(2012){Kann}, {Stecklum}, \&
  {Laux}}]{Kann2012GCN13337}
{Kann}, D.~A., {Stecklum}, B., \& {Laux}, U. 2012, GRB Coordinates Network,
  13337

\bibitem[{{Kann} {et~al.}(2019{\natexlab{c}}){Kann}, {Thoene}, {Selsing},
  {Izzo}, {de Ugarte Postigo}, {Pugliese}, {Sbarufatti}, {Heintz}, {D'Elia},
  {Covino}, {Wiersema}, {Perley}, {Vergani}, {Fynbo}, {Watson}, {Tanvir},
  {Hartmann}, {Xu}, {Schulze}, \& {Bolmer}}]{Kann2019GCN23710}
{Kann}, D.~A., {Thoene}, C.~C., {Selsing}, J., {et~al.} 2019{\natexlab{c}}, GRB
  Coordinates Network, 23710

\bibitem[{{Kawai} {et~al.}(2006){Kawai}, {Kosugi}, {Aoki}, {Yamada}, {Totani},
  {Ohta}, {Iye}, {Hattori}, {Aoki}, {Furusawa}, {Hurley}, {Kawabata},
  {Kobayashi}, {Komiyama}, {Mizumoto}, {Nomoto}, {Noumaru}, {Ogasawara},
  {Sato}, {Sekiguchi}, {Shirasaki}, {Suzuki}, {Takata}, {Tamagawa}, {Terada},
  {Watanabe}, {Yatsu}, \& {Yoshida}}]{Kawai2006Nature}
{Kawai}, N., {Kosugi}, G., {Aoki}, K., {et~al.} 2006, \nat, 440, 184

\bibitem[{{Keel} {et~al.}(2013){Keel}, {Hartmann}, \&
  {Kaur}}]{Keel2013GCN14507}
{Keel}, W.~C., {Hartmann}, D., \& {Kaur}, A. 2013, GCN Circulars, 14507

\bibitem[{{Kelemen}(2009)}]{Kelemen2009GCN10028}
{Kelemen}, J. 2009, GRB Coordinates Network, 1028

\bibitem[{{Kennedy}(2014)}]{Kennedy2014GCN16196}
{Kennedy}, M. 2014, GRB Coordinates Network, 16196

\bibitem[{{Kennedy} \& {Garnavich}(2014)}]{Kennedy2014GCN16201}
{Kennedy}, M. \& {Garnavich}, P. 2014, GRB Coordinates Network, 16201

\bibitem[{{Khorunzhev} {et~al.}(2013){Khorunzhev}, {Burenin}, {Pavlinsky},
  {Sunyaev}, {Bikmaev}, {Sakhibullin}, {Khamitov}, \&
  {Kirbiyik}}]{Khorunzhev2013GCN15244}
{Khorunzhev}, G., {Burenin}, R., {Pavlinsky}, M., {et~al.} 2013, GRB
  Coordinates Network, 15244

\bibitem[{{Kim} \& {Im}(2019)}]{Kim2019GCN23732}
{Kim}, J. \& {Im}, M. 2019, GRB Coordinates Network, 23732

\bibitem[{{Kim} {et~al.}(2019){Kim}, {Im}, {Lee}, {Kim}, {de Ugrate Postigo},
  \& {Castro-Tirado}}]{Kim2019GCN23734}
{Kim}, J., {Im}, M., {Lee}, C.~U., {et~al.} 2019, GRB Coordinates Network,
  23734

\bibitem[{{King} {et~al.}(2014){King}, {Blinov}, {Giannios}, {Papadakis},
  {Angelakis}, {Balokovic}, {Fuhrmann}, {Hovatta}, {Khodade}, {Kiehlmann},
  {Kylafis}, {Kus}, {Myserlis}, {Modi}, {Panopoulou}, {Papamastorakis},
  {Pavlidou}, {Pazderska}, {Pazderski}, {Pearson}, {Rajarshi}, {Ramaprakash},
  {Readhead}, {Reig}, {Tassis}, \& {Zensus}}]{King2014MNRAS}
{King}, O.~G., {Blinov}, D., {Giannios}, D., {et~al.} 2014, \mnras, 445, L114

\bibitem[{{Kinugasa} {et~al.}(2009{\natexlab{a}}){Kinugasa}, {Honda},
  {Hashimoto}, {Takahashi}, \& {Taguchi}}]{Kinugasa2009GCN9292}
{Kinugasa}, K., {Honda}, S., {Hashimoto}, O., {Takahashi}, H., \& {Taguchi}, H.
  2009{\natexlab{a}}, GRB Coordinates Network, 9292

\bibitem[{{Kinugasa} {et~al.}(2009{\natexlab{b}}){Kinugasa}, {Honda},
  {Hashimoto}, {Takahashi}, \& {Taguchi}}]{Kinugasa2009GCN10248}
{Kinugasa}, K., {Honda}, S., {Hashimoto}, O., {Takahashi}, H., \& {Taguchi}, H.
  2009{\natexlab{b}}, GCN Circulars, 10248

\bibitem[{{Kinugasa} {et~al.}(2009{\natexlab{c}}){Kinugasa}, {Honda},
  {Takahashi}, {Taguchi}, \& {Hashimoto}}]{Kinugasa2009GCN10275}
{Kinugasa}, K., {Honda}, S., {Takahashi}, H., {Taguchi}, H., \& {Hashimoto}, O.
  2009{\natexlab{c}}, GRB Coordinates Network, 10275

\bibitem[{{Kinugasa} {et~al.}(2010){Kinugasa}, {Honda}, {Takahashi}, {Taguchi},
  \& {Hashimoto}}]{Kinugasa2010GCN10452}
{Kinugasa}, K., {Honda}, S., {Takahashi}, H., {Taguchi}, H., \& {Hashimoto}, O.
  2010, GRB Coordinates Network, 10452

\bibitem[{{Kinugawa} {et~al.}(2019){Kinugawa}, {Harikane}, \&
  {Asano}}]{Kinugawa2019ApJ}
{Kinugawa}, T., {Harikane}, Y., \& {Asano}, K. 2019, \apj, 878, 128

\bibitem[{{Kistler} {et~al.}(2008){Kistler}, {Y{\"u}ksel}, {Beacom}, \&
  {Stanek}}]{Kistler2008ApJ}
{Kistler}, M.~D., {Y{\"u}ksel}, H., {Beacom}, J.~F., \& {Stanek}, K.~Z. 2008,
  \apjl, 673, L119

\bibitem[{{Klein} {et~al.}(2010){Klein}, {Morgan}, {Perley}, \&
  {Bloom}}]{Klein2010GCN10627}
{Klein}, C.~R., {Morgan}, A.~N., {Perley}, D.~A., \& {Bloom}, J.~S. 2010, GRB
  Coordinates Network, 10627

\bibitem[{{Klose} {et~al.}(2019){Klose}, {Schmidl}, {Kann}, {Nicuesa
  Guelbenzu}, {Schulze}, {Greiner}, {Olivares E.}, {Kr{\"u}hler}, {Schady},
  {Afonso}, {Filgas}, {Fynbo}, {Rau}, {Rossi}, {Takats}, {Tanga}, {Updike}, \&
  {Varela}}]{Klose2019AA}
{Klose}, S., {Schmidl}, S., {Kann}, D.~A., {et~al.} 2019, \aap, 622, A138

\bibitem[{{Klotz} {et~al.}(2008{\natexlab{a}}){Klotz}, {Boer}, \&
  {Atteia}}]{Klotz2008GCN7595}
{Klotz}, A., {Boer}, M., \& {Atteia}, J.~L. 2008{\natexlab{a}}, GRB Coordinates
  Network, 7595

\bibitem[{{Klotz} {et~al.}(2008{\natexlab{b}}){Klotz}, {Boer}, \&
  {Atteia}}]{Klotz2008GCN7799}
{Klotz}, A., {Boer}, M., \& {Atteia}, J.~L. 2008{\natexlab{b}}, GRB Coordinates
  Network, 7799

\bibitem[{{Klotz} {et~al.}(2008{\natexlab{c}}){Klotz}, {Boer}, \&
  {Atteia}}]{Klotz2008GCN7795}
{Klotz}, A., {Boer}, M., \& {Atteia}, J.~L. 2008{\natexlab{c}}, GRB Coordinates
  Network, 7795

\bibitem[{{Klotz} {et~al.}(2012{\natexlab{a}}){Klotz}, {Gendre}, \&
  {Boer}}]{Klotz2012GCN13132}
{Klotz}, A., {Gendre}, B., \& {Boer}, M. 2012{\natexlab{a}}, GRB Coordinates
  Network, 13132

\bibitem[{{Klotz} {et~al.}(2009){Klotz}, {Gendre}, {Boer}, \&
  {Atteia}}]{Klotz2009GCN10200}
{Klotz}, A., {Gendre}, B., {Boer}, M., \& {Atteia}, J.~L. 2009, GCN Circulars,
  10200

\bibitem[{{Klotz} {et~al.}(2011){Klotz}, {Gendre}, {Boer}, \&
  {Atteia}}]{Klotz2011GCN12758}
{Klotz}, A., {Gendre}, B., {Boer}, M., \& {Atteia}, J.~L. 2011, GRB Coordinates
  Network, 12758

\bibitem[{{Klotz} {et~al.}(2012{\natexlab{b}}){Klotz}, {Gendre}, {Boer}, \&
  {Atteia}}]{Klotz2012GCN12883}
{Klotz}, A., {Gendre}, B., {Boer}, M., \& {Atteia}, J.~L. 2012{\natexlab{b}},
  GRB Coordinates Network, 12883

\bibitem[{{Klotz} {et~al.}(2012{\natexlab{c}}){Klotz}, {Gendre}, {Boer}, \&
  {Atteia}}]{Klotz2012GCN13108}
{Klotz}, A., {Gendre}, B., {Boer}, M., \& {Atteia}, J.~L. 2012{\natexlab{c}},
  GRB Coordinates Network, 13108

\bibitem[{{Klotz} {et~al.}(2012{\natexlab{d}}){Klotz}, {Gendre}, {Boer}, \&
  {Atteia}}]{Klotz2012GCN13887}
{Klotz}, A., {Gendre}, B., {Boer}, M., \& {Atteia}, J.~L. 2012{\natexlab{d}},
  GRB Coordinates Network, 13887

\bibitem[{{Klotz} {et~al.}(2013){Klotz}, {Gendre}, {Boer}, {Siellez}, {Dereli},
  {Bardho}, \& {Atteia}}]{Klotz2013GCN14818}
{Klotz}, A., {Gendre}, B., {Boer}, M., {et~al.} 2013, GRB Coordinates Network,
  14818

\bibitem[{{Klotz} {et~al.}(2014){Klotz}, {Turpin}, {MacPherson}, {Coward},
  {Boer}, {Gendre}, {Siellez}, {Dereli}, {Bardho}, {Williams}, \&
  {Martin}}]{Klotz2014GCN15952}
{Klotz}, A., {Turpin}, D., {MacPherson}, D., {et~al.} 2014, GRB Coordinates
  Network, 15952

\bibitem[{{Klunko} \& {Pozanenko}(2008)}]{Klunko2008GCN7890}
{Klunko}, E. \& {Pozanenko}, A. 2008, GRB Coordinates Network, 7890

\bibitem[{{Kocevski} {et~al.}(2006{\natexlab{a}}){Kocevski}, {Bloom}, \&
  {McGrath}}]{Kocevski2006GCN4528}
{Kocevski}, D., {Bloom}, J.~S., \& {McGrath}, E.~J. 2006{\natexlab{a}}, GRB
  Coordinates Network, 4528

\bibitem[{{Kocevski} {et~al.}(2006{\natexlab{b}}){Kocevski}, {Bloom}, \&
  {McGrath}}]{Kocevski2006GCN4540}
{Kocevski}, D., {Bloom}, J.~S., \& {McGrath}, E.~J. 2006{\natexlab{b}}, GRB
  Coordinates Network, 4540

\bibitem[{{Kong}(2018)}]{Kong2018GCN23475}
{Kong}, A.~K.~H. 2018, GRB Coordinates Network, 23475

\bibitem[{{Kopac} {et~al.}(2010){Kopac}, {Dintinjana}, \&
  {Gomboc}}]{Kopac2010GCN11177}
{Kopac}, D., {Dintinjana}, B., \& {Gomboc}, A. 2010, GRB Coordinates Network,
  11177

\bibitem[{{Kopa{\v{c}}} {et~al.}(2015){Kopa{\v{c}}}, {Mundell}, {Japelj},
  {Arnold}, {Steele}, {Guidorzi}, {Dichiara}, {Kobayashi}, {Gomboc},
  {Harrison}, {Lamb}, {Melandri}, {Smith}, {Virgili}, {Castro-Tirado},
  {Gorosabel}, {J{\"a}rvinen}, {S{\'a}nchez-Ram{\'\i}rez}, {Oates}, \&
  {Jel{\'\i}nek}}]{Kopac2015ApJ}
{Kopa{\v{c}}}, D., {Mundell}, C.~G., {Japelj}, J., {et~al.} 2015, \apj, 813, 1

\bibitem[{{Kouveliotou} {et~al.}(1993){Kouveliotou}, {Meegan}, {Fishman},
  {Bhat}, {Briggs}, {Koshut}, {Paciesas}, \& {Pendleton}}]{Kouveliotou1993ApJ}
{Kouveliotou}, C., {Meegan}, C.~A., {Fishman}, G.~J., {et~al.} 1993, \apjl,
  413, L101

\bibitem[{{Kr{\"u}hler} {et~al.}(2012){Kr{\"u}hler}, {Fynbo}, {Geier},
  {Hjorth}, {Malesani}, {Milvang-Jensen}, {Levan}, {Sparre}, {Watson}, \&
  {Zafar}}]{Kruhler2012AA}
{Kr{\"u}hler}, T., {Fynbo}, J.~P.~U., {Geier}, S., {et~al.} 2012, \aap, 546, A8

\bibitem[{{Kr{\"u}hler} {et~al.}(2011{\natexlab{a}}){Kr{\"u}hler}, {Greiner},
  {Schady}, {Savaglio}, {Afonso}, {Clemens}, {Elliott}, {Filgas}, {Gruber},
  {Kann}, {Klose}, {K{\"u}pc{\"u}-Yolda{\c{s}}}, {McBreen}, {Olivares},
  {Pierini}, {Rau}, {Rossi}, {Nardini}, {Nicuesa Guelbenzu}, {Sudilovsky}, \&
  {Updike}}]{Kruehler2011AA}
{Kr{\"u}hler}, T., {Greiner}, J., {Schady}, P., {et~al.} 2011{\natexlab{a}},
  \aap, 534, A108

\bibitem[{{Kr{\"u}hler} {et~al.}(2013){Kr{\"u}hler}, {Ledoux}, {Fynbo},
  {Vreeswijk}, {Schmidl}, {Malesani}, {Christensen}, {De Cia}, {Hjorth},
  {Jakobsson}, {Kann}, {Kaper}, {Vergani}, {Afonso}, {Covino}, {de Ugarte
  Postigo}, {D'Elia}, {Filgas}, {Goldoni}, {Greiner}, {Hartoog},
  {Milvang-Jensen}, {Nardini}, {Piranomonte}, {Rossi},
  {S{\'a}nchez-Ram{\'\i}rez}, {Schady}, {Schulze}, {Sudilovsky}, {Tanvir},
  {Tagliaferri}, {Watson}, {Wiersema}, {Wijers}, \& {Xu}}]{Kruhler2013AA}
{Kr{\"u}hler}, T., {Ledoux}, C., {Fynbo}, J.~P.~U., {et~al.} 2013, \aap, 557,
  A18

\bibitem[{{Kr{\"u}hler} {et~al.}(2011{\natexlab{b}}){Kr{\"u}hler}, {Schady},
  {Greiner}, {Afonso}, {Bottacini}, {Clemens}, {Filgas}, {Klose}, {Koch},
  {K{\"u}pc{\"u}-Yolda{\c{s}}}, {Oates}, {Olivares E.}, {Page}, {McBreen},
  {Nardini}, {Nicuesa Guelbenzu}, {Rau}, {Roming}, {Rossi}, {Updike}, \&
  {Yolda{\c{s}}}}]{Kruhler2011AA}
{Kr{\"u}hler}, T., {Schady}, P., {Greiner}, J., {et~al.} 2011{\natexlab{b}},
  \aap, 526, A153

\bibitem[{{Krushinski} {et~al.}(2011){Krushinski}, {Zalozhnich}, {Popov},
  {Sinykov}, {Yurkov}, {Sergienko}, {Varda}, {Gorbovskoy}, {Lipunov},
  {Kornilov}, {Kuvshinov}, {Belinski}, {Tyurina}, {Shatskiy}, {Balanutsa},
  {Chazov}, {Kuznetsov}, {Zimnukhov}, {Kornilov}, {Sankovich}, {Shurpakov},
  {Ivanov}, {Poleshchuk}, {Yazev}, {Budnev}, {Gres}, {Chuvalaev},
  {Konstantinov}, {Tlatov}, {Parhomenko}, {Dormidontov}, \&
  {Sennik}}]{Krushinski2011GCN12789}
{Krushinski}, V., {Zalozhnich}, I., {Popov}, A., {et~al.} 2011, GRB Coordinates
  Network, 12789

\bibitem[{{Kubanek} {et~al.}(2013){Kubanek}, {Topinka}, {Hanhol}, \&
  {Meehan}}]{Kubanek2013GCN15403}
{Kubanek}, P., {Topinka}, M., {Hanhol}, L., \& {Meehan}, S. 2013, GRB
  Coordinates Network, 15403

\bibitem[{{Kugel}(2011)}]{Kugel2011GCN11647}
{Kugel}, F. 2011, GRB Coordinates Network, 11647

\bibitem[{{Kuin} \& {D'Elia}(2011)}]{Kuin2011GCN11718}
{Kuin}, N.~P.~M. \& {D'Elia}, V. 2011, GRB Coordinates Network, 11718

\bibitem[{{Kuin} {et~al.}(2009){Kuin}, {Landsman}, {Page}, {Schady}, {Still},
  {Breeveld}, {de Pasquale}, {Roming}, {Brown}, {Carter}, {James}, {Curran},
  {Cucchiara}, {Gronwall}, {Holland}, {Hoversten}, {Hunsberger}, {Kennedy},
  {Koch}, {Lamoureux}, {Marshall}, {Oates}, {Parsons}, {Palmer}, \&
  {Smith}}]{Kuin2009MNRAS}
{Kuin}, N.~P.~M., {Landsman}, W., {Page}, M.~J., {et~al.} 2009, \mnras, 395,
  L21

\bibitem[{{Kuin} \& {Mangano}(2008)}]{Kuin2008GCN7808}
{Kuin}, N.~P.~M. \& {Mangano}, V. 2008, GRB Coordinates Network, 7808

\bibitem[{{Kuin} \& {Marshall}(2014)}]{Kuin2014GCN16130}
{Kuin}, N.~P.~M. \& {Marshall}, F.~E. 2014, GRB Coordinates Network, 16130

\bibitem[{{Kuin} {et~al.}(2008){Kuin}, {Sbarufatti}, {Marshall}, \&
  {Schady}}]{Kuin2008GCN7844}
{Kuin}, N.~P.~M., {Sbarufatti}, B., {Marshall}, F., \& {Schady}, P. 2008, GRB
  Coordinates Network, 7844

\bibitem[{{Kuin} \& {Swift/UVOT Team}(2019)}]{Kuin2019GCN26538}
{Kuin}, N.~P.~M. \& {Swift/UVOT Team}. 2019, GRB Coordinates Network, 26538

\bibitem[{{Kuin} \& {Swift/UVOT Team}(2021)}]{Kuin2021GCN30355}
{Kuin}, N.~P.~M. \& {Swift/UVOT Team}. 2021, GRB Coordinates Network, 30355

\bibitem[{{Kumar} {et~al.}(2019){Kumar}, {Singh}, {Sahu}, \&
  {Anupama}}]{Kumar2019ATel12383}
{Kumar}, B., {Singh}, A., {Sahu}, D.~K., \& {Anupama}, G.~C. 2019, The
  Astronomer's Telegram, 12383

\bibitem[{{Kumar} \& {Zhang}(2015)}]{Kumar2015PhR}
{Kumar}, P. \& {Zhang}, B. 2015, \physrep, 561, 1

\bibitem[{{Kuroda} {et~al.}(2011{\natexlab{a}}){Kuroda}, {Hanayama}, {Miyaji},
  {Watanabe}, {Yanagisawa}, {Nagayama}, {Yoshida}, {Ohta}, \&
  {Kawai}}]{Kuroda2011GCN11972}
{Kuroda}, D., {Hanayama}, H., {Miyaji}, T., {et~al.} 2011{\natexlab{a}}, GRB
  Coordinates Network, 11972

\bibitem[{{Kuroda} {et~al.}(2011{\natexlab{b}}){Kuroda}, {Hanayama}, {Miyaji},
  {Watanabe}, {Yanagisawa}, {Nagayama}, {Yoshida}, {Ohta}, \&
  {Kawai}}]{Kuroda2011GCN12204}
{Kuroda}, D., {Hanayama}, H., {Miyaji}, T., {et~al.} 2011{\natexlab{b}}, GRB
  Coordinates Network, 12204

\bibitem[{{Kuroda} {et~al.}(2013{\natexlab{a}}){Kuroda}, {Hanayama}, {Miyaji},
  {Watanabe}, {Yanagisawa}, {Nagayama}, {Yoshida}, {Ohta}, \&
  {Kawai}}]{Kuroda2013GCN14513}
{Kuroda}, D., {Hanayama}, H., {Miyaji}, T., {et~al.} 2013{\natexlab{a}}, GCN
  Circulars, 14513

\bibitem[{{Kuroda} {et~al.}(2014{\natexlab{a}}){Kuroda}, {Hanayama}, {Miyaji},
  {Watanabe}, {Yanagisawa}, {Nagayama}, {Yoshida}, {Ohta}, \&
  {Kawai}}]{Kuroda2014GCN16132}
{Kuroda}, D., {Hanayama}, H., {Miyaji}, T., {et~al.} 2014{\natexlab{a}}, GRB
  Coordinates Network, 16132

\bibitem[{{Kuroda} {et~al.}(2014{\natexlab{b}}){Kuroda}, {Hanayama}, {Miyaji},
  {Watanabe}, {Yanagisawa}, {Nagayama}, {Yoshida}, {Ohta}, \&
  {Kawai}}]{Kuroda2014GCN16488}
{Kuroda}, D., {Hanayama}, H., {Miyaji}, T., {et~al.} 2014{\natexlab{b}}, GRB
  Coordinates Network, 16488

\bibitem[{{Kuroda} {et~al.}(2015{\natexlab{a}}){Kuroda}, {Hanayama}, {Miyaji},
  {Watanabe}, {Yanagisawa}, {Nagayama}, {Yoshida}, {Ohta}, \&
  {Kawai}}]{Kuroda2015GCN18276}
{Kuroda}, D., {Hanayama}, H., {Miyaji}, T., {et~al.} 2015{\natexlab{a}}, GRB
  Coordinates Network, 18276

\bibitem[{{Kuroda} {et~al.}(2015{\natexlab{b}}){Kuroda}, {Hanayama}, {Miyaji},
  {Watanabe}, {Yanagisawa}, {Nagayama}, {Yoshida}, {Ohta}, \&
  {Kawai}}]{Kuroda2015GCN18301}
{Kuroda}, D., {Hanayama}, H., {Miyaji}, T., {et~al.} 2015{\natexlab{b}}, GRB
  Coordinates Network, 18301

\bibitem[{{Kuroda} {et~al.}(2010{\natexlab{a}}){Kuroda}, {Hanayama}, {Miyaji},
  {Watanabe}, {Yanagisawa}, {Yoshida}, {Ohta}, \& {Kawai}}]{Kuroda2010GCN11249}
{Kuroda}, D., {Hanayama}, H., {Miyaji}, T., {et~al.} 2010{\natexlab{a}}, GRB
  Coordinates Network, 11249

\bibitem[{{Kuroda} {et~al.}(2010{\natexlab{b}}){Kuroda}, {Hanayama}, {Miyaji},
  {Watanabe}, {Yanagisawa}, {Yoshida}, {Ohta}, \& {Kawai}}]{Kuroda2010GCN11205}
{Kuroda}, D., {Hanayama}, H., {Miyaji}, T., {et~al.} 2010{\natexlab{b}}, GRB
  Coordinates Network, 11205

\bibitem[{{Kuroda} {et~al.}(2011{\natexlab{c}}){Kuroda}, {Hanayama}, {Miyaji},
  {Watanabe}, {Yanagisawa}, {Yoshida}, {Ohta}, \& {Kawai}}]{Kuroda2011GCN11652}
{Kuroda}, D., {Hanayama}, H., {Miyaji}, T., {et~al.} 2011{\natexlab{c}}, GRB
  Coordinates Network, 11652

\bibitem[{{Kuroda} {et~al.}(2010{\natexlab{c}}){Kuroda}, {Yanagisawa},
  {Shimizu}, {Toda}, {Nagayama}, {Yoshida}, {Ohta}, \&
  {Kawai}}]{Kuroda2010GCN11172}
{Kuroda}, D., {Yanagisawa}, K., {Shimizu}, Y., {et~al.} 2010{\natexlab{c}}, GRB
  Coordinates Network, 11172

\bibitem[{{Kuroda} {et~al.}(2010{\natexlab{d}}){Kuroda}, {Yanagisawa},
  {Shimizu}, {Toda}, {Nagayama}, {Yoshida}, {Ohta}, \&
  {Kawai}}]{Kuroda2010GCN11189}
{Kuroda}, D., {Yanagisawa}, K., {Shimizu}, Y., {et~al.} 2010{\natexlab{d}}, GRB
  Coordinates Network, 11189

\bibitem[{{Kuroda} {et~al.}(2010{\natexlab{e}}){Kuroda}, {Yanagisawa},
  {Shimizu}, {Toda}, {Nagayama}, {Yoshida}, {Ohta}, \&
  {Kawai}}]{Kuroda2010GCN11241}
{Kuroda}, D., {Yanagisawa}, K., {Shimizu}, Y., {et~al.} 2010{\natexlab{e}}, GRB
  Coordinates Network, 11241

\bibitem[{{Kuroda} {et~al.}(2011{\natexlab{d}}){Kuroda}, {Yanagisawa},
  {Shimizu}, {Toda}, {Nagayama}, {Yoshida}, {Ohta}, \&
  {Kawai}}]{Kuroda2011GCN11651}
{Kuroda}, D., {Yanagisawa}, K., {Shimizu}, Y., {et~al.} 2011{\natexlab{d}}, GRB
  Coordinates Network, 11651

\bibitem[{{Kuroda} {et~al.}(2011{\natexlab{e}}){Kuroda}, {Yanagisawa},
  {Shimizu}, {Toda}, {Nagayama}, {Yoshida}, {Ohta}, \&
  {Kawai}}]{Kuroda2011GCN11719}
{Kuroda}, D., {Yanagisawa}, K., {Shimizu}, Y., {et~al.} 2011{\natexlab{e}}, GRB
  Coordinates Network, 11719

\bibitem[{{Kuroda} {et~al.}(2011{\natexlab{f}}){Kuroda}, {Yanagisawa},
  {Shimizu}, {Toda}, {Nagayama}, {Yoshida}, {Ohta}, \&
  {Kawai}}]{Kuroda2011GCN12743}
{Kuroda}, D., {Yanagisawa}, K., {Shimizu}, Y., {et~al.} 2011{\natexlab{f}}, GRB
  Coordinates Network, 12743

\bibitem[{{Kuroda} {et~al.}(2013{\natexlab{b}}){Kuroda}, {Yanagisawa},
  {Shimizu}, {Toda}, {Nagayama}, {Yoshida}, {Ohta}, \&
  {Kawai}}]{Kuroda2013GCN14465}
{Kuroda}, D., {Yanagisawa}, K., {Shimizu}, Y., {et~al.} 2013{\natexlab{b}}, GCN
  Circulars, 14465

\bibitem[{{Kuroda} {et~al.}(2014{\natexlab{c}}){Kuroda}, {Yanagisawa},
  {Shimizu}, {Toda}, {Nagayama}, {Yoshida}, {Ohta}, \&
  {Kawai}}]{Kuroda2014GCN16131}
{Kuroda}, D., {Yanagisawa}, K., {Shimizu}, Y., {et~al.} 2014{\natexlab{c}}, GRB
  Coordinates Network, 16131

\bibitem[{{Kuroda} {et~al.}(2014{\natexlab{d}}){Kuroda}, {Yanagisawa},
  {Shimizu}, {Toda}, {Nagayama}, {Yoshida}, {Ohta}, \&
  {Kawai}}]{Kuroda2014GCN16160}
{Kuroda}, D., {Yanagisawa}, K., {Shimizu}, Y., {et~al.} 2014{\natexlab{d}}, GRB
  Coordinates Network, 16160

\bibitem[{{Kuroda} {et~al.}(2015{\natexlab{c}}){Kuroda}, {Yanagisawa},
  {Shimizu}, {Toda}, {Nagayama}, {Yoshida}, {Ohta}, \&
  {Kawai}}]{Kuroda2015GCN18267}
{Kuroda}, D., {Yanagisawa}, K., {Shimizu}, Y., {et~al.} 2015{\natexlab{c}}, GRB
  Coordinates Network, 18267

\bibitem[{{Kuroda} {et~al.}(2010{\natexlab{f}}){Kuroda}, {Yanagisawa},
  {Shimizu}, {Toda}, {Nakajima}, {Yatsu}, {Mori}, {Endo}, {Shimokawabe},
  {Kawai}, {Nagayama}, {Yoshida}, \& {Ohta}}]{Kuroda2010GCN10440}
{Kuroda}, D., {Yanagisawa}, K., {Shimizu}, Y., {et~al.} 2010{\natexlab{f}}, GRB
  Coordinates Network, 10440

\bibitem[{{LaCluyze} {et~al.}(2012){LaCluyze}, {Haislip}, {Ivarsen}, {Maturi},
  {Reichart}, {Moore}, {Cromartie}, {Egger}, {Foster}, {Frank}, {Nysewander},
  {Oza}, {Speckhard}, {Trotter}, \& {Crain}}]{LaCluyze2012GCN13109}
{LaCluyze}, A., {Haislip}, J., {Ivarsen}, K., {et~al.} 2012, GRB Coordinates
  Network, 13109

\bibitem[{{LaCluyze} {et~al.}(2009{\natexlab{a}}){LaCluyze}, {Reichart},
  {Haislip}, {Ivarsen}, {Egger}, {Foster}, {Moore}, {Oza}, {Schubel},
  {Styblova}, {Trotter}, {Crain}, \& {Nysewander}}]{LaCluyze2009GCN10046}
{LaCluyze}, A., {Reichart}, D., {Haislip}, J., {et~al.} 2009{\natexlab{a}}, GRB
  Coordinates Network, 10046

\bibitem[{{LaCluyze} {et~al.}(2009{\natexlab{b}}){LaCluyze}, {Reichart},
  {Haislip}, {Ivarsen}, {Egger}, {Foster}, {Moore}, {Oza}, {Schubel},
  {Styblova}, {Trotter}, {Crain}, \& {Nysewander}}]{LaCluyze2009GCN10107}
{LaCluyze}, A., {Reichart}, D., {Haislip}, J., {et~al.} 2009{\natexlab{b}}, GRB
  Coordinates Network, 10107

\bibitem[{{LaCluyze} {et~al.}(2009{\natexlab{c}}){LaCluyze}, {Reichart},
  {Smith}, {Caton}, {Hawkins}, {Haislip}, {Ivarsen}, {Egger}, {Foster},
  {Moore}, {Oza}, {Schubel}, {Styblova}, {Trotter}, {Crain}, \&
  {Nysewander}}]{LaCluyze2009GCN10109}
{LaCluyze}, A., {Reichart}, D., {Smith}, A., {et~al.} 2009{\natexlab{c}}, GRB
  Coordinates Network, 10109

\bibitem[{{Lamb} {et~al.}(2021){Lamb}, {Kann}, {Fern{\'a}ndez}, {Mandel},
  {Levan}, \& {Tanvir}}]{Lamb2021MNRAS}
{Lamb}, G.~P., {Kann}, D.~A., {Fern{\'a}ndez}, J.~J., {et~al.} 2021, \mnras,
  506, 4163

\bibitem[{{Landsman} \& {Page}(2009)}]{Landsman2009GCN}
{Landsman}, W.~B. \& {Page}, K. 2009, GRB Coordinates Network, 9717

\bibitem[{{Landsman} \& {Stamatikos}(2009)}]{Landsman2009GCN10041}
{Landsman}, W.~B. \& {Stamatikos}, M. 2009, GRB Coordinates Network, 1041

\bibitem[{{Laskar} {et~al.}(2018{\natexlab{a}}){Laskar}, {Alexander}, {Berger},
  {Guidorzi}, {Margutti}, {Fong}, {Kilpatrick}, {Milne}, {Drout}, {Mundell},
  {Kobayashi}, {Lunnan}, {Barniol Duran}, {Menten}, {Ioka}, \&
  {Williams}}]{Laskar2018ApJ}
{Laskar}, T., {Alexander}, K.~D., {Berger}, E., {et~al.} 2018{\natexlab{a}},
  \apj, 862, 94

\bibitem[{{Laskar} {et~al.}(2018{\natexlab{b}}){Laskar}, {Berger}, {Chornock},
  {Margutti}, {Fong}, \& {Zauderer}}]{Laskar2018ApJ2}
{Laskar}, T., {Berger}, E., {Chornock}, R., {et~al.} 2018{\natexlab{b}}, \apj,
  858, 65

\bibitem[{{Laskar} {et~al.}(2015){Laskar}, {Berger}, {Margutti}, {Perley},
  {Zauderer}, {Sari}, \& {Fong}}]{Laskar2015ApJ}
{Laskar}, T., {Berger}, E., {Margutti}, R., {et~al.} 2015, \apj, 814, 1

\bibitem[{{Laskar} {et~al.}(2014){Laskar}, {Berger}, {Tanvir}, {Zauderer},
  {Margutti}, {Levan}, {Perley}, {Fong}, {Wiersema}, {Menten}, \&
  {Hrudkova}}]{Laskar2014ApJ}
{Laskar}, T., {Berger}, E., {Tanvir}, N., {et~al.} 2014, \apj, 781, 1

\bibitem[{{Laskar} {et~al.}(2013){Laskar}, {Berger}, {Zauderer}, {Margutti},
  {Soderberg}, {Chakraborti}, {Lunnan}, {Chornock}, {Chandra}, \&
  {Ray}}]{Laskar2013ApJ}
{Laskar}, T., {Berger}, E., {Zauderer}, B.~A., {et~al.} 2013, \apj, 776, 119

\bibitem[{{Laskar} {et~al.}(2019){Laskar}, {van Eerten}, {Schady}, {Mundell},
  {Alexander}, {Barniol Duran}, {Berger}, {Bolmer}, {Chornock}, {Coppejans},
  {Fong}, {Gomboc}, {Jordana-Mitjans}, {Kobayashi}, {Margutti}, {Menten},
  {Sari}, {Yamazaki}, {Lipunov}, {Gorbovskoy}, {Kornilov}, {Tyurina},
  {Zimnukhov}, {Podesta}, {Levato}, {Buckley}, {Tlatov}, {Rebolo}, \&
  {Serra-Ricart}}]{Laskar2019ApJ}
{Laskar}, T., {van Eerten}, H., {Schady}, P., {et~al.} 2019, \apj, 884, 121

\bibitem[{{Lee} {et~al.}(2010){Lee}, {Im}, \& {Urata}}]{LeeI2010}
{Lee}, I.-D., {Im}, M.-S., \& {Urata}, Y. 2010, Journal of Korean Astronomical
  Society, 43, 95

\bibitem[{{Leonini} {et~al.}(2013){Leonini}, {Guerrini}, {Rosi}, \& {Tinjaca
  Ramirez}}]{Leonini2013GCN15150}
{Leonini}, S., {Guerrini}, G., {Rosi}, P., \& {Tinjaca Ramirez}, L.~M. 2013,
  GRB Coordinates Network, 15150

\bibitem[{{Levan} {et~al.}(2014){Levan}, {Tanvir}, {Starling}, {Wiersema},
  {Page}, {Perley}, {Schulze}, {Wynn}, {Chornock}, {Hjorth}, {Cenko},
  {Fruchter}, {O'Brien}, {Brown}, {Tunnicliffe}, {Malesani}, {Jakobsson},
  {Watson}, {Berger}, {Bersier}, {Cobb}, {Covino}, {Cucchiara}, {de Ugarte
  Postigo}, {Fox}, {Gal-Yam}, {Goldoni}, {Gorosabel}, {Kaper}, {Kr{\"u}hler},
  {Karjalainen}, {Osborne}, {Pian}, {S{\'a}nchez-Ram{\'{\i}}rez}, {Schmidt},
  {Skillen}, {Tagliaferri}, {Th{\"o}ne}, {Vaduvescu}, {Wijers}, \&
  {Zauderer}}]{Levan2014ApJ}
{Levan}, A.~J., {Tanvir}, N.~R., {Starling}, R.~L.~C., {et~al.} 2014, \apj,
  781, 13

\bibitem[{{Levato} {et~al.}(2012){Levato}, {Saffe}, {Mallamaci}, {Lopez},
  {Podest}, {Denisenko}, {Kuznetsov}, {Lipunov}, {Gorbovskoy}, {Kornilov},
  {Kuvshinov}, {Belinski}, {Tyurina}, {Shatskiy}, {Balanutsa}, {Zimnukhov},
  {Chazov}, {Tlatov}, {Parhomenko}, {Dormidontov}, {Sennik}, {Ivanov}, {Yazev},
  {Budnev}, {Gres}, {Chuvalaev}, {Poleshchuk}, {Yurkov}, {Sergienko}, {Varda},
  {Sinyakov}, {Krushinski}, {Zalozhnich}, {Popov}, {Bourdanov}, \&
  {Punanova}}]{Levato2012GCN13443}
{Levato}, H., {Saffe}, C., {Mallamaci}, C., {et~al.} 2012, GRB Coordinates
  Network, 13443

\bibitem[{{Levesque} {et~al.}(2010){Levesque}, {Bloom}, {Butler}, {Perley},
  {Cenko}, {Prochaska}, {Kewley}, {Bunker}, {Chen}, {Chornock}, {Filippenko},
  {Glazebrook}, {Lopez}, {Masiero}, {Modjaz}, {Morgan}, \&
  {Poznanski}}]{Levesque2010MNRAS}
{Levesque}, E.~M., {Bloom}, J.~S., {Butler}, N.~R., {et~al.} 2010, \mnras, 401,
  963

\bibitem[{{Li} {et~al.}(2020){Li}, {Wang}, {Zheng}, {Pozanenko}, {Filippenko},
  {Qin}, {Wang}, {Jiang}, {Li}, {Lin}, {Liang}, {Volnova}, {Elenin}, {Klunko},
  {Inasaridze}, {Kusakin}, \& {Lu}}]{Li2020ApJ}
{Li}, L., {Wang}, X.-G., {Zheng}, W., {et~al.} 2020, \apj, 900, 176

\bibitem[{{Li}(2008)}]{Li2008MNRAS}
{Li}, L.-X. 2008, \mnras, 388, 1487

\bibitem[{{Liang} \& {Li}(2009)}]{LiangLi2009ApJ}
{Liang}, S.~L. \& {Li}, A. 2009, \apjl, 690, L56

\bibitem[{Lidz {et~al.}(2021)Lidz, Chang, Mas-Ribas, \& Sun}]{Lidz2021}
Lidz, A., Chang, T.-C., Mas-Ribas, L., \& Sun, G. 2021, The Astrophysical
  Journal, 917, 58

\bibitem[{{Lien} {et~al.}(2014){Lien}, {Sakamoto}, {Gehrels}, {Palmer},
  {Barthelmy}, {Graziani}, \& {Cannizzo}}]{Lien2014ApJ}
{Lien}, A., {Sakamoto}, T., {Gehrels}, N., {et~al.} 2014, \apj, 783, 24

\bibitem[{{Lipunov} {et~al.}(2018{\natexlab{a}}){Lipunov}, {Gorbovskoy},
  {Tiurina}, {Vlasenko}, {Kornilov}, {Kuznetsov}, {Chazov}, {Gorbunov},
  {Zimnukhov}, {Kuvshinov}, {Balanutsa}, {Vladimirov}, {Buckley}, {Rebolo},
  {Serra}, {Lodieu}, {Israelian}, {Suarez-Andres}, {Tlatov}, {Senik},
  {Dormidontov}, {Podesta}, {Podesta}, {Lopez}, {Francile}, {Levato}, {Gres},
  {Budnev}, {Ishmuhametova}, {Gabovich}, {Yurkov}, \&
  {Sergienko}}]{Lipunov2018GCN23023}
{Lipunov}, V., {Gorbovskoy}, E., {Tiurina}, N., {et~al.} 2018{\natexlab{a}},
  GRB Coordinates Network, 23023

\bibitem[{{Lipunov} {et~al.}(2018{\natexlab{b}}){Lipunov}, {Gorbovskoy},
  {Tyurina}, {Kornilov}, {Zimnukhov}, {Vladimirov}, {Krylov}, {Gorbunov},
  {Balanutsa}, {Kuznetsov}, {Chazov}, {Kuvshinov}, {Podesta}, {Lopez},
  {Podesta}, {Levato}, {Saffe}, {Juan}, {Rebolo}, {Serra-Ricart}, {Israelyan},
  {Buckley}, {Potter}, {Gres}, {Budnev}, {Tlatov}, {Senik}, {Dormidontov},
  {Yurkov}, {Gabovich}, \& {Sergienko}}]{Lipunov2018GCN22543}
{Lipunov}, V., {Gorbovskoy}, E., {Tyurina}, N., {et~al.} 2018{\natexlab{b}},
  GRB Coordinates Network, 22543

\bibitem[{{Lipunov} {et~al.}(2016){Lipunov}, {Gorosabel}, {Pruzhinskaya}, {de
  Ugarte Postigo}, {Pelassa}, {Tsvetkova}, {Sokolov}, {Kann}, {Xu},
  {Gorbovskoy}, {Krushinski}, {Kornilov}, {Balanutsa}, {Boronina}, {Budnev},
  {Cano}, {Castro-Tirado}, {Chazov}, {Connaughton}, {Delvaux}, {Frederiks},
  {Fynbo}, {Gabovich}, {Goldstein}, {Greiner}, {Gress}, {Ivanov}, {Jakobsson},
  {Klose}, {Knust}, {Komarova}, {Konstantinov}, {Krylov}, {Kuvshinov},
  {Kuznetsov}, {Lipunova}, {Moskvitin}, {Pal'shin}, {Pandey}, {Poleshchuk},
  {Schmidl}, {Sergienko}, {Sinyakov}, {Schulze}, {Sokolov}, {Sokolova},
  {Sparre}, {Th{\"o}ne}, {Tlatov}, {Tyurina}, {Ulanov}, {Yazev}, \&
  {Yurkov}}]{Lipunov2016MNRAS}
{Lipunov}, V.~M., {Gorosabel}, J., {Pruzhinskaya}, M.~V., {et~al.} 2016,
  \mnras, 455, 712

\bibitem[{{Littlefield} \& {Garnavich}(2018)}]{Littlefield2018GCN22538}
{Littlefield}, C. \& {Garnavich}, P. 2018, GRB Coordinates Network, 22538

\bibitem[{{Littlejohns} {et~al.}(2015){Littlejohns}, {Butler}, {Cucchiara},
  {Watson}, {Fox}, {Lee}, {Kutyrev}, {Richer}, {Klein}, {Prochaska}, {Bloom},
  {Troja}, {Ramirez-Ruiz}, {de Diego}, {Georgiev}, {Gonz{\'a}lez},
  {Rom{\'a}n-Z{\'u}{\~n}iga}, {Gehrels}, \& {Moseley}}]{Littlejohns2015MNRAS}
{Littlejohns}, O.~M., {Butler}, N.~R., {Cucchiara}, A., {et~al.} 2015, \mnras,
  449, 2919

\bibitem[{{Littlejohns} {et~al.}(2014){Littlejohns}, {Butler}, {Cucchiara},
  {Watson}, {Kutyrev}, {Lee}, {Richer}, {Klein}, {Fox}, {Prochaska}, {Bloom},
  {Troja}, {Ramirez-Ruiz}, {de Diego}, {Georgiev}, {Gonz{\'a}lez},
  {Rom{\'a}n-Z{\'u}{\~n}iga}, {Gehrels}, \& {Moseley}}]{Littlejohns2014AJ}
{Littlejohns}, O.~M., {Butler}, N.~R., {Cucchiara}, A., {et~al.} 2014, \aj,
  148, 2

\bibitem[{{Littlejohns} {et~al.}(2013){Littlejohns}, {Tanvir}, {Willingale},
  {Evans}, {O'Brien}, \& {Levan}}]{Littlejohns2013MNRAS}
{Littlejohns}, O.~M., {Tanvir}, N.~R., {Willingale}, R., {et~al.} 2013, \mnras,
  436, 3640

\bibitem[{{Liu} {et~al.}(2008){Liu}, {Wang}, {Xin}, {Qiu}, {Wei}, {Hu}, {Deng},
  {Zheng}, \& {Urata}}]{Liu2008GCN8618}
{Liu}, H., {Wang}, J., {Xin}, L.~P., {et~al.} 2008, GRB Coordinates Network,
  8618

\bibitem[{{Lloyd-Ronning} {et~al.}(2002){Lloyd-Ronning}, {Fryer}, \&
  {Ramirez-Ruiz}}]{Lloyd-Ronning2002ApJ}
{Lloyd-Ronning}, N.~M., {Fryer}, C.~L., \& {Ramirez-Ruiz}, E. 2002, \apj, 574,
  554

\bibitem[{{L{\"u}} {et~al.}(2017){L{\"u}}, {Wang}, {Lu}, {Lan}, {Gao}, {Liang},
  {Graham}, {Zheng}, {Filippenko}, \& {Zhang}}]{Lue2017ApJ}
{L{\"u}}, H., {Wang}, X., {Lu}, R., {et~al.} 2017, \apj, 843, 114

\bibitem[{{Lyman} {et~al.}(2017){Lyman}, {Levan}, {Tanvir}, {Fynbo}, {McGuire},
  {Perley}, {Angus}, {Bloom}, {Conselice}, {Fruchter}, {Hjorth}, {Jakobsson},
  \& {Starling}}]{Lyman2017MNRAS}
{Lyman}, J.~D., {Levan}, A.~J., {Tanvir}, N.~R., {et~al.} 2017, \mnras, 467,
  1795

\bibitem[{{Ma} {et~al.}(2015){Ma}, {Maio}, {Ciardi}, \& {Salvaterra}}]{Ma2015}
{Ma}, Q., {Maio}, U., {Ciardi}, B., \& {Salvaterra}, R. 2015, \mnras, 449, 3006

\bibitem[{{Ma} {et~al.}(2017){Ma}, {Maio}, {Ciardi}, \& {Salvaterra}}]{Ma2017}
{Ma}, Q., {Maio}, U., {Ciardi}, B., \& {Salvaterra}, R. 2017, \mnras, 466, 1140

\bibitem[{{Maehara}(2014)}]{Maehara2014GCN16484}
{Maehara}, H. 2014, GRB Coordinates Network, 16484

\bibitem[{{MAGIC Collaboration} {et~al.}(2019{\natexlab{a}}){MAGIC
  Collaboration}, {Acciari}, {Ansoldi}, {Antonelli}, {Arbet Engels}, {Baack},
  {Babi{\'c}}, {Banerjee}, {Barres de Almeida}, {Barrio}, {Becerra
  Gonz{\'a}lez}, {Bednarek}, {Bellizzi}, {Bernardini}, {Berti}, {Besenrieder},
  {Bhattacharyya}, {Bigongiari}, {Biland}, {Blanch}, {Bonnoli},
  {Bo{\v{s}}njak}, {Busetto}, {Carosi}, {Carosi}, {Ceribella}, {Chai},
  {Chilingaryan}, {Cikota}, {Colak}, {Colin}, {Colombo}, {Contreras},
  {Cortina}, {Covino}, {D'Amico}, {D'Elia}, {da Vela}, {Dazzi}, {de Angelis},
  {de Lotto}, {Delfino}, {Delgado}, {Depaoli}, {di Pierro}, {di Venere}, {Do
  Souto Espi{\~n}eira}, {Dominis Prester}, {Donini}, {Dorner}, {Doro},
  {Elsaesser}, {Fallah Ramazani}, {Fattorini}, {Fern{\'a}ndez-Barral},
  {Ferrara}, {Fidalgo}, {Foffano}, {Fonseca}, {Font}, {Fruck}, {Fukami},
  {Gallozzi}, {Garc{\'\i}a L{\'o}pez}, {Garczarczyk}, {Gasparyan}, {Gaug},
  {Giglietto}, {Giordano}, {Godinovi{\'c}}, {Green}, {Guberman}, {Hadasch},
  {Hahn}, {Herrera}, {Hoang}, {Hrupec}, {H{\"u}tten}, {Inada}, {Inoue},
  {Ishio}, {Iwamura}, {Jouvin}, {Kerszberg}, {Kubo}, {Kushida}, {Lamastra},
  {Lelas}, {Leone}, {Lindfors}, {Lombardi}, {Longo}, {L{\'o}pez},
  {L{\'o}pez-Coto}, {L{\'o}pez-Oramas}, {Loporchio}, {Machado de Oliveira
  Fraga}, {Maggio}, {Majumdar}, {Makariev}, {Mallamaci}, {Maneva}, {Manganaro},
  {Mannheim}, {Maraschi}, {Mariotti}, {Mart{\'\i}nez}, {Masuda}, {Mazin},
  {Mi{\'c}anovi{\'c}}, {Miceli}, {Minev}, {Miranda}, {Mirzoyan}, {Molina},
  {Moralejo}, {Morcuende}, {Moreno}, {Moretti}, {Munar-Adrover}, {Neustroev},
  {Nigro}, {Nilsson}, {Ninci}, {Nishijima}, {Noda}, {Nogu{\'e}s}, {N{\"o}the},
  {Nozaki}, {Paiano}, {Palacio}, {Palatiello}, {Paneque}, {Paoletti},
  {Paredes}, {Pe{\~n}il}, {Peresano}, {Persic}, {Prada Moroni}, {Prandini},
  {Puljak}, {Rhode}, {Rib{\'o}}, {Rico}, {Righi}, {Rugliancich}, {Saha},
  {Sahakyan}, {Saito}, {Sakurai}, {Satalecka}, {Schmidt}, {Schweizer},
  {Sitarek}, {{\v{S}}nidari{\'c}}, {Sobczynska}, {Somero}, {Stamerra}, {Strom},
  {Strzys}, {Suda}, {Suri{\'c}}, {Takahashi}, {Tavecchio}, {Temnikov},
  {Terzi{\'c}}, {Teshima}, {Torres-Alb{\`a}}, {Tosti}, {Tsujimoto}, {Vagelli},
  {van Scherpenberg}, {Vanzo}, {Vazquez Acosta}, {Vigorito}, {Vitale}, {Vovk},
  {Will}, {Zari{\'c}}, \& {Nava}}]{MAGIC2019Nature1}
{MAGIC Collaboration}, {Acciari}, V.~A., {Ansoldi}, S., {et~al.}
  2019{\natexlab{a}}, \nat, 575, 455

\bibitem[{{MAGIC Collaboration} {et~al.}(2019{\natexlab{b}}){MAGIC
  Collaboration}, {Acciari}, {Ansoldi}, {Antonelli}, {Engels}, {Baack},
  {Babi{\'c}}, {Banerjee}, {Barres de Almeida}, {Barrio}, {Becerra
  Gonz{\'a}lez}, {Bednarek}, {Bellizzi}, {Bernardini}, {Berti}, {Besenrieder},
  {Bhattacharyya}, {Bigongiari}, {Biland}, {Blanch}, {Bonnoli},
  {Bo{\v{s}}njak}, {Busetto}, {Carosi}, {Ceribella}, {Chai}, {Chilingaryan},
  {Cikota}, {Colak}, {Colin}, {Colombo}, {Contreras}, {Cortina}, {Covino},
  {D'Elia}, {da Vela}, {Dazzi}, {de Angelis}, {de Lotto}, {Delfino}, {Delgado},
  {Depaoli}, {di Pierro}, {di Venere}, {Do Souto Espi{\~n}eira}, {Dominis
  Prester}, {Donini}, {Dorner}, {Doro}, {Elsaesser}, {Fallah Ramazani},
  {Fattorini}, {Ferrara}, {Fidalgo}, {Foffano}, {Fonseca}, {Font}, {Fruck},
  {Fukami}, {Garc{\'\i}a L{\'o}pez}, {Garczarczyk}, {Gasparyan}, {Gaug},
  {Giglietto}, {Giordano}, {Godinovi{\'c}}, {Green}, {Guberman}, {Hadasch},
  {Hahn}, {Herrera}, {Hoang}, {Hrupec}, {H{\"u}tten}, {Inada}, {Inoue},
  {Ishio}, {Iwamura}, {Jouvin}, {Kerszberg}, {Kubo}, {Kushida}, {Lamastra},
  {Lelas}, {Leone}, {Lindfors}, {Lombardi}, {Longo}, {L{\'o}pez},
  {L{\'o}pez-Coto}, {L{\'o}pez-Oramas}, {Loporchio}, {Machado de Oliveira
  Fraga}, {Maggio}, {Majumdar}, {Makariev}, {Mallamaci}, {Maneva}, {Manganaro},
  {Mannheim}, {Maraschi}, {Mariotti}, {Mart{\'\i}nez}, {Mazin},
  {Mi{\'c}anovi{\'c}}, {Miceli}, {Minev}, {Miranda}, {Mirzoyan}, {Molina},
  {Moralejo}, {Morcuende}, {Moreno}, {Moretti}, {Munar-Adrover}, {Neustroev},
  {Nigro}, {Nilsson}, {Ninci}, {Nishijima}, {Noda}, {Nogu{\'e}s}, {Nozaki},
  {Paiano}, {Palatiello}, {Paneque}, {Paoletti}, {Paredes}, {Pe{\~n}il},
  {Peresano}, {Persic}, {Moroni}, {Prandini}, {Puljak}, {Rhode}, {Rib{\'o}},
  {Rico}, {Righi}, {Rugliancich}, {Saha}, {Sahakyan}, {Saito}, {Sakurai},
  {Satalecka}, {Schmidt}, {Schweizer}, {Sitarek}, {{\v{S}}nidari{\'c}},
  {Sobczynska}, {Somero}, {Stamerra}, {Strom}, {Strzys}, {Suda}, {Suri{\'c}},
  {Takahashi}, {Tavecchio}, {Temnikov}, {Terzi{\'c}}, {Teshima},
  {Torres-Alb{\`a}}, {Tosti}, {Vagelli}, {van Scherpenberg}, {Vanzo}, {Vazquez
  Acosta}, {Vigorito}, {Vitale}, {Vovk}, {Will}, {Zari{\'c}}, {Nava}, {Veres},
  {Bhat}, {Briggs}, {Cleveland}, {Hamburg}, {Hui}, {Mailyan}, {Preece},
  {Roberts}, {von Kienlin}, {Wilson-Hodge}, {Kocevski}, {Arimoto}, {Tak},
  {Asano}, {Axelsson}, {Barbiellini}, {Bissaldi}, {Dirirsa}, {Gill}, {Granot},
  {McEnery}, {Omodei}, {Razzaque}, {Piron}, {Racusin}, {Thompson}, {Campana},
  {Bernardini}, {Kuin}, {Siegel}, {Cenko}, {O'Brien}, {Capalbi}, {Da{\i}}, {de
  Pasquale}, {Gropp}, {Klingler}, {Osborne}, {Perri}, {Starling},
  {Tagliaferri}, {Tohuvavohu}, {Ursi}, {Tavani}, {Cardillo}, {Casentini},
  {Piano}, {Evangelista}, {Verrecchia}, {Pittori}, {Lucarelli}, {Bulgarelli},
  {Parmiggiani}, {Anderson}, {Anderson}, {Bernardi}, {Bolmer},
  {Caballero-Garc{\'\i}a}, {Carrasco}, {Castell{\'o}n}, {Castro Segura},
  {Castro-Tirado}, {Cherukuri}, {Cockeram}, {D'Avanzo}, {di Dato}, {Diretse},
  {Fender}, {Fern{\'a}ndez-Garc{\'\i}a}, {Fynbo}, {Fruchter}, {Greiner},
  {Gromadzki}, {Heintz}, {Heywood}, {van der Horst}, {Hu}, {Inserra}, {Izzo},
  {Jaiswal}, {Jakobsson}, {Japelj}, {Kankare}, {Kann}, {Kouveliotou}, {Klose},
  {Levan}, {Li}, {Lotti}, {Maguire}, {Malesani}, {Manulis}, {Marongiu},
  {Martin}, {Melandri}, {Micha{\l}owski}, {Miller-Jones}, {Misra}, {Moin},
  {Mooley}, {Nasri}, {Nicholl}, {Noschese}, {Novara}, {Pandey}, {Peretti},
  {P{\'e}rez Del Pulgar}, {P{\'e}rez-Torres}, {Perley}, {Piro}, {Ragosta},
  {Resmi}, {Ricci}, {Rossi}, {S{\'a}nchez-Ram{\'\i}rez}, {Selsing}, {Schulze},
  {Smartt}, {Smith}, {Sokolov}, {Stevens}, {Tanvir}, {Th{\"o}ne}, {Tiengo},
  {Tremou}, {Troja}, {de Ugarte Postigo}, {Valeev}, {Vergani}, {Wieringa},
  {Woudt}, {Xu}, {Yaron}, \& {Young}}]{MAGIC2019Nature2}
{MAGIC Collaboration}, {Acciari}, V.~A., {Ansoldi}, S., {et~al.}
  2019{\natexlab{b}}, \nat, 575, 459

\bibitem[{{Malesani} {et~al.}(2006){Malesani}, {D'Avanzo}, {Chincarini},
  {Tagliaferri}, {Campana}, {Covino}, {Fugazza}, {Fernandez-Soto}, {Della
  Valle}, {Antonelli}, {Fiore}, {Piranomonte}, \&
  {Stella}}]{Malesani2006GCN4541}
{Malesani}, D., {D'Avanzo}, P., {Chincarini}, G., {et~al.} 2006, GRB
  Coordinates Network, 4541

\bibitem[{{Malesani} {et~al.}(2014{\natexlab{a}}){Malesani}, {D'Avanzo}, \&
  {Martinez Osorio}}]{Malesani2014GCN15980}
{Malesani}, D., {D'Avanzo}, P., \& {Martinez Osorio}, Y.~F. 2014{\natexlab{a}},
  GRB Coordinates Network, 15980

\bibitem[{{Malesani} \& {Fynbo}(2018)}]{Malesani2018GCN22551}
{Malesani}, D. \& {Fynbo}, J.~P.~U. 2018, GRB Coordinates Network, 22551

\bibitem[{{Malesani} {et~al.}(2014{\natexlab{b}}){Malesani}, {Kruehler},
  {Santiago}, {de Ugarte Postigo}, {Dahle}, \& {Soto}}]{Malesani2014GCN16193}
{Malesani}, D., {Kruehler}, T., {Santiago}, E., {et~al.} 2014{\natexlab{b}},
  GRB Coordinates Network, 16193

\bibitem[{{Malesani} {et~al.}(2011){Malesani}, {Leloudas}, {Xu}, {de Ugarte
  Postigo}, {Hjorth}, {Jakobsson}, \& {Nielsen}}]{Malesani2011GCN12220}
{Malesani}, D., {Leloudas}, G., {Xu}, D., {et~al.} 2011, GRB Coordinates
  Network, 12220

\bibitem[{{Malesani} {et~al.}(2012){Malesani}, {Schulze}, {de Ugarte Postigo},
  {Xu}, {D'Elia}, {Fynbo}, \& {Tanvir}}]{Malesani2012GCN13649}
{Malesani}, D., {Schulze}, S., {de Ugarte Postigo}, A., {et~al.} 2012, GRB
  Coordinates Network, 13649

\bibitem[{{Malesani} {et~al.}(2020){Malesani}, {Vielfaure}, {Izzo}, {Vergani},
  {Burgarella}, {D'Elia}, {de Ugarte Postigo}, {Fynbo}, {Levan},
  {Milvang-Jensen}, {Kann}, {Palazzi}, {Pugliese}, {Tanvir}, {Wiersema}, \&
  {Stargate Consortium}}]{Malesani2020GCN29100}
{Malesani}, D.~B., {Vielfaure}, J.~B., {Izzo}, L., {et~al.} 2020, GRB
  Coordinates Network, 29100

\bibitem[{{Mangano} {et~al.}(2007){Mangano}, {Holland}, {Malesani}, {Troja},
  {Chincarini}, {Zhang}, {La Parola}, {Brown}, {Burrows}, {Campana}, {Capalbi},
  {Cusumano}, {Della Valle}, {Gehrels}, {Giommi}, {Grupe}, {Guidorzi}, {Mineo},
  {Moretti}, {Osborne}, {Pandey}, {Perri}, {Romano}, {Roming}, \&
  {Tagliaferri}}]{mangano2007}
{Mangano}, V., {Holland}, S.~T., {Malesani}, D., {et~al.} 2007, \aap, 470, 105

\bibitem[{{Mao} {et~al.}(2009{\natexlab{a}}){Mao}, {Cha}, \&
  {Bai}}]{Mao2009GCN9285}
{Mao}, J., {Cha}, G., \& {Bai}, J. 2009{\natexlab{a}}, GRB Coordinates Network,
  9285

\bibitem[{{Mao} {et~al.}(2009{\natexlab{b}}){Mao}, {Li}, \&
  {Bai}}]{Mao2009GCN10092}
{Mao}, J., {Li}, S., \& {Bai}, J. 2009{\natexlab{b}}, GRB Coordinates Network,
  1092

\bibitem[{{Mao} {et~al.}(2012){Mao}, {Malesani}, {D'Avanzo}, {Covino}, {Li},
  {Jakobsson}, \& {Bai}}]{Mao2012AA}
{Mao}, J., {Malesani}, D., {D'Avanzo}, P., {et~al.} 2012, \aap, 538, A1

\bibitem[{{Marshall} {et~al.}(2011){Marshall}, {Antonelli}, {Burrows},
  {Covino}, {de Pasquale}, {Evans}, {Fugazza}, {Holland}, {Liang}, {O'Brien},
  {Oates}, {Osborne}, {Pagani}, {Sakamoto}, {Siegel}, {Wu}, \&
  {Zhang}}]{Marshall2011ApJ}
{Marshall}, F.~E., {Antonelli}, L.~A., {Burrows}, D.~N., {et~al.} 2011, \apj,
  727, 132

\bibitem[{{Marshall} {et~al.}(2008){Marshall}, {Barthelmy}, {Burrows}, {Evans},
  {Oates}, {Stamatikos}, \& {Gehrels}}]{Marshall2008GCNR129}
{Marshall}, F.~E., {Barthelmy}, S.~D., {Burrows}, D.~N., {et~al.} 2008, GCN
  Report, 129

\bibitem[{{Marshall} \& {Sonbas}(2014)}]{Marshall2014GCN17219}
{Marshall}, F.~E. \& {Sonbas}, E. 2014, GRB Coordinates Network, 17219

\bibitem[{{Marshall} \& {Troja}(2018)}]{Marshall2018GCN22549}
{Marshall}, F.~E. \& {Troja}, E. 2018, GRB Coordinates Network, 22549

\bibitem[{{Martin-Carrillo} {et~al.}(2014){Martin-Carrillo}, {Hanlon},
  {Topinka}, {LaCluyz{\'e}}, {Savchenko}, {Kann}, {Trotter}, {Covino},
  {Kr{\"u}hler}, {Greiner}, {McGlynn}, {Murphy}, {Tisdall}, {Meehan}, {Wade},
  {McBreen}, {Reichart}, {Fugazza}, {Haislip}, {Rossi}, {Schady}, {Elliott}, \&
  {Klose}}]{Martin-Carrillo2014AA}
{Martin-Carrillo}, A., {Hanlon}, L., {Topinka}, M., {et~al.} 2014, \aap, 567,
  A84

\bibitem[{{Martin-Carrillo} {et~al.}(2016){Martin-Carrillo}, {Murphy},
  {Hanlon}, {van Heerden}, {van Soelen}, \&
  {Meintjes}}]{Martin-Carrillo2016GCN20305}
{Martin-Carrillo}, A., {Murphy}, D., {Hanlon}, L., {et~al.} 2016, GRB
  Coordinates Network, 20305

\bibitem[{{Martone} {et~al.}(2018){Martone}, {Guidorzi}, {Kobayashi},
  {Mundell}, {Gomboc}, {Steele}, {Cucchiara}, \&
  {Morris}}]{Martone2018GCN22976}
{Martone}, R., {Guidorzi}, C., {Kobayashi}, S., {et~al.} 2018, GRB Coordinates
  Network, 22976

\bibitem[{{Maselli} {et~al.}(2014){Maselli}, {Melandri}, {Nava}, {Mundell},
  {Kawai}, {Campana}, {Covino}, {Cummings}, {Cusumano}, {Evans}, {Ghirlanda},
  {Ghisellini}, {Guidorzi}, {Kobayashi}, {Kuin}, {La Parola}, {Mangano},
  {Oates}, {Sakamoto}, {Serino}, {Virgili}, {Zhang}, {Barthelmy}, {Beardmore},
  {Bernardini}, {Bersier}, {Burrows}, {Calderone}, {Capalbi}, {Chiang},
  {D'Avanzo}, {D'Elia}, {De Pasquale}, {Fugazza}, {Gehrels}, {Gomboc},
  {Harrison}, {Hanayama}, {Japelj}, {Kennea}, {Kopac}, {Kouveliotou}, {Kuroda},
  {Levan}, {Malesani}, {Marshall}, {Nousek}, {O'Brien}, {Osborne}, {Pagani},
  {Page}, {Page}, {Perri}, {Pritchard}, {Romano}, {Saito}, {Sbarufatti},
  {Salvaterra}, {Steele}, {Tanvir}, {Vianello}, {Weigand}, {Wiersema}, {Yatsu},
  {Yoshii}, \& {Tagliaferri}}]{Maselli2014Science}
{Maselli}, A., {Melandri}, A., {Nava}, L., {et~al.} 2014, Science, 343, 48

\bibitem[{{Masi} \& {Nocentini}(2013)}]{Masi2013GCN14789}
{Masi}, G. \& {Nocentini}, F. 2013, GRB Coordinates Network, 14789

\bibitem[{{Maticic} \& {Skvarc}(2009)}]{Maticic2009GCN9715}
{Maticic}, S. \& {Skvarc}, J. 2009, GRB Coordinates Network, 9715

\bibitem[{{Mazaeva} {et~al.}(2015{\natexlab{a}}){Mazaeva}, {Inasaridze},
  {Kvaratskhelia}, {Molotov}, \& {Pozanenko}}]{Mazaeva2015GCN18327}
{Mazaeva}, E., {Inasaridze}, R., {Kvaratskhelia}, O., {Molotov}, I., \&
  {Pozanenko}, A. 2015{\natexlab{a}}, GRB Coordinates Network, 18327

\bibitem[{{Mazaeva} {et~al.}(2015{\natexlab{b}}){Mazaeva}, {Inasaridze},
  {Zhuzhunadze}, {Molotov}, \& {Pozanenko}}]{Mazaeva2015GCN18289}
{Mazaeva}, E., {Inasaridze}, R., {Zhuzhunadze}, V., {Molotov}, I., \&
  {Pozanenko}, A. 2015{\natexlab{b}}, GRB Coordinates Network, 18289

\bibitem[{{Mazaeva} {et~al.}(2019){Mazaeva}, {Pozanenko}, {Volnova}, {Belkin},
  \& {Krugov}}]{Mazaeva2019GCN23741}
{Mazaeva}, E., {Pozanenko}, A., {Volnova}, A., {Belkin}, S., \& {Krugov}, M.
  2019, GRB Coordinates Network, 23741

\bibitem[{{Mazaeva} {et~al.}(2015{\natexlab{c}}){Mazaeva}, {Reva}, {Kusakin},
  {Volnova}, \& {Pozanenko}}]{Mazaeva2015GCN18281}
{Mazaeva}, E., {Reva}, I., {Kusakin}, A., {Volnova}, A., \& {Pozanenko}, A.
  2015{\natexlab{c}}, GRB Coordinates Network, 18281

\bibitem[{{Mazets} {et~al.}(1981){Mazets}, {Golenetskii}, {Ilinskii}, {Panov},
  {Aptekar}, {Gurian}, {Proskura}, {Sokolov}, {Sokolova}, \&
  {Kharitonova}}]{Mazets1981ApSS}
{Mazets}, E.~P., {Golenetskii}, S.~V., {Ilinskii}, V.~N., {et~al.} 1981, \apss,
  80, 3

\bibitem[{{Mazzali} {et~al.}(2016){Mazzali}, {Sullivan}, {Pian}, {Greiner}, \&
  {Kann}}]{Mazzali2016MNRAS}
{Mazzali}, P.~A., {Sullivan}, M., {Pian}, E., {Greiner}, J., \& {Kann}, D.~A.
  2016, \mnras, 458, 3455

\bibitem[{{McCauley} \& {Pagani}(2015)}]{McCauley2015GCN18270}
{McCauley}, L.~M. \& {Pagani}, C. 2015, GRB Coordinates Network, 18270

\bibitem[{{McGuire} {et~al.}(2016){McGuire}, {Tanvir}, {Levan}, {Trenti},
  {Stanway}, {Shull}, {Wiersema}, {Perley}, {Starling}, {Bremer}, {Stocke},
  {Hjorth}, {Rhoads}, {Curtis-Lake}, {Schulze}, {Levesque}, {Robertson},
  {Fynbo}, {Ellis}, \& {Fruchter}}]{McGuire2016ApJ}
{McGuire}, J.~T.~W., {Tanvir}, N.~R., {Levan}, A.~J., {et~al.} 2016, \apj, 825,
  135

\bibitem[{{Melandri} {et~al.}(2015){Melandri}, {Bernardini}, {D'Avanzo},
  {S{\'a}nchez-Ram{\'\i}rez}, {Nappo}, {Nava}, {Japelj}, {de Ugarte Postigo},
  {Oates}, {Campana}, {Covino}, {D'Elia}, {Ghirlanda}, {Gafton}, {Ghisellini},
  {Gnedin}, {Goldoni}, {Gorosabel}, {Libbrecht}, {Malesani}, {Salvaterra},
  {Th{\"o}ne}, {Vergani}, {Xu}, \& {Tagliaferri}}]{Melandri2015AA}
{Melandri}, A., {Bernardini}, M.~G., {D'Avanzo}, P., {et~al.} 2015, \aap, 581,
  A86

\bibitem[{{Melandri} {et~al.}(2014{\natexlab{a}}){Melandri}, {Covino},
  {Rogantini}, {Salvaterra}, {Sbarufatti}, {Bernardini}, {Campana}, {D'Avanzo},
  {D'Elia}, {Fugazza}, {Ghirlanda}, {Ghisellini}, {Nava}, {Vergani}, \&
  {Tagliaferri}}]{Melandri2014AABAT6}
{Melandri}, A., {Covino}, S., {Rogantini}, D., {et~al.} 2014{\natexlab{a}},
  \aap, 565, A72

\bibitem[{{Melandri} {et~al.}(2017){Melandri}, {Covino}, {Zaninoni}, {Campana},
  {Bolmer}, {Cobb}, {Gorosabel}, {Kim}, {Kuin}, {Kuroda}, {Malesani},
  {Mundell}, {Nappo}, {Sbarufatti}, {Smith}, {Steele}, {Topinka}, {Trotter},
  {Virgili}, {Bernardini}, {D'Avanzo}, {D'Elia}, {Fugazza}, {Ghirlanda},
  {Gomboc}, {Greiner}, {Guidorzi}, {Haislip}, {Hanayama}, {Hanlon}, {Im},
  {Ivarsen}, {Japelj}, {Jel{\'\i}nek}, {Kawai}, {Kobayashi}, {Kopac},
  {LaCluyz{\'e}}, {Martin-Carrillo}, {Murphy}, {Reichart}, {Salvaterra},
  {Salafia}, {Tagliaferri}, \& {Vergani}}]{Melandri2017AA}
{Melandri}, A., {Covino}, S., {Zaninoni}, E., {et~al.} 2017, \aap, 607, A29

\bibitem[{{Melandri} {et~al.}(2011){Melandri}, {D'Avanzo}, {Fugazza}, \&
  {Palazzi}}]{Melandri2011GCN11963}
{Melandri}, A., {D'Avanzo}, P., {Fugazza}, D., \& {Palazzi}, E. 2011, GRB
  Coordinates Network, 11963

\bibitem[{{Melandri} {et~al.}(2008){Melandri}, {Gomboc}, {Guidorzi}, {Smith},
  {Steele}, {Bersier}, {Mundell}, {Carter}, {Kobayashi}, {Burgdorf}, {Bode},
  {Rol}, {O'Brien}, {Bannister}, \& {Tanvir}}]{Melandri2008GCN7813}
{Melandri}, A., {Gomboc}, A., {Guidorzi}, C., {et~al.} 2008, GRB Coordinates
  Network, 7813

\bibitem[{{Melandri} {et~al.}(2019){Melandri}, {Izzo}, {D'Avanzo}, {Malesani},
  {Valle}, {Pian}, {Tanvir}, {Ragosta}, {Olivares}, {Carini}, {Palazzi},
  {Piranomonte}, {Jonker}, {Rossi}, {Kann}, {Hartmann}, {Inserra}, {Kankare},
  {Maguire}, {Smartt}, {Yaron}, {Young}, \& {Manulis}}]{Melandri2019GCN23983}
{Melandri}, A., {Izzo}, L., {D'Avanzo}, P., {et~al.} 2019, GRB Coordinates
  Network, 23983

\bibitem[{{Melandri} {et~al.}(2022){Melandri}, {Izzo}, {Pian}, {Malesani},
  {Della Valle}, {Rossi}, {D'Avanzo}, {Guetta}, {Mazzali}, {Benetti},
  {Masetti}, {Palazzi}, {Savaglio}, {Amati}, {Antonelli}, {Ashall},
  {Bernardini}, {Campana}, {Carini}, {Covino}, {D'Elia}, {de Ugarte Postigo},
  {De Pasquale}, {Filippenko}, {Fruchter}, {Fynbo}, {Giunta}, {Hartmann},
  {Jakobsson}, {Japelj}, {Jonker}, {Kann}, {Lamb}, {Levan}, {Martin-Carrillo},
  {M{\o}ller}, {Piranomonte}, {Pugliese}, {Salvaterra}, {Schulze}, {Starling},
  {Stella}, {Tagliaferri}, {Tanvir}, \& {Watson}}]{Melandri2022AA}
{Melandri}, A., {Izzo}, L., {Pian}, E., {et~al.} 2022, \aap, 659, A39

\bibitem[{{Melandri} {et~al.}(2010){Melandri}, {Kopac}, \&
  {Cano}}]{Melandri2010GCN11229}
{Melandri}, A., {Kopac}, D., \& {Cano}, Z. 2010, GRB Coordinates Network, 11229

\bibitem[{{Melandri} {et~al.}(2014{\natexlab{b}}){Melandri}, {Pian}, {D'Elia},
  {D'Avanzo}, {Della Valle}, {Mazzali}, {Tagliaferri}, {Cano}, {Levan},
  {M{$\Delta$}oller}, {Amati}, {Bernardini}, {Bersier}, {Bufano}, {Campana},
  {Castro-Tirado}, {Covino}, {Ghirlanda}, {Hurley}, {Malesani}, {Masetti},
  {Palazzi}, {Piranomonte}, {Rossi}, {Salvaterra}, {Starling}, {Tanaka},
  {Tanvir}, \& {Vergani}}]{Melandri2014AA}
{Melandri}, A., {Pian}, E., {D'Elia}, V., {et~al.} 2014{\natexlab{b}}, \aap,
  567, A29

\bibitem[{{Melandri} {et~al.}(2012){Melandri}, {Sbarufatti}, {D'Avanzo},
  {Salvaterra}, {Campana}, {Covino}, {Vergani}, {Nava}, {Ghisellini},
  {Ghirlanda}, {Fugazza}, {Mangano}, {Capalbi}, \&
  {Tagliaferri}}]{Melandri2012}
{Melandri}, A., {Sbarufatti}, B., {D'Avanzo}, P., {et~al.} 2012, \mnras, 421,
  1265

\bibitem[{{Melandri} {et~al.}(2014{\natexlab{c}}){Melandri}, {Virgili},
  {Guidorzi}, {Bernardini}, {Kobayashi}, {Mundell}, {Gomboc}, {Dintinjana},
  {Hentunen}, {Japelj}, {Kopa{\v{c}}}, {Kuroda}, {Morgan}, {Steele}, {Quadri},
  {Arici}, {Arnold}, {Girelli}, {Hanayama}, {Kawai}, {Miku{\v{z}}}, {Nissinen},
  {Salmi}, {Smith}, {Strabla}, {Tonincelli}, \& {Quadri}}]{Melandri2014AA2}
{Melandri}, A., {Virgili}, F.~J., {Guidorzi}, C., {et~al.} 2014{\natexlab{c}},
  \aap, 572, A55

\bibitem[{{Mereghetti} {et~al.}(2021){Mereghetti}, {Gotz}, {Ferrigno}, {Bozzo},
  {Savchenko}, {Ducci}, \& {Borkowski}}]{Mereghetti2021GCN29650}
{Mereghetti}, S., {Gotz}, D., {Ferrigno}, C., {et~al.} 2021, GRB Coordinates
  Network, 29650

\bibitem[{{M{\'e}sz{\'a}ros} \& {Rees}(1997)}]{Meszaros1997ApJ}
{M{\'e}sz{\'a}ros}, P. \& {Rees}, M.~J. 1997, \apj, 476, 232

\bibitem[{{Micha{\l}owski} {et~al.}(2021){Micha{\l}owski}, {Kami{\'n}ski},
  {Kami{\'n}ska}, \& {Wnuk}}]{Michalowski2021NatAs}
{Micha{\l}owski}, M.~J., {Kami{\'n}ski}, K., {Kami{\'n}ska}, M.~K., \& {Wnuk},
  E. 2021, Nature Astronomy, 5, 995

\bibitem[{{Miller} {et~al.}(2008){Miller}, {Bloom}, \&
  {Perley}}]{Miller2008GCN7827}
{Miller}, A.~A., {Bloom}, J.~S., \& {Perley}, D.~A. 2008, GRB Coordinates
  Network, 7827

\bibitem[{{Minaev} \& {Pozanenko}(2019)}]{Minaev2019GCN23714}
{Minaev}, P. \& {Pozanenko}, A. 2019, GRB Coordinates Network, 23714

\bibitem[{{Misra} {et~al.}(2021){Misra}, {Resmi}, {Kann}, {Marongiu}, {Moin},
  {Klose}, {Bernardi}, {de Ugarte Postigo}, {Jaiswal}, {Schulze}, {Perley},
  {Ghosh}, {Dimple}, {Kumar}, {Gupta}, {Micha{\l}owski}, {Mart{\'\i}n},
  {Cockeram}, {Cherukuri}, {Bhalerao}, {Anderson}, {Pandey}, {Anupama},
  {Th{\"o}ne}, {Barway}, {Wieringa}, {Fynbo}, \& {Habeeb}}]{Misra2021MNRAS}
{Misra}, K., {Resmi}, L., {Kann}, D.~A., {et~al.} 2021, \mnras, 504, 5685

\bibitem[{{Moody} {et~al.}(2010){Moody}, {Laney}, {Pearson}, \&
  {Pace}}]{Moody2010GCN10665}
{Moody}, J.~W., {Laney}, D., {Pearson}, R., \& {Pace}, C. 2010, GRB Coordinates
  Network, 10665

\bibitem[{{Moon} {et~al.}(2008){Moon}, {Jin}, {Yuk}, {Lee}, {Nam}, {Cha},
  {Cho}, {Kyeong}, {Park}, {Mock}, {Han}, {Lee}, {Park}, {Han}, {Pak}, {Kim},
  \& {Kim}}]{MoonB2008}
{Moon}, B., {Jin}, H., {Yuk}, I.-S., {et~al.} 2008, \pasj, 60, 849

\bibitem[{{Morgan}(2011)}]{Morgan2011GCN12760}
{Morgan}, A.~N. 2011, GRB Coordinates Network, 12760

\bibitem[{{Morgan}(2013{\natexlab{a}})}]{Morgan2013GCN14453}
{Morgan}, A.~N. 2013{\natexlab{a}}, GCN Circulars, 14453

\bibitem[{{Morgan}(2013{\natexlab{b}})}]{Morgan2013GCN14802}
{Morgan}, A.~N. 2013{\natexlab{b}}, GRB Coordinates Network, 14802

\bibitem[{{Morgan} {et~al.}(2014){Morgan}, {Perley}, {Cenko}, {Bloom},
  {Cucchiara}, {Richards}, {Filippenko}, {Haislip}, {LaCluyze}, {Corsi},
  {Melandri}, {Cobb}, {Gomboc}, {Horesh}, {James}, {Li}, {Mundell}, {Reichart},
  \& {Steele}}]{Morgan2014MNRAS}
{Morgan}, A.~N., {Perley}, D.~A., {Cenko}, S.~B., {et~al.} 2014, \mnras, 440,
  1810

\bibitem[{{Morgan} {et~al.}(2010){Morgan}, {Perley}, {Klein}, \&
  {Bloom}}]{Morgan2010GCN10648}
{Morgan}, A.~N., {Perley}, D.~A., {Klein}, C.~R., \& {Bloom}, J.~S. 2010, GRB
  Coordinates Network, 10648

\bibitem[{{Mori} {et~al.}(2008){Mori}, {Nakajima}, {Shimokawabe}, {Kawai},
  {Kuroda}, {Yoshida}, {Yanagisawa}, {Shimizu}, {Toda}, \&
  {Nagayama}}]{Mori2008GCN8619}
{Mori}, Y.~A., {Nakajima}, H., {Shimokawabe}, T., {et~al.} 2008, GRB
  Coordinates Network, 8619

\bibitem[{{Moskvitin} {et~al.}(2014{\natexlab{a}}){Moskvitin}, {Burenin},
  {Uklein}, {Sokolov}, \& {Sokolova}}]{Moskvitin2014GCN16499}
{Moskvitin}, A., {Burenin}, R., {Uklein}, R., {Sokolov}, V., \& {Sokolova}, T.
  2014{\natexlab{a}}, GRB Coordinates Network, 16499

\bibitem[{{Moskvitin} \& {Fatkhullin}(2009)}]{Moskvitin2009GCN10101}
{Moskvitin}, A. \& {Fatkhullin}, T. 2009, GRB Coordinates Network, 10101

\bibitem[{{Moskvitin} {et~al.}(2009){Moskvitin}, {Fatkhullin}, \&
  {Valeev}}]{Moskvitin2009GCN9709}
{Moskvitin}, A., {Fatkhullin}, T., \& {Valeev}, A. 2009, GRB Coordinates
  Network, 9709

\bibitem[{{Moskvitin}(2011)}]{Moskvitin2011GCN11962}
{Moskvitin}, A.~S. 2011, GRB Coordinates Network, 11962

\bibitem[{{Moskvitin}(2013)}]{Moskvitin2013GCN15412}
{Moskvitin}, A.~S. 2013, GRB Coordinates Network, 15412

\bibitem[{{Moskvitin} \& {Goranskij}(2015)}]{Moskvitin2015GCN18275}
{Moskvitin}, A.~S. \& {Goranskij}, V.~P. 2015, GRB Coordinates Network, 18275

\bibitem[{{Moskvitin} {et~al.}(2014{\natexlab{b}}){Moskvitin}, {Komarova},
  {Sokolova}, \& {Sokolov}}]{Moskvitin2014GCN16518}
{Moskvitin}, A.~S., {Komarova}, V.~N., {Sokolova}, T.~N., \& {Sokolov}, V.~V.
  2014{\natexlab{b}}, GRB Coordinates Network, 16518

\bibitem[{{Moskvitin} \& {Sokolov}(2011)}]{Moskvitin2011GCN12251}
{Moskvitin}, A.~S. \& {Sokolov}, V.~V. 2011, GRB Coordinates Network, 12251, 1

\bibitem[{{Motohara} {et~al.}(2009){Motohara}, {Konishi}, {Toshikawa},
  {Mitani}, {Minezaki}, {Koshida}, {Kato}, {Yoshii}, \&
  {Ita}}]{Motohara2009GCN10047}
{Motohara}, K., {Konishi}, M., {Toshikawa}, K., {et~al.} 2009, GRB Coordinates
  Network, 1047

\bibitem[{{Nagayama}(2013)}]{Nagayama2013GCN14793}
{Nagayama}, T. 2013, GRB Coordinates Network, 14793

\bibitem[{{Naito} {et~al.}(2010){Naito}, {Sako}, {Suzuki}, {Kobara}, {Omori},
  {Nagayama}, {Kurita}, \& {Oi}}]{Naito2010GCN10881}
{Naito}, H., {Sako}, T., {Suzuki}, D., {et~al.} 2010, GRB Coordinates Network,
  10881

\bibitem[{{Nakajima} {et~al.}(2011){Nakajima}, {Yatsu}, {Enomoto}, {Kawakami},
  {Tokoyoda}, {Ohkawa}, \& {Kawai}}]{Nakajima2011GCN11724}
{Nakajima}, H., {Yatsu}, Y., {Enomoto}, T., {et~al.} 2011, GRB Coordinates
  Network, 11724

\bibitem[{{Nakajima} {et~al.}(2009){Nakajima}, {Yatsu}, {Mori}, {Endo},
  {Shimokawabe}, \& {Kawai}}]{Nakajima2009GCN10260}
{Nakajima}, H., {Yatsu}, Y., {Mori}, Y.~A., {et~al.} 2009, GRB Coordinates
  Network, 10260

\bibitem[{{Nappo} {et~al.}(2014){Nappo}, {Ghisellini}, {Ghirlanda}, {Melandri},
  {Nava}, \& {Burlon}}]{Nappo2014MNRAS}
{Nappo}, F., {Ghisellini}, G., {Ghirlanda}, G., {et~al.} 2014, \mnras, 445,
  1625

\bibitem[{{Nardini} {et~al.}(2014){Nardini}, {Elliott}, {Filgas}, {Schady},
  {Greiner}, {Kr{\"u}hler}, {Klose}, {Afonso}, {Kann}, {Nicuesa Guelbenzu},
  {Olivares E.}, {Rau}, {Rossi}, {Sudilovsky}, \& {Schmidl}}]{Nardini2014AA}
{Nardini}, M., {Elliott}, J., {Filgas}, R., {et~al.} 2014, \aap, 562, A29

\bibitem[{{Nardini} {et~al.}(2011){Nardini}, {Greiner}, {Kr{\"u}hler},
  {Filgas}, {Klose}, {Afonso}, {Clemens}, {Guelbenzu}, {Olivares E.}, {Rau},
  {Rossi}, {Updike}, {K{\"u}pc{\"u} Yolda{\c s}}, {Yolda{\c s}}, {Burlon},
  {Elliott}, \& {Kann}}]{Nardini2011AA}
{Nardini}, M., {Greiner}, J., {Kr{\"u}hler}, T., {et~al.} 2011, \aap, 531, A39

\bibitem[{{Nava} {et~al.}(2012){Nava}, {Salvaterra}, {Ghirlanda}, {Ghisellini},
  {Campana}, {Covino}, {Cusumano}, {D'Avanzo}, {D'Elia}, {Fugazza}, {Melandri},
  {Sbarufatti}, {Vergani}, \& {Tagliaferri}}]{nava2012}
{Nava}, L., {Salvaterra}, R., {Ghirlanda}, G., {et~al.} 2012, \mnras, 421, 1256

\bibitem[{Nekola {et~al.}(2010)Nekola, Hudec, Jelínek, Kubánek, Strobl, \&
  Polášek}]{bart}
Nekola, M., Hudec, R., Jelínek, M., {et~al.} 2010, Advances in Astronomy, 2010

\bibitem[{{Nelson}(2011)}]{Nelson2011GCN12174}
{Nelson}, P. 2011, GRB Coordinates Network, 12174

\bibitem[{{Nicuesa Guelbenzu} {et~al.}(2011){Nicuesa Guelbenzu}, {Klose},
  {Rossi}, {Kann}, {Kr{\"u}hler}, {Greiner}, {Rau}, {Olivares E.}, {Afonso},
  {Filgas}, {K{\"u}pc{\"u} Yolda{\c{s}}}, {McBreen}, {Nardini}, {Schady},
  {Schmidl}, {Updike}, \& {Yolda{\c{s}}}}]{NicuesaGuelbenzu2011AA}
{Nicuesa Guelbenzu}, A., {Klose}, S., {Rossi}, A., {et~al.} 2011, \aap, 531, L6

\bibitem[{{Niino} {et~al.}(2012){Niino}, {Hashimoto}, {Aoki}, {Hattori},
  {Yabe}, \& {Nomoto}}]{Niino2012PASJ}
{Niino}, Y., {Hashimoto}, T., {Aoki}, K., {et~al.} 2012, \pasj, 64, 115

\bibitem[{{Nir} {et~al.}(2021{\natexlab{a}}){Nir}, {Ofek}, {Ben-Ami}, {Segev},
  {Polishook}, \& {Manulis}}]{Nir2020MNRAS}
{Nir}, G., {Ofek}, E.~O., {Ben-Ami}, S., {et~al.} 2021{\natexlab{a}}, \mnras,
  505, 2477

\bibitem[{{Nir} {et~al.}(2021{\natexlab{b}}){Nir}, {Ofek}, \&
  {Gal-Yam}}]{Nir2021RNAAS}
{Nir}, G., {Ofek}, E.~O., \& {Gal-Yam}, A. 2021{\natexlab{b}}, Research Notes
  of the American Astronomical Society, 5, 27

\bibitem[{{Norris} \& {Macomb}(2013)}]{Norris2013GCN14511}
{Norris}, J. \& {Macomb}, D. 2013, GCN Circulars, 14511

\bibitem[{{Nysewander} {et~al.}(2011{\natexlab{a}}){Nysewander}, {Haislip},
  {Ivarsen}, {Lacluyze}, {Maturi}, {Reichart}, {Moore}, {Cromartie}, {Egger},
  {Foster}, {Frank}, {Oza}, {Speckhard}, {Trotter}, \&
  {Crain}}]{Nysewander2011GCN12751}
{Nysewander}, M., {Haislip}, J., {Ivarsen}, K., {et~al.} 2011{\natexlab{a}},
  GRB Coordinates Network, 12751

\bibitem[{{Nysewander} {et~al.}(2011{\natexlab{b}}){Nysewander}, {Haislip},
  {Lacluyze}, {Ivarsen}, {Reichart}, {Moore}, {Trotter}, {Egger}, {Foster},
  {Oza}, {Cromartie}, {Speckhard}, \& {Crain}}]{Nysewander2011GCN}
{Nysewander}, M., {Haislip}, J., {Lacluyze}, A., {et~al.} 2011{\natexlab{b}},
  GCN Circulars, 12645

\bibitem[{{Oates} \& {Cummings}(2009)}]{Oates2009GCN9265}
{Oates}, S.~R. \& {Cummings}, J.~R. 2009, GRB Coordinates Network, 9265

\bibitem[{{Oates} \& {Marshall}(2008)}]{Oates2008GCN7607}
{Oates}, S.~R. \& {Marshall}, F.~E. 2008, GRB Coordinates Network, 7607

\bibitem[{{Oates} {et~al.}(2009){Oates}, {Page}, {Schady}, {de Pasquale},
  {Koch}, {Breeveld}, {Brown}, {Chester}, {Holland}, {Hoversten}, {Kuin},
  {Marshall}, {Roming}, {Still}, {vanden Berk}, {Zane}, \&
  {Nousek}}]{Oates09MNRAS}
{Oates}, S.~R., {Page}, M.~J., {Schady}, P., {et~al.} 2009, \mnras, 395, 490

\bibitem[{{Oates} \& {Racusin}(2009)}]{Oates2009GCN10054}
{Oates}, S.~R. \& {Racusin}, J.~L. 2009, GRB Coordinates Network, 1054, 1

\bibitem[{{Oates} \& {Stamatikos}(2008)}]{Oates2008GCN7611}
{Oates}, S.~R. \& {Stamatikos}, M. 2008, GRB Coordinates Network, 7611

\bibitem[{{Olivares E.} {et~al.}(2015){Olivares E.}, {Greiner}, {Schady},
  {Klose}, {Kr{\"u}hler}, {Rau}, {Savaglio}, {Kann}, {Pignata}, {Elliott},
  {Rossi}, {Nardini}, {Afonso}, {Filgas}, {Nicuesa Guelbenzu}, {Schmidl}, \&
  {Sudilovsky}}]{Olivares2015AA}
{Olivares E.}, F., {Greiner}, J., {Schady}, P., {et~al.} 2015, \aap, 577, A44

\bibitem[{{Padmanabhan} \& {Loeb}(2021)}]{Padmanabhan2021arXiv}
{Padmanabhan}, H. \& {Loeb}, A. 2021, arXiv e-prints, arXiv:2101.12222

\bibitem[{{Page} {et~al.}(2011){Page}, {Starling}, {Fitzpatrick}, {Pandey},
  {Osborne}, {Schady}, {McBreen}, {Campana}, {Ukwatta}, {Pagani}, {Beardmore},
  \& {Evans}}]{Page2011MNRAS}
{Page}, K.~L., {Starling}, R.~L.~C., {Fitzpatrick}, G., {et~al.} 2011, \mnras,
  416, 2078

\bibitem[{{Page} {et~al.}(2019){Page}, {Oates}, {De Pasquale}, {Breeveld},
  {Emery}, {Kuin}, {Marshall}, {Siegel}, \& {Roming}}]{Page2019MNRAS}
{Page}, M.~J., {Oates}, S.~R., {De Pasquale}, M., {et~al.} 2019, \mnras, 488,
  2855

\bibitem[{{Palmerio} \& {Daigne}(2021)}]{Palmerio2020arXiv}
{Palmerio}, J.~T. \& {Daigne}, F. 2021, \aap, 649, A166

\bibitem[{{Panaitescu} \& {Kumar}(2000)}]{Panaitescu2000ApJ}
{Panaitescu}, A. \& {Kumar}, P. 2000, \apj, 543, 66

\bibitem[{{Pandey} \& {Kumar}(2012)}]{Pandey2012GCN13904}
{Pandey}, S.~B. \& {Kumar}, B. 2012, GRB Coordinates Network, 13904

\bibitem[{{Pandey} \& {Kumar}(2014{\natexlab{a}})}]{Pandey2014GCN16133}
{Pandey}, S.~B. \& {Kumar}, B. 2014{\natexlab{a}}, GRB Coordinates Network,
  16133, 1

\bibitem[{{Pandey} \& {Kumar}(2014{\natexlab{b}})}]{Pandey2014GCN16517}
{Pandey}, S.~B. \& {Kumar}, B. 2014{\natexlab{b}}, GRB Coordinates Network,
  16517

\bibitem[{{Pandey} {et~al.}(2014){Pandey}, {Kumar}, \&
  {Agarwal}}]{Pandey2014GCN16164}
{Pandey}, S.~B., {Kumar}, B., \& {Agarwal}, A. 2014, GRB Coordinates Network,
  16164

\bibitem[{{Pandey} {et~al.}(2013){Pandey}, {Kumar}, \&
  {Joshi}}]{Pandey2013GCN15501}
{Pandey}, S.~B., {Kumar}, B., \& {Joshi}, Y.~C. 2013, GRB Coordinates Network,
  15501

\bibitem[{{Pandey} {et~al.}(2011){Pandey}, {Yadav}, {Sagar}, \&
  {Stalin}}]{Pandey2011GCN12792}
{Pandey}, S.~B., {Yadav}, R.~K.~S., {Sagar}, R., \& {Stalin}, C.~S. 2011, GRB
  Coordinates Network, 12792

\bibitem[{{Park} {et~al.}(2012){Park}, {Pak}, {Im}, {Choi}, {Jeon}, {Chang},
  {Jeong}, {Lim}, \& {Kim}}]{ParkWK2012}
{Park}, W.-K., {Pak}, S., {Im}, M., {et~al.} 2012, \pasp, 124, 839

\bibitem[{{Patel} {et~al.}(2010){Patel}, {Warren}, {Mortlock}, \&
  {Fynbo}}]{Patel2010AA}
{Patel}, M., {Warren}, S.~J., {Mortlock}, D.~J., \& {Fynbo}, J.~P.~U. 2010,
  \aap, 512, L3

\bibitem[{{Paul} {et~al.}(2011){Paul}, {Wei}, {Basa}, \&
  {Zhang}}]{Paul2011CRPhy}
{Paul}, J., {Wei}, J., {Basa}, S., \& {Zhang}, S.-N. 2011, Comptes Rendus
  Physique, 12, 298

\bibitem[{{Pearson} {et~al.}(2010){Pearson}, {Moody}, \&
  {Pace}}]{Pearson2010GCN10626}
{Pearson}, R., {Moody}, J.~W., \& {Pace}, C. 2010, GRB Coordinates Network,
  10626

\bibitem[{{Pe'er}(2015)}]{Pe'er2015AdAst}
{Pe'er}, A. 2015, Advances in Astronomy, 2015, 907321

\bibitem[{{Pei}(1992)}]{Pei1992ApJ}
{Pei}, Y.~C. 1992, \apj, 395, 130

\bibitem[{{P{\'e}rez-Ram{\'\i}rez} {et~al.}(2010){P{\'e}rez-Ram{\'\i}rez}, {de
  Ugarte Postigo}, {Gorosabel}, {Aloy}, {J{\'o}hannesson}, {Guerrero},
  {Osborne}, {Page}, {Warwick}, {Horv{\'a}th}, {Veres}, {Jel{\'\i}nek},
  {Kub{\'a}nek}, {Guziy}, {Bremer}, {Winters}, {Riva}, \&
  {Castro-Tirado}}]{Perez2010AA}
{P{\'e}rez-Ram{\'\i}rez}, D., {de Ugarte Postigo}, A., {Gorosabel}, J.,
  {et~al.} 2010, \aap, 510, A105

\bibitem[{{Perley} {et~al.}(2005){Perley}, {Bloom}, {Cooper}, {Newman},
  {Guhathakurta}, {Prochaska}, \& {Chen}}]{Perley2005GCN3932}
{Perley}, D., {Bloom}, J.~S., {Cooper}, M., {et~al.} 2005, GRB Coordinates
  Network, 3932

\bibitem[{{Perley}(2009{\natexlab{a}})}]{Perley2009GCN10060}
{Perley}, D.~A. 2009{\natexlab{a}}, GRB Coordinates Network, 10060

\bibitem[{{Perley}(2009{\natexlab{b}})}]{Perley2009GCN10058}
{Perley}, D.~A. 2009{\natexlab{b}}, GRB Coordinates Network, 10058

\bibitem[{{Perley} {et~al.}(2010){Perley}, {Bloom}, {Klein}, {Covino},
  {Minezaki}, {Wo{\'z}niak}, {Vestrand}, {Williams}, {Milne}, {Butler},
  {Updike}, {Kr{\"u}hler}, {Afonso}, {Antonelli}, {Cowie}, {Ferrero},
  {Greiner}, {Hartmann}, {Kakazu}, {K{\"u}pc{\"u} Yolda{\c{s}}}, {Morgan},
  {Price}, {Prochaska}, \& {Yoshii}}]{Perley2010MNRAS}
{Perley}, D.~A., {Bloom}, J.~S., {Klein}, C.~R., {et~al.} 2010, \mnras, 406,
  2473

\bibitem[{{Perley} {et~al.}(2014{\natexlab{a}}){Perley}, {Cao}, \&
  {Cenko}}]{Perley2014GCN17228}
{Perley}, D.~A., {Cao}, Y., \& {Cenko}, S.~B. 2014{\natexlab{a}}, GRB
  Coordinates Network, 17228

\bibitem[{{Perley} \& {Cenko}(2013{\natexlab{a}})}]{Perley2013GCN14804}
{Perley}, D.~A. \& {Cenko}, S.~B. 2013{\natexlab{a}}, GRB Coordinates Network,
  14804

\bibitem[{{Perley} \& {Cenko}(2013{\natexlab{b}})}]{Perley2013GCN15423}
{Perley}, D.~A. \& {Cenko}, S.~B. 2013{\natexlab{b}}, GRB Coordinates Network,
  15423

\bibitem[{{Perley} \& {Cenko}(2014{\natexlab{a}})}]{Perley2014GCN16498}
{Perley}, D.~A. \& {Cenko}, S.~B. 2014{\natexlab{a}}, GRB Coordinates Network,
  16498

\bibitem[{{Perley} \& {Cenko}(2014{\natexlab{b}})}]{Perley2014GCN16491}
{Perley}, D.~A. \& {Cenko}, S.~B. 2014{\natexlab{b}}, GRB Coordinates Network,
  16491

\bibitem[{{Perley} \& {Cenko}(2015)}]{Perley2015GCN18295}
{Perley}, D.~A. \& {Cenko}, S.~B. 2015, GRB Coordinates Network, 18295

\bibitem[{{Perley} {et~al.}(2014{\natexlab{b}}){Perley}, {Cenko}, {Corsi},
  {Tanvir}, {Levan}, {Kann}, {Sonbas}, {Wiersema}, {Zheng}, {Zhao}, {Bai},
  {Bremer}, {Castro-Tirado}, {Chang}, {Clubb}, {Frail}, {Fruchter}, {G{\"o}{\u
  g}{\"u}{\c s}}, {Greiner}, {G{\"u}ver}, {Horesh}, {Filippenko}, {Klose},
  {Mao}, {Morgan}, {Pozanenko}, {Schmidl}, {Stecklum}, {Tanga}, {Volnova},
  {Volvach}, {Wang}, {Winters}, \& {Xin}}]{Perley2014ApJ}
{Perley}, D.~A., {Cenko}, S.~B., {Corsi}, A., {et~al.} 2014{\natexlab{b}},
  \apj, 781, 37

\bibitem[{{Perley} {et~al.}(2016{\natexlab{a}}){Perley}, {Kr{\"u}hler},
  {Schulze}, {de Ugarte Postigo}, {Hjorth}, {Berger}, {Cenko}, {Chary},
  {Cucchiara}, {Ellis}, {Fong}, {Fynbo}, {Gorosabel}, {Greiner}, {Jakobsson},
  {Kim}, {Laskar}, {Levan}, {Micha{\l}owski}, {Milvang-Jensen}, {Tanvir},
  {Th{\"o}ne}, \& {Wiersema}}]{Perley2016shoals1}
{Perley}, D.~A., {Kr{\"u}hler}, T., {Schulze}, S., {et~al.} 2016{\natexlab{a}},
  \apj, 817, 7

\bibitem[{{Perley} {et~al.}(2011){Perley}, {Morgan}, {Updike}, {Yuan},
  {Akerlof}, {Miller}, {Bloom}, {Cenko}, {Li}, {Filippenko}, {Prochaska},
  {Kann}, {Tanvir}, {Levan}, {Butler}, {Christian}, {Hartmann}, {Milne},
  {Rykoff}, {Rujopakarn}, {Wheeler}, \& {Williams}}]{Perley2011AJ}
{Perley}, D.~A., {Morgan}, A.~N., {Updike}, A., {et~al.} 2011, \aj, 141, 36

\bibitem[{{Perley} {et~al.}(2009){Perley}, {Prochaska}, {Kalas}, {Howard},
  {Fitzgerald}, {Marcy}, \& {Graham}}]{Perley2009GCN10272}
{Perley}, D.~A., {Prochaska}, J.~X., {Kalas}, P., {et~al.} 2009, GRB
  Coordinates Network, 10272

\bibitem[{{Perley} {et~al.}(2016{\natexlab{b}}){Perley}, {Quimby}, {Yan},
  {Vreeswijk}, {De Cia}, {Lunnan}, {Gal-Yam}, {Yaron}, {Filippenko}, {Graham},
  {Laher}, \& {Nugent}}]{Perley2016ApJ}
{Perley}, D.~A., {Quimby}, R.~M., {Yan}, L., {et~al.} 2016{\natexlab{b}}, \apj,
  830, 13

\bibitem[{{Piranomonte} {et~al.}(2006){Piranomonte}, {D'Elia}, {D'Avanzo},
  {Malesani}, {Grazian}, {Fugazza}, {Antonelli}, {Campana}, {Chincarini},
  {Covino}, {Della Valle}, {Fernandez-Soto}, {Fiore}, {Stella}, {Tagliaferri},
  \& {Testa}}]{Piranomonte2006GCN4583}
{Piranomonte}, S., {D'Elia}, V., {D'Avanzo}, P., {et~al.} 2006, GRB Coordinates
  Network, 4583

\bibitem[{{Piranomonte} {et~al.}(2011){Piranomonte}, {Vergani}, {Malesani},
  {Fynbo}, {Wiersema}, \& {Kaper}}]{Piranomonte2011GCN12164}
{Piranomonte}, S., {Vergani}, S.~D., {Malesani}, D., {et~al.} 2011, GRB
  Coordinates Network, 12164

\bibitem[{{Planck Collaboration} {et~al.}(2020){Planck Collaboration},
  {Aghanim}, {Akrami}, {Ashdown}, {Aumont}, {Baccigalupi}, {Ballardini},
  {Banday}, {Barreiro}, {Bartolo}, {Basak}, {Battye}, {Benabed}, {Bernard},
  {Bersanelli}, {Bielewicz}, {Bock}, {Bond}, {Borrill}, {Bouchet}, {Boulanger},
  {Bucher}, {Burigana}, {Butler}, {Calabrese}, {Cardoso}, {Carron},
  {Challinor}, {Chiang}, {Chluba}, {Colombo}, {Combet}, {Contreras}, {Crill},
  {Cuttaia}, {de Bernardis}, {de Zotti}, {Delabrouille}, {Delouis}, {Di
  Valentino}, {Diego}, {Dor{\'e}}, {Douspis}, {Ducout}, {Dupac}, {Dusini},
  {Efstathiou}, {Elsner}, {En{\ss}lin}, {Eriksen}, {Fantaye}, {Farhang},
  {Fergusson}, {Fernandez-Cobos}, {Finelli}, {Forastieri}, {Frailis},
  {Fraisse}, {Franceschi}, {Frolov}, {Galeotta}, {Galli}, {Ganga},
  {G{\'e}nova-Santos}, {Gerbino}, {Ghosh}, {Gonz{\'a}lez-Nuevo}, {G{\'o}rski},
  {Gratton}, {Gruppuso}, {Gudmundsson}, {Hamann}, {Handley}, {Hansen},
  {Herranz}, {Hildebrandt}, {Hivon}, {Huang}, {Jaffe}, {Jones}, {Karakci},
  {Keih{\"a}nen}, {Keskitalo}, {Kiiveri}, {Kim}, {Kisner}, {Knox},
  {Krachmalnicoff}, {Kunz}, {Kurki-Suonio}, {Lagache}, {Lamarre}, {Lasenby},
  {Lattanzi}, {Lawrence}, {Le Jeune}, {Lemos}, {Lesgourgues}, {Levrier},
  {Lewis}, {Liguori}, {Lilje}, {Lilley}, {Lindholm}, {L{\'o}pez-Caniego},
  {Lubin}, {Ma}, {Mac{\'\i}as-P{\'e}rez}, {Maggio}, {Maino}, {Mandolesi},
  {Mangilli}, {Marcos-Caballero}, {Maris}, {Martin}, {Martinelli},
  {Mart{\'\i}nez-Gonz{\'a}lez}, {Matarrese}, {Mauri}, {McEwen}, {Meinhold},
  {Melchiorri}, {Mennella}, {Migliaccio}, {Millea}, {Mitra},
  {Miville-Desch{\^e}nes}, {Molinari}, {Montier}, {Morgante}, {Moss}, {Natoli},
  {N{\o}rgaard-Nielsen}, {Pagano}, {Paoletti}, {Partridge}, {Patanchon},
  {Peiris}, {Perrotta}, {Pettorino}, {Piacentini}, {Polastri}, {Polenta},
  {Puget}, {Rachen}, {Reinecke}, {Remazeilles}, {Renzi}, {Rocha}, {Rosset},
  {Roudier}, {Rubi{\~n}o-Mart{\'\i}n}, {Ruiz-Granados}, {Salvati}, {Sandri},
  {Savelainen}, {Scott}, {Shellard}, {Sirignano}, {Sirri}, {Spencer},
  {Sunyaev}, {Suur-Uski}, {Tauber}, {Tavagnacco}, {Tenti}, {Toffolatti},
  {Tomasi}, {Trombetti}, {Valenziano}, {Valiviita}, {Van Tent}, {Vibert},
  {Vielva}, {Villa}, {Vittorio}, {Wandelt}, {Wehus}, {White}, {White},
  {Zacchei}, \& {Zonca}}]{Planck2020AA}
{Planck Collaboration}, {Aghanim}, N., {Akrami}, Y., {et~al.} 2020, \aap, 641,
  A6

\bibitem[{{Poole} {et~al.}(2008){Poole}, {Breeveld}, {Page}, {Landsman},
  {Holland}, {Roming}, {Kuin}, {Brown}, {Gronwall}, {Hunsberger}, {Koch},
  {Mason}, {Schady}, {vanden Berk}, {Blustin}, {Boyd}, {Broos}, {Carter},
  {Chester}, {Cucchiara}, {Hancock}, {Huckle}, {Immler}, {Ivanushkina},
  {Kennedy}, {Marshall}, {Morgan}, {Pandey}, {de Pasquale}, {Smith}, \&
  {Still}}]{Poole2008MNRAS}
{Poole}, T.~S., {Breeveld}, A.~A., {Page}, M.~J., {et~al.} 2008, \mnras, 383,
  627

\bibitem[{{Porterfield} {et~al.}(2011){Porterfield}, {Siegel}, \&
  {Wolf}}]{Porterfield2011GCN12203}
{Porterfield}, B.~L., {Siegel}, M.~H., \& {Wolf}, C.~A. 2011, GRB Coordinates
  Network, 12203, 1

\bibitem[{{Pozanenko} {et~al.}(2017){Pozanenko}, {Volnova}, {Mazaeva},
  {Inasaridze}, {Ayvazyan}, {Inasaridze}, {Zhuzhunadze}, {Reva}, {Burkhonov},
  {Rumyantsev}, {Krugly}, \& {Molotov}}]{Pozanenko2017AAC}
{Pozanenko}, A., {Volnova}, A., {Mazaeva}, E., {et~al.} 2017, Astronomy \&
  Astrophysics (Caucasus), 1, 8

\bibitem[{{Preece} {et~al.}(2014){Preece}, {Burgess}, {von Kienlin}, {Bhat},
  {Briggs}, {Byrne}, {Chaplin}, {Cleveland}, {Collazzi}, {Connaughton},
  {Diekmann}, {Fitzpatrick}, {Foley}, {Gibby}, {Giles}, {Goldstein}, {Greiner},
  {Gruber}, {Jenke}, {Kippen}, {Kouveliotou}, {McBreen}, {Meegan}, {Paciesas},
  {Pelassa}, {Tierney}, {van der Horst}, {Wilson-Hodge}, {Xiong}, {Younes},
  {Yu}, {Ackermann}, {Ajello}, {Axelsson}, {Baldini}, {Barbiellini}, {Baring},
  {Bastieri}, {Bellazzini}, {Bissaldi}, {Bonamente}, {Bregeon}, {Brigida},
  {Bruel}, {Buehler}, {Buson}, {Caliandro}, {Cameron}, {Caraveo}, {Cecchi},
  {Charles}, {Chekhtman}, {Chiang}, {Chiaro}, {Ciprini}, {Claus},
  {Cohen-Tanugi}, {Cominsky}, {Conrad}, {D'Ammando}, {de Angelis}, {de Palma},
  {Dermer}, {Desiante}, {Digel}, {Di Venere}, {Drell}, {Drlica-Wagner},
  {Favuzzi}, {Franckowiak}, {Fukazawa}, {Fusco}, {Gargano}, {Gehrels},
  {Germani}, {Giglietto}, {Giordano}, {Giroletti}, {Godfrey}, {Granot},
  {Grenier}, {Guiriec}, {Hadasch}, {Hanabata}, {Harding}, {Hayashida},
  {Iyyani}, {Jogler}, {J{\'o}hannesson}, {Kawano}, {Kn{\"o}dlseder},
  {Kocevski}, {Kuss}, {Lande}, {Larsson}, {Larsson}, {Latronico}, {Longo},
  {Loparco}, {Lovellette}, {Lubrano}, {Mayer}, {Mazziotta}, {Michelson},
  {Mizuno}, {Monzani}, {Moretti}, {Morselli}, {Murgia}, {Nemmen}, {Nuss},
  {Nymark}, {Ohno}, {Ohsugi}, {Okumura}, {Omodei}, {Orienti}, {Paneque},
  {Perkins}, {Pesce-Rollins}, {Piron}, {Pivato}, {Porter}, {Racusin},
  {Rain{\`o}}, {Rando}, {Razzano}, {Razzaque}, {Reimer}, {Reimer}, {Ritz},
  {Roth}, {Ryde}, {Sartori}, {Scargle}, {Schulz}, {Sgr{\`o}}, {Siskind},
  {Spandre}, {Spinelli}, {Suson}, {Tajima}, {Takahashi}, {Thayer}, {Thayer},
  {Tibaldo}, {Tinivella}, {Torres}, {Tosti}, {Troja}, {Usher}, {Vandenbroucke},
  {Vasileiou}, {Vianello}, {Vitale}, {Werner}, {Winer}, {Wood}, \&
  {Zhu}}]{Preece2014Science}
{Preece}, R., {Burgess}, J.~M., {von Kienlin}, A., {et~al.} 2014, Science, 343,
  51

\bibitem[{{Price} {et~al.}(2006){Price}, {Cowie}, {Minezaki}, {Schmidt},
  {Songaila}, \& {Yoshii}}]{Price2006ApJ}
{Price}, P.~A., {Cowie}, L.~L., {Minezaki}, T., {et~al.} 2006, \apj, 645, 851

\bibitem[{{Prochaska} {et~al.}(2009){Prochaska}, {Sheffer}, {Perley}, {Bloom},
  {Lopez}, {Dessauges-Zavadsky}, {Chen}, {Filippenko}, {Ganeshalingam}, {Li},
  {Miller}, \& {Starr}}]{Prochaska2009ApJ}
{Prochaska}, J.~X., {Sheffer}, Y., {Perley}, D.~A., {et~al.} 2009, \apjl, 691,
  L27

\bibitem[{{Pruzhinskaya} {et~al.}(2014){Pruzhinskaya}, {Krushinsky},
  {Lipunova}, {Gorbovskoy}, {Balanutsa}, {Kuznetsov}, {Denisenko}, {Kornilov},
  {Tyurina}, {Lipunov}, {Tlatov}, {Parkhomenko}, {Budnev}, {Yazev}, {Ivanov},
  {Gress}, {Yurkov}, {Gabovich}, {Sergienko}, \&
  {Sinyakov}}]{Pruzhinskaya2014NewA}
{Pruzhinskaya}, M.~V., {Krushinsky}, V.~V., {Lipunova}, G.~V., {et~al.} 2014,
  \na, 29, 65

\bibitem[{{Qin} {et~al.}(2010){Qin}, {Liang}, {Lu}, {Wei}, \&
  {Zhang}}]{Qin2010MNRAS}
{Qin}, S.-F., {Liang}, E.-W., {Lu}, R.-J., {Wei}, J.-Y., \& {Zhang}, S.-N.
  2010, \mnras, 406, 558

\bibitem[{{Racusin} {et~al.}(2008){Racusin}, {Karpov}, {Sokolowski}, {Granot},
  {Wu}, {Pal'Shin}, {Covino}, {van der Horst}, {Oates}, {Schady}, {Smith},
  {Cummings}, {Starling}, {Piotrowski}, {Zhang}, {Evans}, {Holland}, {Malek},
  {Page}, {Vetere}, {Margutti}, {Guidorzi}, {Kamble}, {Curran}, {Beardmore},
  {Kouveliotou}, {Mankiewicz}, {Melandri}, {O'Brien}, {Page}, {Piran},
  {Tanvir}, {Wrochna}, {Aptekar}, {Barthelmy}, {Bartolini}, {Beskin}, {Bondar},
  {Bremer}, {Campana}, {Castro-Tirado}, {Cucchiara}, {Cwiok}, {D'Avanzo},
  {D'Elia}, {Della Valle}, {de Ugarte Postigo}, {Dominik}, {Falcone}, {Fiore},
  {Fox}, {Frederiks}, {Fruchter}, {Fugazza}, {Garrett}, {Gehrels},
  {Golenetskii}, {Gomboc}, {Gorosabel}, {Greco}, {Guarnieri}, {Immler},
  {Jelinek}, {Kasprowicz}, {La Parola}, {Levan}, {Mangano}, {Mazets},
  {Molinari}, {Moretti}, {Nawrocki}, {Oleynik}, {Osborne}, {Pagani}, {Pandey},
  {Paragi}, {Perri}, {Piccioni}, {Ramirez-Ruiz}, {Roming}, {Steele}, {Strom},
  {Testa}, {Tosti}, {Ulanov}, {Wiersema}, {Wijers}, {Winters}, {Zarnecki},
  {Zerbi}, {M{\'e}sz{\'a}ros}, {Chincarini}, \& {Burrows}}]{Racusin2008Nature}
{Racusin}, J.~L., {Karpov}, S.~V., {Sokolowski}, M., {et~al.} 2008, \nat, 455,
  183

\bibitem[{{Ramsay} {et~al.}(2018){Ramsay}, {Lyman}, {Ulaczyk}, {Steeghs},
  {Wiersema}, {Dyer}, {Gompertz}, {Levan}, {Cutter}, {Ackley}, {Galloway},
  {Rol}, {Dhillon}, {O'Brien}, {Starling}, {Poshyachinda}, {Pollacco},
  {Thrane}, \& {Palle}}]{Ramsay2018GCN23503}
{Ramsay}, G., {Lyman}, J., {Ulaczyk}, K., {et~al.} 2018, GRB Coordinates
  Network, 23503

\bibitem[{{Rhodes} {et~al.}(2020){Rhodes}, {van der Horst}, {Fender},
  {Monageng}, {Anderson}, {Antoniadis}, {Bietenholz}, {B{\"o}ttcher}, {Bright},
  {Green}, {Kouveliotou}, {Kramer}, {Motta}, {Wijers}, {Williams}, \&
  {Woudt}}]{Rhodes2020MNRAS}
{Rhodes}, L., {van der Horst}, A.~J., {Fender}, R., {et~al.} 2020, \mnras, 496,
  3326

\bibitem[{{Robertson} \& {Ellis}(2012)}]{Robertson2012ApJ}
{Robertson}, B.~E. \& {Ellis}, R.~S. 2012, \apj, 744, 95

\bibitem[{{Rol} {et~al.}(2005){Rol}, {Wijers}, {Kouveliotou}, {Kaper}, \&
  {Kaneko}}]{Rol2005ApJ}
{Rol}, E., {Wijers}, R. A.~M.~J., {Kouveliotou}, C., {Kaper}, L., \& {Kaneko},
  Y. 2005, \apj, 624, 868

\bibitem[{{Roming} {et~al.}(2012){Roming}, {Bil{\'e}n}, {Burrows}, {Falcone},
  {Fox}, {Herter}, {Kennea}, {McConnell}, \& {Nousek}}]{Roming2012MSAIS}
{Roming}, P.~W.~A., {Bil{\'e}n}, S.~G., {Burrows}, D.~N., {et~al.} 2012,
  Memorie della Societa Astronomica Italiana Supplementi, 21, 155

\bibitem[{{Roming} {et~al.}(2005){Roming}, {Kennedy}, {Mason}, {Nousek}, {Ahr},
  {Bingham}, {Broos}, {Carter}, {Hancock}, {Huckle}, {Hunsberger}, {Kawakami},
  {Killough}, {Koch}, {McLelland}, {Smith}, {Smith}, {Soto}, {Boyd},
  {Breeveld}, {Holland}, {Ivanushkina}, {Pryzby}, {Still}, \&
  {Stock}}]{Roming2005SSRv}
{Roming}, P.~W.~A., {Kennedy}, T.~E., {Mason}, K.~O., {et~al.} 2005, \ssr, 120,
  95

\bibitem[{{Ronchi} {et~al.}(2020){Ronchi}, {Fumagalli}, {Ravasio}, {Oganesyan},
  {Toffano}, {Salafia}, {Nava}, {Ascenzi}, {Ghirlanda}, \&
  {Ghisellini}}]{Ronchi2020AA}
{Ronchi}, M., {Fumagalli}, F., {Ravasio}, M.~E., {et~al.} 2020, \aap, 636, A55

\bibitem[{{Rossi} {et~al.}(2022){Rossi}, {Frederiks}, {Kann}, {De Pasquale},
  {Pian}, {Lamb}, {D'Avanzo}, {Izzo}, {Levan}, {Malesani}, {Melandri}, {Nicuesa
  Guelbenzu}, {Schulze}, {Strausbaugh}, {Tanvir}, {Amati}, {Campana},
  {Cucchiara}, {Ghirlanda}, {Della Valle}, {Klose}, {Salvaterra}, {Starling},
  {Stratta}, {Tsvetkova}, {Vergani}, {D'A{\`\i}}, {Burgarella}, {Covino},
  {D'Elia}, {de Ugarte Postigo}, {Fausey}, {Fynbo}, {Frontera}, {Guidorzi},
  {Heintz}, {Masetti}, {Maiorano}, {Mundell}, {Oates}, {Page}, {Palazzi},
  {Palmerio}, {Pugliese}, {Rau}, {Saccardi}, {Sbarufatti}, {Svinkin},
  {Tagliaferri}, {van der Horst}, {Watson}, {Ulanov}, {Wiersema}, {Xu}, \&
  {Zhang}}]{Rossi2022AA}
{Rossi}, A., {Frederiks}, D.~D., {Kann}, D.~A., {et~al.} 2022, \aap, 665, A125

\bibitem[{{Rossi} {et~al.}(2011){Rossi}, {Schulze}, {Klose}, {Kann}, {Rau},
  {Krimm}, {J{\'o}hannesson}, {Panaitescu}, {Yuan}, {Ferrero}, {Kr{\"u}hler},
  {Greiner}, {Schady}, {Pandey}, {Amati}, {Afonso}, {Akerlof}, {Arnold},
  {Clemens}, {Filgas}, {Hartmann}, {K{\"u}pc{\"u} Yolda{\c s}}, {McBreen},
  {McKay}, {Nicuesa Guelbenzu}, {Olivares}, {Paciesas}, {Rykoff}, {Szokoly},
  {Updike}, \& {Yolda{\c s}}}]{Rossi2011AA}
{Rossi}, A., {Schulze}, S., {Klose}, S., {et~al.} 2011, \aap, 529, A142

\bibitem[{{Rujopakarn} {et~al.}(2008){Rujopakarn}, {Guver}, \&
  {Smith}}]{Rujopakarn2008GCN7792}
{Rujopakarn}, W., {Guver}, T., \& {Smith}, D.~A. 2008, GRB Coordinates Network,
  7792

\bibitem[{{Rujopakarn} {et~al.}(2011){Rujopakarn}, {Schaefer}, \&
  {Rykoff}}]{Rujopakarn2011GCN11707}
{Rujopakarn}, W., {Schaefer}, B.~E., \& {Rykoff}, E.~S. 2011, GRB Coordinates
  Network, 11707

\bibitem[{{Rumyantsev} {et~al.}(2008{\natexlab{a}}){Rumyantsev}, {Antoniuk}, \&
  {Pozanenko}}]{Rumyantsev2008GCN7974}
{Rumyantsev}, V., {Antoniuk}, K., \& {Pozanenko}, A. 2008{\natexlab{a}}, GRB
  Coordinates Network, 7974

\bibitem[{{Rumyantsev} {et~al.}(2011{\natexlab{a}}){Rumyantsev}, {Antoniuk}, \&
  {Pozanenko}}]{Rumyantsev2011GCN11973}
{Rumyantsev}, V., {Antoniuk}, K., \& {Pozanenko}, A. 2011{\natexlab{a}}, GRB
  Coordinates Network, 11973

\bibitem[{{Rumyantsev} {et~al.}(2011{\natexlab{b}}){Rumyantsev}, {Antoniuk}, \&
  {Pozanenko}}]{Rumyantsev2011GCN11979}
{Rumyantsev}, V., {Antoniuk}, K., \& {Pozanenko}, A. 2011{\natexlab{b}}, GRB
  Coordinates Network, 11979

\bibitem[{{Rumyantsev} {et~al.}(2008{\natexlab{b}}){Rumyantsev}, {Antonyuk},
  {Andreev}, \& {Pozanenko}}]{Rumyantsev2008GCN8645}
{Rumyantsev}, V., {Antonyuk}, K., {Andreev}, M., \& {Pozanenko}, A.
  2008{\natexlab{b}}, GRB Coordinates Network, 8645

\bibitem[{{Rumyantsev} {et~al.}(2009){Rumyantsev}, {Biryukov}, \&
  {Pozanenko}}]{Rumyantsev2009GCN10116}
{Rumyantsev}, V., {Biryukov}, V., \& {Pozanenko}, A. 2009, GRB Coordinates
  Network, 10116

\bibitem[{{Rumyantsev} {et~al.}(2015){Rumyantsev}, {Mazaeva}, {Volnova}, \&
  {Pozanenko}}]{Rumyantsev2015GCN18556}
{Rumyantsev}, V., {Mazaeva}, E., {Volnova}, A., \& {Pozanenko}, A. 2015, GRB
  Coordinates Network, 18556

\bibitem[{{Rumyantsev} \& {Pozanenko}(2008)}]{Rumyantsev2008GCN7857}
{Rumyantsev}, V. \& {Pozanenko}, A. 2008, GRB Coordinates Network, 7857

\bibitem[{{Rumyantsev} \&
  {Pozanenko}(2010{\natexlab{a}})}]{Rumyantsev2010GCN10783}
{Rumyantsev}, V. \& {Pozanenko}, A. 2010{\natexlab{a}}, GRB Coordinates
  Network, 10783

\bibitem[{{Rumyantsev} \&
  {Pozanenko}(2010{\natexlab{b}})}]{Rumyantsev2010GCN10883}
{Rumyantsev}, V. \& {Pozanenko}, A. 2010{\natexlab{b}}, GRB Coordinates
  Network, 10883

\bibitem[{{Rumyantsev} {et~al.}(2011{\natexlab{c}}){Rumyantsev}, {Pozanenko},
  \& {Klunko}}]{Rumyantsev2011GCN11986}
{Rumyantsev}, V., {Pozanenko}, A., \& {Klunko}, E. 2011{\natexlab{c}}, GRB
  Coordinates Network, 11986

\bibitem[{{Rumyantsev} {et~al.}(2010){Rumyantsev}, {Shakhovkoy}, \&
  {Pozanenko}}]{Rumyantsev2010GCN10634}
{Rumyantsev}, V., {Shakhovkoy}, D., \& {Pozanenko}, A. 2010, GRB Coordinates
  Network, 10634

\bibitem[{{Ryan} {et~al.}(2020){Ryan}, {van Eerten}, {Piro}, \&
  {Troja}}]{Ryan2020ApJ}
{Ryan}, G., {van Eerten}, H., {Piro}, L., \& {Troja}, E. 2020, \apj, 896, 166

\bibitem[{{Saccardi} {et~al.}(2023){Saccardi}, {Vergani}, {De Cia}, {D'Elia},
  {Heintz}, {Izzo}, {Palmerio}, {Petitjean}, {Rossi}, {de Ugarte Postigo},
  {Christensen}, {Konstantopoulou}, {Levan}, {Malesani}, {M{\o}ller},
  {Ramburuth-Hurt}, {Salvaterra}, {Tanvir}, {Th{\"o}ne}, {Vejlgaard}, {Fynbo},
  {Kann}, {Schady}, {Watson}, {Wiersema}, {Campana}, {Covino}, {De Pasquale},
  {Fausey}, {Hartmann}, {van der Horst}, {Jakobsson}, {Palazzi}, {Pugliese},
  {Savaglio}, {Starling}, {Stratta}, \& {Zafar}}]{Saccardi2023}
{Saccardi}, A., {Vergani}, S.~D., {De Cia}, A., {et~al.} 2023, \aap, 671, A84

\bibitem[{{Sahu}(2014)}]{Sahu2014GCN16272}
{Sahu}, D.~K. 2014, GRB Coordinates Network, 16272

\bibitem[{{Sahu} {et~al.}(2012){Sahu}, {Anupama}, \&
  {Pandey}}]{Sahu2012GCN13185}
{Sahu}, D.~K., {Anupama}, G.~C., \& {Pandey}, S.~B. 2012, GRB Coordinates
  Network, 13185

\bibitem[{{Sahu} {et~al.}(2010{\natexlab{a}}){Sahu}, {Arora}, {Singh}, \&
  {Kartha}}]{Sahu2010GCN11197}
{Sahu}, D.~K., {Arora}, S., {Singh}, N.~S., \& {Kartha}, S.~S.
  2010{\natexlab{a}}, GRB Coordinates Network, 11197

\bibitem[{{Sahu} {et~al.}(2010{\natexlab{b}}){Sahu}, {Bhatt}, \&
  {Arora}}]{Sahu2010GCN11175}
{Sahu}, D.~K., {Bhatt}, B.~C., \& {Arora}, S. 2010{\natexlab{b}}, GRB
  Coordinates Network, 11175

\bibitem[{{Sahu} \& {Fort{\'\i}n}(2020)}]{Sahu2020ApJ}
{Sahu}, S. \& {Fort{\'\i}n}, C. E.~L. 2020, \apjl, 895, L41

\bibitem[{{Sakamoto} {et~al.}(2012){Sakamoto}, {Donato}, {Gehrels}, {Okajima},
  \& {Urata}}]{Sakamoto2012GCN12894}
{Sakamoto}, T., {Donato}, D., {Gehrels}, N., {Okajima}, T., \& {Urata}, Y.
  2012, GRB Coordinates Network, 12894

\bibitem[{{Salvaterra}(2015)}]{Salvaterra2015}
{Salvaterra}, R. 2015, Journal of High Energy Astrophysics, 7, 35

\bibitem[{{Salvaterra} {et~al.}(2012){Salvaterra}, {Campana}, {Vergani},
  {Covino}, {D'Avanzo}, {Fugazza}, {Ghirlanda}, {Ghisellini}, {Melandri},
  {Nava}, {Sbarufatti}, {Flores}, {Piranomonte}, \&
  {Tagliaferri}}]{Salvaterra2012ApJ}
{Salvaterra}, R., {Campana}, S., {Vergani}, S.~D., {et~al.} 2012, \apj, 749, 68

\bibitem[{{Salvaterra} {et~al.}(2009){Salvaterra}, {Della Valle}, {Campana},
  {Chincarini}, {Covino}, {D'Avanzo}, {Fern{\'a}ndez-Soto}, {Guidorzi},
  {Mannucci}, {Margutti}, {Th{\"o}ne}, {Antonelli}, {Barthelmy}, {de Pasquale},
  {D'Elia}, {Fiore}, {Fugazza}, {Hunt}, {Maiorano}, {Marinoni}, {Marshall},
  {Molinari}, {Nousek}, {Pian}, {Racusin}, {Stella}, {Amati}, {Andreuzzi},
  {Cusumano}, {Fenimore}, {Ferrero}, {Giommi}, {Guetta}, {Holland}, {Hurley},
  {Israel}, {Mao}, {Markwardt}, {Masetti}, {Pagani}, {Palazzi}, {Palmer},
  {Piranomonte}, {Tagliaferri}, \& {Testa}}]{Salvaterra2009Nature}
{Salvaterra}, R., {Della Valle}, M., {Campana}, S., {et~al.} 2009, \nat, 461,
  1258

\bibitem[{{Salvaterra} {et~al.}(2013){Salvaterra}, {Maio}, {Ciardi}, \&
  {Campisi}}]{Salvaterra2013}
{Salvaterra}, R., {Maio}, U., {Ciardi}, B., \& {Campisi}, M.~A. 2013, \mnras,
  429, 2718

\bibitem[{{S{\'a}nchez-Ram{\'\i}rez} {et~al.}(2017){S{\'a}nchez-Ram{\'\i}rez},
  {Hancock}, {J{\'o}hannesson}, {Murphy}, {de Ugarte Postigo}, {Gorosabel},
  {Kann}, {Kr{\"u}hler}, {Oates}, {Japelj}, {Th{\"o}ne}, {Lundgren}, {Perley},
  {Malesani}, {de Gregorio Monsalvo}, {Castro-Tirado}, {D'Elia}, {Fynbo},
  {Garcia-Appadoo}, {Goldoni}, {Greiner}, {Hu}, {Jel{\'\i}nek}, {Jeong},
  {Kamble}, {Klose}, {Kuin}, {Llorente}, {Mart{\'\i}n}, {Nicuesa Guelbenzu},
  {Rossi}, {Schady}, {Sparre}, {Sudilovsky}, {Tello}, {Updike}, {Wiersema}, \&
  {Zhang}}]{Sanchez-Ramirez2017MNRAS}
{S{\'a}nchez-Ram{\'\i}rez}, R., {Hancock}, P.~J., {J{\'o}hannesson}, G.,
  {et~al.} 2017, \mnras, 464, 4624

\bibitem[{{Sanchez-Ramirez} {et~al.}(2010){Sanchez-Ramirez}, {Tello}, {Sota},
  {Gorosabel}, \& {Castro-Tirado}}]{Sanchez-Ramirez2010GCN11180}
{Sanchez-Ramirez}, R., {Tello}, J.~C., {Sota}, A., {Gorosabel}, J., \&
  {Castro-Tirado}, A.~J. 2010, GRB Coordinates Network, 11180

\bibitem[{{Sari} {et~al.}(1998){Sari}, {Piran}, \& {Narayan}}]{Sari1998ApJ}
{Sari}, R., {Piran}, T., \& {Narayan}, R. 1998, \apjl, 497, L17

\bibitem[{{Sasada} {et~al.}(2018){Sasada}, {Nakaoka}, {Kawabata}, {Uchida},
  {Yamazaki}, \& {Kawabata}}]{Sasada2018GCN22977}
{Sasada}, M., {Nakaoka}, T., {Kawabata}, M., {et~al.} 2018, GRB Coordinates
  Network, 22977

\bibitem[{{Schady} {et~al.}(2007){Schady}, {Mason}, {Page}, {de Pasquale},
  {Morris}, {Romano}, {Roming}, {Immler}, \& {vanden Berk}}]{Schady2007MNRAS}
{Schady}, P., {Mason}, K.~O., {Page}, M.~J., {et~al.} 2007, \mnras, 377, 273

\bibitem[{{Schady} {et~al.}(2010){Schady}, {Page}, {Oates}, {Still}, {de
  Pasquale}, {Dwelly}, {Kuin}, {Holland}, {Marshall}, \&
  {Roming}}]{Schady2010MNRAS}
{Schady}, P., {Page}, M.~J., {Oates}, S.~R., {et~al.} 2010, \mnras, 401, 2773

\bibitem[{{Schaefer} \& {Pandey}(2009)}]{Schaefer2009GCN10036}
{Schaefer}, B.~E. \& {Pandey}, S.~B. 2009, GRB Coordinates Network, 1036

\bibitem[{{Schaefer} {et~al.}(2011){Schaefer}, {Zheng}, \&
  {Flewelling}}]{Schaefer2011GCN12197}
{Schaefer}, B.~E., {Zheng}, W., \& {Flewelling}, H. 2011, GRB Coordinates
  Network, 12197, 1

\bibitem[{{Schlafly} \& {Finkbeiner}(2011)}]{Schlafly2011ApJ}
{Schlafly}, E.~F. \& {Finkbeiner}, D.~P. 2011, \apj, 737, 103

\bibitem[{{Schmalz} {et~al.}(2018){Schmalz}, {Graziani}, {Pozanenko},
  {Volnova}, {Mazaeva}, \& {Molotov}}]{Schmalz2018GCN23020}
{Schmalz}, S., {Graziani}, F., {Pozanenko}, A., {et~al.} 2018, GRB Coordinates
  Network, 23020

\bibitem[{{Schmidl} {et~al.}(2015){Schmidl}, {Knust}, \&
  {Greiner}}]{Schmidl2015GCN18277}
{Schmidl}, S., {Knust}, F., \& {Greiner}, J. 2015, GRB Coordinates Network,
  18277

\bibitem[{{Schulze} {et~al.}(2015){Schulze}, {Chapman}, {Hjorth}, {Levan},
  {Jakobsson}, {Bj{\"o}rnsson}, {Perley}, {Kr{\"u}hler}, {Gorosabel}, {Tanvir},
  {de Ugarte Postigo}, {Fynbo}, {Milvang-Jensen}, {M{\o}ller}, \&
  {Watson}}]{Schulze2015ApJ}
{Schulze}, S., {Chapman}, R., {Hjorth}, J., {et~al.} 2015, \apj, 808, 73

\bibitem[{{Schulze} {et~al.}(2011){Schulze}, {Klose}, {Bj{\"o}rnsson},
  {Jakobsson}, {Kann}, {Rossi}, {Kr{\"u}hler}, {Greiner}, \&
  {Ferrero}}]{Schulze2011}
{Schulze}, S., {Klose}, S., {Bj{\"o}rnsson}, G., {et~al.} 2011, \aap, 526, A23

\bibitem[{{Schweyer} {et~al.}(2014){Schweyer}, {Wiseman}, {Schady}, \&
  {Greiner}}]{Schweyer2014GCN17212}
{Schweyer}, T., {Wiseman}, P., {Schady}, P., \& {Greiner}, J. 2014, GRB
  Coordinates Network, 17212

\bibitem[{Seiffert {et~al.}(2021)Seiffert, Balady, Chang, Dyer, Fausey,
  Guiriec, Hart, Morris, Rodriguez, Roming, Rud, Russel, Sambruna, Terrile,
  Torossian, van~der Horst, White, Willems, Woodmansee, \&
  Young}]{Seiffert_2021}
Seiffert, M., Balady, A., Chang, T.-C., {et~al.} 2021, in {UV}/Optical/{IR}
  Space Telescopes and Instruments: Innovative Technologies and Concepts X, ed.
  J.~B. Breckinridge, H.~P. Stahl, \& A.~A. Barto ({SPIE})

\bibitem[{{Selsing} {et~al.}(2019){Selsing}, {Malesani}, {Goldoni}, {Fynbo},
  {Kr{\"u}hler}, {Antonelli}, {Arabsalmani}, {Bolmer}, {Cano}, {Christensen},
  {Covino}, {D'Avanzo}, {D'Elia}, {De Cia}, {de Ugarte Postigo}, {Flores},
  {Friis}, {Gomboc}, {Greiner}, {Groot}, {Hammer}, {Hartoog}, {Heintz},
  {Hjorth}, {Jakobsson}, {Japelj}, {Kann}, {Kaper}, {Ledoux}, {Leloudas},
  {Levan}, {Maiorano}, {Melandri}, {Milvang-Jensen}, {Palazzi}, {Palmerio},
  {Perley}, {Pian}, {Piranomonte}, {Pugliese}, {S{\'a}nchez-Ram{\'{\i}}rez},
  {Savaglio}, {Schady}, {Schulze}, {Sollerman}, {Sparre}, {Tagliaferri},
  {Tanvir}, {Th{\"o}ne}, {Vergani}, {Vreeswijk}, {Watson}, {Wiersema},
  {Wijers}, {Xu}, \& {Zafar}}]{Selsing2019AA}
{Selsing}, J., {Malesani}, D., {Goldoni}, P., {et~al.} 2019, \aap, 623, A92

\bibitem[{{Sheffer} {et~al.}(2009){Sheffer}, {Prochaska}, {Draine}, {Perley},
  \& {Bloom}}]{Sheffer2009ApJ}
{Sheffer}, Y., {Prochaska}, J.~X., {Draine}, B.~T., {Perley}, D.~A., \&
  {Bloom}, J.~S. 2009, \apjl, 701, L63

\bibitem[{{Siegel} \& {Markwardt}(2010)}]{Siegel2010GCN11242}
{Siegel}, M.~H. \& {Markwardt}, C.~B. 2010, GRB Coordinates Network, 11242

\bibitem[{{Siegel} \& {Marshall}(2010)}]{Siegel2010GCN10625}
{Siegel}, M.~H. \& {Marshall}, F. 2010, GRB Coordinates Network, 10625

\bibitem[{{Siegel} \& {Rowlinson}(2009)}]{Siegel2009GCN9377}
{Siegel}, M.~H. \& {Rowlinson}, B.~A. 2009, GRB Coordinates Network, 9377

\bibitem[{{Siegel} \& {Wolf}(2011)}]{Siegel2011GCN12199}
{Siegel}, M.~H. \& {Wolf}, C.~A. 2011, GRB Coordinates Network, 12199

\bibitem[{{Skrutskie} {et~al.}(2006){Skrutskie}, {Cutri}, {Stiening},
  {Weinberg}, {Schneider}, {Carpenter}, {Beichman}, {Capps}, {Chester},
  {Elias}, {Huchra}, {Liebert}, {Lonsdale}, {Monet}, {Price}, {Seitzer},
  {Jarrett}, {Kirkpatrick}, {Gizis}, {Howard}, {Evans}, {Fowler}, {Fullmer},
  {Hurt}, {Light}, {Kopan}, {Marsh}, {McCallon}, {Tam}, {Van Dyk}, \&
  {Wheelock}}]{Skrutskie2006AJ}
{Skrutskie}, M.~F., {Cutri}, R.~M., {Stiening}, R., {et~al.} 2006, \aj, 131,
  1163

\bibitem[{{Sonbas} {et~al.}(2015){Sonbas}, {Basturk}, {Guver}, {Gogus},
  {Eryilmaz}, {Erece}, \& {Kirbiyik}}]{Sonbas2015GCN18314}
{Sonbas}, E., {Basturk}, O., {Guver}, T., {et~al.} 2015, GRB Coordinates
  Network, 18314

\bibitem[{{Sonbas} {et~al.}(2013){Sonbas}, {Temiz}, {Guver}, {Eker}, {Kaynar},
  {Gogus}, \& {Kirbiyik}}]{Sonbas2013GCN15161}
{Sonbas}, E., {Temiz}, U., {Guver}, T., {et~al.} 2013, GRB Coordinates Network,
  15161

\bibitem[{{Sonbas} {et~al.}(2014){Sonbas}, {Ukwatta}, {Amaral-Rogers}, \&
  {Chester}}]{Sonbas2014GCNR470}
{Sonbas}, E., {Ukwatta}, T.~N., {Amaral-Rogers}, A., \& {Chester}, M.~M. 2014,
  GCN Report, 470

\bibitem[{{Sparre} {et~al.}(2014){Sparre}, {Hartoog}, {Kr{\"u}hler}, {Fynbo},
  {Watson}, {Wiersema}, {D'Elia}, {Zafar}, {Afonso}, {Covino}, {de Ugarte
  Postigo}, {Flores}, {Goldoni}, {Greiner}, {Hjorth}, {Jakobsson}, {Kaper},
  {Klose}, {Levan}, {Malesani}, {Milvang-Jensen}, {Nardini}, {Piranomonte},
  {Sollerman}, {S{\'a}nchez-Ram{\'\i}rez}, {Schulze}, {Tanvir}, {Vergani}, \&
  {Wijers}}]{Sparre2014}
{Sparre}, M., {Hartoog}, O.~E., {Kr{\"u}hler}, T., {et~al.} 2014, \apj, 785,
  150

\bibitem[{{Sparre} {et~al.}(2011){Sparre}, {Sollerman}, {Fynbo}, {Malesani},
  {Goldoni}, {de Ugarte Postigo}, {Covino}, {D'Elia}, {Flores}, {Hammer},
  {Hjorth}, {Jakobsson}, {Kaper}, {Leloudas}, {Levan}, {Milvang-Jensen},
  {Schulze}, {Tagliaferri}, {Tanvir}, {Watson}, {Wiersema}, \&
  {Wijers}}]{Sparre2011ApJ}
{Sparre}, M., {Sollerman}, J., {Fynbo}, J.~P.~U., {et~al.} 2011, \apjl, 735,
  L24

\bibitem[{{Spergel} {et~al.}(2003){Spergel}, {Verde}, {Peiris}, {Komatsu},
  {Nolta}, {Bennett}, {Halpern}, {Hinshaw}, {Jarosik}, {Kogut}, {Limon},
  {Meyer}, {Page}, {Tucker}, {Weiland}, {Wollack}, \&
  {Wright}}]{Spergel2003ApJS}
{Spergel}, D.~N., {Verde}, L., {Peiris}, H.~V., {et~al.} 2003, \apjs, 148, 175

\bibitem[{{Sposetti} \& {Immler}(2010)}]{Sposetti2010GCN11213}
{Sposetti}, S. \& {Immler}, S. 2010, GRB Coordinates Network, 11213

\bibitem[{{Srivastava} {et~al.}(2018){Srivastava}, {Kumar}, {Otzer}, {de},
  {Bhalerao}, {Anupama}, \& {Kasliwal}}]{Srivastava2018GCN23510}
{Srivastava}, S., {Kumar}, H., {Otzer}, S., {et~al.} 2018, GRB Coordinates
  Network, 23510

\bibitem[{{Starling} {et~al.}(2009){Starling}, {Rol}, {van der Horst}, {Yoon},
  {Pal'Shin}, {Ledoux}, {Page}, {Fynbo}, {Wiersema}, {Tanvir}, {Jakobsson},
  {Guidorzi}, {Curran}, {Levan}, {O'Brien}, {Osborne}, {Svinkin}, {de Ugarte
  Postigo}, {Oosting}, \& {Howarth}}]{Starling2009MNRAS}
{Starling}, R.~L.~C., {Rol}, E., {van der Horst}, A.~J., {et~al.} 2009, \mnras,
  400, 90

\bibitem[{{Starling} {et~al.}(2007){Starling}, {Wijers}, {Wiersema}, {Rol},
  {Curran}, {Kouveliotou}, {van der Horst}, \& {Heemskerk}}]{Starling2007ApJ}
{Starling}, R.~L.~C., {Wijers}, R.~A.~M.~J., {Wiersema}, K., {et~al.} 2007,
  \apj, 661, 787

\bibitem[{{Steele} {et~al.}(2017){Steele}, {Kopa{\v{c}}}, {Arnold}, {Smith},
  {Kobayashi}, {Jermak}, {Mundell}, {Gomboc}, {Guidorzi}, {Melandri}, \&
  {Japelj}}]{Steele2017ApJ}
{Steele}, I.~A., {Kopa{\v{c}}}, D., {Arnold}, D.~M., {et~al.} 2017, \apj, 843,
  143

\bibitem[{{Steele} {et~al.}(2004){Steele}, {Smith}, {Rees}, {Baker}, {Bates},
  {Bode}, {Bowman}, {Carter}, {Etherton}, {Ford}, {Fraser}, {Gomboc}, {Lett},
  {Mansfield}, {Marchant}, {Medrano-Cerda}, {Mottram}, {Raback}, {Scott},
  {Tomlinson}, \& {Zamanov}}]{Steele2004SPIE}
{Steele}, I.~A., {Smith}, R.~J., {Rees}, P.~C., {et~al.} 2004, in Society of
  Photo-Optical Instrumentation Engineers (SPIE) Conference Series, Vol. 5489,
  Ground-based Telescopes, ed. J.~{Oschmann}, Jacobus~M., 679--692

\bibitem[{{Steidel} \& {Hamilton}(1992)}]{Steidel1992}
{Steidel}, C.~C. \& {Hamilton}, D. 1992, \aj, 104, 941

\bibitem[{{Steinhardt} {et~al.}(2021){Steinhardt}, {Andersen}, {Brammer},
  {Christensen}, {Fynbo}, {Laursen}, {Milvang-Jensen}, {Oesch}, \&
  {Toft}}]{Steinhardt2021arXiv}
{Steinhardt}, C.~L., {Andersen}, M.~I., {Brammer}, G.~B., {et~al.} 2021, Nature
  Astronomy, 5, 993

\bibitem[{{Stratta} {et~al.}(2011){Stratta}, {Gallerani}, \&
  {Maiolino}}]{Stratta2011AA}
{Stratta}, G., {Gallerani}, S., \& {Maiolino}, R. 2011, \aap, 532, A45

\bibitem[{{Stratta} {et~al.}(2013){Stratta}, {Gendre}, {Atteia}, {Bo{\"e}r},
  {Coward}, {De Pasquale}, {Howell}, {Klotz}, {Oates}, \&
  {Piro}}]{Stratta2013ApJ}
{Stratta}, G., {Gendre}, B., {Atteia}, J.~L., {et~al.} 2013, \apj, 779, 66

\bibitem[{{Stratta} {et~al.}(2007){Stratta}, {Maiolino}, {Fiore}, \&
  {D'Elia}}]{Stratta2007ApJ}
{Stratta}, G., {Maiolino}, R., {Fiore}, F., \& {D'Elia}, V. 2007, \apjl, 661,
  L9

\bibitem[{{Strobl} {et~al.}(2010){Strobl}, {Blazek}, {Jelinek}, {Polasek},
  {Kubanek}, {Nekola}, {Kocka}, \& {Hudec}}]{Strobl2010GCN11340}
{Strobl}, J., {Blazek}, M., {Jelinek}, M., {et~al.} 2010, GRB Coordinates
  Network, 11340

\bibitem[{{Strobl} {et~al.}(2018){Strobl}, {Jelinek}, \&
  {Hudec}}]{Strobl2018GCN22541}
{Strobl}, J., {Jelinek}, M., \& {Hudec}, R. 2018, GRB Coordinates Network,
  22541

\bibitem[{{Strobl} {et~al.}(2013){Strobl}, {Jelinek}, {Polasek}, {Jakubec},
  {Skala}, \& {Hudec}}]{Strobl2013GCN15410}
{Strobl}, J., {Jelinek}, M., {Polasek}, C., {et~al.} 2013, GRB Coordinates
  Network, 15410

\bibitem[{{Sudilovsky} {et~al.}(2012){Sudilovsky}, {Nicuesa Guelbenzu}, \&
  {Greiner}}]{Sudilovsky2012GCN13129}
{Sudilovsky}, V., {Nicuesa Guelbenzu}, A., \& {Greiner}, J. 2012, GRB
  Coordinates Network, 13129

\bibitem[{{Sugita} {et~al.}(2009){Sugita}, {Yamaoka}, {Ohno}, {Tashiro},
  {Nakagawa}, {Urata}, {Pal'Shin}, {Golenetskii}, {Sakamoto}, {Cummings},
  {Krimm}, {Stamatikos}, {Parsons}, {Barthelmy}, \& {Gehrels}}]{Sugita2009PASJ}
{Sugita}, S., {Yamaoka}, K., {Ohno}, M., {et~al.} 2009, \pasj, 61, 521

\bibitem[{{Tagliaferri} {et~al.}(2005){Tagliaferri}, {Antonelli}, {Chincarini},
  {Fern{\'a}ndez-Soto}, {Malesani}, {Della Valle}, {D'Avanzo}, {Grazian},
  {Testa}, {Campana}, {Covino}, {Fiore}, {Stella}, {Castro-Tirado},
  {Gorosabel}, {Burrows}, {Capalbi}, {Cusumano}, {Conciatore}, {D'Elia},
  {Filliatre}, {Fugazza}, {Gehrels}, {Goldoni}, {Guetta}, {Guziy}, {Held},
  {Hurley}, {Israel}, {Jel{\'\i}nek}, {Lazzati}, {L{\'o}pez-Echarri},
  {Melandri}, {Mirabel}, {Moles}, {Moretti}, {Mason}, {Nousek}, {Osborne},
  {Pellizza}, {Perna}, {Piranomonte}, {Piro}, {de Ugarte Postigo}, \&
  {Romano}}]{Tagliaferri2005AA}
{Tagliaferri}, G., {Antonelli}, L.~A., {Chincarini}, G., {et~al.} 2005, \aap,
  443, L1

\bibitem[{{Takahashi} \& {Arai}(2014)}]{Takahashi2014GCN16167}
{Takahashi}, J. \& {Arai}, A. 2014, GRB Coordinates Network, 16167

\bibitem[{{Takahashi} {et~al.}(2013){Takahashi}, {Morihana}, {Honda}, \&
  {Takagi}}]{Takahashi2013GCN14495}
{Takahashi}, J., {Morihana}, K., {Honda}, S., \& {Takagi}, Y. 2013, GCN
  Circulars, 14495

\bibitem[{{Takaki} {et~al.}(2014){Takaki}, {Nakaoka}, {Moritani}, \&
  {Kawabata}}]{Takaki2014GCN16487}
{Takaki}, K., {Nakaoka}, T., {Moritani}, Y., \& {Kawabata}, K.~S. 2014, GRB
  Coordinates Network, 16487

\bibitem[{{Tanigawa} {et~al.}(2013){Tanigawa}, {Yoshii}, {Ito}, {Saito},
  {Yano}, {Usui}, {Tachibana}, {Kurita}, {Yatsu}, \&
  {Kawai}}]{Tanigawa2013GCN15481}
{Tanigawa}, T., {Yoshii}, T., {Ito}, K., {et~al.} 2013, GRB Coordinates
  Network, 15481

\bibitem[{{Tanvir} \& {Ball}(2012)}]{Tanvir2012GCN13532}
{Tanvir}, N.~R. \& {Ball}, J. 2012, GRB Coordinates Network, 13532

\bibitem[{{Tanvir} {et~al.}(2009){Tanvir}, {Fox}, {Levan}, {Berger},
  {Wiersema}, {Fynbo}, {Cucchiara}, {Kr{\"u}hler}, {Gehrels}, {Bloom},
  {Greiner}, {Evans}, {Rol}, {Olivares}, {Hjorth}, {Jakobsson}, {Farihi},
  {Willingale}, {Starling}, {Cenko}, {Perley}, {Maund}, {Duke}, {Wijers},
  {Adamson}, {Allan}, {Bremer}, {Burrows}, {Castro-Tirado}, {Cavanagh}, {de
  Ugarte Postigo}, {Dopita}, {Fatkhullin}, {Fruchter}, {Foley}, {Gorosabel},
  {Kennea}, {Kerr}, {Klose}, {Krimm}, {Komarova}, {Kulkarni}, {Moskvitin},
  {Mundell}, {Naylor}, {Page}, {Penprase}, {Perri}, {Podsiadlowski}, {Roth},
  {Rutledge}, {Sakamoto}, {Schady}, {Schmidt}, {Soderberg}, {Sollerman},
  {Stephens}, {Stratta}, {Ukwatta}, {Watson}, {Westra}, {Wold}, \&
  {Wolf}}]{Tanvir2009Nature}
{Tanvir}, N.~R., {Fox}, D.~B., {Levan}, A.~J., {et~al.} 2009, \nat, 461, 1254

\bibitem[{{Tanvir} {et~al.}(2019){Tanvir}, {Fynbo}, {de Ugarte Postigo},
  {Japelj}, {Wiersema}, {Malesani}, {Perley}, {Levan}, {Selsing}, {Cenko},
  {Kann}, {Milvang-Jensen}, {Berger}, {Cano}, {Chornock}, {Covino},
  {Cucchiara}, {D'Elia}, {Gargiulo}, {Goldoni}, {Gomboc}, {Heintz}, {Hjorth},
  {Izzo}, {Jakobsson}, {Kaper}, {Kr{\"u}hler}, {Laskar}, {Myers},
  {Piranomonte}, {Pugliese}, {Rossi}, {S{\'a}nchez-Ram{\'\i}rez}, {Schulze},
  {Sparre}, {Stanway}, {Tagliaferri}, {Th{\"o}ne}, {Vergani}, {Vreeswijk},
  {Wijers}, {Watson}, \& {Xu}}]{Tanvir2019MNRAS}
{Tanvir}, N.~R., {Fynbo}, J.~P.~U., {de Ugarte Postigo}, A., {et~al.} 2019,
  \mnras, 483, 5380

\bibitem[{{Tanvir} {et~al.}(2018){Tanvir}, {Laskar}, {Levan}, {Perley}, {Zabl},
  {Fynbo}, {Rhoads}, {Cenko}, {Greiner}, {Wiersema}, {Hjorth}, {Cucchiara},
  {Berger}, {Bremer}, {Cano}, {Cobb}, {Covino}, {D'Elia}, {Fong}, {Fruchter},
  {Goldoni}, {Hammer}, {Heintz}, {Jakobsson}, {Kann}, {Kaper}, {Klose},
  {Knust}, {Kr{\"u}hler}, {Malesani}, {Misra}, {Nicuesa Guelbenzu}, {Pugliese},
  {S{\'a}nchez-Ram{\'\i}rez}, {Schulze}, {Stanway}, {de Ugarte Postigo},
  {Watson}, {Wijers}, \& {Xu}}]{Tanvir2018ApJ}
{Tanvir}, N.~R., {Laskar}, T., {Levan}, A.~J., {et~al.} 2018, \apj, 865, 107

\bibitem[{{Tanvir} {et~al.}(2021){Tanvir}, {Le Floc'h}, {Christensen},
  {Caruana}, {Salvaterra}, {Ghirlanda}, {Ciardi}, {Maio}, {D'Odorico},
  {Piedipalumbo}, {Campana}, {Noterdaeme}, {Graziani}, {Amati}, {Bagoly},
  {Bal{\'a}zs}, {Basa}, {Behar}, {De Cia}, {Valle}, {De Pasquale}, {Frontera},
  {Gomboc}, {G{\"o}tz}, {Horvath}, {Hudec}, {Mereghetti}, {O'Brien}, {Osborne},
  {Paltani}, {Rosati}, {Sergijenko}, {Stanway}, {Sz{\'e}csi}, {Toth}, {Urata},
  {Vergani}, \& {Zane}}]{Tanvir2021THESEUS}
{Tanvir}, N.~R., {Le Floc'h}, E., {Christensen}, L., {et~al.} 2021,
  Experimental Astronomy [\eprint[arXiv]{2104.09532}]

\bibitem[{{Tanvir} {et~al.}(2012{\natexlab{a}}){Tanvir}, {Levan}, {Fruchter},
  {Fynbo}, {Hjorth}, {Wiersema}, {Bremer}, {Rhoads}, {Jakobsson}, {O'Brien},
  {Stanway}, {Bersier}, {Natarajan}, {Greiner}, {Watson}, {Castro-Tirado},
  {Wijers}, {Starling}, {Misra}, {Graham}, \& {Kouveliotou}}]{Tanvir2012ApJ}
{Tanvir}, N.~R., {Levan}, A.~J., {Fruchter}, A.~S., {et~al.}
  2012{\natexlab{a}}, \apj, 754, 46

\bibitem[{{Tanvir} {et~al.}(2006){Tanvir}, {Levan}, {Priddey}, {Fruchter}, \&
  {Hjorth}}]{Tanvir2006GCN4602}
{Tanvir}, N.~R., {Levan}, A.~J., {Priddey}, R.~S., {Fruchter}, A.~S., \&
  {Hjorth}, J. 2006, GRB Coordinates Network, 4602

\bibitem[{{Tanvir} {et~al.}(2010){Tanvir}, {Wiersema}, \&
  {Levan}}]{Tanvir2010GCN11230}
{Tanvir}, N.~R., {Wiersema}, K., \& {Levan}, A.~J. 2010, GRB Coordinates
  Network, 11230

\bibitem[{{Tanvir} {et~al.}(2011){Tanvir}, {Wiersema}, {Levan}, {Cenko}, \&
  {Geballe}}]{Tanvir2011GCN12225}
{Tanvir}, N.~R., {Wiersema}, K., {Levan}, A.~J., {Cenko}, S.~B., \& {Geballe},
  T. 2011, GRB Coordinates Network, 12225

\bibitem[{{Tanvir} {et~al.}(2012{\natexlab{b}}){Tanvir}, {Wiersema}, {Levan},
  {Fox}, {Fruchter}, \& {Krogsrud}}]{Tanvir2012GCN13441}
{Tanvir}, N.~R., {Wiersema}, K., {Levan}, A.~J., {et~al.} 2012{\natexlab{b}},
  GRB Coordinates Network, 13441

\bibitem[{{Tello} {et~al.}(2012){Tello}, {Sanchez-Ramirez}, {Gorosabel},
  {Castro-Tirado}, {Rivero}, {Gomez-Velarde}, \& {Klotz}}]{Tello2012GCN13118}
{Tello}, J.~C., {Sanchez-Ramirez}, R., {Gorosabel}, J., {et~al.} 2012, GRB
  Coordinates Network, 13118

\bibitem[{{Terron} {et~al.}(2013){Terron}, {Fernandez}, \&
  {Gorosabel}}]{Terron2013GCN15411}
{Terron}, V., {Fernandez}, M., \& {Gorosabel}, J. 2013, GRB Coordinates
  Network, 15411

\bibitem[{{Th{\"o}ne} {et~al.}(2011){Th{\"o}ne}, {Campana}, {Lazzati}, {de
  Ugarte Postigo}, {Fynbo}, {Christensen}, {Levan}, {Aloy}, {Hjorth},
  {Jakobsson}, {Levesque}, {Malesani}, {Milvang-Jensen}, {Roming}, {Tanvir},
  {Wiersema}, {Gladders}, {Wuyts}, \& {Dahle}}]{Thoene2011MNRAS}
{Th{\"o}ne}, C.~C., {Campana}, S., {Lazzati}, D., {et~al.} 2011, \mnras, 414,
  479

\bibitem[{{Th{\"o}ne} {et~al.}(2013){Th{\"o}ne}, {Fynbo}, {Goldoni}, {de Ugarte
  Postigo}, {Campana}, {Vergani}, {Covino}, {Kr{\"u}hler}, {Kaper}, {Tanvir},
  {Zafar}, {D'Elia}, {Gorosabel}, {Greiner}, {Groot}, {Hammer}, {Jakobsson},
  {Klose}, {Levan}, {Milvang-Jensen}, {Nicuesa}, {Palazzi}, {Piranomonte},
  {Tagliaferri}, {Watson}, {Wiersema}, \& {Wijers}}]{Thoene2013MNRAS}
{Th{\"o}ne}, C.~C., {Fynbo}, J.~P.~U., {Goldoni}, P., {et~al.} 2013, \mnras,
  428, 3590

\bibitem[{{Th{\"o}ne} {et~al.}(2010){Th{\"o}ne}, {Kann}, {J{\'o}hannesson},
  {Selj}, {Jaunsen}, {Fynbo}, {Akerlof}, {Baliyan}, {Bartolini}, {Bikmaev},
  {Bloom}, {Burenin}, {Cobb}, {Covino}, {Curran}, {Dahle}, {Ferrero}, {Foley},
  {French}, {Fruchter}, {Ganesh}, {Graham}, {Greco}, {Guarnieri}, {Hanlon},
  {Hjorth}, {Ibrahimov}, {Israel}, {Jakobsson}, {Jel{\'{\i}}nek}, {Jensen},
  {J{\o}rgensen}, {Khamitov}, {Koch}, {Levan}, {Malesani}, {Masetti}, {Meehan},
  {Melady}, {Nanni}, {N{\"a}r{\"a}nen}, {Pakstiene}, {Pavlinsky}, {Perley},
  {Piccioni}, {Pizzichini}, {Pozanenko}, {Roming}, {Rujopakarn}, {Rumyantsev},
  {Rykoff}, {Sharapov}, {Starr}, {Sunyaev}, {Swan}, {Tanvir}, {Terra}, {de
  Ugarte Postigo}, {Vreeswijk}, {Wilson}, {Yost}, \& {Yuan}}]{Thoene2010AA}
{Th{\"o}ne}, C.~C., {Kann}, D.~A., {J{\'o}hannesson}, G., {et~al.} 2010, \aap,
  523, A70

\bibitem[{{Tkachenko} {et~al.}(2010){Tkachenko}, {Khamitov}, {Burenin},
  {Pavlinsky}, {Sunyaev}, {Bikmaev}, {Sakhibullin}, {Eker}, {Kiziloglu}, \&
  {Gogus}}]{Tkachenko2010GCN11254}
{Tkachenko}, A., {Khamitov}, I., {Burenin}, R., {et~al.} 2010, GRB Coordinates
  Network, 11254

\bibitem[{{Totani} {et~al.}(2016){Totani}, {Aoki}, {Hattori}, \&
  {Kawai}}]{Totani2016PASJ}
{Totani}, T., {Aoki}, K., {Hattori}, T., \& {Kawai}, N. 2016, \pasj, 68, 15

\bibitem[{{Totani} {et~al.}(2014){Totani}, {Aoki}, {Hattori}, {Kosugi},
  {Niino}, {Hashimoto}, {Kawai}, {Ohta}, {Sakamoto}, \&
  {Yamada}}]{Totani2014PASJ}
{Totani}, T., {Aoki}, K., {Hattori}, T., {et~al.} 2014, \pasj, 66, 63

\bibitem[{{Totani} {et~al.}(2006){Totani}, {Kawai}, {Kosugi}, {Aoki}, {Yamada},
  {Iye}, {Ohta}, \& {Hattori}}]{Totani2006PASJ}
{Totani}, T., {Kawai}, N., {Kosugi}, G., {et~al.} 2006, \pasj, 58, 485

\bibitem[{{Tristram} {et~al.}(2012){Tristram}, {Fukui}, \&
  {Sako}}]{Tristram2012GCN13228}
{Tristram}, P.~J., {Fukui}, A., \& {Sako}, T. 2012, GRB Coordinates Network,
  13228

\bibitem[{{Troja} {et~al.}(2007){Troja}, {Cusumano}, {O'Brien}, {Zhang},
  {Sbarufatti}, {Mangano}, {Willingale}, {Chincarini}, {Osborne}, {Marshall},
  {Burrows}, {Campana}, {Gehrels}, {Guidorzi}, {Krimm}, {La Parola}, {Liang},
  {Mineo}, {Moretti}, {Page}, {Romano}, {Tagliaferri}, {Zhang}, {Page}, \&
  {Schady}}]{Troja2007ApJ}
{Troja}, E., {Cusumano}, G., {O'Brien}, P.~T., {et~al.} 2007, \apj, 665, 599

\bibitem[{{Troja} {et~al.}(2012){Troja}, {Sakamoto}, {Guidorzi}, {Norris},
  {Panaitescu}, {Kobayashi}, {Omodei}, {Brown}, {Burrows}, {Evans}, {Gehrels},
  {Marshall}, {Mawson}, {Melandri}, {Mundell}, {Oates}, {Pal'shin}, {Preece},
  {Racusin}, {Steele}, {Tanvir}, {Vasileiou}, {Wilson-Hodge}, \&
  {Yamaoka}}]{Troja2012ApJ}
{Troja}, E., {Sakamoto}, T., {Guidorzi}, C., {et~al.} 2012, \apj, 761, 50

\bibitem[{{Trotter} {et~al.}(2013{\natexlab{a}}){Trotter}, {Lacluyze},
  {Reichart}, {Haislip}, {Berger}, {Carroll}, {Cromartie}, {Egger}, {Foster},
  {Foster}, {Frank}, {Ivarsen}, {James}, {Maples}, {Moore}, {Nysewander},
  {Speckhard}, {Taylor}, \& {Crain}}]{Trotter2013GCN14815}
{Trotter}, A., {Lacluyze}, A., {Reichart}, D., {et~al.} 2013{\natexlab{a}}, GRB
  Coordinates Network, 14815

\bibitem[{{Trotter} {et~al.}(2013{\natexlab{b}}){Trotter}, {Lacluyze},
  {Reichart}, {Haislip}, {Berger}, {Carroll}, {Cromartie}, {Egger}, {Foster},
  {Foster}, {Frank}, {Ivarsen}, {James}, {Maples}, {Moore}, {Nysewander},
  {Speckhard}, {Taylor}, \& {Crain}}]{Trotter2013GCN14826}
{Trotter}, A., {Lacluyze}, A., {Reichart}, D., {et~al.} 2013{\natexlab{b}}, GRB
  Coordinates Network, 14826

\bibitem[{{Uehara} {et~al.}(2012){Uehara}, {Toma}, {Kawabata}, {Chiyonobu},
  {Fukazawa}, {Ikejiri}, {Inoue}, {Itoh}, {Komatsu}, {Miyamoto}, {Mizuno},
  {Nagae}, {Nakaya}, {Ohsugi}, {Sakimoto}, {Sasada}, {Tanaka}, {Uemura},
  {Yamanaka}, {Yamashita}, {Yamazaki}, \& {Yoshida}}]{Uehara2012ApJ}
{Uehara}, T., {Toma}, K., {Kawabata}, K.~S., {et~al.} 2012, \apjl, 752, L6

\bibitem[{{Ukwatta} {et~al.}(2010){Ukwatta}, {Sakamoto}, {Gehrels}, \&
  {Dhuga}}]{Ukwatta2010GCN11198}
{Ukwatta}, T.~N., {Sakamoto}, T., {Gehrels}, N., \& {Dhuga}, K.~S. 2010, GRB
  Coordinates Network, 11198

\bibitem[{{Updike} {et~al.}(2009{\natexlab{a}}){Updike}, {Brittain},
  {Hartmann}, {Colson}, {Cumbee}, {Hackett}, {Lewis}, \&
  {Kronberg}}]{Updike2009GCN9529}
{Updike}, A., {Brittain}, S., {Hartmann}, D., {et~al.} 2009{\natexlab{a}}, GCN
  Circulars, 9529

\bibitem[{{Updike} {et~al.}(2009{\natexlab{b}}){Updike}, {Brittain},
  {Hartmann}, {Colson}, {Cumbee}, {Hackett}, {Lewis}, \&
  {Kronberg}}]{Updike2009GCN9575}
{Updike}, A., {Brittain}, S., {Hartmann}, D., {et~al.} 2009{\natexlab{b}}, GCN
  Circulars, 9575

\bibitem[{{Updike} {et~al.}(2009{\natexlab{c}}){Updike}, {Rossi}, \&
  {Greiner}}]{Updike2009GCN10271}
{Updike}, A., {Rossi}, A., \& {Greiner}, J. 2009{\natexlab{c}}, GRB Coordinates
  Network, 10271

\bibitem[{{Updike} {et~al.}(2010{\natexlab{a}}){Updike}, {Hartmann}, \& {de
  Pree}}]{Updike2010GCN10637}
{Updike}, A.~C., {Hartmann}, D.~H., \& {de Pree}, C. 2010{\natexlab{a}}, GRB
  Coordinates Network, 10637

\bibitem[{{Updike} {et~al.}(2010{\natexlab{b}}){Updike}, {Hartmann}, {Keel}, \&
  {Darnell}}]{Updike2010GCN11174}
{Updike}, A.~C., {Hartmann}, D.~H., {Keel}, W., \& {Darnell}, E.
  2010{\natexlab{b}}, GRB Coordinates Network, 11174

\bibitem[{{Updike} {et~al.}(2010{\natexlab{c}}){Updike}, {Hartmann}, \&
  {Murphy}}]{Updike2010GCN10619}
{Updike}, A.~C., {Hartmann}, D.~H., \& {Murphy}, B. 2010{\natexlab{c}}, GRB
  Coordinates Network, 10619

\bibitem[{{Urata} {et~al.}(2014){Urata}, {Huang}, {Takahashi}, {Im}, {Yamaoka},
  {Tashiro}, {Kim}, {Jang}, \& {Pak}}]{Urata2014ApJ}
{Urata}, Y., {Huang}, K., {Takahashi}, S., {et~al.} 2014, \apj, 789, 146

\bibitem[{{Ursi} {et~al.}(2020){Ursi}, {Tavani}, {Frederiks}, {Romani},
  {Verrecchia}, {Marisaldi}, {Aptekar}, {Antonelli}, {Argan}, {Bulgarelli},
  {Barbiellini}, {Caraveo}, {Cardillo}, {Casentini}, {Cattaneo}, {Chen},
  {Costa}, {Donnarumma}, {Evangelista}, {Feroci}, {Ferrari}, {Fuschino},
  {Galli}, {Giuliani}, {Labanti}, {Lazzarotto}, {Longo}, {Lucarelli},
  {Morselli}, {Paoletti}, {Parmiggiani}, {Piano}, {Pilia}, {Pittori},
  {Svinkin}, {Trois}, {Tsvetkova}, {Vercellone}, \& {Vittorini}}]{Ursi2020ApJ}
{Ursi}, A., {Tavani}, M., {Frederiks}, D.~D., {et~al.} 2020, \apj, 904, 133

\bibitem[{{Usui} {et~al.}(2011){Usui}, {Aoki}, {Song}, {Hayashi}, {Kawakami},
  {Tokoyoda}, {Saito}, {Yatsu}, \& {Kawai}}]{Usui2011GCN12768}
{Usui}, R., {Aoki}, Y., {Song}, S., {et~al.} 2011, GRB Coordinates Network,
  12768

\bibitem[{{van de Stadt} {et~al.}(2013){van de Stadt}, {Wiersema}, {Bekkers},
  {Seynen}, \& {Nieuwenhout}}]{vandeStadt2013GCN14521}
{van de Stadt}, I., {Wiersema}, K., {Bekkers}, T., {Seynen}, M., \&
  {Nieuwenhout}, F. 2013, GCN Circulars, 14521

\bibitem[{{Vangioni} {et~al.}(2015){Vangioni}, {Olive}, {Prestegard}, {Silk},
  {Petitjean}, \& {Mandic}}]{Vangioni2015MNRAS}
{Vangioni}, E., {Olive}, K.~A., {Prestegard}, T., {et~al.} 2015, \mnras, 447,
  2575

\bibitem[{{Varela} {et~al.}(2016){Varela}, {van Eerten}, {Greiner}, {Schady},
  {Elliott}, {Sudilovsky}, {Kr{\"u}hler}, {van der Horst}, {Bolmer}, {Knust},
  {Agurto}, {Azagra}, {Belloche}, {Bertoldi}, {De Breuck}, {Delvaux}, {Filgas},
  {Graham}, {Kann}, {Klose}, {Menten}, {Nicuesa Guelbenzu}, {Rau}, {Rossi},
  {Schmidl}, {Schuller}, {Schweyer}, {Tanga}, {Weiss}, {Wiseman}, \&
  {Wyrowski}}]{Varela2016AA}
{Varela}, K., {van Eerten}, H., {Greiner}, J., {et~al.} 2016, \aap, 589, A37

\bibitem[{{Vergani} {et~al.}(2011){Vergani}, {Flores}, {Covino}, {Fugazza},
  {Gorosabel}, {Levan}, {Puech}, {Salvaterra}, {Tello}, {de Ugarte Postigo},
  {D'Avanzo}, {D'Elia}, {Fern{\'a}ndez}, {Fynbo}, {Ghirlanda},
  {Jel{\'{\i}}nek}, {Lundgren}, {Malesani}, {Palazzi}, {Piranomonte},
  {Rodrigues}, {S{\'a}nchez-Ram{\'{\i}}rez}, {Terr{\'o}n}, {Th{\"o}ne},
  {Antonelli}, {Campana}, {Castro-Tirado}, {Goldoni}, {Hammer}, {Hjorth},
  {Jakobsson}, {Kaper}, {Melandri}, {Milvang-Jensen}, {Sollerman},
  {Tagliaferri}, {Tanvir}, {Wiersema}, \& {Wijers}}]{Vergani2011AA}
{Vergani}, S.~D., {Flores}, H., {Covino}, S., {et~al.} 2011, \aap, 535, A127

\bibitem[{{Vergani} {et~al.}(2015){Vergani}, {Salvaterra}, {Japelj}, {Le
  Floc'h}, {D'Avanzo}, {Fernandez-Soto}, {Kr{\"u}hler}, {Melandri}, {Boissier},
  {Covino}, {Puech}, {Greiner}, {Hunt}, {Perley}, {Petitjean}, {Vinci},
  {Hammer}, {Levan}, {Mannucci}, {Campana}, {Flores}, {Gomboc}, \&
  {Tagliaferri}}]{Vergani2014AA}
{Vergani}, S.~D., {Salvaterra}, R., {Japelj}, J., {et~al.} 2015, \aap, 581,
  A102

\bibitem[{{Vestrand} {et~al.}(2014){Vestrand}, {Wren}, {Panaitescu}, {Wozniak},
  {Davis}, {Palmer}, {Vianello}, {Omodei}, {Xiong}, {Briggs}, {Elphick},
  {Paciesas}, \& {Rosing}}]{Vestrand2014Science}
{Vestrand}, W.~T., {Wren}, J.~A., {Panaitescu}, A., {et~al.} 2014, Science,
  343, 38

\bibitem[{{Vielfaure} {et~al.}(2020){Vielfaure}, {Vergani}, {Japelj}, {Fynbo},
  {Gronke}, {Heintz}, {Malesani}, {Petitjean}, {Tanvir}, {D'Elia}, {Kann},
  {Palmerio}, {Salvaterra}, {Wiersema}, {Arabsalmani}, {Campana}, {Covino}, {De
  Pasquale}, {de Ugarte Postigo}, {Hammer}, {Hartmann}, {Jakobsson},
  {Kouveliotou}, {Laskar}, {Levan}, \& {Rossi}}]{Vielfaure2020AA}
{Vielfaure}, J.~B., {Vergani}, S.~D., {Japelj}, J., {et~al.} 2020, \aap, 641,
  A30

\bibitem[{{Virgili} {et~al.}(2013{\natexlab{a}}){Virgili}, {Mundell}, {Japelj},
  {Gomboc}, {Smith}, \& {Melandri}}]{Virgili2013GCN15406}
{Virgili}, F.~J., {Mundell}, C.~G., {Japelj}, J., {et~al.} 2013{\natexlab{a}},
  GRB Coordinates Network, 15406

\bibitem[{{Virgili} {et~al.}(2013{\natexlab{b}}){Virgili}, {Mundell}, \&
  {Melandri}}]{Virgili2013GCN14785}
{Virgili}, F.~J., {Mundell}, C.~G., \& {Melandri}, A. 2013{\natexlab{b}}, GRB
  Coordinates Network, 14785

\bibitem[{{Virgili} {et~al.}(2013{\natexlab{c}}){Virgili}, {Mundell},
  {Melandri}, \& {Gomboc}}]{Virgili2013GCN14840}
{Virgili}, F.~J., {Mundell}, C.~G., {Melandri}, A., \& {Gomboc}, A.
  2013{\natexlab{c}}, GRB Coordinates Network, 14840

\bibitem[{{Virgili} {et~al.}(2013{\natexlab{d}}){Virgili}, {Mundell},
  {Pal'shin}, {Guidorzi}, {Margutti}, {Melandri}, {Harrison}, {Kobayashi},
  {Chornock}, {Henden}, {Updike}, {Cenko}, {Tanvir}, {Steele}, {Cucchiara},
  {Gomboc}, {Levan}, {Cano}, {Mottram}, {Clay}, {Bersier}, {Kopa{\v c}},
  {Japelj}, {Filippenko}, {Li}, {Svinkin}, {Golenetskii}, {Hartmann}, {Milne},
  {Williams}, {O'Brien}, {Fox}, \& {Berger}}]{Virgili2013ApJ}
{Virgili}, F.~J., {Mundell}, C.~G., {Pal'shin}, V., {et~al.}
  2013{\natexlab{d}}, \apj, 778, 54

\bibitem[{{Virgili} {et~al.}(2011){Virgili}, {Zhang}, {Nagamine}, \&
  {Choi}}]{Virgili2011MNRAS}
{Virgili}, F.~J., {Zhang}, B., {Nagamine}, K., \& {Choi}, J.-H. 2011, \mnras,
  417, 3025

\bibitem[{{Volkov}(2008)}]{Volkov2008GCN8604}
{Volkov}, I. 2008, GRB Coordinates Network, 8604

\bibitem[{{Volnova} {et~al.}(2014{\natexlab{a}}){Volnova}, {Klunko},
  {Eselevich}, {Korobtsev}, \& {Pozanenko}}]{Volnova2014GCN16168}
{Volnova}, A., {Klunko}, E., {Eselevich}, M., {Korobtsev}, I., \& {Pozanenko},
  A. 2014{\natexlab{a}}, GRB Coordinates Network, 16168

\bibitem[{{Volnova} {et~al.}(2014{\natexlab{b}}){Volnova}, {Klunko},
  {Eselevich}, {Korobtsev}, \& {Pozanenko}}]{Volnova2014GCN16281}
{Volnova}, A., {Klunko}, E., {Eselevich}, M., {Korobtsev}, I., \& {Pozanenko},
  A. 2014{\natexlab{b}}, GRB Coordinates Network, 16281

\bibitem[{{Volnova} {et~al.}(2014{\natexlab{c}}){Volnova}, {Klunko},
  {Eselevich}, {Korobtsev}, \& {Pozanenko}}]{Volnova2014GCN16141}
{Volnova}, A., {Klunko}, E., {Eselevich}, M., {Korobtsev}, I., \& {Pozanenko},
  A. 2014{\natexlab{c}}, GRB Coordinates Network, 16141

\bibitem[{{Volnova} {et~al.}(2012{\natexlab{a}}){Volnova}, {Klunko}, \&
  {Pozanenko}}]{Volnova2012GCN13236}
{Volnova}, A., {Klunko}, E., \& {Pozanenko}, A. 2012{\natexlab{a}}, GRB
  Coordinates Network, 13236

\bibitem[{{Volnova} {et~al.}(2010{\natexlab{a}}){Volnova}, {Msu}, {Ibrahimov},
  {Karimov}, \& {Pozanenko}}]{Volnova2010GCN10821}
{Volnova}, A., {Msu}, S., {Ibrahimov}, M., {Karimov}, R., \& {Pozanenko}, A.
  2010{\natexlab{a}}, GRB Coordinates Network, 10821

\bibitem[{{Volnova} {et~al.}(2009){Volnova}, {Pavlenko}, {Sklyanov},
  {Antoniuk}, \& {Pozanenko}}]{Volnova2009GCN9741}
{Volnova}, A., {Pavlenko}, E., {Sklyanov}, A., {Antoniuk}, O., \& {Pozanenko},
  A. 2009, GRB Coordinates Network, 9741

\bibitem[{{Volnova} {et~al.}(2010{\natexlab{b}}){Volnova}, {Pozanenko},
  {Andreev}, {Sergeev}, {Rumyantsev}, {Grankin}, {Erofeeva}, {Kornienko},
  {Molotov}, {Klunko}, \& {Satovski}}]{Volnova2010GCN11395}
{Volnova}, A., {Pozanenko}, A., {Andreev}, M., {et~al.} 2010{\natexlab{b}}, GRB
  Coordinates Network, 11395

\bibitem[{{Volnova} {et~al.}(2010{\natexlab{c}}){Volnova}, {Pozanenko},
  {Vozyakova}, {Satovski}, {Im}, \& {Ibrahimov}}]{Volnova2010GCN11153}
{Volnova}, A., {Pozanenko}, A., {Vozyakova}, O., {et~al.} 2010{\natexlab{c}},
  GRB Coordinates Network, 11153

\bibitem[{{Volnova} {et~al.}(2015{\natexlab{a}}){Volnova}, {Reva}, {Kusakin},
  {Mazaeva}, \& {Pozanenko}}]{Volnova2015GCN18319}
{Volnova}, A., {Reva}, I., {Kusakin}, A., {Mazaeva}, E., \& {Pozanenko}, A.
  2015{\natexlab{a}}, GRB Coordinates Network, 18319

\bibitem[{{Volnova} {et~al.}(2015{\natexlab{b}}){Volnova}, {Sergeev},
  {Andreev}, {Mazaeva}, \& {Pozanenko}}]{Volnova2015GCN18320}
{Volnova}, A., {Sergeev}, A., {Andreev}, M., {Mazaeva}, E., \& {Pozanenko}, A.
  2015{\natexlab{b}}, GRB Coordinates Network, 18320

\bibitem[{{Volnova} {et~al.}(2012{\natexlab{b}}){Volnova}, {Varda}, {Sinyakov},
  {Litvinenko}, {Kouprianov}, {Molotov}, \& {Pozanenko}}]{Volnova2012GCN13235}
{Volnova}, A., {Varda}, D., {Sinyakov}, E., {et~al.} 2012{\natexlab{b}}, GRB
  Coordinates Network, 13235

\bibitem[{{Vreeswijk} {et~al.}(2011{\natexlab{a}}){Vreeswijk}, {Groot},
  {Carter}, {Warwick}, {Xu}, {De Cia}, {Jakobsson}, \&
  {Fynbo}}]{Vreeswijk2011GCN11640}
{Vreeswijk}, P., {Groot}, P., {Carter}, P., {et~al.} 2011{\natexlab{a}}, GRB
  Coordinates Network, 11640

\bibitem[{{Vreeswijk} {et~al.}(2018){Vreeswijk}, {Kann}, {Heintz}, {de Ugarte
  Postigo}, {Milvang-Jensen}, {Malesani}, {Covino}, {Levan}, \&
  {Pugliese}}]{Vreeswijk2018GCN22996}
{Vreeswijk}, P.~M., {Kann}, D.~A., {Heintz}, K.~E., {et~al.} 2018, GRB
  Coordinates Network, 22996

\bibitem[{{Vreeswijk} {et~al.}(2007){Vreeswijk}, {Ledoux}, {Smette}, {Ellison},
  {Jaunsen}, {Andersen}, {Fruchter}, {Fynbo}, {Hjorth}, {Kaufer}, {M{\o}ller},
  {Petitjean}, {Savaglio}, \& {Wijers}}]{Vreeswijk2007AA}
{Vreeswijk}, P.~M., {Ledoux}, C., {Smette}, A., {et~al.} 2007, \aap, 468, 83

\bibitem[{{Vreeswijk} {et~al.}(2011{\natexlab{b}}){Vreeswijk}, {Ledoux},
  {Smette}, {Ellison}, {Jaunsen}, {Andersen}, {Fruchter}, {Fynbo}, {Hjorth},
  {Kaufer}, {M{\o}ller}, {Petitjean}, {Savaglio}, \&
  {Wijers}}]{Vreeswijk2011AA}
{Vreeswijk}, P.~M., {Ledoux}, C., {Smette}, A., {et~al.} 2011{\natexlab{b}},
  \aap, 532, C3

\bibitem[{{{\v{S}}imon} {et~al.}(2010){{\v{S}}imon}, {Pol{\'a}{\v{s}}ek},
  {Jel{\'\i}nek}, {Hudec}, \& {{\v{S}}trobl}}]{Simon2010AA}
{{\v{S}}imon}, V., {Pol{\'a}{\v{s}}ek}, C., {Jel{\'\i}nek}, M., {Hudec}, R., \&
  {{\v{S}}trobl}, J. 2010, \aap, 510, A49

\bibitem[{{Walker} {et~al.}(2012){Walker}, {Court}, {Duffy}, {Edwards},
  {Herath}, {Kirk}, {Patel}, {Prajs}, {Tunbridge}, {Williams}, {Wood},
  {Wright}, {Munoz-Darias}, {Knigge}, {Coriat}, {Gimeno}, \&
  {Gorosabel}}]{Walker2012GCN13112}
{Walker}, C., {Court}, J., {Duffy}, R., {et~al.} 2012, GRB Coordinates Network,
  13112

\bibitem[{{Wang}(2013)}]{Wang2013AA}
{Wang}, F.~Y. 2013, \aap, 556, A90

\bibitem[{{Wang} \& {Dai}(2009)}]{Wang2009MNRAS}
{Wang}, F.~Y. \& {Dai}, Z.~G. 2009, \mnras, 400, L10

\bibitem[{{Wang} \& {Dai}(2011)}]{Wang2011ApJ}
{Wang}, F.~Y. \& {Dai}, Z.~G. 2011, \apjl, 727, L34

\bibitem[{{Wang} {et~al.}(2020){Wang}, {Qiu}, \& {Wei}}]{Wang2020RAA}
{Wang}, J., {Qiu}, Y.-L., \& {Wei}, J.-Y. 2020, Research in Astronomy and
  Astrophysics, 20, 124

\bibitem[{{Wang} {et~al.}(2015){Wang}, {Zheng}, {Filippenko}, \&
  {Cenko}}]{Wang2015GCN18184}
{Wang}, X., {Zheng}, W., {Filippenko}, A.~V., \& {Cenko}, S.~B. 2015, GRB
  Coordinates Network, 18184

\bibitem[{{Wang} {et~al.}(2022){Wang}, {Chen}, {Huang}, {Chen}, {Zheng},
  {D'Elia}, {De Pasquale}, {Pozanenko}, {Xin}, {Stratta}, {Ukwatta}, {Akerlof},
  {Geng}, {Han}, {Hentunen}, {Klunko}, {Kuin}, {Nissinen}, {Rujopakarn},
  {Rumyantsev}, {Rykoff}, {Salmi}, {Schaefer}, {Volnova}, {Wu}, {Wei}, {Liang},
  {Zhang}, \& {Filippenko}}]{Wang2022ApJ}
{Wang}, X.-G., {Chen}, Y.-Z., {Huang}, X.-L., {et~al.} 2022, \apj, 939, 39

\bibitem[{{Wang} {et~al.}(2019){Wang}, {Liu}, {Zhang}, {Xi}, \&
  {Zhang}}]{Wang2019ApJ}
{Wang}, X.-Y., {Liu}, R.-Y., {Zhang}, H.-M., {Xi}, S.-Q., \& {Zhang}, B. 2019,
  \apj, 884, 117

\bibitem[{{Ward} {et~al.}(2008){Ward}, {Holland}, \&
  {Marshall}}]{Ward2008GCN7996}
{Ward}, P., {Holland}, S.~T., \& {Marshall}, F.~E. 2008, GRB Coordinates
  Network, 7996

\bibitem[{{Watson} {et~al.}(2015){Watson}, {Butler}, {Kutyrev}, {Lee},
  {Richer}, {Fox}, {Prochaska}, {Bloom}, {Cucchiara}, {Troja}, {Littlejohns},
  {Ramirez-Ruiz}, {de Diego}, {Georgiev}, {Gonzalez}, {Roman-Zuniga},
  {Gehrels}, {Moseley}, {Capone}, {Golkhou}, \& {Toy}}]{Watson2015GCN18512}
{Watson}, A.~M., {Butler}, N., {Kutyrev}, A., {et~al.} 2015, GRB Coordinates
  Network, 18512

\bibitem[{{Watson} {et~al.}(2019){Watson}, {Butler}, {Kutyrev}, {Lee},
  {Richer}, {Fox}, {Prochaska}, {Bloom}, {Cucchiara}, {Troja}, {Littlejohns},
  {Ramirez-Ruiz}, {Gonzalez}, {Roman-Zuniga}, {Moseley}, {Capone}, {Golkhou},
  \& {Toy}}]{Watson2019GCN23751}
{Watson}, A.~M., {Butler}, N., {Kutyrev}, A., {et~al.} 2019, GRB Coordinates
  Network, 23751

\bibitem[{{Watson} {et~al.}(2018){Watson}, {Pereyra}, {Butler}, {Becerra},
  {Lee}, {Roman-Zuniga}, {Kutyrev}, \& {Troja}}]{Watson2018GCN23481}
{Watson}, A.~M., {Pereyra}, M., {Butler}, N., {et~al.} 2018, GRB Coordinates
  Network, 23481

\bibitem[{{Watson} {et~al.}(2006){Watson}, {Reeves}, {Hjorth}, {Fynbo},
  {Jakobsson}, {Pedersen}, {Sollerman}, {Castro Cer{\'o}n}, {McBreen}, \&
  {Foley}}]{Watson2006ApJ}
{Watson}, D., {Reeves}, J.~N., {Hjorth}, J., {et~al.} 2006, \apjl, 637, L69

\bibitem[{{Wei} {et~al.}(2006){Wei}, {Yan}, \& {Fan}}]{Wei2006ApJ}
{Wei}, D.~M., {Yan}, T., \& {Fan}, Y.~Z. 2006, \apjl, 636, L69

\bibitem[{{Wei} {et~al.}(2016){Wei}, {Cordier}, {Antier}, {Antilogus},
  {Atteia}, {Bajat}, {Basa}, {Beckmann}, {Bernardini}, {Boissier}, {Bouchet},
  {Burwitz}, {Claret}, {Dai}, {Daigne}, {Deng}, {Dornic}, {Feng}, {Foglizzo},
  {Gao}, {Gehrels}, {Godet}, {Goldwurm}, {Gonzalez}, {Gosset}, {G{\"o}tz},
  {Gouiffes}, {Grise}, {Gros}, {Guilet}, {Han}, {Huang}, {Huang}, {Jouret},
  {Klotz}, {La Marle}, {Lachaud}, {Le Floch}, {Lee}, {Leroy}, {Li}, {Li}, {Li},
  {Liang}, {Lyu}, {Mercier}, {Migliori}, {Mochkovitch}, {O'Brien}, {Osborne},
  {Paul}, {Perinati}, {Petitjean}, {Piron}, {Qiu}, {Rau}, {Rodriguez},
  {Schanne}, {Tanvir}, {Vangioni}, {Vergani}, {Wang}, {Wang}, {Wang}, {Wang},
  {Watson}, {Webb}, {Wei}, {Willingale}, {Wu}, {Wu}, {Xin}, {Xu}, {Yu}, {Yu},
  {Yu}, {Zhang}, {Zhang}, {Zhang}, \& {Zhou}}]{Wei2016arXiv}
{Wei}, J., {Cordier}, B., {Antier}, S., {et~al.} 2016, arXiv e-prints,
  arXiv:1610.06892

\bibitem[{{West} {et~al.}(2008{\natexlab{a}}){West}, {Haislip}, {Brennan},
  {Reichart}, {Nysewander}, {Lacluyze}, {Ivarsen}, {Crain}, {Foster}, {Holmes},
  {Schubel}, {Rhine}, {Styblova}, {Trotter}, \& {Weaver}}]{West2008GCN8449}
{West}, J.~P., {Haislip}, J., {Brennan}, T., {et~al.} 2008{\natexlab{a}}, GRB
  Coordinates Network, 8449

\bibitem[{{West} {et~al.}(2008{\natexlab{b}}){West}, {McLin}, {Brennan},
  {Haislip}, {Reichart}, {Cominsky}, {Graves}, {Spear}, {Ivarsen}, {Crain},
  {Foster}, {Holmes}, {Lacluyze}, {Schubel}, {Styblova}, {Trotter}, \&
  {Weaver}}]{West2008GCN8617}
{West}, J.~P., {McLin}, K., {Brennan}, T., {et~al.} 2008{\natexlab{b}}, GRB
  Coordinates Network, 8617

\bibitem[{{White} {et~al.}(2021){White}, {Bauer}, {Baumgartner}, {Bautz},
  {Berger}, {Cenko}, {Chang}, {Falcone}, {Fausey}, {Feldman}, {Fox}, {Fox},
  {Fruchter}, {Fryer}, {Ghirlanda}, {Gorski}, {Grant}, {Guiriec}, {Hart},
  {Hartmann}, {Hennawi}, {Kann}, {Kaplan}, {Kennea}, {Kocevski}, {Kouveliotou},
  {Lawrence}, {Levan}, {Lidz}, {Lien}, {Littenberg}, {Mas-Ribas}, {Moss},
  {O'Brien}, {O'Meara}, {Palmer}, {Pasham}, {Racusin}, {Remillard}, {Roberts},
  {Roming}, {Rud}, {Salvaterra}, {Sambruna}, {Seiffert}, {Sun}, {Tanvir},
  {Terrile}, {Thomas}, {van der Horst}, {Verstrand}, {Willems}, {Wilson-Hodge},
  {Young}, {Amati}, {Bozzo}, {Karczewski}, {Hernandez-Monteagudo}, {Rebolo
  Lopez}, {Genova-Santos}, {Martin}, {Granot}, {Bemiamini}, {Gil}, \&
  {Burns}}]{White2021SPIE}
{White}, N.~E., {Bauer}, F.~E., {Baumgartner}, W., {et~al.} 2021, in Society of
  Photo-Optical Instrumentation Engineers (SPIE) Conference Series, Vol. 11821,
  Society of Photo-Optical Instrumentation Engineers (SPIE) Conference Series,
  1182109

\bibitem[{{Wiersema} {et~al.}(2014){Wiersema}, {Covino}, {Toma}, {van der
  Horst}, {Varela}, {Min}, {Greiner}, {Starling}, {Tanvir}, {Wijers},
  {Campana}, {Curran}, {Fan}, {Fynbo}, {Gorosabel}, {Gomboc}, {G{\"o}tz},
  {Hjorth}, {Jin}, {Kobayashi}, {Kouveliotou}, {Mundell}, {O'Brien}, {Pian},
  {Rowlinson}, {Russell}, {Salvaterra}, {di Serego Alighieri}, {Tagliaferri},
  {Vergani}, {Elliott}, {Fari{\~n}a}, {Hartoog}, {Karjalainen}, {Klose},
  {Knust}, {Levan}, {Schady}, {Sudilovsky}, \&
  {Willingale}}]{Wiersema2014Nature}
{Wiersema}, K., {Covino}, S., {Toma}, K., {et~al.} 2014, \nat, 509, 201

\bibitem[{{Wiersema} {et~al.}(2012){Wiersema}, {Curran}, {Kr{\"u}hler},
  {Melandri}, {Rol}, {Starling}, {Tanvir}, {van der Horst}, {Covino}, {Fynbo},
  {Goldoni}, {Gorosabel}, {Hjorth}, {Klose}, {Mundell}, {O'Brien}, {Palazzi},
  {Wijers}, {D'Elia}, {Evans}, {Filgas}, {Gomboc}, {Greiner}, {Guidorzi},
  {Kaper}, {Kobayashi}, {Kouveliotou}, {Levan}, {Rossi}, {Rowlinson}, {Steele},
  {de Ugarte Postigo}, \& {Vergani}}]{Wiersema2012MNRAS}
{Wiersema}, K., {Curran}, P.~A., {Kr{\"u}hler}, T., {et~al.} 2012, \mnras, 426,
  2

\bibitem[{{Wiggins}(2013)}]{Wiggins2013GCN14490}
{Wiggins}, P. 2013, GCN Circulars, 14490

\bibitem[{{Woosley} \& {Bloom}(2006)}]{Woosley2006ARAA}
{Woosley}, S.~E. \& {Bloom}, J.~S. 2006, \araa, 44, 507

\bibitem[{{Wren} {et~al.}(2011){Wren}, {Vestrand}, {Wozniak}, \&
  {Davis}}]{Wren2011GCN11730}
{Wren}, J., {Vestrand}, W.~T., {Wozniak}, P.~R., \& {Davis}, H. 2011, GRB
  Coordinates Network, 11730

\bibitem[{{Wren} {et~al.}(2012){Wren}, {Wozniak}, {Davis}, \&
  {Vestrand}}]{Wren2012GCN13545}
{Wren}, J., {Wozniak}, P., {Davis}, H., \& {Vestrand}, W.~T. 2012, GRB
  Coordinates Network, 13545

\bibitem[{{Xiao} {et~al.}(2019){Xiao}, {Li}, {Li}, {Li}, {Liao}, {Xiong},
  {Liu}, {Li}, {Li}, {Chang}, {Lu}, {Zhao}, {Zhang}, {Zhang}, {Zou}, {Jin},
  {Zhang}, {Li}, {Lu}, {Song}, {Wu}, {Xu}, \& {Zhang}}]{Xiao2019GCN23716}
{Xiao}, S., {Li}, C.~K., {Li}, X.~B., {et~al.} 2019, GRB Coordinates Network,
  23716

\bibitem[{{Xie} {et~al.}(2020){Xie}, {Wang}, {Zheng}, {Filippenko}, {Qin},
  {Li}, {Zheng}, {Zou}, {Lin}, {Zhu}, {Yuk}, {Lu}, \& {Liang}}]{Xie2020ApJ}
{Xie}, L., {Wang}, X.-G., {Zheng}, W., {et~al.} 2020, \apj, 896, 4

\bibitem[{{Xin} {et~al.}(2008){Xin}, {Feng}, {Zhai}, {Qiu}, {Wei}, {Hu},
  {Deng}, {Wang}, {Urata}, \& {Zheng}}]{Xin2008GCN7814}
{Xin}, L.~P., {Feng}, Q.~C., {Zhai}, M., {et~al.} 2008, GRB Coordinates
  Network, 7814

\bibitem[{{Xin} {et~al.}(2013){Xin}, {Han}, {Qiu}, {Wei}, {Wang}, {Deng}, \&
  {Wu}}]{Xin2013GCN15146}
{Xin}, L.~P., {Han}, X.~H., {Qiu}, Y.~L., {et~al.} 2013, GRB Coordinates
  Network, 15146

\bibitem[{{Xin} {et~al.}(2011){Xin}, {Liang}, {Wei}, {Zhang}, {Lv}, {Zheng},
  {Urata}, {Im}, {Wang}, {Qiu}, {Deng}, {Huang}, {Hu}, {Jeon}, {Li}, \&
  {Han}}]{Xin2011MNRAS}
{Xin}, L.-P., {Liang}, E.-W., {Wei}, J.-Y., {et~al.} 2011, \mnras, 410, 27

\bibitem[{{Xin} {et~al.}(2009){Xin}, {Qian}, {Qiu}, {Wang}, {Wei}, {Zheng},
  {Deng}, \& {Hu}}]{Xin2009GCN10279}
{Xin}, L.~P., {Qian}, S.~B., {Qiu}, Y.~L., {et~al.} 2009, GRB Coordinates
  Network, 10279

\bibitem[{{Xin} {et~al.}(2016){Xin}, {Wang}, {Lin}, {Liang}, {L{\"u}}, {Zhong},
  {Urata}, {Zhao}, {Wu}, {Wei}, {Huang}, {Qiu}, \& {Deng}}]{Xin2016ApJ}
{Xin}, L.-P., {Wang}, Y.-Z., {Lin}, T.-T., {et~al.} 2016, \apj, 817, 152

\bibitem[{{Xin} {et~al.}(2012){Xin}, {Wei}, {Qiu}, {Wang}, {Deng}, {Wu}, \&
  {Han}}]{Xin2012GCN13221}
{Xin}, L.~P., {Wei}, J.~Y., {Qiu}, Y.~L., {et~al.} 2012, GRB Coordinates
  Network, 13221

\bibitem[{{Xin} {et~al.}(2018){Xin}, {Zhong}, {Liang}, {Wang}, {Liu}, {Zhang},
  {Huang}, {Li}, {Qiu}, {Han}, \& {Wei}}]{Xin2018ApJ}
{Xin}, L.-P., {Zhong}, S.-Q., {Liang}, E.-W., {et~al.} 2018, \apj, 860, 8

\bibitem[{{Xu}(2014)}]{Xu2014GCN16140}
{Xu}, D. 2014, GRB Coordinates Network, 16140

\bibitem[{{Xu} {et~al.}(2014){Xu}, {Bai}, {Zhang}, {Esamdin}, \&
  {Ma}}]{Xu2014GCN15956}
{Xu}, D., {Bai}, C.~H., {Zhang}, X., {Esamdin}, A., \& {Ma}, L. 2014, GRB
  Coordinates Network, 15956

\bibitem[{{Xu} {et~al.}(2013{\natexlab{a}}){Xu}, {de Ugarte Postigo},
  {Leloudas}, {Kr{\"u}hler}, {Cano}, {Hjorth}, {Malesani}, {Fynbo},
  {Th{\"o}ne}, {S{\'a}nchez-Ram{\'{\i}}rez}, {Schulze}, {Jakobsson}, {Kaper},
  {Sollerman}, {Watson}, {Cabrera-Lavers}, {Cao}, {Covino}, {Flores}, {Geier},
  {Gorosabel}, {Hu}, {Milvang-Jensen}, {Sparre}, {Xin}, {Zhang}, {Zheng}, \&
  {Zou}}]{Xu2013ApJ}
{Xu}, D., {de Ugarte Postigo}, A., {Leloudas}, G., {et~al.} 2013{\natexlab{a}},
  \apj, 776, 98

\bibitem[{{Xu} {et~al.}(2013{\natexlab{b}}){Xu}, {Fynbo}, {Jakobsson}, {Cano},
  {Milvang-Jensen}, {Malesani}, {de Ugarte Postigo}, \&
  {Hayes}}]{Xu2013GCN15407}
{Xu}, D., {Fynbo}, J.~P.~U., {Jakobsson}, P., {et~al.} 2013{\natexlab{b}}, GRB
  Coordinates Network, 15407

\bibitem[{{Xu} {et~al.}(2011{\natexlab{a}}){Xu}, {Fynbo}, {Nielsen}, \&
  {Jakobsson}}]{Xu2011GCN11970}
{Xu}, D., {Fynbo}, J.~P.~U., {Nielsen}, M., \& {Jakobsson}, P.
  2011{\natexlab{a}}, GRB Coordinates Network, 11970

\bibitem[{{Xu} {et~al.}(2009{\natexlab{a}}){Xu}, {Fynbo}, {Tanvir}, {Hjorth},
  {Leloudas}, {Malesani}, {Jakobsson}, {Wilson}, \&
  {Andersen}}]{Xu2009GCN10053}
{Xu}, D., {Fynbo}, J.~P.~U., {Tanvir}, N.~R., {et~al.} 2009{\natexlab{a}}, GRB
  Coordinates Network, 1053

\bibitem[{{Xu} {et~al.}(2011{\natexlab{b}}){Xu}, {Kankare}, {Kangas}, \&
  {Jakobsson}}]{Xu2011GCN11974}
{Xu}, D., {Kankare}, E., {Kangas}, T., \& {Jakobsson}, P. 2011{\natexlab{b}},
  GRB Coordinates Network, 11974

\bibitem[{{Xu} {et~al.}(2009{\natexlab{b}}){Xu}, {Leloudas}, {Malesani},
  {Jakobsson}, {Lindberg}, \& {Andersen}}]{Xu2009GCN10269}
{Xu}, D., {Leloudas}, G., {Malesani}, D., {et~al.} 2009{\natexlab{b}}, GRB
  Coordinates Network, 10269

\bibitem[{{Xu} {et~al.}(2016){Xu}, {Malesani}, {Fynbo}, {Tanvir}, {Levan}, \&
  {Perley}}]{Xu2016GCN19600}
{Xu}, D., {Malesani}, D., {Fynbo}, J.~P.~U., {et~al.} 2016, GRB Coordinates
  Network, 19600

\bibitem[{{Xu} {et~al.}(2009{\natexlab{c}}){Xu}, {Malesani}, {Hjorth},
  {Djupvik}, {Datson}, {Jakobsson}, {Carmona}, \&
  {Baldovin-Saavedra}}]{Xu2009GCN10196}
{Xu}, D., {Malesani}, D., {Hjorth}, J., {et~al.} 2009{\natexlab{c}}, GCN
  Circulars, 10196

\bibitem[{{Xu} {et~al.}(2009{\natexlab{d}}){Xu}, {Malesani}, {Hjorth},
  {Jakobsson}, {Carmona}, \& {Baldovin-Saavedra}}]{Xu2009GCN10205}
{Xu}, D., {Malesani}, D., {Hjorth}, J., {et~al.} 2009{\natexlab{d}}, GCN
  Circulars, 10205

\bibitem[{{Xu} {et~al.}(2015){Xu}, {Qin}, {Hu}, {Han}, {Zhang}, {Esamdin}, \&
  {Ma}}]{Xu2015GCN18269}
{Xu}, D., {Qin}, Y., {Hu}, Y.~D., {et~al.} 2015, GRB Coordinates Network, 18269

\bibitem[{{Xu} {et~al.}(2011{\natexlab{c}}){Xu}, {Thygesen}, {Kiaee}, \&
  {Jakobsson}}]{Xu2011GCN11961}
{Xu}, D., {Thygesen}, A., {Kiaee}, F., \& {Jakobsson}, P. 2011{\natexlab{c}},
  GRB Coordinates Network, 11961

\bibitem[{{Xu} {et~al.}(2013{\natexlab{c}}){Xu}, {Zhang}, {Cao}, \&
  {Hu}}]{Xu2013GCN15142}
{Xu}, D., {Zhang}, C.~M., {Cao}, C., \& {Hu}, S.~M. 2013{\natexlab{c}}, GRB
  Coordinates Network, 15142

\bibitem[{{Yanagisawa} {et~al.}(2015){Yanagisawa}, {Kuroda}, {Shimizu},
  {Izumiura}, {Yoshida}, {Ohta}, \& {Kawai}}]{Yanagisawa2015GCN18278}
{Yanagisawa}, K., {Kuroda}, D., {Shimizu}, Y., {et~al.} 2015, GRB Coordinates
  Network, 18278

\bibitem[{{Yano} {et~al.}(2014){Yano}, {Yoshii}, {Saito}, {Tachibana},
  {Ohuchi}, {Kurita}, {Ono}, {Fujiwara}, {Yatsu}, \&
  {Kawai}}]{Yano2014GCN16501}
{Yano}, Y., {Yoshii}, T., {Saito}, Y., {et~al.} 2014, GRB Coordinates Network,
  16501

\bibitem[{{Yatsu} {et~al.}(2013){Yatsu}, {Yano}, {Usui}, {Tachibana}, {Ito},
  {Yoshii}, {Kurita}, {Saito}, \& {Kawai}}]{Yatsu2013GCN14454}
{Yatsu}, Y., {Yano}, Y., {Usui}, R., {et~al.} 2013, GCN Circulars, 14454

\bibitem[{{Yonetoku} {et~al.}(2014){Yonetoku}, {Mihara}, {Sawano}, {Ikeda},
  {Harayama}, {Takata}, {Yoshida}, {Seta}, {Toyanago}, {Kagawa}, {Kawai},
  {Kawai}, {Sakamoto}, {Serino}, {Kurosawa}, {Gunji}, {Tanimori}, {Murakami},
  {Yatsu}, {Yamaoka}, {Yoshida}, {Kawabata}, {Matsumoto}, {Tsumura},
  {Matsuura}, {Shirahata}, {Okita}, {Yanagisawa}, {Yoshida}, \&
  {Motohara}}]{Yonetoku2014SPIE}
{Yonetoku}, D., {Mihara}, T., {Sawano}, T., {et~al.} 2014, in Society of
  Photo-Optical Instrumentation Engineers (SPIE) Conference Series, Vol. 9144,
  Space Telescopes and Instrumentation 2014: Ultraviolet to Gamma Ray, ed.
  T.~{Takahashi}, J.-W.~A. {den Herder}, \& M.~{Bautz}, 91442S

\bibitem[{{Yonetoku} {et~al.}(2004){Yonetoku}, {Murakami}, {Nakamura},
  {Yamazaki}, {Inoue}, \& {Ioka}}]{yonetoku2004}
{Yonetoku}, D., {Murakami}, T., {Nakamura}, T., {et~al.} 2004, \apj, 609, 935

\bibitem[{{Yoshida} {et~al.}(2014){Yoshida}, {Itoh}, {Moritani}, {Ali},
  {Essam}, {Takey}, \& {Hamed}}]{Yoshida2014GCN15957}
{Yoshida}, M., {Itoh}, R., {Moritani}, Y., {et~al.} 2014, GRB Coordinates
  Network, 15957

\bibitem[{{Yoshida} {et~al.}(2009{\natexlab{a}}){Yoshida}, {Kuroda},
  {Yanagisawa}, {Shimizu}, {Nagayama}, {Toda}, \& {Kawai}}]{Yoshida2009GCN9267}
{Yoshida}, M., {Kuroda}, D., {Yanagisawa}, K., {et~al.} 2009{\natexlab{a}}, GRB
  Coordinates Network, 9267

\bibitem[{{Yoshida} {et~al.}(2009{\natexlab{b}}){Yoshida}, {Kuroda},
  {Yanagisawa}, {Shimizu}, {Nagayama}, {Toda}, \& {Kawai}}]{Yoshida2009GCN9266}
{Yoshida}, M., {Kuroda}, D., {Yanagisawa}, K., {et~al.} 2009{\natexlab{b}}, GRB
  Coordinates Network, 9266

\bibitem[{{Yoshida} {et~al.}(2010{\natexlab{a}}){Yoshida}, {Sasada}, {Komatsu},
  \& {Kawabata}}]{Yoshida2010GCN11190}
{Yoshida}, M., {Sasada}, M., {Komatsu}, T., \& {Kawabata}, K.~S.
  2010{\natexlab{a}}, GRB Coordinates Network, 11190

\bibitem[{{Yoshida} {et~al.}(2010{\natexlab{b}}){Yoshida}, {Yanagisawa},
  {Kuroda}, {Ohta}, \& {Kawai}}]{Yoshida2010AIPC}
{Yoshida}, M., {Yanagisawa}, K., {Kuroda}, D., {Ohta}, K., \& {Kawai}, N.
  2010{\natexlab{b}}, in American Institute of Physics Conference Series, Vol.
  1279, Deciphering the Ancient Universe with Gamma-ray Bursts, ed. N.~{Kawai}
  \& S.~{Nagataki}, 469--471

\bibitem[{{Yuan} {et~al.}(2008){Yuan}, {Rykoff}, {Schaefer}, {Rujopakarn},
  {G{\"u}ver}, {Aharonian}, {Akerlof}, {Ashley}, {Barthelmy}, {Gehrels},
  {G{\"o}{\v{g}}a}, {{\"u}{\c{s}}}, {Horns}, {Kizilo{\v{g}}lu}, {Krimm},
  {McKay}, {{\"O}zel}, {Phillips}, {Quimby}, {Rowell}, {Swan}, {Vestrand},
  {Wheeler}, \& {Wren}}]{Yuan2008AIPC}
{Yuan}, F., {Rykoff}, E.~S., {Schaefer}, B.~E., {et~al.} 2008, in American
  Institute of Physics Conference Series, Vol. 1065, 2008 Nanjing Gamma-ray
  Burst Conference, ed. Y.-F. {Huang}, Z.-G. {Dai}, \& B.~{Zhang}, 103--106

\bibitem[{{Yuan} {et~al.}(2016){Yuan}, {Amati}, {Cannizzo}, {Cordier},
  {Gehrels}, {Ghirlanda}, {G{\"o}tz}, {Produit}, {Qiu}, {Sun}, {Tanvir}, {Wei},
  \& {Zhang}}]{Yuan2016SSRv}
{Yuan}, W., {Amati}, L., {Cannizzo}, J.~K., {et~al.} 2016, \ssr, 202, 235

\bibitem[{{Zafar} {et~al.}(2018{\natexlab{a}}){Zafar}, {Heintz}, {Fynbo},
  {Malesani}, {Bolmer}, {Ledoux}, {Arabsalmani}, {Kaper}, {Campana},
  {Starling}, {Selsing}, {Kann}, {de Ugarte Postigo}, {Schweyer},
  {Christensen}, {M{\o}ller}, {Japelj}, {Perley}, {Tanvir}, {D'Avanzo},
  {Hartmann}, {Hjorth}, {Covino}, {Sbarufatti}, {Jakobsson}, {Izzo},
  {Salvaterra}, {D'Elia}, \& {Xu}}]{Zafar2018ApJ}
{Zafar}, T., {Heintz}, K.~E., {Fynbo}, J.~P.~U., {et~al.} 2018{\natexlab{a}},
  \apjl, 860, L21

\bibitem[{{Zafar} {et~al.}(2012){Zafar}, {Watson}, {El{\'{\i}}asd{\'o}ttir},
  {Fynbo}, {Kr{\"u}hler}, {Schady}, {Leloudas}, {Jakobsson}, {Th{\"o}ne},
  {Perley}, {Morgan}, {Bloom}, \& {Greiner}}]{Zafar2012ApJ}
{Zafar}, T., {Watson}, D., {El{\'{\i}}asd{\'o}ttir}, {\'A}., {et~al.} 2012,
  \apj, 753, 82

\bibitem[{{Zafar} {et~al.}(2011{\natexlab{a}}){Zafar}, {Watson}, {Fynbo},
  {Malesani}, {Jakobsson}, \& {de Ugarte Postigo}}]{Zafar2011AA}
{Zafar}, T., {Watson}, D., {Fynbo}, J.~P.~U., {et~al.} 2011{\natexlab{a}},
  \aap, 532, A143

\bibitem[{{Zafar} {et~al.}(2018{\natexlab{b}}){Zafar}, {Watson}, {M{\o}ller},
  {Selsing}, {Fynbo}, {Schady}, {Wiersema}, {Levan}, {Heintz}, {de Ugarte
  Postigo}, {D'Elia}, {Jakobsson}, {Bolmer}, {Japelj}, {Covino}, {Gomboc}, \&
  {Cano}}]{Zafar2018MNRAS}
{Zafar}, T., {Watson}, D., {M{\o}ller}, P., {et~al.} 2018{\natexlab{b}},
  \mnras, 479, 1542

\bibitem[{{Zafar} {et~al.}(2010){Zafar}, {Watson}, {Malesani}, {Vreeswijk},
  {Fynbo}, {Hjorth}, {Levan}, \& {Micha{\l}owski}}]{Zafar2010AA}
{Zafar}, T., {Watson}, D.~J., {Malesani}, D., {et~al.} 2010, \aap, 515, A94

\bibitem[{{Zafar} {et~al.}(2011{\natexlab{b}}){Zafar}, {Watson}, {Tanvir},
  {Fynbo}, {Starling}, \& {Levan}}]{Zafar2011ApJ}
{Zafar}, T., {Watson}, D.~J., {Tanvir}, N.~R., {et~al.} 2011{\natexlab{b}},
  \apj, 735, 2

\bibitem[{{Zeh} {et~al.}(2004){Zeh}, {Klose}, \& {Hartmann}}]{Zeh2004Apj}
{Zeh}, A., {Klose}, S., \& {Hartmann}, D.~H. 2004, \apj, 609, 952

\bibitem[{{Zeh} {et~al.}(2006){Zeh}, {Klose}, \& {Kann}}]{Zeh2006ApJ}
{Zeh}, A., {Klose}, S., \& {Kann}, D.~A. 2006, \apj, 637, 889

\bibitem[{{Zerbi} {et~al.}(2001){Zerbi}, {Chincarini}, {Ghisellini},
  {Rondon{\'o}}, {Tosti}, {Antonelli}, {Conconi}, {Covino}, {Cutispoto},
  {Molinari}, {Nicastro}, {Palazzi}, {Akerlof}, {Burderi}, {Campana}, {Crimi},
  {Danzinger}, {di Paola}, {Fernandez-Soto}, {Fiore}, {Frontera}, {Fugazza},
  {Gentile}, {Goldoni}, {Israel}, {Jordan}, {Lorenzetti}, {McBreen},
  {Martinetti}, {Mazzoleni}, {Masetti}, {Messina}, {Meurs}, {Monfardini},
  {Nucciarelli}, {Orlandini}, {Paul}, {Pian}, {Saracco}, {Sardone}, {Stella},
  {Tagliaferri}, {Tavani}, {Testa}, \& {Vitali}}]{Zerbi2001AN}
{Zerbi}, R.~M., {Chincarini}, G., {Ghisellini}, G., {et~al.} 2001,
  Astronomische Nachrichten, 322, 275

\bibitem[{Zhang(2018)}]{zhang_2018}
Zhang, B. 2018, The Physics of Gamma-Ray Bursts (Cambridge University Press)

\bibitem[{{Zhang} {et~al.}(2009){Zhang}, {Zhang}, {Virgili}, {Liang}, {Kann},
  {Wu}, {Proga}, {Lv}, {Toma}, {M{\'e}sz{\'a}ros}, {Burrows}, {Roming}, \&
  {Gehrels}}]{Zhang2009ApJ}
{Zhang}, B., {Zhang}, B.-B., {Virgili}, F.~J., {et~al.} 2009, \apj, 703, 1696

\bibitem[{{Zhao} {et~al.}(2011){Zhao}, {Bai}, \& {Mao}}]{Zhao2011GCN11733}
{Zhao}, X.~H., {Bai}, J.~M., \& {Mao}, J. 2011, GRB Coordinates Network, 11733

\bibitem[{{Zhao} {et~al.}(2012{\natexlab{a}}){Zhao}, {Mao}, {Xu}, \&
  {Bai}}]{Zhao2012GCN13122}
{Zhao}, X.~H., {Mao}, J., {Xu}, D., \& {Bai}, J.~M. 2012{\natexlab{a}}, GRB
  Coordinates Network, 13122

\bibitem[{{Zhao} {et~al.}(2012{\natexlab{b}}){Zhao}, {Mao}, \&
  {Bai}}]{Zhao2012GCN13898}
{Zhao}, X.~H., {Mao}, J.~R., \& {Bai}, J.~M. 2012{\natexlab{b}}, GRB
  Coordinates Network, 13898

\bibitem[{{Zheng} \& {Filippenko}(2018)}]{Zheng2018GCN23033}
{Zheng}, W. \& {Filippenko}, A.~V. 2018, GRB Coordinates Network, 23033

\bibitem[{{Zheng} {et~al.}(2014){Zheng}, {Filippenko}, {Morgan}, \&
  {Cenko}}]{Zheng2014GCN16137}
{Zheng}, W., {Filippenko}, A.~V., {Morgan}, A., \& {Cenko}, S.~B. 2014, GRB
  Coordinates Network, 16137

\bibitem[{{Zheng} {et~al.}(2011){Zheng}, {Schaefer}, \&
  {Flewelling}}]{Zheng2011GCN12205}
{Zheng}, W., {Schaefer}, B.~E., \& {Flewelling}, H. 2011, GRB Coordinates
  Network, 12205

\bibitem[{{Zheng} {et~al.}(2012){Zheng}, {Shen}, {Sakamoto}, {Beardmore}, {De
  Pasquale}, {Wu}, {Gorosabel}, {Urata}, {Sugita}, {Zhang}, {Pozanenko},
  {Nissinen}, {Sahu}, {Im}, {Ukwatta}, {Andreev}, {Klunko}, {Volnova},
  {Akerlof}, {Anto}, {Barthelmy}, {Breeveld}, {Carsenty},
  {Castillo-Carri{\'o}n}, {Castro-Tirado}, {Chester}, {Chuang}, {Cunniffe}, {De
  Ugarte Postigo}, {Duffard}, {Flewelling}, {Gehrels}, {G{\"u}ver}, {Guziy},
  {Hentunen}, {Huang}, {Jel{\'\i}nek}, {Koch}, {Kub{\'a}nek}, {Kuin}, {McKay},
  {Mottola}, {Oates}, {O'Brien}, {Ohno}, {Page}, {Pandey}, {P{\'e}rez del
  Pulgar}, {Rujopakarn}, {Rykoff}, {Salmi}, {S{\'a}nchez-Ram{\'\i}rez},
  {Schaefer}, {Sergeev}, {Sonbas}, {Sota}, {Tello}, {Yamaoka}, {Yost}, \&
  {Yuan}}]{Zheng2012ApJ}
{Zheng}, W., {Shen}, R.~F., {Sakamoto}, T., {et~al.} 2012, \apj, 751, 90

\bibitem[{{Zhuchkov} {et~al.}(2008){Zhuchkov}, {Bikmaev}, {Sakhibullin},
  {Khamitov}, {Eker}, {Kiziloglu}, {Gogus}, {Burenin}, {Pavlinsky}, \&
  {Sunyaev}}]{Zhuchkov2008GCN7803}
{Zhuchkov}, R., {Bikmaev}, I., {Sakhibullin}, N., {et~al.} 2008, GRB
  Coordinates Network, 7803

\bibitem[{{Zou} {et~al.}(2006){Zou}, {Dai}, \& {Xu}}]{Zou2006ApJ}
{Zou}, Y.~C., {Dai}, Z.~G., \& {Xu}, D. 2006, \apj, 646, 1098

\end{thebibliography}

\begin{appendix}

\section{Redshift Measurement Delays} 
\label{redshift delays}

The question might be asked {\it Why not do this from the ground using \eg, the existing assets utilized in this work?} The answer to this can be found by analysing the \emph{Swift} percentage of redshifts recovered {\it and} the delay time to broadcast them for follow-up spectroscopy by large telescopes.

While \emph{Swift} has proved exceptionally efficient at providing prompt gamma-ray ($\sim 2\arcmin-3\arcmin$), X-ray ($\sim 2\arcsec-3\arcsec$) and often UV/optical ($\sim 0\farcs5$) localization of GRBs, it has in the vast majority of cases been reliant on ground-based follow-up to determine the GRB redshift with only $\sim 30\%$ recovered \citep{Perley2016ApJ}. 

In some cases a detection by \emph{Swift} UVOT does provide rapid limits on the redshift. The white light filter typically used for finding charts is most sensitive in the UV, and cuts out at around 7000 {\AA}. Hence, a UVOT detection is immediately indicative of $z\lesssim5$, with the most distant GRB detected by UVOT being GRB 060522 \citep{Fox2006GCN5150,Holland2006GCN5158} at redshift $z=5.11$ \citep{Cenko2006GCN5155}. However, the converse is not true. While $>50\%$ of \emph{Swift} detected GRBs are undetected by the UVOT, only a small fraction of these lie at $z>5$. The majority of non-detections are the result of either observations of insufficient depth, where the extrapolation of the X-ray flux to the optical lies below the available upper limits, where appropriate ground-based telescopes were unavailable because of weather, downtime, sky location or lack of time allocation to GRB science, or where intrinsic extinction renders the optical afterglow too faint, so called dark GRBs \citep{Jakobsson2004ApJ,Rol2005ApJ,Greiner2011AA,Melandri2012}.

A consequence of this is that any robust redshift measurements are substantially delayed from the time of the burst trigger. Such delays have significant implications for the ability to plan follow-up. It is much more straightforward to recover the GRB redshift with earlier data when the afterglow is brighter. The situation for high-$z$ bursts is even more acute, since NIR spectroscopy is not routinely attempted. Hence, the afterglow must first be identified as a high-$z$ candidate via imaging observations and then targeted for NIR spectroscopy. 

\begin{figure}[!t]
         \centering 
         \includegraphics[width=\columnwidth]{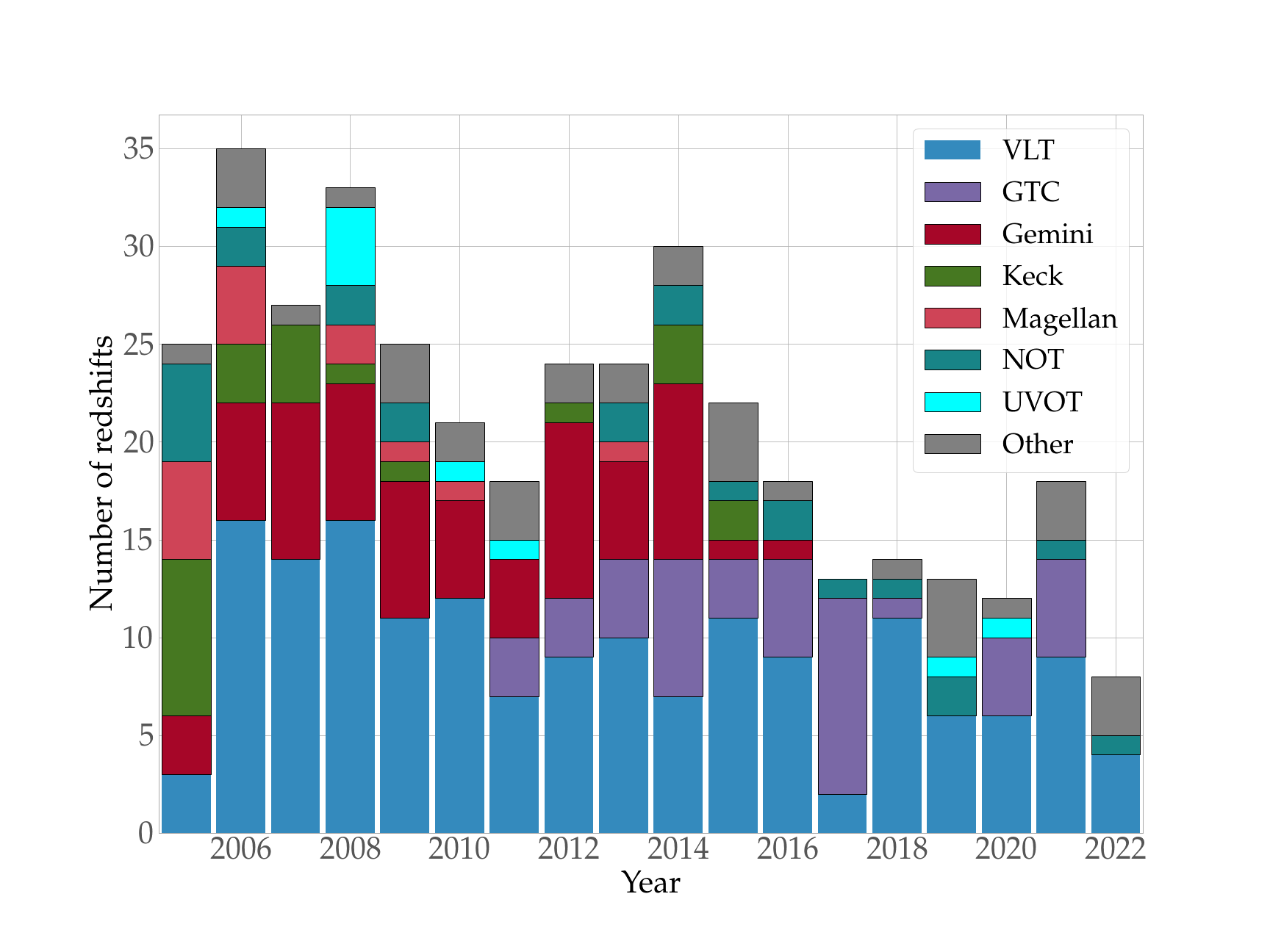} 
         \caption{Number of redshifts obtained each year since the launch of \emph{Swift}. The histograms are color-coded by the telescope that obtained the redshift measurement. }
         \label{sources}
\end{figure}

To quantify this effect we retrieved relevant times via a search of the GCN archives for all \emph{Swift} bursts. Fig. \ref{sources} gives an overview of the number of redshifts determined each year since the launch of Swift and the telescopes used to obtain them. The number of redshifts retrieved has clearly decreased as the mission has continued, while the GRB detection rate has maintained a comparable rate of between $\sim$ 80 and 100 GRBs yr$^{-1}$.

Where possible we record both the time of observation either from the GCN or available archives, and the time of dissemination to the community. Where recovery of the observation time is not possible we set it equal to the report time. We did not include observations obtained within the first days but only reported via papers that appear much later (months to years after the GRB) because they were not, for whatever reason, promptly reported to the community. 

\begin{figure}[!t]
         \centering 
         \includegraphics[width=\columnwidth]{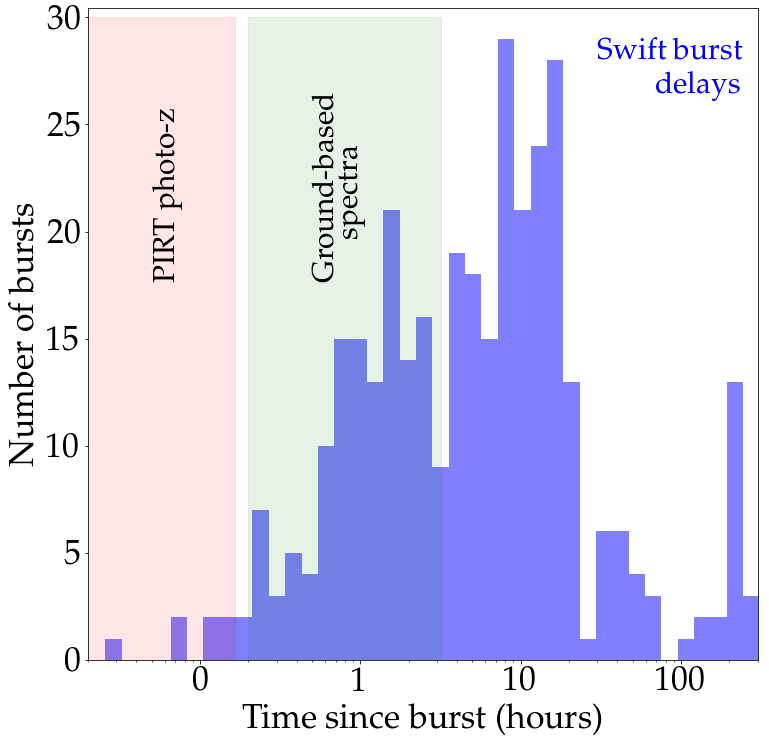} 
         \caption{Delays in obtaining redshifts for GRBs in the \emph{Swift} era. Highlighted are the time spans during which PIRT is expected to measure photo-$z$s on-board and transmit them to the ground, as well as our planned rapid follow-up observations.}
         \label{redshift_delay}
\end{figure}

\begin{figure}[!t]
         \centering 
         \includegraphics[width=\columnwidth]{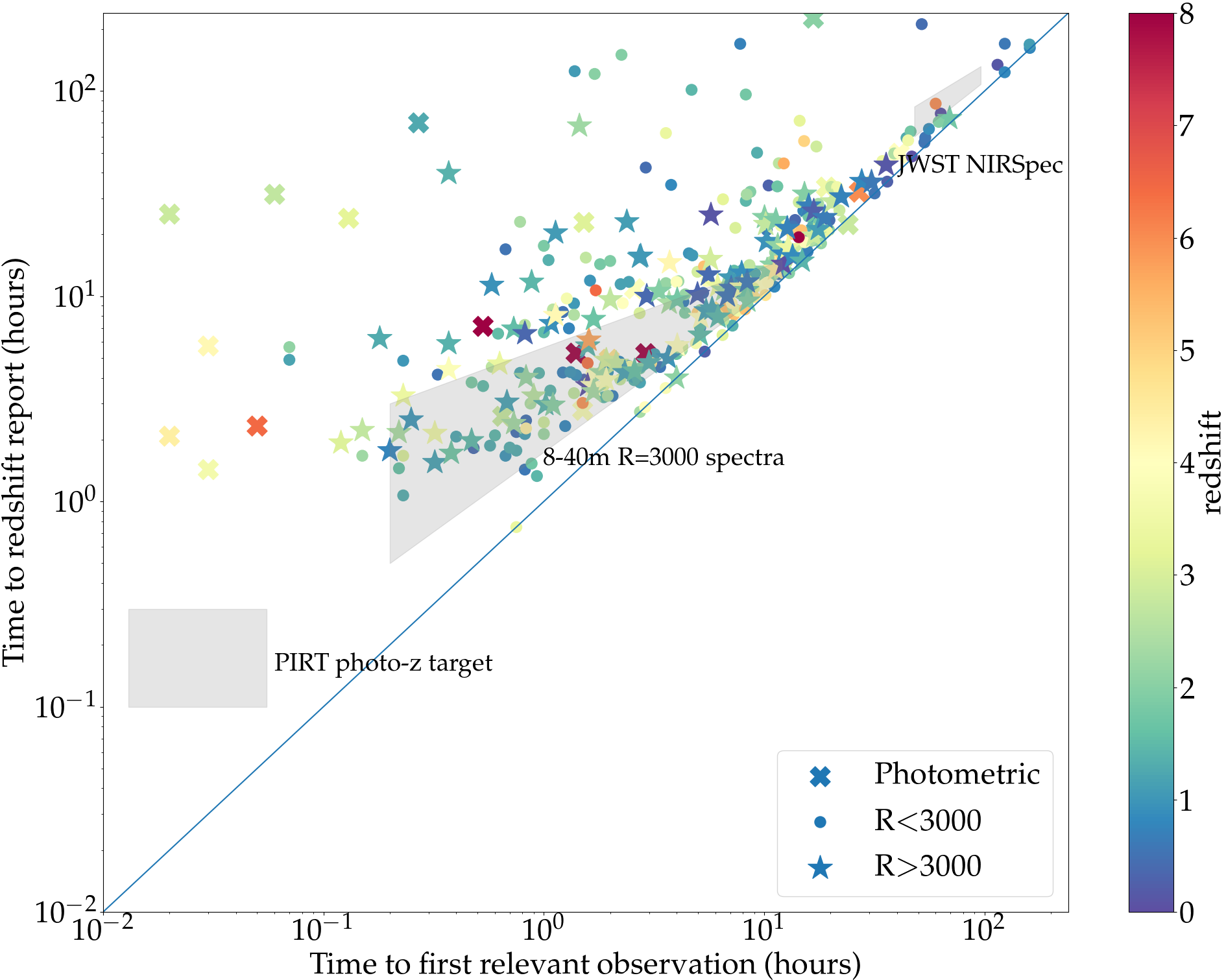} 
         \caption{Delays between the time when a redshift was recorded vs. when the first relevant observation was obtained. We differentiate between redshifts that are only photometric, those that have been obtained by low-to-medium resolution spectrographs (resolution R$<3000$) and those that have been obtained by medium-to-high resolution spectrographs (R$>3000$). For each observation, the redshift itself is color-coded. Finally, with gray areas, we highlight the expected observation and report times for the initial PIRT observation, our planned follow-up with large ground-based facilities, and planned follow-up with JWST NIRSpec.}
         \label{zhist2}
\end{figure}

Fig.~\ref{redshift_delay} and Fig.~\ref{zhist2} show the delay times for all \emph{Swift} bursts, both in terms of the time to observation, and of reporting to the community. The mean and median delays to observations are 29 and 5.5 hours (with the mean significantly skewed by a handful of very long delay times), and the 90\% range $0.3-200$ hours. For reports to the community the corresponding mean and median times are 45 and 11.8 hours with 90\% of the reports coming within $2-200$ hours. It is important to note that the earliest robust indications of a redshift (photometric or spectroscopic) are not reported to the community for $>1$ hour after the burst. For the high redshift bursts, the situation is further complicated by the requirement to first identify IR counterparts before acquiring spectroscopy. 

\begin{figure}[!t]
    \centering
         \includegraphics[width=\columnwidth]{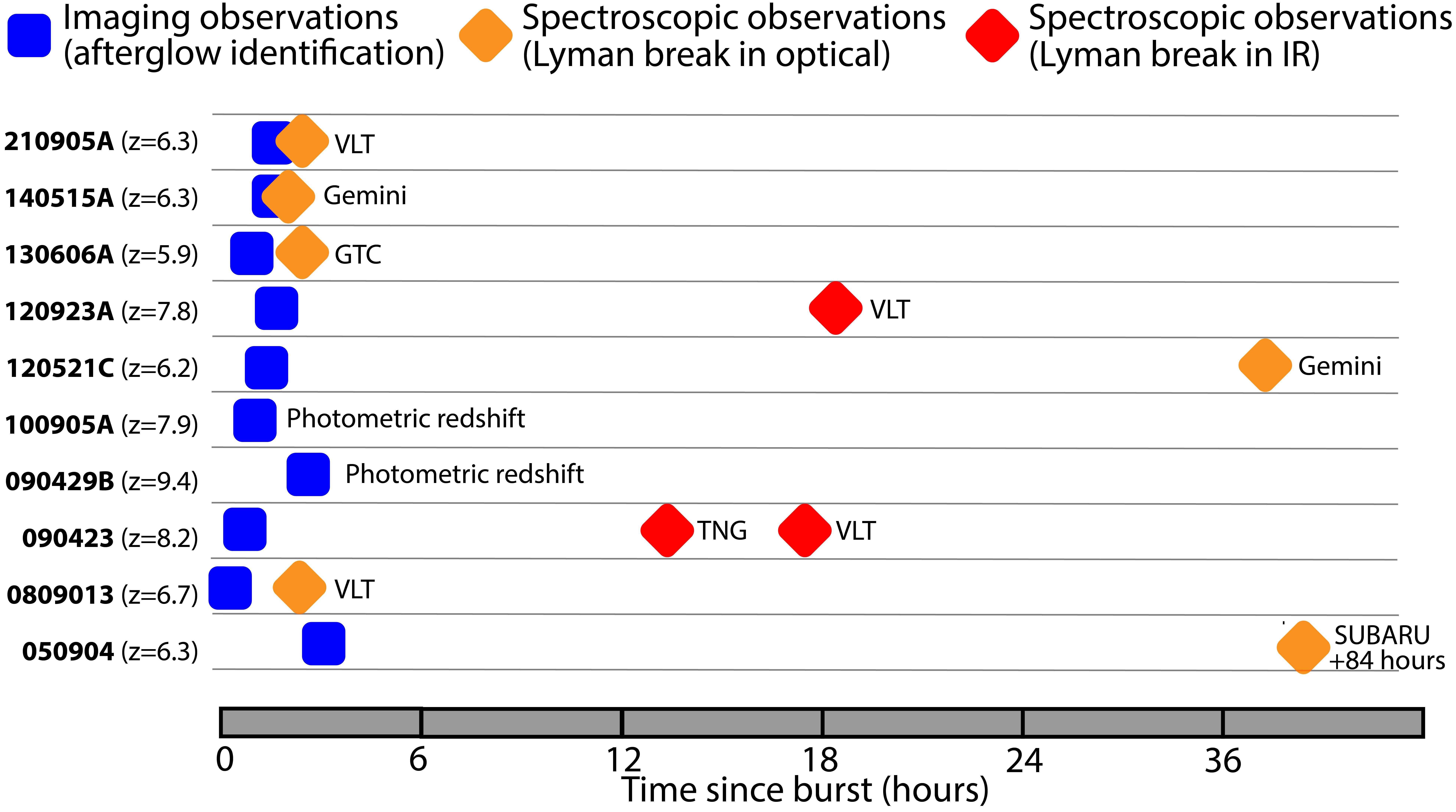} 
         \caption{Timelines of the redshift determination for ten high-$z$ GRBs in the $z\gtrsim6$ sample. Two of these only have relatively insecure photometric redshifts. Three further ones were so distant that the Lyman-$\alpha$ line lies in the infrared and only low-S/N spectra allowing a redshift determination and little else could be obtained. In the other cases the Lyman-$\alpha$ line still lies in the optical regime and higher-quality spectra were usually obtained (especially for GRBs 130606A, 140515A, and 210905A).}
         \label{timelines}
\end{figure}

In Fig.~\ref{timelines} we show timelines for all bursts in our high-$z$ sample with spectroscopic or photometric redshifts of $z\gtrsim6$. The mean and median delays here are not substantially different to those for the population of bursts as a whole. However, this is skewed low because for bursts at $z\sim6$ $z^\prime$-band detections are still possible, enabling direct spectroscopy, in some cases with optical spectrographs (\eg, \citealt{Kawai2006Nature}). For example, the X-shooter instrument has provided more GRB redshifts than any other, but acquires with an optical camera. In the case of $z \sim 6$ bursts it can therefore, on occasion, obtain rapid spectroscopy, as was the case for GRB 210905A \citep{Rossi2022AA}. For bursts which require IR spectroscopy none was attempted at $<12$ hours, and in each case, not at the same site as the initial photometric identification as a high$-z$ candidate.

Having quantified the actual delays in obtaining observations, it is also relevant to consider their origin. The majority of the redshifts for GRBs reported in the literature arise from ground-based observations. Only in a handful of cases were there {\em Swift} UVOT grism observations (three events, \citealt{Kuin2009MNRAS,Kuin2019GCN26538,Kuin2021GCN30355}), or UVOT determined photometric redshifts ($\sim 15$ events\footnote{The absence of a GCN reporting an UVOT photometric redshift does not imply that no constraints were possible from the UVOT data, only that they were not reported. See \cite{Kruhler2011AA} for a study using \emph{Swift} UVOT and GROND to determine photometric redshifts.}). The delay to observations can therefore be decomposed into physical constraints and operational constraints. Once a burst is detected it takes some time before it becomes visible to a given observatory -- such delays are unavoidable and can be many hours. The operational constraints then refer to the time to begin observations once the target is visible. A robust quantification of these issues is difficult because different telescopes have individual pointing constraints and the reasons for delays during visibility can also arise from varying sources (technical issues, weather, visiting astronomers, etc.). 

\begin{figure}[!t]
         \centering 
         \includegraphics[width=\columnwidth]{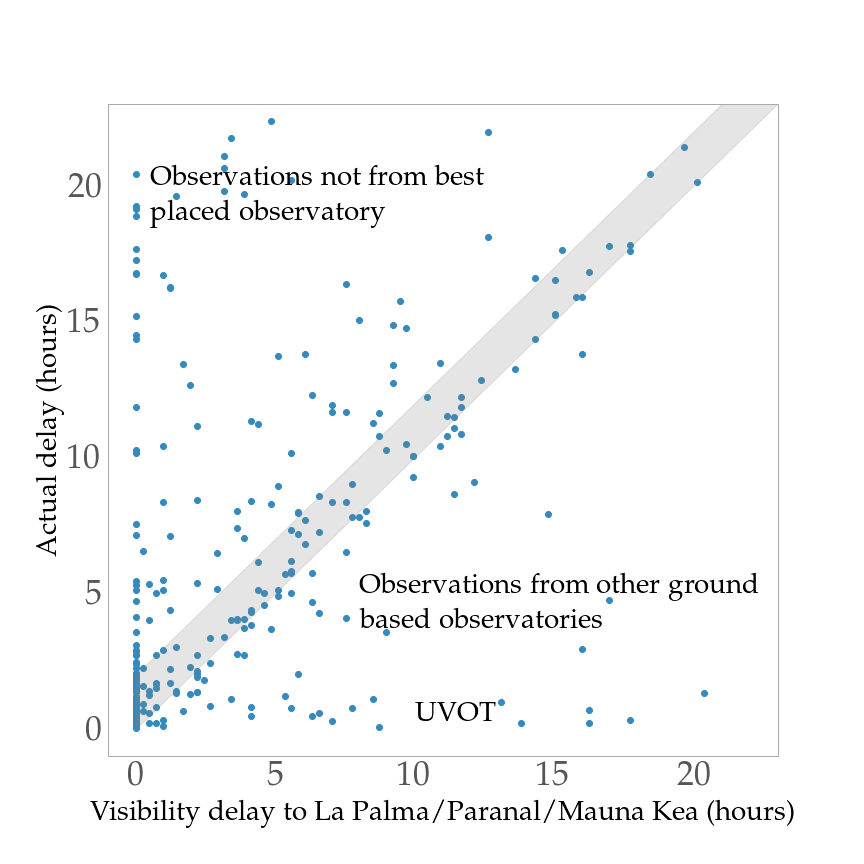} 
         \caption{Actual delay time in obtaining redshift observations vs. the minimum possible delay. These are from the three premier observing sites La Palma, Chile, and Mauna Kea. GRBs apparently observed \emph{before} the minimum possible delay (lower right) were observed by other observatories (\eg, in China or Russia) or by UVOT. Large delays despite being well-placed (top left) are usually down to weather or technical issues, instrumentation not being mounted or the lack of a redshift programme.}
         \label{threeobservatories}
\end{figure}

To approximately understand the origin of the delays for \emph{Swift} bursts we calculate the delay from the burst time until the source reaches 40 degrees of elevation with the Sun 18 degrees below the horizon at one of the following three: 1.) La Palma (Canary Islands, Spain), 2.) Chile (for which we adopt Cerro Paranal as a location) or 3.) Mauna Kea (Hawaii, USA), since these three sites are responsible for the vast majority of GRB redshift measurements. The resulting outcome is shown in Fig. \ref{threeobservatories}. A handful of bursts are apparently observed \emph{before} the earliest possible observations. These cases are predominantly GRBs that had redshift measurements from other observatories including \emph{Swift} UVOT. In some cases they are also cases where the burst was promptly visible at one of the three sites, but in twilight, or with an elevation $<40^\circ$. One feature that is clear from Fig. \ref{threeobservatories} is that having redshifts and arc second positions determined onboard \emph{Swift} by UVOT, reduces the delay time, so that when the afterglow does become visible to big glass these observatories can be ready to obtain the required spectra for high$-z$ GRBs, when the afterglow is still bright.

The China-France Space Variable Objects Monitor (\emph{SVOM}) will launch in the near future and is predicted to detect $\approx5$ GRBs per year at $z>5$ by using a coded mask GRB detector in the X-ray band \citep{Gotz2009AIPC,Paul2011CRPhy,Wei2016arXiv,Wang2020RAA}. SVOM carries an onboard Visible Telescope (VT) that is sensitive from 400-950\,nm, considerably redder than the UV/Optical Telescope (VT) on  \emph{Swift},  which will provide rapid identification of optically dark afterglows. 
\emph{SVOM} will rely on a network of ground-based telescopes to identify and obtain high resolution NIR spectra of the high redshift events. While the performance of this network is yet to be seen, it very likely will have the same challenges as \emph{Swift} in obtaining rapid ground-based follow-up (telescope availability, weather, visibility, etc.) which means the rare high$-z$ events may be missed or observed too late to obtain high quality spectra. 

Improvements in the speed of obtaining redshifts, and hence in redshift completion since bursts are brighter earlier, can therefore be obtained by three key improvements in the design of future programmes. 
\begin{enumerate}
\item \emph{GRB detectors should point at regions of sky that are promptly visible to large ground-based observatories.} Although the duty cycle of large ground-based telescopes is not 100\% there are relatively few gaps in the night time regions between the three major sites considered here. Therefore a pointing strategy which prioritises such visibility could potentially significantly improve delay times. 
\item Future GRB observatories should provide arc sec positions, accurate enough for direct spectroscopic measurements in the majority of cases. 
\item Since ground-based follow-up of premier facilities (\eg, ELTs) is very limited, missions should provide prompt and reliable indications of high redshift. 
\end{enumerate}
\GE{} aims to maximize use of these strategies.

\section{Details on the new sample}
\label{sampledetails}

\subsection{Additional data added to the \emph{Swift}-era sample}

The following GRBs have data added that changes their light curves (usually at very early times) without changing the analyses presented in K10. For GRB 080319B, we added the high time resolution data of the prompt flare as measured by the TORTORA wide field camera presented by \cite{Beskin2010ApJL}. We add $V$-band data presented by \cite{Brivio2022AA} to the light curve of GRB 080928. \cite{Page2019MNRAS} present an analysis of \emph{Swift} UVOT ``settling images'', very earlytime $v$-band detections. In the context of our sample, we add these data points to the light curves of GRBs 050922C, 060418, 060908, and 081008. Further GRBs are mentioned below in the context of our early-time sample.

\subsection{The high$-z$ sample}
\label{hizgrbsdetails}
\subsubsection{GRB 050904, $z=6.295\pm0.002$}
\label{GRB050904}
Data are taken from \cite{Haislip2006Nature,Tagliaferri2005AA,Kawai2006Nature,Boer2006ApJ,Price2006ApJ,Berger2007ApJ,Gendre2007AA}, and GCN Circulars \citep{Perley2005GCN3932}. Deep host galaxy observations are presented in \cite{Berger2007ApJ,Tanvir2012ApJ,Schulze2015ApJ,McGuire2016ApJ,Blanchard2016ApJ}. The redshift is given by \cite{Kawai2006Nature}. The full analysis of this GRB afterglow has been done by \cite{Kann2007AJ}.

This GRB afterglow is characterized by a very early, extremely luminous \citep{Kann2007AJ} prompt flash and a reverse shock \citep{Boer2006ApJ,Tagliaferri2005AA,Wei2006ApJ}. The very luminous and extremely variable X-ray afterglow \citep{Cusumano2006Nature,Cusumano2007AA,Watson2006ApJ} has led to the interpretation that it belongs to the class of ultralong GRBs \citep{Zou2006ApJ,Kann2018AA}. It was also highly luminous in the radio bands \citep{Frail2006ApJ}, and generally one of the most luminous GRBs observed to date \citep[][K10]{Sugita2009PASJ}. The presence or absence of dust along the line of sight to this GRB has been controversially discussed \citep{Stratta2007ApJ,Stratta2011AA,Zafar2010AA,Zafar2011ApJ,LiangLi2009ApJ}, an interesting contrast to the very high column density in X-rays and evidence for dense surroundings \citep{Campana2007ApJ,Gou2007ApJ}.

\subsubsection{GRB 080913, $z=6.733$}

Data are taken from \citet[][see also \citealt{Bolmer2018AA}]{Greiner2009ApJ}, \cite{Perez2010AA}. Deep host galaxy observations are presented in \cite{Tanvir2012ApJ,Basa2012AA,Blanchard2016ApJ}. The redshift is given by \cite{Patel2010AA}. The full analysis of this GRB afterglow has been done by K10. The afterglow is relatively faint and shows a steady decay from early on \citep{Greiner2009ApJ}, which is followed by a very strong rebrightening \citep{Greiner2009ApJ,Perez2010AA}. The GRB itself was quite short temporally, in contrast to the expected long lasting time dilated light curves such as in the case of GRB 050904. This led \cite{Zhang2009ApJ} to discuss the long/short nature of such GRBs.

\subsubsection{GRB 090423, $z=8.23^{+0.06}_{-0.07}$}

Data are taken from \citet[][see also \citealt{Bolmer2018AA}]{Tanvir2009Nature}, \cite{Yoshida2010AIPC,Laskar2014ApJ}. Deep host galaxy observations are presented in \cite{Tanvir2012ApJ,Laskar2014ApJ,Blanchard2016ApJ}. The redshift is given by \citet[][see also \citealt{Salvaterra2009Nature}]{Tanvir2009Nature}. The full analysis of this GRB afterglow has been done by K10. The afterglow shows an extended early plateau phase \citep{Tanvir2009Nature,Yoshida2010AIPC}. Similar to GRB 080913, it was temporally short \citep{Zhang2009ApJ}.

\subsubsection{090429B: $z=9.4^{+0.12}_{-0.36}$}

Data are taken from \citet[][see also \citealt{Bolmer2018AA}]{Cucchiara2011ApJ}. Deep host-galaxy observations are presented in \cite{Cucchiara2011ApJ,Tanvir2012ApJ}. The photometric redshift is given by \cite{Cucchiara2011ApJ}. This is the most distant GRB known to date. Data are very sparse. From the two $K$-band detections, we derive a decay slope $\alpha\approx0.6$, in full agreement with \cite{Cucchiara2011ApJ}. The $J$ band is strongly suppressed by Lyman damping, leaving only two filters to measure the SED. Without dust, we find $\beta\approx1.14$. While this value is not extraordinary, \cite{Cucchiara2011ApJ} show via a joint NIR to X-ray fit that the cooling break must lie between the two spectral regimes, and that the intrinsic slope in the NIR range is $\beta=0.51\pm0.17$. Fixing this value, we derive $A_V=0.23,0.15,$ and 0.11 mag for MW, LMC, and SMC dust, respectively. Owing to the high-$z$ nature of the GRB, we prefer SMC dust, our value being in excellent agreement with the more detailed fitting procedure from \cite{Cucchiara2011ApJ}. We find the afterglow luminosity is very close to that of the GRB 090423 afterglow.

\subsubsection{GRB 100905A, $z=7.88^{+0.75}_{-0.94}$}

Data are taken from \cite{Bolmer2018AA}, and GCN Circulars \citep{Im2010GCN11222}. Deep host-galaxy observations are presented in \cite{Bolmer2018AA}. The (photometric) redshift is given by \cite{Bolmer2018AA}. This GRB has very sparse data, with only two epochs of $JHK$ detections. \cite{Bolmer2018AA} derive a photometric redshift from GROND data. We find the afterglow exhibits a shallow decay, $\alpha=0.29\pm0.09$, which is shallower than that reported by \cite{Bolmer2018AA}. It is indeed so shallow that the late upper limit on a host galaxy also implies that a light curve break must have occurred. Assuming a fixed break immediately after the GROND epoch at 0.658 d, we find $\alpha_2\gtrsim0.88$, which does not yield strong evidence that this break is a jet break. The large errors in the data lead to an insecure spectral slope, we find the $J_G H_G K_G$ SED can be fit by a simple power-law with no extinction and $\beta=0.48\pm0.73$, in excellent agreement with \cite{Bolmer2018AA}.

\subsubsection{GRB 120521C, $z=6.15$}

Data are taken from \cite{Laskar2014ApJ}, where we follow their correction of the $z^\prime$-band flux. We also present early $R_C I_C Z$ upper limits from observations with the \emph{Tautenberg} Schmidt, first reported in \cite{Kann2012GCN13337}. The redshift is given by \cite{Laskar2014ApJ}, based on a very low-S/N spectrum which shows a Lyman-$\alpha$ cutoff. The light curve shows a long rise and likely a smooth rollover to the decay. We fix $n=2$ and find $\alpha_1=-0.84\pm0.26$, $\alpha_2=0.96\pm0.11$, $t_b=0.248\pm0.049$ d. We find a very similar $\alpha_1$ to \cite{Laskar2014ApJ}, but a somewhat earlier break time, and shallower post-break decay slope $\alpha_2$, however, they still agree within errors. We follow the best-fit of \cite{Laskar2014ApJ} and fix $\beta=0.34$, and derive a small amount of extinction, $A_V=0.13\pm0.03$ mag from the $z^\prime JHK$ SED, using SMC dust.

\subsubsection{GRB 120923A, $z=7.84^{+0.06}_{-0.12}$}

Data are taken from \cite{Tanvir2018ApJ}. We also present upper limits from the 1.8m (Bohyunsan/KASINICS) in Korea, first reported in \citep{Im2012GCN13823}. The redshift is given by \cite{Tanvir2018ApJ}, based on a very low S/N spectrum which shows a Lyman-$\alpha$ cutoff. Deep host-galaxy observations are presented in \cite{Tanvir2018ApJ,Blanchard2016ApJ}. We find that similar to several other high-$z$ GRB afterglows, that of GRB 120923A exhibits an early plateau phase. We find $\alpha_1=0.06\pm0.09$, $\alpha_2=1.96\pm0.29$, $t_b=1.59\pm0.33$ d. This implies $\Delta\alpha=1.89\pm0.30$, an unusually large value \citep{Zeh2006ApJ}. The $JHK$ SED is modeled perfectly well by a simple power-law with no dust extinction and $\beta=0.54\pm0.48$.

\subsubsection{GRB 130606A, $z=5.91285\pm0.00002$}

Data are taken from \cite{Castro-Tirado2013arXiv,Littlejohns2014AJ,Littlejohns2015MNRAS,Hartoog2015AA,Bolmer2018AA}, from GCN Circulars \citep{Masi2013GCN14789,Nagayama2013GCN14793,Morgan2013GCN14802,Perley2013GCN14804,Trotter2013GCN14815,Trotter2013GCN14826,Klotz2013GCN14818}, as well the 2.0 m LT/FTN (first reported in \citealt{Virgili2013GCN14785,Virgili2013GCN14840}) and 1.0m LOAO observations (first reported in \citealt{Im2013GCN14800}). Deep host galaxy observations are presented in \cite{McGuire2016ApJ}. The redshift is given by \citet[][see also \citealt{Chornock2013ApJ,Castro-Tirado2013arXiv,Totani2014PASJ,Totani2016PASJ}]{Hartoog2015AA}. This GRB has the second most luminous afterglow among the high-$z$ sample, and was studied in great detail spectroscopically \citep{Chornock2013ApJ,Castro-Tirado2013arXiv,Hartoog2015AA,Totani2014PASJ,Totani2016PASJ}. The optical/NIR light curve is also, by far, the richest of the high-$z$ GRBs, and, similar to GRB 050904 \citep{Boer2006ApJ}, it was detected at early times by small (aperture $<0.5$ m) robotic telescopes.

We find that the afterglow initially features a rise to an early peak, and a rollover into a decay phase. It is $\alpha_{\rm rise}=0.972\pm0.088$, $\alpha_1=1.591\pm0.097$, $t_b=0.0076\pm0.0009$ d ($657\pm79$ s), with $n=1$ fixed. This fit yields a $r^\prime R_C i^\prime z^\prime JHK_S$ SED. At $t\approx0.039$ d, the afterglow transitions into a shallow decay phase before breaking again into a ``normal'' decay. It is $\alpha_{\rm plateau}=0.502\pm0.29$, $\alpha_2=1.807\pm0.017$, $t_{b,2}=0.160\pm0.008$ d, and $n=2$ was fixed (we found $n=2.08\pm0.53$ when leaving it free, however, this leads to very large errors in the normalizations, so we fixed $n$ when deriving the SED). This decay is $2.2\sigma$ steeper than $\alpha_1$, however, it is still likely that the plateau represents an energy injection and the decay after the end of the injection continues as before. The final $z^\prime$ data point lies significantly above the extrapolation of the earlier decay slope, indicating another rebrightening has probably taken place.

The second fit yields a $g_G r_G r^\prime R_C i^\prime Z z^\prime YJ_G JH_G HK_S K_G$ SED ($g_G$ being an upper limit only). We fit both SEDs simultaneously, leaving the normalizations free for each SED but sharing the other parameters. We find no evidence for any color evolution between the two SEDs. A fit without extinction is perfectly acceptable, and we find $\beta=0.886\pm0.092$. Fits with extinction in all cases yield similar spectral slopes and $A_V=0$ mag within errors. This is in full agreement with ``Solution 3'' of \cite{Littlejohns2014AJ} and in good agreement ($1.4\sigma$) with the X-ray-to-optical fit shown in \cite{Hartoog2015AA}, who also set an upper limit on the extinction of $A_V\lesssim0.2$ mag. At $\approx0.05$ d, this is among the most luminous GRB afterglows ever detected, comparable to the dusty moderately high-$z$ GRB afterglows of GRB 080607 and GRB 140311A, as well as the strong rebrightening of GRB 081029.

\subsubsection{GRB 140515A, $z=6.3298\pm0.0004$}

Data are taken from \cite{Melandri2015AA,Bolmer2018AA}, and GCN Circulars \citep{Fong2014GCN16274}. Deep host galaxy observations are presented in \cite{McGuire2016ApJ}. The redshift is given by \citet[][see also \citealt{Chornock2014arXiv}]{Melandri2015AA}.

The available data are fit by a simple power-law decay, however, later upper limits indicate a light curve break must have occurred. Furthermore, the earliest detections \citep{Fong2014GCN16274} lie somewhat above the back extrapolation of the later decay and were not included in the fit. Fixing the break time to $t_b=1$ d, we find $\alpha_1=0.48\pm0.23$, $\alpha_2\gtrsim1.67$. This is qualitatively in agreement with the X-ray results presented in \cite{Melandri2015AA}, who find $t_b\geq10^5$ s and a very steep decay $\alpha_2=3.9\pm0.6$. The prebreak slope we find is shallower than that given by \cite{Melandri2015AA}, as they include the early detection \citep{Fong2014GCN16274}, and use only their data, while we perform a joint fit together with the GROND data \citep{Bolmer2018AA}. The $i_Gz_GJ_GH_GK_G$ SED ($i_Gz_G$ are clearly affected by Lyman damping and are not included) is fit perfectly well by a simple power-law with no dust extinction and $\beta=0.22\pm0.35$, a very shallow spectral slope. From an optical to X-ray fit using their X-shooter spectrum, \cite{Melandri2015AA} derive $\beta=0.33\pm0.02$ and $A_V=0.11\pm0.02$ mag.

\subsubsection{GRB 210905A, $z=6.312$}

This was a very energetic and long high-$z$ event with many similarities to GRB 050904. All data, the redshift, and the full analysis are presented in \cite{Rossi2022AA,Saccardi2023}. At late times, the afterglow is one of the most luminous ever detected (along with that of GRB 160625B \cite{Xu2016GCN19600}).

\subsubsection{Erratum on GNz-11 Flash}

\cite{Jiang2021NatAs2} reported on a potential $K$-band flash detected in the spectral sequence of the lensed high-$z$ galaxy GN-z11, which they find to be at $z\approx11$ \citep{Jiang2021NatAs1}. They interpreted this flash as associated with the prompt emission of a GRB. This led to intense discussion in the literature \citep{Nir2020MNRAS,Nir2021RNAAS,Padmanabhan2021arXiv,Steinhardt2021arXiv,Jiang2021arXiv}, before \cite{Michalowski2021NatAs} convincingly showed that it resulted from a satellite in Earth orbit crossing the spectrograph slit.
In \cite{Kann2020RNAAS}, we showed that, assuming GN-z11 Flash is located at $z=11$, its brightness would be in agreement with the early UV/optical/NIR luminosity of transients associated with GRBs. While the association has been shown to very likely be incorrect, we here wish to point out one incorrect statement in that work. The first upper limit after the ``flash'' was also put into context, and it was stated the afterglow of this potential GRB had to be among the least luminous known so far. However, this continued to assume the spectral slope derived for the flash, which was extremely blue and not in agreement with usual GRB afterglow spectra. Applying a spectral slope in the typical range $\beta\approx0.5-1.1$, as expected for GRB afterglows, leads to a significantly larger $dRc$ and therefore significantly \emph{shallower} upper limits, \eg, for $\beta=0.6$, it is $dRc=-5.47$ mag, whereas it is $dRc=-1.98$ mag for the spectral slope $\beta=-1.2$ which \cite{Jiang2021NatAs2} measure. Therefore, the upper limits do not impose a strong constraint on potential afterglow emission if it were to exist.

\subsection{The early time afterglow sample}
\label{ELCsampledetails}

\subsubsection{GRB 080413A, $z=2.4330$}

Data are taken from \cite{Yuan2008AIPC}, K10, from GCN Circulars \citep{Klotz2008GCN7595,Fukui2008GCN7622,Marshall2008GCNR129}, as well as our own \emph{Swift} UVOT data (first reported in \citealt{Oates2008GCN7607}), 0.6 m REM optical and NIR data (first reported in \citealt{Antonelli2008GCN7597}), and 2.0 m LT and FTS data (first reported in \citealt{Gomboc2008GCN7625}), which are mostly upper limits with one unpublished low-significance detection. The redshift is given by \cite{Fynbo2009ApJS}.

The analysis of this GRB afterglow was first presented in K10. Here, with expanded data from \emph{Swift} UVOT, REM, and LT, we reanalyze the event. The early afterglow shows variability superposed on the general decay. Data beginning at 0.005 d can be fit well ($\chi^2/d.o.f.=1.29$) with a smoothly broken power-law with $\alpha_1=0.824\pm0.038$, $\alpha_2=1.520\pm0.030$, $t_b=0.015\pm0.002$ d, $n=10$ fixed and no host galaxy contribution. The post break decay slope is in agreement with K10, whereas a different choice of starting time makes the prebreak slope different.

The SED ($uBVR_CI_Cz^\prime JHK$) is significantly broader than before, and now shows clear curvature. A fit without dust finds it to be moderately red, $\beta=1.17\pm0.10$ ($\chi^2/d.o.f.=1.58$). A fit with MW dust shows that the 2175 {\AA} is not detected, and therefore ruled out. LMC dust, with a smaller bump, is able to be accommodated, however, the fit yields high extinction and a strongly negative intrinsic spectral slope. K10 stated that SMC dust yielded the best result, which we confirm, however, a free fit yields an intrinsic spectral slope $\approx0$ ($\chi^2/d.o.f.=0.28$). We fix the intrinsic spectral slope to a value derived from X-rays under the assumption $\nu_c$ lies between optical and X-rays, $\beta=0.55$, and derive $A_V=0.228\pm0.037$ mag ($\chi^2/d.o.f.=0.49$). This value is larger than that derived for SMC dust in K10 ($\beta=0.52\pm0.37$, $A_V=0.13\pm0.17$ mag), but consistent within errors. The somewhat larger value also leads to a larger correction, we find $dRc=-2.904^{+0.178}_{-0.179}$ mag. As already stated in K10, the early afterglow is one of the most luminous known, reaching nearly $9^{\textnormal{th}}$ magnitude in the $z=1$ system.

\subsubsection{GRB 080413B, $z=1.1014$}

Data are taken from \cite{Filgas2011AA2}, from GCN Circulars \citep{Brennan2008GCN7629}, as well as our own \emph{Swift} UVOT data, first reported in \cite{Oates2008GCN7611}, and FTS data, first reported in \cite{Gomboc2008GCN7626}. The data from \cite{Brennan2008GCN7629} were made fainter by 0.7 mag to bring them into agreement with those of \cite{Filgas2011AA2}. The redshift is given by \cite{Fynbo2009ApJS}.

This burst was first analyzed in K11 with GCN and preliminary \emph{Swift} UVOT data. It shows a decay turning into a long plateau phase, followed by a steep decay, and has been explained by a double-jet model \citep{Filgas2011AA2}. It also shows strong evidence for spectral evolution from a very flat to a ``normal'' spectral slope \citep[][K11]{Filgas2011AA2}.

With the full \emph{Swift} UVOT data set, we confirm and expand these results, and fill the large data gap found in \cite{Filgas2011AA2}. Our first fit covers data up to 0.1d. We find the early afterglow is best fit with a broken power-law decay, with a steep to shallow transition. It is $\alpha_1=0.817\pm0.022$, $\alpha_2=0.523\pm0.011$, $t_b=0.0098\pm0.0011$ d ($850.2\pm96.8$ s), $n=-10$ fixed and here we do not add a host contribution which is several magnitudes fainter. The fit is excellent, $\chi^2/d.o.f.=0.62$. The second fit uses data beginning at 0.14 d (there is no data between 0.1 and 0.14 d) and we find: $\alpha_3=-0.009\pm0.058$, $\alpha_2=2.39\pm0.08$, $t_b=2.28\pm0.14$d, $n=1$ fixed (a soft rollover), the host galaxy magnitudes for $g^\prime r^\prime$ were left free. Only these two bands show a transition into the host. Scatter in the data leads to a worse fit than the initial one ($\chi^2/d.o.f.=2.33$). The first fit agrees very well with the result from K11, and the second fit yields the same $\alpha_4$ within errors, but the additional \emph{Swift} UVOT data coverage shows a different behavior for the earlier part of this data set (a flat plateau instead of a decay similar to the earliest decay), and the break time is over a day earlier.

In both cases, the SED is very broad\footnote{$uvw2uvm2uvw1ubg_Gv$ $r_G i_G z_G J_G H_G K_G$}, but the two show completely different behavior, as found before. The first SED shows some scatter, with especially $b$ being too faint, the source of this offset is unclear as the second SED does not show this effect. The $uvw2\,uvm2\,uvw1$ bands are affected by Lyman dropout and are not included. We fit both SEDs simultaneously, fixing $A_V$ as a shared parameter, but not $\beta$. Without dust, we find $\beta_1=0.409\pm0.047$, $\beta_2=0.772\pm0.046$, with $\chi^2/d.o.f.=2.65$. The result for the second SED is very similar to that found in K11, but the result for the early SED is steeper. This stems from the addition of the full UVOT $u$ data set, which shows some curvature in the SEDs. For MW dust, we find slightly negative extinction (0 within errors). For LMC dust, the extinction is higher, but a 2175 {\AA} bump is ruled out by the bright $g^\prime$ band. The best fit is found for SMC dust, with $\beta_1=0.03\pm0.14$, $\beta_2=0.39\pm0.14$, $A_V=0.23\pm0.08$ mag. The latter slope value agrees perfectly with the late-time XRT spectrum presented on the XRT repository if the cooling break lies between optical and X-rays. We find $dRc_1=-0.657^{+0.156}_{-0.153}$ mag, $dRc_2=-0.674^{+0.152}_{-0.153}$ mag, $\approx0.44$ mag brighter than the value found in K11.

\subsubsection{GRB 080603B, $z=2.6893$}

Data are taken from \cite{Jelinek2012AcPol}, from GCN Circulars \citep{Rujopakarn2008GCN7792,Klotz2008GCN7795,Klotz2008GCN7799,Zhuchkov2008GCN7803,Kuin2008GCN7808,Xin2008GCN7814,Miller2008GCN7827,Klunko2008GCN7890,Rumyantsev2008GCN7974,Ibrahimov2008GCN7975}, as well as our own observations from \emph{Swift} UVOT (first reported in \citealt{Kuin2008GCN7808}), from the \emph{Tautenberg} (first reported in \citealt{Kann2008GCN7865,Kann2008GCN7823}), and the LT (first reported in \citealt{Melandri2008GCN7813}). The redshift is given by \cite{Fynbo2009ApJS}.

This GRB afterglow features multiple phases and breaks in its light curve. The initial afterglow is reported to be ``rapidly fading'' \citep{Rujopakarn2008GCN7792} from an initial detection at mid-time 25.3 s. However, we find the afterglow to fade rapidly from the first detection of the $white$ finding chart on, at 79 s, with a slope $\alpha_{steep}=1.89\pm0.10$, a good indicator of reverse-shock flash behavior. During the first $v$ observations, the afterglow transitions into a plateau phase, a behavior noted before \citep{Klotz2008GCN7795,Klotz2008GCN7799}. This is followed by dense multicolor coverage showing a further break, first reported by \cite{Zhuchkov2008GCN7803} and confirmed by \cite{Jelinek2012AcPol}. We find $\alpha_1=0.436\pm0.019$, $\alpha_2=1.147\pm0.046$, $t_b=0.109\pm0.006$ d, $n=10$ fixed and no host-galaxy contribution. Within errors, this is fully in agreement with \cite{Jelinek2012AcPol} and also similar to the results of \cite{Zhuchkov2008GCN7803}. The afterglow then breaks for a final time during a gap in data coverage, and later observations show it to decay steeply, with $\alpha_3=2.35\pm0.20$, as first noted by \cite{Kann2008GCN7823}. This latter break very likely represents a jet break. The earlier break may stem from the cessation of energy injection into the afterglow.

The SED is broad ($uvw1\,ubg^\prime vr^\prime R_C i^\prime J H K_S$) and straight. $uvw1\,ubg^\prime$ are affected by Lyman dropout and are not included. Fits with dust find very small values, all 0 within errors (negative for SMC dust). We prefer a fit without dust and a spectral slope $\beta=0.621\pm0.068$ ($\chi^2/d.o.f.=0.74$). This agrees fully within errors with the value derived by \cite{Jelinek2012AcPol}. For this fit, we find $dRc=-2.413^{+0.046}_{-0.045}$ mag.

\subsubsection{GRB 080605, $z=1.6403$}

Data are taken from \cite{Zafar2012ApJ,Jelinek2013EAS}, from GCN Circulars \citep{Clemens2008GCN7851,Rumyantsev2008GCN7857}, as well as our own analysis of \emph{Swift} UVOT data (first reported in \citealt{Holland2008GCN7830,Kuin2008GCN7844}), Tautenburg  observations (first reported in \citealt{Kann2008GCN7829,Kann2008GCN7845,Kann2008GCN7864}, see Sect. \ref{ExtSample} for the special analysis done for this data), 2.0 m LT and 2.0 m FTN observations (first reported in \citealt{Gomboc2008GCN7831}), and 8.2m VLT/FORS2 spectroscopy acquisition images (first reported in \citealt{Jakobsson2008GCN7832}). Host galaxy observations have been taken from \cite{Kruhler2012AA,Blanchard2016ApJ,Lyman2017MNRAS}. The redshift is given by \cite{Fynbo2009ApJS}.

This was a bright burst in a crowded field with high line-of-sight extinction and a clear 2175 {\AA} bump in the spectrum \citep{Zafar2012ApJ}. The afterglow shows a steep-shallow-steep evolution. We find $\alpha_{steep}=1.344\pm0.017$, $\alpha_1=0.620\pm0.005$, $t_b=0.00575\pm0.00021$ d ($t_b=487\pm18$ s), $n=-10$ fixed ($\chi^2/d.o.f.=3.23$ owing to some scatter). The early decay compares well with the result \cite{Jelinek2013EAS} find, $\alpha_{steep}=1.27\pm0.04$, based on a smaller data set. Using data after 0.0076 d, we find $\alpha_1=0.535\pm0.015$, $\alpha_2=0.782\pm0.022$, $t_b=0.069\pm0.013$ d ($t_b=5981\pm1137$ s), $n=10$ fixed ($\chi^2/d.o.f.=2.26$). This break is clearly not a jet break. To our knowledge, it has not been reported in the literature yet.

The SED derived from the second fit is broad ($bg_Gg^\prime vr_Gr^\prime R_Ci_Gi^\prime I_Cz_GJHK$) and clearly red but not strongly curved, without dust we find $\beta=1.482\pm0.079$ ($\chi^2/d.o.f.=1.41$). A fit with MW dust finds very little extinction and an intrinsic spectral slope that is still too red: $\beta=1.37\pm0.21$, $A_V=0.085\pm0.146$ mag ($\chi^2/d.o.f.=1.51$). LMC dust leads to a significantly bluer intrinsic spectral slope and higher extinction: $\beta=0.31\pm0.45$, $A_V=0.78\pm0.29$ mag ($\chi^2/d.o.f.=0.92$). We point out that the $v$ band lies in the 2175 {\AA} but does not support its detection. It is clearly seen in the X-shooter spectrum shown by \cite{Zafar2012ApJ}, but not very deep, and their photometry does not actually cover it. Therefore, it may come as no surprise that our SED is also well-fit by SMC dust: $\beta=0.57\pm0.35$, $A_V=0.51\pm0.19$ mag ($\chi^2/d.o.f.=0.84$). \cite{Zafar2011AA} find a large extinction value $A_V=1.20^{+0.09}_{-0.10}$ mag with a full Fitzpatrick-Massa parametrization. Later studies do not support this large extinction, \cite{Zafar2012ApJ} find $\beta=0.60\pm0.03$, $A_V=0.52^{+0.13}_{-0.16}$ mag, and $\beta=0.60\pm0.02$, $A_V=0.50^{+0.13}_{-0.10}$ mag, for two different SEDs at different times, with $R_V=3.24\pm1.05$ and $R_V=3.19^{+0.86}_{-0.89}$, respectively. \cite{Greiner2011AA} derive $\beta=0.67\pm0.01$, $A_V=0.47\pm0.03$ mag. These results are in good agreement with our SMC dust result, which we will continue to use henceforth. For this result, we find $dRc=-2.473\pm0.489$ mag. The early afterglow in the $z=1$ system is luminous, with $R_C\approx11.9$ mag.

\subsubsection{GRB 080721, $z=2.5914$}

Data are taken from \cite{Starling2009MNRAS,Page2019MNRAS}, from GCN Circulars \citep{Chen2008GCN7990,Huang2008GCN7999}, as well as our own \emph{Swift} UVOT analysis (first reported in \citealt{Holland2008GCN7991,Ward2008GCN7996}). The redshift is given by \cite{Fynbo2009ApJS}.

The GRB was already presented in K10, however, we here add the large and complete UVOT data set. We reaffirm a fit with a broken power-law as presented in K10, we find $\alpha_1=1.192\pm0.009$, $\alpha_2=1.53\pm0.09$, $t_b=2.08\pm0.89$ d, $n=10$ fixed and no host-galaxy contribution. This implies $\Delta\alpha=0.34\pm0.09$, in agreement with the result of K10 within errors and still conform with a cooling-break passage.

The SED ($uvw1ubvR_CI_C$) is significantly improved compared to the analysis of K10. It is clearly Lyman-damped in $uvw1u$, so these bands are not included. The slope without dust is still red ($\beta=1.48\pm0.12$, $\chi^2/d.o.f.=0.83$), but less so than in K10 ($\beta=2.36\pm0.30$). MW dust is ruled out as there is no indication of a 2175 {\AA} bump in the $I_C$ band. A fit with LMC dust yields a viable result but the intrinsic spectral slope is significantly steeper than the X-ray slope $\beta_X=0.86$ derived by \cite{Starling2009MNRAS}, so we prefer the fit with SMC dust, just as K10 did, finding $\beta=0.65\pm0.79$, $A_V=0.18\pm0.17$ mag, $\chi^2/d.o.f.=0.50$. As a direct comparison, a fit with $\beta=0.86$ fixed yields $A_V=0.137\pm0.025$ mag, significantly less than the result of K10, $A_V=0.35\pm0.07$ mag.

Using our new SMC result, we derive $dRc=-3.002^{+0.709}_{-0.759}$ mag, less than the K10 value of $dRc=-3.68$ mag. The earlier detection from \cite{Page2019MNRAS} leads to a similar early peak magnitude of $R_C\approx9.2$ mag, however. Therefore, GRB 080721 still has one of the most luminous early afterglows discovered so far.

\subsubsection{GRB 081007, $z=0.5295\pm0.0001$}

Data are taken from \cite{Jin2013ApJ}, \citet[][including unpublished higher time-resolution early GROND observations]{Olivares2015AA}. Host-galaxy observations are taken from \cite{Vergani2014AA,Blanchard2016ApJ,Lyman2017MNRAS}. The redshift is given by \cite{Berger2008GCN8335}.

This was a moderately low-redshift GRB with dense follow-up and an associated, spectroscopically confirmed SN. The SED shows a blue intrinsic spectrum and a quite large amount of SMC dust extinction, $A_V=0.82\pm0.09$. For this fit, we derive $dRc=0.377\pm0.128$ mag, the early afterglow peaks at $R_C=15.7$ mag.

\subsubsection{GRB 081029, $z=3.8479\pm0.0002$}

Data are taken from \cite{Nardini2011AA,Holland2012ApJ}, and from a GCN circular \citep{West2008GCN8449}. The redshift is given by \cite{Holland2012ApJ}.

This GRB afterglow shows a very complex evolution which was followed up in detail by GROND \citep{Nardini2011AA}. The initial afterglow shows a shallow decay with a break to a more regular value, we find: $\alpha_{shallow}=0.420\pm0.027$, $\alpha_{1,a}=0.789\pm0.017$, $t_b=0.01118\pm0.00097$ d ($t_b=966\pm84$ s), $n=10$ fixed ($\chi^2/d.o.f.=0.85$). This is followed by a very strong, fast rebrightening which turns over into a decay phase with superposed small-scale substructure. For this phase, we measure: $\alpha_{rise}=-4.19\pm0.18$, $\alpha_{1,b}=0.707\pm0.0069$, $t_b=0.05442\pm0.00063$ d, $n=1.68\pm0.13$ ($\chi^2/d.o.f.=5.32$ resulting from the above-mentioned deviations). The decay after the rise is similar to the decay preceding it. After a sharp break at 0.195 d, the afterglow decays steeply before turning more shallow again - this slope decrease would indicate a transition to the host galaxy, however, deep late-time observations \citep{Nardini2011AA} do not reveal any host at the GRB position. We find: $\alpha_2=2.019\pm0.011$, $\alpha_3=0.97\pm0.16$, $t_b=1.90\pm0.25$ d, $n=-5$ fixed ($\chi^2/d.o.f.=2.61$). As the afterglow is dominated by GROND data, \cite{Nardini2011AA} find values mostly in agreement with ours. For the early phase, they find $\alpha_{shallow}=0.38\pm0.05$, $\alpha_{1,a}=1.12\pm0.06$, the latter value being significantly steeper than our result. For the rise, they determine $\alpha_{rise}\approx-4.7$, in general agreement with our value. They exclude flaring activity from the following decay and derive an underlying $\alpha_{1,b}\approx0.47$, shallower than our result. For the steep decay, they find $\alpha_3=2.3\pm0.2$. When fitting the entirety of the light curve with a sum of broken power-law models, they find similar values except for the rise, which becomes much steeper, $\alpha_{rise}=-8.2\pm0.4$. They explain the final flattening with the renewed detection of the early component $\alpha_{1,a}$. Together with color evolution, \cite{Nardini2011AA} interpret the light curve as the superposition of two separated and spectrally different components, with the first one becoming dominant again at late times as the second decays more rapidly.

As \cite{Nardini2011AA} detect a color change in the SED between the first and second component, we derive two SEDs and fit them separately. The first ($g_G r_G R_C i_G z_G J J_G H_G H K_G$) is determined from the first fit derived above, whereas the second ($uvw2 uvm2 uvw1 u b g_G v r_G R_C i_G I_C z_G J J_G H_G H K_G K$ are upper limits only) is derived from the fit the strong rise and following decay. Data blueward of the $R$ bands is affected by Lyman-damping and is excluded. The SEDs show some scatter and clear curvature. Without dust, we find $\beta_1=0.869\pm0.044$ ($\chi^2/d.o.f.=3.76$) and $\beta_2=1.062\pm0.011$ ($\chi^2/d.o.f.=30.54$). The latter value is in perfect agreement with that of \cite{Nardini2011AA}. The two SEDs are markedly different by $4.2\sigma$, confirming the result of \cite{Nardini2011AA}. A fit with MW dust finds negative extinction, whereas fits with LMC dust are generally viable, but find extremely blue (partially negative) intrinsic spectral slopes. Fits with SMC dust yield the best result, with $\beta_1=0.11\pm0.20$, $A_{V,1}=0.25\pm0.06$ mag ($\chi^2/d.o.f.=1.89$), 
$\beta_2=0.33\pm0.05$, $A_{V,2}=0.240\pm0.016$ mag ($\chi^2/d.o.f.=5.00$), respectively. The intrinsic spectral slopes agree within the large errors, however, $\Delta\beta$ is identical to the ft without dust. The derived $A_V$ for the two fits is in excellent agreement with each other, remaining constant as expected whereas the underlying slope changes. \cite{Nardini2011AA}, using an X-ray-to-optical analysis, find a less-curved SED and therefore a redder intrinsic result and less extinction: $\beta=1.00\pm0.01$, $A_V=0.03^{+0.02}_{-0.03}$ mag for SMC dust.

For the second, more precise SMC fit, we derive a large $dRc=-4.222^{+0.099}_{-0.096}$ mag. At $z=1$, the early afterglow is luminous ($R_C=11.6$ mag). The strong rebrightening makes this one of the most luminous afterglows known at later times.

\subsubsection{GRB 081203A, $z=2.05\pm0.01$}

Data are taken from \cite{Page2019MNRAS}, from GCN circulars \citep{Andreev2008GCN8596,Andreev2008GCN8615,Volkov2008GCN8604,West2008GCN8617,Liu2008GCN8618,Mori2008GCN8619,Isogai2008GCN8629,Rumyantsev2008GCN8645,Fatkhullin2008GCN8695} as well as our own \emph{Swift} UVOT analysis (first reported in \citealt{DePasquale2008GCN8603}). The redshift is given by \cite{Kuin2009MNRAS}.

This was a very bright GRB afterglow which yielded the first \emph{Swift} UVOT grism spectrum which allowed the measurement of the redshift. It was already presented in K10 but here we add the full UVOT data set which improves especially the characterization of the early afterglow. It shows a well-detailed rise to a rollover peak followed by a simple power-law decay, we find: $\alpha_{rise}=-3.13\pm0.08$, $\alpha_1=1.549\pm0.009$, $t_b=0.00434\pm0.00006$ d ($t_b=375.0\pm5.2$ s), $n=1$ fixed, and no host-galaxy contribution ($\chi^2/d.o.f.=4.56$ resulting from scatter and small error bars).

The SED ($uvw2uvm2uvw1ubg^\prime vR_CI_C$) is significantly improved compared to the one used in K10. The UV filters and $u$ are Lyman-damped and excluded. Also, $g^\prime $ is anomalously low and not included. The goodness of the afterglow fit leads to small errors in the SED and increased $\chi^2$ values. For no extinction, we find $\beta=0.92\pm0.05$ ($\chi^2/d.o.f.=5.17$.) MW dust leads to negative extinction, and SMC dust yields an excellent fit but with a negative intrinsic spectral slope. LMC dust yields a viable result, $\beta=0.72\pm0.17$, $A_V=0.10\pm0.08$ mag, but with a worse fit than without dust ($\chi^2/d.o.f.=.95$). We follow K10 and use SMC dust, fixing the spectral slope to $\beta_O=\Gamma_X-1-0.5=0.45$, for this value we find $A_V=0.127\pm0.017$ mag, fully in agreement with K10. For this fit, we derive $dRc=-2.065^{+0.042}_{-0.043}$ mag. The early peak rises to $R_C=10.4$ mag in the $z=1$ system.

\subsubsection{GRB 090426, $z=2.609$}

Data are taken from \cite{Antonelli2009AA,Xin2011MNRAS,Thoene2011MNRAS,NicuesaGuelbenzu2011AA}, from GCN circulars \citep{Yoshida2009GCN9266,Yoshida2009GCN9267,Mao2009GCN9285,Kinugasa2009GCN9292}, and from our own \emph{Swift} UVOT analysis (first reported in \citealt{Oates2009GCN9265}). The redshift is given by \cite{Levesque2010MNRAS}.

This GRB was discussed widely in the literature as it has a very short duration, especially in the rest-frame, but is likely to be a Type II GRB \citep[][K11]{Antonelli2009AA,Zhang2009ApJ,Levesque2010MNRAS,Xin2011MNRAS,Thoene2011MNRAS,NicuesaGuelbenzu2011AA}. \cite{NicuesaGuelbenzu2011AA} already showed the optical luminosity of the afterglow is compatible with a Type II origin. We therefore include this GRB in our sample.

The afterglow shows a complex evolution. After a potential initial sharp rise, the afterglow decays at a shallow rate before smoothly turning over into a steeper day \citep{Xin2011MNRAS}. We find: $\alpha_{shallow}=0.20\pm0.07$, $\alpha_1=1.177\pm0.038$, $t_b=0.00236\pm0.00026$ d ($t_b=203.9\pm22.5$ s), $n=3$ fixed ($\chi^2/d.o.f.=1.07$). These values are fully in agreement with those of \cite{Xin2011MNRAS}, who find $\alpha_{shallow}=0.26\pm0.07$, $\alpha_1=1.22\pm0.04$, $t_b=227\pm27$ s. \cite{NicuesaGuelbenzu2011AA} also find values that agree well, except for a steeper initial decay: $\alpha_{shallow}=0.48\pm0.04$, $\alpha_1=1.22\pm0.05$, $t_b=290\pm20$ s (with $n=3$ fixed).
After a short plateau phase, at $\approx0.06$ d (\citealt{Xin2011MNRAS} find $t_b=0.082\pm0.042$ d), the decay steepens again. The afterglow, following a data gap, shows complex behavior, which is likely comprise of a short, steep decay, another sharp rise, and then a clear steep decay that is likely post-jet-break \citep{NicuesaGuelbenzu2011AA}. This variability, however, is not well-defined. Fitting data from 0.06 d onward, including the multicolor host-galaxy detections, we find $\alpha_2=0.664\pm0.020$, $\alpha_3=2.64\pm0.11$, $t_b=0.497\pm0.009$ d, $n=10$ fixed ($\chi^2/d.o.f.=1.67$). Our results also generally agree with those of \cite{NicuesaGuelbenzu2011AA}, who found: $\alpha_2=0.46\pm0.15$, $\alpha_3=2.43\pm0.19$, $t_b=0.39\pm0.02$ d.

The SED is very broad ($uvw2uvw1ubg_Gg^\prime Vr_GR_Ci_GI_Cz_G$ $J_GH_GK_S$), is straight and shows some scatter. Without dust, we find $\beta=0.850\pm0.046$. Fits with dust models yield negative extinction (SMC) or only small amounts of extinction compatible with 0 (LMC, MW). We therefore use the fit without extinction, This is fully compatible with \cite{NicuesaGuelbenzu2011AA}, who find $\beta=0.76\pm0.14$ and no evidence for extinction either.

Using the dustless fit, we find $dRc=-2.488\pm0.030$ mag. The early peaks at $R_C=13.8$ mag.

\subsubsection{GRB 090516, $z=4.111\pm0.006$}

Data are taken from \cite{Bolmer2018AA}, GCN Circulars \citep{Gorosabel2009GCN9379,Christie2009GCN9396}, the automatic \emph{Swift} UVOT analysis page (offline at time of writing, see \citealt{Siegel2009GCN9377} for a preliminary report), and our own FTS data (first reported in \citealt{Guidorzi2009GCN9375}), as well as VLT/FORS2 data (first reported in \citealt{deUgartePostigo2009GCN9381}). A late-time host galaxy observation has been taken from \cite{Greiner2015ApJ}. The redshift is given by \cite{deUgartePostigo2012AA}.

This light curve has only been studied at late times so far. At early times, there is likely a rise following the first, faint $white$ detection. Our afterglow discovery data from FTN shows the afterglow to be achromatically decaying with $\alpha_1=1.03\pm0.06$. This is followed by a large data gap, during which few observations were taken, however, the behavior reported by \cite{Coward2010PASA} is contrary to known afterglow behavior (it would be an extreme flare with a very steep decay), so we do not implement this data. The afterglow is significantly recovered by the Liverpool Telescope and NOT \citep{Gorosabel2009GCN9379}, and then GROND and VLT, at a magnitude only 0.5 mag fainter than the end of FTS observations and those of Stardome \citep{Christie2009GCN9396}. It actually rebrightens a bit before going over into a steady decay with $\alpha_2=1.551\pm0.024$. The final GROND data show it potentially flattening off to a host \citep{Bolmer2018AA}, which, however, would be exceedingly luminous at this high redshift. \cite{Bolmer2018AA} add this host component to their fit and therefore find a somewhat steeper decay than we do $\alpha_2\approx1.7$.

The SED ($g_G r_G R_C i_G i^\prime z_G J_G H_G K_G$) shows clear signs of a Lyman dropout affecting the bands blueward of $i_G$. It furthermore shows curvature, and a red spectral slope without dust, $\beta=1.253\pm0.034$, $\chi^2/d.o.f.=22.1$. A fit with MW dust yields negative extinction, whereas free fits with LMC and SMC dust find negative spectral slopes and large extinction. Fixing the intrinsic spectral slope to a value derived from the XRT spectrum ($\beta=0.52$), we find an improved but still formally unacceptable ($\chi^2/d.o.f.=6.87$) fit with SMC dust and $A_V=0.250\pm0.011$ mag. The bad $\chi^2$ is a result of small errors, the fit looks satisfactory visually. \cite{Bolmer2018AA} find an intrinsically steeper slope ($\beta=0.97$) but similar extinction ($A_V=0.19\pm0.03$ mag). The moderately large redshift combined with the extinction leads to a large $dRc=-4.681^{+0.097}_{-0.101}$ mag. This large correction combined with the long plateau phase leads the afterglow of GRB 090516 to be one of the most intrinsically luminous at the late peak time of 0.124 d in the rest-frame.

\subsubsection{GRB 090618, $z=0.54$}

Data are taken from \cite{Cano2011MNRAS,Page2011MNRAS}, and from GCN circulars \citep{Updike2009GCN9529,Updike2009GCN9575,Galeev2009GCN9548}. The redshift is given by \cite{Cenko2009GCN9518}. This was a very bright GRB at moderate redshift with a well observed, bright afterglow, also extensively studied in X-rays \citep{Page2011MNRAS,Campana2011MNRAS}. It has a prominent SN bump, but without spectroscopic confirmation \citep{Cano2011MNRAS}. Using a fit with no dust extinction, we find $dRc=+1.571\pm0.006$ mag, with a moderately luminous afterglow peak of $R_C\approx15.1$ mag.

\subsubsection{GRB 090726, $z=2.713$}

Data were taken from \cite{Simon2010AA}, from GCN circulars \citep{Moskvitin2009GCN9709,Maticic2009GCN9715,Volnova2009GCN9741,Kelemen2009GCN10028}, and from our own \emph{Swift} UVOT analysis (first reported in \citealt{Landsman2009GCN}), as well as our own rereduction of the Ond\v rejov D50 data from \cite{Simon2010AA} at a higher time resolution. This reanalysis does not recover the short flare \cite{Simon2010AA} report. The redshift is given by \cite{Tanvir2019MNRAS}.

The afterglow is densely observed at early times but there is almost no data after 0.1 d. The early afterglow shows a rise and a long rollover to a relatively shallow decay before breaking to a steeper decay. We find $\alpha_{rise}=-1.60\pm0.12$, $\alpha_1=0.61\pm0.03$, $t_b=0.00381\pm0.00022$ d ($t_b=329\pm19$ s), $n=1$ fixed ($\chi2/d.o.f.=2.21$) for the early afterglow, and $\alpha_1=0.55\pm0.02$, $\alpha_2=1.49\pm0.08$, $t_b=0.047\pm0.0026$ d, and $n=10$ fixed, for the later afterglow. \cite{Simon2010AA} undertake a different fit to their data but qualitatively find the same behavior.

The SED ($ubvR_C$) is very red, and also $u$ is affected by Lyman-damping, and we find $\beta=3.41\pm0.60$. \cite{Simon2010AA} already pointed out that the $R_C$ band lies significantly under an extrapolation of the X-ray slope. The \emph{Swift} XRT repository gives an X-ray slope of $\Gamma-1=1.30$, which we adopt. The short baseline does not allow us to distinguish between dust models (the 2175 {\AA} bump is insufficiently covered), and we conservatively use the SMC dust solution which has the lowest extinction, $A_V=0.39\pm0.11$ mag ($\chi^2/d.o.f.=0.00006$). For this fit, we find a large $dRc=-4.310^{+0.410}_{-0.388}$ mag. The corrected afterglow is luminous and peaks at $R_C=13.1$ mag.

\subsubsection{GRB 091018, $z=0.9710\pm0.0003$}

Data are taken from \cite{Wiersema2012MNRAS,Page2019MNRAS}, from GCN circulars \citep{Schaefer2009GCN10036,deUgartePostigo2009GCN10043,LaCluyze2009GCN10046,Motohara2009GCN10047,Cobb2009GCN10110}, and from our own \emph{Swift} UVOT analysis (first reported in \citealt{Landsman2009GCN10041}, our analysis includes early time-reseolved event mode data which was not given in \citealt{Wiersema2012MNRAS}), as well as FRAM/PAO observations. Host galaxy observations are taken from \cite{Vergani2014AA}. The redshift is given by \cite{Wiersema2012MNRAS}.

We find the afterglow is best fit by a broken power-law with $\alpha_1=0.882\pm0.007$, $\alpha_2=1.542\pm0.013$, $t_b=0.402\pm0.009$ d, $n=10$ fixed, and individual host galaxy magnitudes from late time follow-up ($\chi^2/d.o.f.=3.33$). The high $\chi^2$ results from substructure seen in the dense GROND coverage, as pointed out by \cite{Wiersema2012MNRAS}. They find similar values, with $\alpha_1=0.81\pm0.01$, $\alpha_2=1.33\pm0.02$, $t_b=0.374\pm0.019$ d. 

The SED is very broad and rich ($uvw2uvm2uvw1uBg_G V r_G$ $ R_C i_G I_C z_G YJJ_G H_G H K_S K_G$), however, there are a few outliers and we exclude $Bg_GJHK_S$ (note the NIR is still covered by GROND observations). Furthermore, the UV bands are Lyman-damped. We find the SED is best fit by a simple power-law without dust ($\beta=0.607\pm0.018$, $\chi^2/d.o.f.=3.45$), all dust models yield negative extinction. \cite{Wiersema2012MNRAS} find similar values from a joint X-ray-to-optical fit, with $\beta=0.58\pm0.07$, along with a low amount of SMC extinction, $A_V=0.070^{+0.015}_{-0.018}$ mag. For our extinction-less fit, we find $dRc=+0.074\pm0.001$ mag. The early afterglow peaks at $R_C=13.4$ mag.

\subsubsection{GRB 091020, $z=1.71$}

Data are taken from \cite{Gorbovskoy2013ARep,Page2019MNRAS}, from GCN Circulars \citep{Gorbovskoy2009GCN10052,Xu2009GCN10053,LaCluyze2009GCN10109}, and from our own \emph{Swift} UVOT analysis (first reported in \citealt{Oates2009GCN10054}), \emph{Tautenberg} (first reported in \citealt{Kann2009GCN10076,Kann2009GCN10090}), and Lick Nickel observations (first reported in \citealt{Perley2009GCN10058,Perley2009GCN10060}). The redshift is given by \cite{Xu2009GCN10053}.

A detailed multiwavelength analysis of this GRB has not yet been presented in the literature. We find it exhibits and initial plateau phase (likely preceded by an unobserved rise) that breaks into a simple power-law decay with no further significant break seen in the data until several days after the GRB (the \emph{Tautenberg} datum at 5.2 d lies beneath the extrapolation of the earlier decay, but it is only a single $3\sigma$ detection). We find: $\alpha_{plateau}=0.108\pm0.053$, $\alpha_1=1.061\pm0.012$, $t_b=0.00294\pm0.00015$ d ($t_b=254\pm13$ s), $n=10$ fixed ($\chi^2/d.o.f.=2.10$). \cite{Gorbovskoy2013ARep}, based only on their MASTER data, find $\alpha_1=1.2\pm0.1$, in agreement with our result.

The SED is broad ($uvw2uvm2uvw1ubvR_CI_CZ$) but the UV filters and $u$ are affected by Lyman-damping and are excluded ($uvw2uvm2$ are upper limits only). The SED is red, we find $\beta=1.91\pm0.13$ ($\chi^2/d.o.f.=1.12$). Free dust fits yield an intrinsic spectral slope that is still too red (MW), negative extinction (SMC), and a viable fit, but with very large errors for LMC dust: $\beta=0.82\pm0.99$, $A_V=0.57\pm0.52$ mag ($\chi^2/d.o.f.=1.03$). Using the X-ray spectral slope from the \emph{Swift} XRT repository, $\Gamma_X-1=1.04\pm0.09$, we find a good fit with LMC dust only: $A_V=0.453\pm0.066$ mag ($\chi^2/d.o.f.=0.54$). While there is no direct evidence for the 2175 {\AA} bump, the center lies between $v$ and $R_C$ and a small bump as here cannot be ruled out.

Using this fit, we derive $dRc=-2.631^{+0.166}_{-0.173}$ mag. The early afterglow is luminous, peaking at $R_C=11.7$ mag.

\subsubsection{GRB 091024, $z=1.0924\pm0.0002$}

Data are taken from \cite{Virgili2013ApJ}, and from GCN Circulars \citep{Mao2009GCN10092,Moskvitin2009GCN10101,Rumyantsev2009GCN10116}. Our own observations with \emph{Tautenberg} did not achieve a detection \citep{Kann2009GCN10077}. The redshift is given by \cite{Virgili2013ApJ}.

This was an extremely long duration GRB of over 1000 s length with three well separated episodes \citep{Gruber2011AA} with a complex afterglow showing multiple peaks which has been explained by separate jets launched by a precursor and the main event \citep{Nappo2014MNRAS}. It was well-observed at early times despite lying in a crowded field behind a large amount of Galactic extinction ($A_{V,Gal}=2.6$ mag).

We fit the two peaks separately. For the first, we find: $\alpha_{rise,1}=-1.98\pm0.14$, $\alpha_1=1.32\pm0.07$, $t_b=0.00547\pm0.00012$ d ($t_b=473\pm10$ s), $n=2.83\pm0.72$ ($\chi^2/d.o.f.=1.48$). For the second, we find: $\alpha_{rise,2}=-0.87\pm0.07$, $\alpha_2=1.092\pm0.025$, $t_b=0.0324\pm0.00089$ d ($t_b=2799\pm77$ s), $n=5$ fixed ($\chi^2/d.o.f.=3.45$). The later afterglow shows a steeper decay with at least one plateau phase. \cite{Virgili2013ApJ} present a series of more complex modelling fits not directly comparable with our simple fits.

Even after correction for the large foreground extinction, the short-baseline SED ($BVR_CI_C$) is red, we find $\beta=2.01\pm0.06$ ($\chi^2/d.o.f.=12.1$). The $B$ band is significantly fainter than an extrapolation of the $VR_CI_C$ SED, which cannot be attributed to the redshift. All dust models are able to fit the SED exceedingly well, with $\chi^2/d.o.f.$ values $<0.01$ in all cases. SMC dust finds a negative intrinsic spectral slope. MW and LMC dust find the suppressed $B$ band results from it lying in the 2175 {\AA} bump. For MW dust, we find $\beta=0.99\pm0.22$, $A_V=0.44\pm0.09$ mag, and for LMC dust $\beta=0.64\pm0.29$, $A_V=0.67\pm.14$ mag. \cite{Virgili2013ApJ} find the X-ray slope is very blue, $\Gamma-1=0.49^{+0.23}_{-0.21}$, so we prefer the LMC fit. For this fit, we derive $dRc=-1.437^{+0.253}_{-0.263}$ mag. The early afterglow is luminous, peaking at $R_C\approx12.9$ mag.

\subsubsection{GRB 091029, $z=2.751$}

Data have been taken from \cite{Filgas2012AA}, and from GCN Circulars \citep{LaCluyze2009GCN10107,
Cobb2009GCN10111}. The redshift is given by \cite{Tanvir2019MNRAS}.

This afterglow shows a complex evolution. It begins with a steep rise which turns over into a standard decay, followed by a shallower decay which then breaks to a steeper day. While such behavior is not unprecedented, if the X-ray afterglow is taken into account it reaches ``the limit of the fireball scenario'' \citep{Filgas2012AA}. We fit data up to 0.06 d with a broken power-law: $\alpha_{rise}=-3.10\pm0.54$, $\alpha_1=0.571\pm0.004$, $t_b=0.00358\pm0.00018$ d ($t_b=309.3=\pm15.6$ s), $n=1.87\pm0.46$ ($\chi^2/d.o.f.=1.39$). Data after 0.06 d are fit with a second broken power-law: $\alpha_{shallow}=0.097\pm0.022$, $\alpha_2=1.211\pm0.014$, $t_b=0.2007\pm0.0096$ d, $n=2$ fixed ($\chi^2/d.o.f.=0.86$). \cite{Filgas2012AA} find $\alpha_{rise}=-2.97\pm0.67$, $\alpha_1=0.576\pm0.004$, and from a narrow- plus wide-jet model, $\alpha_{plateau}=-0.12\pm0.07$, $\alpha_2=1.14\pm0.02$, $t_b=0.161\pm0.009$ d. These values compare well with our results.

We derive two SEDs from the two light curve fits, both of which are very broad ($uvw2uvm2uvw1ubg_Gvr_GR_Ci_GI_Cz_GJ_G$ $H_GK_G$ and $ubg_Gr_GR_C
i_GI_Cz_GJ_GH_GK_G$).  Filters bluer than $g_G$ are influenced by Lyman-damping and are not included, with the UV filters being upper limits only. Without dust, we find $\beta_1=0.429\pm0.026$ ($\chi^2/d.o.f.=1.97$), and $\beta_2=0.326\pm0.030$ ($\chi^2/d.o.f.=1.41$), respectively. This is in qualitative agreement with \cite{Filgas2012AA}, who also find a very blue spectral slope and color evolution. Both SEDs show some scatter and no curvature. Any fits with dust models yield either slightly negative extinction or very low values that are in agreement with 0 within errors. We therefore use the fits without extinction.

For the first of these fits (as we are interested in the early luminosity), we derive $dRc=-2.335\pm0.018$ mag. The early afterglow peak reaches only a moderate luminosity, $R_C=15.0$ mag.

\subsubsection{GRB 091127, $z=0.49044\pm0.00008$}

Data are taken from \cite{Cobb2010ApJ,Filgas2011AA,Vergani2011AA,Troja2012ApJ,Gorbovskoy2013ARep,Coward2017PASA}, and from GCN Circulars \citep{Xu2009GCN10196,Xu2009GCN10205,Klotz2009GCN10200,Andreev2009GCN10207,Haislip2009GCN10219,Haislip2009GCN10230,Haislip2009GCN10249,Kinugasa2009GCN10248}. The redshift is given by \cite{Vergani2011AA}. The associated SN 2009nz was spectroscopically confirmed \citep{Berger2011ApJ}. This was a well-observed GRB afterglow at moderately low redshift with an associated SN. The early afterglow showed a very flat spectral slope and continuous color evolution \citep{Filgas2011AA}. Correcting for a small amount of SMC extinction, we find $dRc=1.484^{+0.187}_{-0.205}$ mag. The early afterglow is only moderately luminous, $r_c=16.4$ mag.

\subsubsection{GRB 091208B, $z=1.0633\pm0.0003$}

Data are taken from \cite{Uehara2012ApJ}, GCN Circulars \citep{Nakajima2009GCN10260,Xu2009GCN10269,Updike2009GCN10271,Andreev2009GCN10273,Kinugasa2009GCN10275,Xin2009GCN10279}, and our own \emph{Swift} UVOT (first reported in \citealt{dePasquale2009GCN10267}), and FTN, FTS, LT data (first reported in \citealt{Cano2009GCN10262}). The redshift is given by \cite{Perley2009GCN10272}.

This GRB was also detected by an improved analysis of \emph{Fermi}/LAT data \citep{Akerlof2011ApJ}. It showed early optical polarization \citep{Uehara2012ApJ}. The afterglow, initially discovered by \citet[][who give no magnitude in their report]{deUgartePostigo2009GCN10255}, shows a simple light curve evolution well fitted by a broken power-law, we find $\alpha_1=0.621\pm0.008$, $\alpha_2=1.554\pm0.185$, $t_b=0.286\pm0.037$ d, $n=10$ fixed, and no host galaxy contribution. Hereby, we excluded our last two detections each in $R_C$ and $I_C$ as they show what seems to be a short-lived rebrightening or flare. The prebreak decay is somewhat shallower than that found by \cite{Uehara2012ApJ}, who, however, included later, more steeply decaying data, and had significantly less early data to measure the slope.

The SED we derive is very broad ($uvw2\,uvm2\,uvw1\,$ $u B g_G g^\prime v r_G R_C i_G I_C z_G J_G H_G$), red, and shows clear curvature. The $uvw2\,uvm2\,uvw1$ bands are affected by Lyman dropout and are not included. For no dust, we find $\beta=1.991\pm0.077$ ($\chi^2/d.o.f.=3.25$). Fits with MW and LMC dust are unable to incorporate the curvature, finding negative extinction and no extinction, respectively. The SMC fit is clearly the best, finding $\beta=0.24\pm0.40$, $A_V=0.85\pm0.19$ mag, and is fully acceptable ($\chi^2/d.o.f.=0.98$). For this fit, we find $dRc=-1.757^{+0.365}_{-0.367}$ mag. The early afterglow is moderately luminous, peaking at $R_C=14.2$ mag.

\subsubsection{GRB 100219A, $z=4.66723\pm0.00037$}

Data are taken from \cite{Mao2012AA,Thoene2013MNRAS,Bolmer2018AA}, and GCN Circulars \citep{Kuroda2010GCN10440,Kinugasa2010GCN10452}, as well as our own \emph{Swift} UVOT data, first reported in \cite{Holland2010GCN10436}. The redshift is given by \cite{Thoene2013MNRAS}. For this GRB, a source was detected \citep{Holland2010GCN10432}, nearby but at a clear offset from the GRB afterglow \citep{Jakobsson2010GCN10438}, which was originally suggested to be the host galaxy \citep{Bloom2010GCN10433}, but was found to be an unrelated foreground galaxy at $z=0.217$ \citep{Cenko2010GCN10443}. This source contaminates the \emph{Swift} UVOT analysis. As a result of the high redshift, the very large $g_{G}-r_{G}\approx3$ mag color, and the low depth of UVOT observations in $v$, we assume that the actual afterglow is undetected in all \emph{Swift} UVOT filters except $white$. In $white$, we use a high S/N detection of this galaxy from the final epoch as a ``host galaxy'' and subtract the value in flux space from the other detections, leaving five actual afterglow detections from the early phase. We give upper limits otherwise.

This GRB afterglow is double peaked \citep{Mao2012AA} and we fit the two peaks separately. For the first peak, we find $\alpha_{rise,1}=-0.49\pm0.18$, $\alpha_{decay,1}=1.29\pm0.06$, $t_{b,1}=0.0126\pm0.0007$ d, $n=10$ fixed. We back extrapolate this rising slope to derive flux densities for the shifts to high-$z$. The second peak gives $\alpha_{rise,2}=-0.58\pm0.43$, $\alpha_{decay,2}=1.36\pm0.03$, $t_{b,2}=0.203\pm0.021$ d, $n=10$ fixed. The post peak decay slope is fully in agreement with \cite{Thoene2013MNRAS}. The host galaxy is detected in $i^\prime$ \citep{Thoene2013MNRAS}, and we find the late $r_{G}$ data exhibit a decreasing decay rate \citep[also remarked upon by][]{Thoene2013MNRAS} indicative of an underlying host, our fit finds $r_{G}^{host}=26.19\pm0.46$ mag (AB magnitude, corrected for Galactic extinction). The high $z$ and large $r_{G}-i_{G}$ color (see below) make this detection somewhat doubtful, however. The similarity of the two post peak decay rates indicates this could be the result of a strong energy injection, as suggested also by \cite{Mao2012AA}. The decay slope also indicates no jet break has taken place, however, this is in strong contrast to the X-ray data, which decay steeply with $\alpha_{X}\gtrsim3$ after $3.5\times10^4$ s \citep{Mao2012AA}. This may indicate a similar case to GRB 070110 \citep{Troja2007ApJ} and GRB 130831A \citep{DePasquale2016MNRAS}.

The SED ($g_{G}R_Ci_{G}z_{G}J_{G}H_{G}K_{G}$) shows clear signs of a Lyman dropout with $g_{G}-R_c=2.92$ mag and also $R_C-i_{G}=1.65$ mag. The data redward of Lyman-$\alpha$ can be well fit ($\chi^2/d.o.f.=0.049$) with an SPL with no dust and $\beta_0=1.33\pm0.18$. However, the X-ray slope is significantly harder, indicating some dust must be reddening the afterglow \citep{Mao2012AA,Japelj2015A&A}. Fixing the spectral slope to the value derived from the XRT repository late time spectrum ($\beta=0.68$), we find good fits with LMC dust and $A_V=0.30\pm0.09$ mag ($\chi^2/d.o.f.=0.20$), as well as SMC dust and $A_V=0.20\pm0.06$ mag ($\chi^2/d.o.f.=0.17$). The 2175 \AA{} bump lies in the $J$ band, and we find no indication of a flux depression, so we prefer the SMC solution. From this fit, we derive a correction for Lyman damping to the $R_C$ composite light curve of 1.19 mag (a factor of almost exactly 3).

Our result agrees very well with that of \cite{Thoene2013MNRAS}, who also used the flux-calibrated X-shooter spectrum and find $\beta=0.66\pm0.13$ $A_V=0.13\pm0.05$ mag. \cite{Japelj2015A&A} find $\beta=0.73\pm0.02$ and $A_V=0.23\pm0.02$ mag for LMC dust (and $\beta\approx0.66$, $A_V\approx0.26$ mag after applying a correction to the X-ray flux), in full agreement with our result. \cite{Zafar2018MNRAS}, fitting a combination of photometric and X-shooter spectroscopic data with a more general model, find a lower spectral slope $\beta=0.55^{+0.06}_{-0.07}$, and a lower extinction of $A_V=0.14\pm0.03$ mag for a blue $R_V=2.65\pm0.09$.

\subsubsection{GRB 100418A, $z=0.6239\pm0.0002$}

Data are taken from \cite{Marshall2011ApJ,Niino2012PASJ,Laskar2015ApJ,Jelinek2016AdAst,deUgartePostigo2018AA}, GROND data from the PhD thesis of K. Varela\footnote{\url{https://www.imprs-astro.mpg.de/sites/default/files/varela_karla.pdf}}, and from GCN circulars \citep{Updike2010GCN10619,Updike2010GCN10637,Siegel2010GCN10625,Pearson2010GCN10626,Klein2010GCN10627,Rumyantsev2010GCN10634,Rumyantsev2010GCN10783,Rumyantsev2010GCN10883,Morgan2010GCN10648,Moody2010GCN10665,Andreev2010GCN10694,Volnova2010GCN10821}. The redshift is given by \cite{deUgartePostigo2018AA}. This was a faint, soft GRB (technically an X-ray Flash, XRF) with a complex light curve evolution. The early afterglow shows a long plateau phase at a faint magnitude level with a strong flare superposed which is only detected in \emph{Swift} UVOT data. This is followed by a very strong rebrightening and a further long plateau phase which goes over into a decay only about 1d after the GRB. The associated SN is likely the faintest found so far for a GRB \citep{Niino2012PASJ,deUgartePostigo2018AA}. A full analysis of the light curve is given in \cite{deUgartePostigo2018AA}. We also use their X-ray t optical SED which finds $\beta=1.061^{+0.024}_{-0.023}$, $A_V=0.086\pm0.039$ for SMC dust. \cite{Japelj2015A&A} find a somewhat bluer intrinsic slope and slightly higher extinction,  $\beta=0.73^{+0.07}_{-0.08}$, $A_V=0.20^{+0.03}_{-0.02}$ mag, for SMC dust. \cite{Zafar2018MNRAS} find very similar values to the ones we use here, $\beta=1.01^{+0.12}_{-0.10}$, $A_V=0.12\pm0.03$ mag, with a blue $R_V=2.42^{+0.08}_{-0.10}$. For this fit, we derive $dRc=+1.161^{+0.055}_{-0.058}$ mag. The early afterglow is of very low luminosity, at a level of $R_C\approx21$ mag. It reaches $R_C\approx19.5$ mag at 0.43 d in the $z=1$ frame.

\subsubsection{GRB 100621A, $z=0.5426$}

Data are taken from \cite{Greiner2013AA}, and from a GCN circular \citep{Naito2010GCN10881}. Late-time host-galaxy observations are taken from \cite{Kruehler2011AA}, see also \cite{Blanchard2016ApJ,Vergani2014AA}. The redshift is given by \cite{Japelj2016AA}. This was a bright but soft GRB at a moderately low redshift. The afterglow is covered very well by early GROND observations and exhibits a peculiar behavior, with a steep rise, a plateau with substructure, and a decay, which is followed by a second, extremely steep rise (\citealt{Greiner2013AA}, see also \citealt{deUgartePostigo2018AA2}), another plateau with substructure, and a final decay. However, there is likely another rebrightening at $\approx1.2$ d, but data density is strongly reduced by this time. We subtract the individual host-galaxy values in each band, and fit data from $\approx0.062$ d onward, the beginning of the second plateau phase. We find $\alpha_1=0.456\pm0.036$, $\alpha_2=2.14\pm0.06$, $t_b=0.1065\pm0.0009$ d, $n=15.3\pm4.4$ ($\chi^2/d.o.f.=0.81$). This compares well with the results from \cite{Greiner2013AA}, $\alpha_1=0.42\pm0.05$, $\alpha_2=2.3\pm0.1$, $t_b\approx0.1$ d.

The afterglow SED ($g_G r_G i_G z_G J_G H_G K_G$) is extremely red, we find $\beta=3.989\pm0.013$ ($\chi^2/d.o.f.=340$, an extremely high result stemming from the very small errors of the SED). Free fits with MW and LMC dust find generally viable results, however, the intrinsic spectral slope is very blue ($\beta=0.05-0.2$). For SMC dust, we find $\beta=0.778\pm0.085$, $A_V=3.715\pm0.098$ mag ($\chi^2/d.o.f.=24.5$, still exceedingly high because of small errors, the fit is perfectly fine visually, however). In all cases, the $g^\prime$ band is too bright and excluded, we caution that the very high extinction implies $g^\prime$ is always close to the host-galaxy level and small effects such as differing apertures for afterglow and host measurements may strongly influence the detection level. \cite{Greiner2013AA} find a very similar extinction value, with $\beta=0.82\pm0.02$, $A_V=3.65\pm0.06$ mag. The same is true for \cite{Kruehler2011AA}, who find $\beta=\Gamma_X-1-0.5=0.79$ and $A_V=3.8\pm0.2$ mag. For the SMC fit, we derive, despite the moderately low redshift, $dRc=-3.595^{+0.140}_{-0.144}$ mag. The afterglow peaks late, during the second peak, at a moderate $R_C\approx15.9$ mag.

\subsubsection{GRB 100814A, $z=1.439$}

Data are taken from \cite{Nardini2014AA,DePasquale2015MNRAS}, and from GCN circulars \citep{Elenin2010GCN11129,Elenin2010GCN11133,Volnova2010GCN11153}. The redshift is given by \cite{Selsing2019AA}. The afterglow evolution is complex and reminiscent of GRB 081029 \citep{Nardini2011AA}. It begins with a sparsely covered flare, a decay and a small rebrightening. After this, it goes over into a smooth decay well-covered by GROND, which turns into a long rise that peaks at $\approx1.2$ d before turning over into a steeper decay, the afterglow after host-galaxy subtraction is detectable until $\approx26$ d. The afterglow shows color evolution across the different components \citep{Nardini2014AA} and we perform two different fits, the first encompassing the first long-term decay and the beginning of the rebrightening, for which we find: $\alpha_1=0.5756\pm0.0026$, $\alpha_{rise,1}=-1.49\pm0.10$, $t_{b,1}=0.340\pm0.014$ d, $n_1=-1.29\pm0.11$ ($\chi^2/d.o.f.=2.33$). For the second fit, our result comes to be: $\alpha_{rise,2}=-0.156\pm0.016$, $\alpha_2=2.162\pm0.013$, $t_{b,2}=2.119\pm0.011$, $n_2=1.86\pm0.08$ ($\chi^2/d.o.f.=5.70$). Hereby, $\alpha_{rise,1}$ and $\alpha_{rise,2}$ cover different parts of the light curve and therefore do not have to be similar. \cite{Nardini2014AA} even fit part of the second peak with a shallow decay. Their other results are in agreement with ours, they find the early decay is slightly chromatic and derive decay indices in the range $\alpha_1=0.40-0.58$. After correcting for the host-galaxy contribution, they find a late decay $\alpha_2=2.25\pm0.08$. Under the assumption of achromatic evolution, \cite{DePasquale2015MNRAS} derive $\alpha_1=0.48\pm0.02$, $\alpha_2=1.97\pm0.02$, $t_b=2.52\pm0.028$ d, also in reasonable agreement with our results.

We also create two SEDs to take the different fits and the color evolution into account. Both SEDs are very broad ($uvw2uvm2uvw1ubg_Gvr_GR_Ci_Gz_GJ_GH_GK_G$), however, we find that $g_G$ is anomalously bright, and $v$ anomalously faint so we exclude them. We also exclude $uvw2uvm2uvw1$ as they are clearly affected by Lyman damping. In accordance with \cite{Nardini2014AA}, the first SED is very blue (while still exhibiting slight curvature), we find $\beta_1=0.348\pm0.020$ ($\chi^2/d.o.f.=3.62$). \cite{Nardini2014AA} generally find $A_V\approx0$ mag, and that the spectral index becomes bluer during the first decay before reaching $\beta=0.18\pm0.08$. The second SED, also exhibiting curvature, is significantly redder and more typical for a GRB afterglow, we find $\beta_2=0.743\pm0.014$ ($\chi^2/d.o.f.=13.9$), in agreement with \cite{Nardini2014AA}, who derive $\beta=0.82\pm0.15$ for a late-time SED at 0.7 Ms. For dust fits, we find that in both cases the MW fit find negative extinction. LMC dust fits are not ruled out, but the intrinsic spectral slope is very blue (negative for the first fit). We prefer the SMC fits, for which we find $\beta=0.10\pm0.06$, $A_V=0.123\pm0.031$ mag, ($\chi^2/d.o.f.=1.43$), and $\beta=0.411\pm0.043$, $A_V=0.158\pm0.020$ mag, ($\chi^2/d.o.f.=4.49$), respectively. The $A_V$ values agree with each other within errors, strongly indicating it is indeed the intrinsic spectrum that changes, without any influence on the environment, in agreement with \cite{Nardini2014AA}. \cite{DePasquale2015MNRAS} find roughly similar results and slightly variable extinction. At 500 s, they find (assuming $\Delta\beta=0.5$ from X-ray-to-optical fits, and LMC extinction) $\beta=0.34\pm0.06$, $A_V=0.139^{+0.079}_{-0.044}$ mag; whereas for 50 ks, they find $\beta=0.50^{+0.02}_{-0.04}$, $A_V=0.044\pm0.016$ mag. \cite{Japelj2015A&A} also find results similar to ours, using SMC dust: $\beta=0.52\pm0.07$, $A_V=0.20\pm0.03$ mag. \cite{Zafar2018MNRAS} find a steeper intrinsic spectral slope, however: $\beta=0.92^{+0.12}_{-0.08}$, $A_V<0.07$ mag, with no $R_V$ value given. Using our second SMC fit, we derive $dRc=-1.218\pm0.045$ mag. The light curve peak at a moderately luminous $R_C=14.5$ mag when shifted to $z=1$.

\subsubsection{GRB 100901A, $z=1.4084$}

Data are taken from \cite{Gorbovskoy2012MNRAS,Hartoog2013MNRAS,Laskar2015ApJ}, from GCN Circulars \citep{DeCia2010GCN11170,Kuroda2010GCN11172,Kuroda2010GCN11189,Kuroda2010GCN11205,Hentunen2010GCN11173,Updike2010GCN11174,Sahu2010GCN11175,Sahu2010GCN11197,Kopac2010GCN11177,Sanchez-Ramirez2010GCN11180,Yoshida2010GCN11190,Andreev2010GCN11191,Andreev2010GCN11200,Ukwatta2010GCN11198,Sposetti2010GCN11213}, as well as our own observations with the \emph{Tautenberg} (first reported in \cite{Kann2010GCN11187,Kann2010GCN11236,Kann2010GCN11246}, the UKIRT/WFCAM (first reported in \citealt{Im2010GCN11208}),
as well as an unpublished GROND data set. The redshift is given by \cite{Hartoog2013MNRAS}. The early afterglow of this GRB shows a remarkable evolution, with multiple rebrightenings, likely stemming from energy injections \citep{Laskar2015ApJ}. After the final peak at $\approx0.38$ d, we find the afterglow can be fitted with a broken power-law, it is $\alpha_1=1.392\pm0.016$, $\alpha_2=1.930\pm0.042$, $t_b=3.26\pm0.22$ d, $n=10$ fixed, and no host galaxy was detected to deep limits in late GROND observations. This is likely a jet break.

The SED is very broad and rich ($uvw2uvm2uvw1uBg_Gvr_G$ $R_Ci_GI_Cz_Gz^\prime YJJ_GH_GHK_GK$). It is well-fit by a simple power-law with $\beta=0.75\pm0.06$ ($\chi^2/d.o.f.=0.81$). All dust models yield viable solutions with low amounts of extinction. For MW dust, we find $\beta=0.57\pm0.24$, $A_V=0.13\pm0.17$ mag ($\chi^2/d.o.f.=0.83$), for LMC dust, we find $\beta=0.40\pm0.26$, $A_V=0.26\pm0.17$ mag ($\chi^2/d.o.f.=0.87$), and for SMC dust, we find $\beta=0.54\pm0.18$, $A_V=0.14\pm0.09$ mag ($\chi^2/d.o.f.=0.86$). While mathematically indistinguishable, the lack of a clear 2175 {\AA} bump makes us prefer the SMC solution. \cite{Japelj2015A&A} also prefer SMC dust, finding a similar spectral slope but higher extinction: $\beta=0.50\pm0.04$, $A_V=0.29\pm0.03$ mag. \cite{Zafar2018MNRAS} find a slightly redder slope and also somewhat higher extinction, $\beta=0.70^{+0.13}_{-0.16}$, $A_V=0.25\pm0.08$ mag, for $R_V=3.01\pm0.11$. Generally, the values are in agreement with our result. For the SMC fit, we derive $dRc=-1.137^{+0.197}_{-0.198}$ mag. The afterglow is only moderately luminous, peaking at $R_C\approx15.8$ mag, however, this peak is until $\approx0.24$ d in the $z=1$ system. At early times, the magnitude is almost always $R_C>16$ mag.

\subsubsection{GRB 100906A, $z=1.727$}

Data are taken from \cite{Gorbovskoy2012MNRAS,Page2019MNRAS}, from GCN Circulars \citep{Kuroda2010GCN11241,Kuroda2010GCN11249,Hentunen2010GCN11253,Tkachenko2010GCN11254,Strobl2010GCN11340,Volnova2010GCN11395}, as well as our own observations with \emph{Swift} UVOT (first reported in \citealt{Siegel2010GCN11242}), \emph{Tautenberg} (first reported in \citealt{Kann2010GCN11238,Kann2010GCN11247}), FTN (first reported in \citealt{Melandri2010GCN11229}), UKIRT/WFCAM (first reported in \citealt{Im2010GCN11232}), and Ond\v rejov D50 data (reported in \citealt{Strobl2010GCN11340}) which are not published in this paper. The redshift is given by \cite{Tanvir2010GCN11230}. The GRB afterglow is very bright and a ``classic'' example of a forward-shock rise, captured very early by MASTER \citep{Gorbovskoy2012MNRAS}. We fit data up to $\approx0.15$ d after the trigger with a smoothly broken power-law which is best fit by an extremely smooth rollover. We find $\alpha_1=-25.27\pm0.47$, $\alpha_2=1.0544\pm0.0026$, $t_b=0.00038\pm0.0000004$ d ($32.83\pm0.32$ s), $n=0.1$ fixed, and no host-galaxy contribution. This extremely steep rise is not realistic, but the extremely smooth rollover implies that the afterglow takes a long logarithmic timespan to asymptotically reach this value. The short baseline of the measured rise leads to an imprecise determination of the rising slope. The post-peak decay index is very precisely measured. After $\approx0.15$ d, the afterglow transitions into a plateau phase with additional variability as well as a following sharp flare overlaid over a break to a steep decay. We find $\alpha_1=0.274\pm0.037$, $\alpha_2=1.988\pm0.020$, $t_b=0.370\pm0.0052$ d, $n=10$ fixed. This is very likely a jet break. \cite{Gorbovskoy2012MNRAS}, using a significantly smaller data set, find generally similar results with $\alpha_1=0.14\pm0.02$, $\alpha_2=2.17^{+0.03}_{-0.04}$, $t_b=0.404^{+0.006}_{-0.002}$ d.

The SED is very broad ($uvw2\,uvm2\,uvw1\,ubg^\prime vR_CI_Cz^\prime YJ$  $HK$). Without dust, we find a moderately red but curved afterglow SED, it is $\beta=0.840\pm0.011$ ($\chi^2/d.o.f.=43.1$). A fit with MW dust yields negative extinction. LMC dust, on the other hand, yields a negative spectral slope and high extinction, we also find no compelling evidence for a 2175 {\AA} bump. SMC dust, finally, yields a viable solutions: $\beta=0.151\pm0.039$, $A_V=0.292\pm0.016$ mag, with $\chi^2/d.o.f.=2.20$. For this fit, we derive $dRc=-1.953\pm0.044$ mag. The early afterglow is luminous, peaking at $R_C\approx10.7$ mag.

\subsubsection{GRB 101219B, $z=0.55185\pm0.00005$}

Data are taken from \cite{Olivares2015AA,Sparre2011ApJ}. The redshift is given by \cite{Sparre2011ApJ}. This was a soft GRB at moderate redshift which was followed-up in detail by \emph{Swift} UVOT and GROND, the latter which discovered an SN bump \citep{Olivares2015AA}, which was confirmed spectroscopically \citep{Sparre2011ApJ}. We find the SED is best fit by a small amount of SMC dust, with $\beta=0.578\pm0.067$, $A_V=0.077\pm0.037$ mag ($\chi^2/d.o.f.=3.05$ as there is some scatter). For this fit, we find $dRc=1.375\pm0.054$ mag. Our results compare well with \cite{Olivares2015AA}, who find $\beta=0.62\pm0.01$, $A_V=0.12\pm0.01$ mag. \cite{Zafar2018MNRAS}, using the X-shooter spectrum and a full Fitzpatrick-Massa parametrization, find a steeper spectral slope $\beta=0.93^{+0.14}_{-0.10}$, but a similar extinction $A_V<0.11$ mag. The steeper spectral slope may be the result of a different $R_V$, but no value is given.

\subsubsection{GRB 110205A, $z=2.21442\pm0.00044$}

Data are taken from \cite{Cucchiara2011ApJ2,Zheng2012ApJ,Gendre2012ApJ,Steele2017ApJ}, and from GCN circulars \citep{Vreeswijk2011GCN11640,Kugel2011GCN11647,Kuroda2011GCN11651,Kuroda2011GCN11652}. The redshift is given by \cite{Cucchiara2011ApJ2}. This GRB afterglow has one of the densest broadband follow-ups obtained to date. The afterglow is characterized by an initial plateau phase with a superposed optical flare linked to prompt-emission activity \citep{Guiriec2016ApJ}, followed by a steep. long rise to a very bright peak. After this, the afterglow decays smoothly, experiences a short plateau phase at $\approx0.5$ d before finally breaking to an even steeper decay. For the rising and decaying phase, before 0.5 d, we find: $\alpha_{rise}=-4.278\pm0.077$, $\alpha_1=1.4873\pm0.0018$, $t_b=0.01074\pm0.00005$ d ($t_b=927.9\pm4.3$ s), $n=2.34\pm0.11$, no host galaxy ($r^\prime\gtrsim27$ mag, \citealt{Cucchiara2011ApJ2}) ($\chi^2/d.o.f.=6.04$ resulting from some scatter and small errors). Data after 0.53 d is fit with a steeper decay, we find $\alpha_2=1.989\pm0.036$ ($\chi^2/d.o.f.=1.76$). We note late-time detections presented in \cite{Cucchiara2011ApJ2} are significantly brighter than the data presented in \cite{Gendre2012ApJ}, the latter being in good agreement with the extrapolation of earlier data after the short plateau phase. 
\cite{Cucchiara2011ApJ2} find $\alpha_{rise}=-6.13\pm0.75$, $\alpha_1=1.71\pm0.28$, $t_b=837^{+51}_{-40}$ s, $\alpha_2=1.74\pm0.28$, roughly in agreement with our results. \cite{Zheng2012ApJ} use a multicomponent model, making comparison difficult, \eg, they fix $\alpha_{rise}=5.5$ in a scenario where the early peak is a pure reverse shock. For a different model, the late-time decay is model-independent and they report $\alpha_1=1.50\pm0.04$, in excellent agreement with our result. Their $t_b=1064\pm42$ s is somewhat later than ours. \cite{Steele2017ApJ} report $\alpha_{rise}=-4.63\pm0.29$, $\alpha_1=1.52\pm0.02$, $t_b=1027\pm8$ s, $n=2.18\pm0.45$, in good agreement with our results.

The SED is very broad ($uvw2 uvm2 uvw1 u b g^\prime v r^\prime R_C i^\prime I_C z^\prime J H K$), shows little scatter but also has very small errors, and is clearly curved. Without dust, we find $\beta=1.302\pm0.011$ ($\chi^2/d.o.f.=32.1$). For MW dut, we find negative extinction, but LMC and SMC dust yield viable solution, with $\beta_{LMC}=0.561\pm0.062$, $A_{V,LMC}=0.396\pm0.033$ mag ($\chi^2/d.o.f.=16.2$), and $\beta_{SMC}=0.738\pm0.042$, $A_{V,SMC}=0.191\pm0.013$ mag ($\chi^2/d.o.f.=6.95$). While the goodness-of-fit is still high due to the small errors, the SMC fit is clearly preferred and represents one of the ``best-behaved'' SEDs in the entire sample.

\cite{Zheng2012ApJ} obtain multiple SEDs at several epochs (their table 6). For SMC dust, they generally find somewhat bluer intrinsic power-law slopes ($\beta\approx0.55$) and therefore somewhat higher extinction ($A_V\approx0.27$ mag), but in general agreement with our result. \cite{Gendre2012ApJ} study the SED at two epochs, initially finding, from a joint X-ray-to-optical fit, $\beta_{OX,1}=0.84\pm0.04$ and no evidence for dust. From later data, they find $\beta_{OX,2}=1.03\pm0.10$, $A_{V,MW}=0.27\pm0.10$ mag, $A_{V,SMC}=0.14\pm0.10$ mag. The SMC result is in reasonable agreement with ours. For our SMC result, we derive $dRc=-2.606^{+0.046}_{-0.047}$ mag, and the early afterglow peak at $z=1$, is luminous, peaking at $R_C\approx11.3$ mag.

\subsubsection{GRB 110213A, $z=1.4607\pm0.0001$}

Data are taken from \cite{Cucchiara2011ApJ2,Jelinek2016AdAst,Wang2022ApJ}, and from GCN circulars \citep{Rujopakarn2011GCN11707,Kuroda2011GCN11719,Nakajima2011GCN11724,Wren2011GCN11730,Zhao2011GCN11733}, and from our own \emph{Swift} UVOT analysis (which completes the partial analysis already presented in \citealt{Cucchiara2011ApJ2} and presents late-time limits beyond the data given in \citealt{Wang2022ApJ}; first reported in \citealt{Kuin2011GCN11718}). The redshift is given by \cite{Cucchiara2011ApJ2}. This GRB had a highly peculiar afterglow. After an initial decay, the afterglow rises slowly to a peak and then decays. We fit data from 0.00086 d to 0.024 d with a smoothly broken power-law and find: $\alpha_{rise,1}=-2.702\pm0.019$, $\alpha_{1,1}=0.7455\pm0.0053$, $t_{b,1}=0.00223\pm0.00001$ d ($t_{b,1}=192.67\pm0.86$ s), $n=1.283\pm0.034$. This is an extremely precise fit, but the very small errors of the dense photometric coverage leads to a very high $\chi^2/d.o.f.=21.2$. This decay then switches to a second rise to a peak with a following decay, in this case, the afterglow evolution is additionally characterized by rapidly variable substructure. In this case, we find: $\alpha_{rise,2}=-0.5253\pm0.0052$, $\alpha_{1,2}=1.4011\pm0.0087$, $t_{b,2}=0.0642\pm0.0001$ d, $n=9.45\pm0.49$. The additional variability leads to an ever worse goodness-of-fit, $\chi^2/d.o.f.=43.2$. The data indicates a plateau phase from $\approx1.2$ d to 2.3 d, after which the afterglow decays sharply, we find $\alpha_2=3.52\pm0.15$. Such a decay is similar to what we find for GRB 120326A (see \S\ref{GRB120326A}) and GRB 060526 \citep{Thoene2010AA}. This steep decay likely represents a post-jet break decay, as already remarked upon by \cite{Cucchiara2011ApJ2}. These authors find $\alpha_{rise,1}=-2.08\pm0.23$, $\alpha_{1,1}=1.10\pm0.24$, $t_{b,1}=263^{+13}_{-19}$ s, $\alpha_{rise,2}=-2.02\pm0.34$, $\alpha_{1,2}=1.80\pm0.15$, $t_{b,2}=0.056$ d. These values are somewhat discrepant to our results, this is because, for one, \cite{Cucchiara2011ApJ2} use a superposition of two Beuermann equations, and secondly, likely, because we let the break smoothness be a free parameter, which influence the temporal slopes.

The SEDs of this GRB afterglow are puzzling. Our temporal fits yield results with very small errors, and the creation of the joint light curve using the normalizations as offsets reveals an extremely densely covered light curve for which even small details, especially during the variability of the second bump, can be traced. This makes it highly unlikely that there are systematic offsets involved which would lead to erroneous scatter in the SEDs. They contain no NIR data but are otherwise broad ($uvw2uvm2uvw1ubg^\prime vr^\prime R_Ci^\prime I_Cz^\prime$, with the first fit containing no $g^\prime$ data, $uvw2uvm2uvw1$ data are affected by Lyman damping and are not included). The SEDs show deviations from a power-law behavior that cannot be explained by the usual dust models, \eg, they are essentially flat between $g^\prime$ and $R_C$, only to the show a large $r^\prime -i^\prime $ color. This behavior negates the possibility to obtain any useful free fit with dust models. Without dust, we find $\beta=1.322\pm0.006$ with an extreme $\chi^2/d.o.f.=742$. Using the ``late'' data fit from the \emph{Swift} XRT repository, we fix $\beta_{opt}=0.90$ and find that an SMC fit yields the ``best'' result, it is $A_V=0.132\pm0.0025$ mag, with $\chi^2/d.o.f.=851$.

However, we note that a dust model exists which shows a similar behavior to the SED derived here, that found in quasars in the high-redshift Universe, and also discovered in a single high-redshift event, GRB 071025 \citep{Perley2010MNRAS,Jang2011ApJ}, which, we note, actually shows a temporal evolution similar to GRB 110213A, with a double bump. A deeper study is beyond the scope of this paper. For the SMC fit, we derive $dRc=-1.306^{+0.013}_{-0.016}$ mag. At $z=1$, the afterglow peaks at $R_C\approx14$ mag.

\subsubsection{GRB 110422A, $z=1.770\pm0.001$}

Data are taken from \cite{Pruzhinskaya2014NewA}, from GCN Circulars \citep{Xu2011GCN11961,Xu2011GCN11970,Xu2011GCN11974,Moskvitin2011GCN11962,Melandri2011GCN11963,Hentunen2011GCN11966,Kuroda2011GCN11972,Rumyantsev2011GCN11973,Rumyantsev2011GCN11979,Rumyantsev2011GCN11986}. as well as our own observations with \emph{Swift} UVOT (first reported in \citealt{Breeveld2011GCN11969}), the Otto Struve/CQUEAN (first reported in \citealt{Jeon2011GCN11967}; including late host galaxy detections in $i^\prime z^\prime$), the LOAO (upper limits only), the AZT-22/SNUCAM, and the Bohyunsan/KASINICS (NIR upper limit only). The redshift is given by \cite{deUgartePostigo2011GCN11978}. The afterglow consists of two parts. Data up to 0.027 d are fit by a single power-law decay, we find $\alpha_1=1.082\pm0.052$ ($\chi^2/d.o.f.=1.84$). Between this time and 0.1 d, a rebrightening must take place which is not covered by observations. Data after 0.1 d are again fit with a simple power-law and a less steep slope ($\alpha_2=0.893\pm0.042$). Hereby, we include the host galaxy as a free parameter in $R_Ci^\prime z^\prime$, the latter two bands have late detections, while $R_C$ shows a flattening at $\approx10$ d, the fit finds $R_C=23.26\pm0.16$ mag (Vega, corrected for Galactic extinction) for the host.

The SED is broad ($uvw2uvm2uvw1ubg^\prime vR_Ci^\prime z^\prime $), and red but straight (the UV filters are Lyman-damped and are not included). It is well-fit by a simple power-law ($\beta=1.79\pm0.13$, $\chi^2/d.o.f.=0.37$), However, this value is steeper than typically found in the fireball model. Free fits with all dust models yield results with little extinction and too-red intrinsic slopes. The \emph{Swift} XRT repository gives $\beta_X=0.84\pm0.09$ for the ``late'' spectrum, and we fix the intrinsic slope to this value. MW dust would require a strong 2175 {\AA} bump which is not observed, and SMC dust shows a too-steep far-UV slope. However, we find an excellent fit with LMC dust, it is $A_V=0.49\pm0.07$ mag ($\chi^2/d.o.f.=0.49$). Using these values, we find we derive $dRc=-2.766^{+0.178}_{-0.181}$ mag. The early afterglow is quite luminous, peaking at $R_C\approx12.1$ mag.

\subsubsection{GRB 110715A, $z=0.8224\pm0.0002$}

Data are taken from \cite{Sanchez-Ramirez2017MNRAS}, and from GCN circulars \citep{Piranomonte2011GCN12164,Nelson2011GCN12174}. The redshift is given by \cite{Sanchez-Ramirez2017MNRAS}. This is the first GRB for which a submillimeter afterglow was detected with ALMA \citep{Sanchez-Ramirez2017MNRAS}. The afterglow shows a complex evolution with multiple rebrightening episodes. The very first detections show evidence for a steep decay which flattens quickly, we find $\alpha_{steep}=1.85\pm0.70$, $\alpha_{1,a}=1.05\pm0.16$, $t_b=0.00133\pm0.00025$ d ($t_b=114.9\pm21.6$ s), $n=-100$ fixed ($\chi^2/d.o.f.=0.39$). After multiple rebrightening episodes which are only sparsely sampled, the afterglow resumes a normal decay at $\approx1.8$ d, from here on, it is covered by multicolor GROND observations. We find it can be fit by a smoothly broken power-law, with $\alpha_{1,b}=1.15\pm0.23$ (in full agreement with $\alpha_{1,b}$ as would be expected for energy injections), $\alpha_2=1.88/pm0.12$, $t_b=3.95\pm0.98$ d, $n=10$ fixed ($\chi^2/d.o.f.=0.53$).

The early (\emph{Swift} UVOT with a few $R_C$ points) and late observations (GROND with a few UVOT points) are almost disconnected in terms of filter coverage, so great care was taken in constructing the SED, which is broad despite the high Galactic foreground extinction ($ubg_Gvr_Gi_Gz_GJ_GH_GK_G$). We find it to be moderately red and clearly curved, $\beta=1.200\pm0.054$ ($\chi^2/d.o.f.=0.99$). All dust models yield good fits, we find for MW dust: $\beta=0.67\pm0.26$, $A_V=0.46\pm0.22$ mag ($\chi^2/d.o.f.=$), for LMC dust: $\beta=0.57\pm0.31$, $A_V=0.54\pm0.26$ mag ($\chi^2/d.o.f.=0.49$), and for SMC dust: $\beta=0.63\pm0.28$, $A_V=0.46\pm0.22$ mag ($\chi^2/d.o.f.=0.50$). These results are statistically indistinguishable, however, the 2175 {\AA} would lie between the $u$ and $b$ bands, both which have large errors, but we see no conclusive evidence for this spectral feature. Therefore, we prefer the SMC fit. This result is different from that of \cite{Sanchez-Ramirez2017MNRAS}, who find $\beta=0.90\pm0.22$, $A_V=0.09\pm0.18$ mag for SMC dust. This leads to us finding a significantly smaller $dRc$ than \cite{Sanchez-Ramirez2017MNRAS} do, $dRc=-0.276^{+0.367}_{-0.370}$ mag compared to $dRc=0.38^{+0.17}_{-0.32}$ mag. While the multiple rebrightenings make the afterglow observationally very bright \citep{Sanchez-Ramirez2017MNRAS}, the early afterglow is luminous but not exceedingly so, with $R_C\approx13.5$ mag.

\subsubsection{GRB 110726A, $z=1.036$}

Data are taken from \cite{Steele2017ApJ}, from GCN circulars \citep{Schaefer2011GCN12197,Kuroda2011GCN12204,Zheng2011GCN12205,Gorosabel2011GCN12206,Moskvitin2011GCN12251}, and from our own \emph{Swift} UVOT analysis (first reported in \citealt{Siegel2011GCN12199,Porterfield2011GCN12203}). We also present late-time $g^\prime r^\prime i^\prime z^\prime $ observations of a bright, resolved host galaxy obtained by the GTC HiPERCAM. The redshift is given by \cite{Cucchiara2011GCN12202}. The initial light curve shows a plateau phase which goes over into a steep decay \citep{Zheng2011GCN12205}, which then transitions into a plateau phase \cite{Steele2017ApJ}. We find $\alpha_{1,1}=1.032\pm0.033$, $\alpha_{plateau}=0.16\pm0.09$, $t_b=0.0189\pm0.0014$ d ($t_b=1629\pm124$ s), $n=-10$ fixed ($\chi^/d.o.f.=0.73$). \cite{Steele2017ApJ} model the light curve with two overlapping components, finding $\alpha_{1,1}=1.03\pm0.05$, in perfect agreement with our value. Data after the plateau are scarce, using only $R_C$ data and our HiPERCAM host magnitude, we find that the late light curve is best fit with a smoothly broken power-law, however, the break time needs to be fixed, as the post-break decay is defined by only one data point. We find, setting the break conservatively at 1 d, right after the next-to-last detection: $\alpha_{1,2}=0.81\pm0.17$, $\alpha_2=1.75\pm0.34$, $t_b=1$ d fixed, $n=10$ fixed ($\chi^2/d.o.f.=0.02$). Using different data, \cite{Steele2017ApJ} find $\alpha_{1,2}=1.13\pm0.33$, in agreement with our value within errors.

The SED is relatively narrow $ubg^\prime vr^\prime R_Ci^\prime $, mildly red and straight (additional UVOT $uvw2uvm2uvw1$ upper limits from  \citealt{Porterfield2011GCN12203} are in agreement with $z>1$), we find without dust: $\beta=1.21\pm0.12$ ($\chi^2/d.o.f.=1.11$). This value is slightly steeper than the X-ray slope ($\Gamma_x=2.00^{+0.23}_{-0.16}$), and the $u$ band shows there is some curvature. Free fits with dust find negative extinction (MW, LMC dust) or a negative intrinsic spectral slope (SMC dust). Fixing $\beta_{opt}=\Gamma_X-1.0-0.5=0.5$ and using SMC dust, we find $A_V=0.297\pm0.052$ mag ($\chi^2/d.o.f.=0.55$). Using this fit, we find $dRc=-0.638\pm0.096$ mag, and the early afterglow is moderately luminous, having $Rc=15.3$ mag.

\subsubsection{GRB 110731A, $z=2.83$}

Data are taken from \cite{Ackermann2013ApJ}, from GCN Circulars \citep{Malesani2011GCN12220,Tanvir2011GCN12225}, as well as our 2.0 m FTN and FTS data (first reported in \citealt{Bersier2011GCN12216}). The host galaxy is well-detected in deep HST observations \citep{Blanchard2016ApJ}, but not in ground-based follow-up \citep{Lue2017ApJ}. The redshift is given by \cite{Tanvir2019MNRAS}. This GRB represented, after several years of \emph{Fermi} operations, the first Type II GRB to be simultaneously detected by \emph{Swift} and \emph{Fermi}/LAT. It was a temporally quite short, very bright event at a moderately high redshift, and has been extensively discussed in \cite{Ackermann2013ApJ}. The initial light curve is found to be decaying rather steeply ($\alpha=1.440\pm0.011$, $\chi^2/d.o.f.=0.70$), in good agreement with \cite{Ackermann2013ApJ}, who find $\alpha=1.37\pm0.03$ from UVOT data alone. After the end of dense data coverage at 0.052 d, the light curve exhibits a flattening, before going over into another steep decay, a rise, and then a further decay. Data are too sparse to categorize the light curve with fits, but \cite{Ackermann2013ApJ} also show the final epoch (from GROND observations) to be significantly brighter than an extrapolation of the early decay.

The SED is broad ($ubg_Gvr_GR_Ci_GI_Cz_GJ_GH_G$), with $ubg_G$ being affected by Lyman damping, showing a clear dropout. The GROND SED values were derived directly versus $r_G$, and for simplicity, we assume $r_G=R_C$. The SED is moderately red ($\beta=1.20\pm0.14$, $\chi^2/d.o.f.=0.48$ for no dust). The XRT spectral slope is actually bluer ($\beta_X=0.78^{+0.15}_{-0.13}$), indicating reddening by dust. Fixing the intrinsic slope to the X-ray value, we find the best fit for LMC dust, with $A_V=0.23\pm0.08$ mag ($\chi^2/d.o.f.=0.47$). For these values, we derive $dRc=-3.358^{+0.259}_{-0.265}$ mag. At $z=1$, the early afterglow is very luminous at $R_C\approx9.3$ mag.

\subsubsection{GRB 111209A, $z=0.67702\pm0.00005$}

Data are taken from \cite{Stratta2013ApJ,Levan2014ApJ,Kann2018AA}, and from GCN Circulars \citep{Nysewander2011GCN}. The redshift is given by \cite{Kann2018AA}. The afterglow of this famous ultralong GRB (\citealt{Gendre2013ApJ,Levan2014ApJ}, see \citealt{Janiuk2013AA} for a potential model) has been analyzed in detail in \cite{Kann2018AA}. GRB 111209A was followed by the highly luminous, spectrally peculiar SN 2011kl \citep{Greiner2015Nat,Mazzali2016MNRAS,Kann2017AA_SN2011kl}. Concerning the SED, we note \cite{Kann2018AA} found a strong color change in the afterglow during a rebrightening episode, with $\beta=0.63\pm0.25$, $A_V=0.25\pm0.11$ mag measured before 0.09 d, and $\beta=1.05\pm0.06$, $A_V=0.12\pm0.04$ mag after the rebrightening. Recently, \cite{Zafar2018MNRAS} derived $\beta=0.68^{+0.12}_{-0.09}$, $A_V=0.18\pm0.08$ mag, and $R_V=2.53^{+0.13}_{-0.15}$ (the ratio of total to selective extinction, for SMC dust it is $R_V=2.93$.) from the X-shooter spectrum, which was taken at the beginning of the color change. These values agree well with the first SED of \cite{Kann2018AA}.

\subsubsection{GRB 111228A, $z=0.71627\pm0.00002$}

Data are taken from \cite{Xin2016ApJ,Klose2019AA}, and from GCN circulars \citep{Kuroda2011GCN12743,Nysewander2011GCN12751,Hentunen2011GCN12754,Klotz2011GCN12758,Morgan2011GCN12760,Usui2011GCN12768,Guver2011GCN12769,Cenko2011GCN12771,Krushinski2011GCN12789,Pandey2011GCN12792}. The redshift is given by \cite{Selsing2019AA}. This was a well-observed afterglow with a clear late-time SN component \citep{Klose2019AA}. The afterglow shows what is likely a very early flare followed by a rise and a very smooth rollover (see also \cite{Klose2019AA}). Correcting for a small amount of SMC dust, we find $dRc=0.603^{+0.062}_{-0.063}$ mag. The early afterglow is not overly luminous, peaking at $R_C=16.7$ mag.

\subsubsection{GRB 120119A, $z=1.72883$}

Data are taken from \cite{Morgan2014MNRAS,Steele2017ApJ,Blanchard2016ApJ}, from GCN Circulars \citep{Elenin2012GCN12871,Klotz2012GCN12883,Sakamoto2012GCN12894}, as well as our own \emph{Swift} UVOT, REM, and 1.0m LOAO data, first reported in \cite{Chester2012GCN12880}, \cite{Fugazza2012GCN12882}, and \cite{Jang2012GCN12898}, respectively. The redshift is given by \cite{Heintz2018MNRAS}. This GRB was followed-up rapidly by multiple telescopes, yielding early multicolor observations which show a clear color change, the colors becoming bluer. This has been interpreted as a sign of dust destruction, the clearest known case for GRB afterglows \citep{Morgan2014MNRAS}. The GRB itself was bright, and also detected by \emph{Fermi} GBM \citep{Gruber2012GCN12874}.

The light curve shows a very early decay (and possibly a peak at earliest times) which we do not fit as the afterglow evolution here is not achromatic resulting from the rapidly variable line-of-sight extinction. At $\approx150$ s, the decay flattens and turns into a shallow rise, which peaks and goes over into a decay again. With the exception of a few NIR data points at the beginning of the rise, this evolution is achromatic. We fit it with a smoothly broken power-law (\citealt{Morgan2014MNRAS} also present host-galaxy magnitudes in many bands) and find $\alpha_{rise}=-0.391\pm0.019$, $\alpha_1=1.467\pm0.009$, $t_b=0.0120\pm0.0001$ d ($1038\pm11$ s), and $n=1.644\pm0.075$, implying a smooth rollover. This is one of the most precise values of $n$ measured so far. \cite{Morgan2014MNRAS} find a decaying slope $\alpha=1.30\pm0.01$, somewhat shallower than our value, possibly stemming from them fitting only the later data with an SPL and not with a smooth break like we do. \cite{Japelj2015A&A} also report $\alpha\approx1.3$.

The high data density leads to a very precise determination of the fit normalizations and to an SED with small error bars and a wide coverage ($uBg^\prime Vr^\prime R_Ci^\prime I_Cz^\prime YJHK$). The SED is very red, we find $\beta=2.60\pm0.01$ as well as clear curvature and an additional deviation (leading to $\chi^2/d.o.f.=47$) indicative of a dust bump. Indeed, we find the SED is best fit (however, it is still $\chi^2/d.o.f.=12.5$ stemming from the small errors and some minor scatter) by LMC dust with $\beta=0.908\pm0.086$, $A_V=1.114\pm0.056$ mag. This represents one of the most secure detections of line-of-sight extinction toward a GRB afterglow. For MW dust, the bump is too shallow compared to the curvature, which yields a very bad fit ($\beta=2.235\pm0.046$, $A_V=0.258\pm0.030$ mag, $\chi^2/d.o.f.=44$), whereas the SMC fit underestimates the data blueward of the dust bump, we find $\beta=1.255\pm0.078$, $A_V=0.799\pm0.046$ mag, $\chi^2/d.o.f.=17.4$. \cite{Morgan2014MNRAS} present an extensive discussion of the SED, we refer to their paper for further details, but note they state that SMC dust fits best of the standard models. Adding X-ray data, they find $\beta=0.92$, $A_V\approx1.15$ mag for two different LMC extinction curves, in full agreement with our value. \cite{Japelj2015A&A} also find results fully agreeing with ours, namely $\beta=0.89\pm0.01$, $A_V=1.07\pm0.03$ mag, and they strongly favor LMC dust. However, they also state LMC dust over-predicts the strength of the 2175 {\AA} bump and that none of the standard models gives a satisfactory fit. \cite{Zafar2018MNRAS} also present results in general agreement with ours, it is $\beta=0.84^{+0.10}_{-0.09}$, $A_V=1.02\pm0.11$ mag, and $R_V=2.99^{+0.24}_{-0.18}$, very similar to the SMC extinction curve. This agrees with the best fit \cite{Morgan2014MNRAS} find with a full Fitzpatrick-Massa parametrization \citep{Fitzpatrick1986ApJ,Fitzpatrick2007ApJ}, they deduce $R_V=4.11\pm1.03$ combined with $\beta=0.92\pm0.02$, $A_V=1.09\pm0.16$ mag. The large extinction leads to a large correction to $z=1$, it is $dRc=-4.355^{+0.147}_{-0.154}$ mag. We find the early afterglow luminosity, despite the evinced capacity to burn dust, is rather unremarkable, reaching $R_C=12.3$ mag; however, the slow rise to the second peak leads to the late light curve lying in the brighter half of the luminosity distribution. The late luminosity and decay rate are similar to the afterglow of GRB 180325A (see also \cite{Capellazzo2022}).

\subsubsection{GRB 120326A, $z=1.798$}
\label{GRB120326A}

Data are taken from \cite{Urata2014ApJ,Melandri2014AA2,Laskar2015ApJ,Jelinek2016AdAst,Jelinek2019AN}, from GCN circulars \citep{Klotz2012GCN13108,LaCluyze2012GCN13109,Walker2012GCN13112,Zhao2012GCN13122,Sahu2012GCN13185}, and from our own 1.0m LOAO observations (first reported in \citealt{Jang2012GCN13139} and partially published in \citealt{Urata2014ApJ}, we present a rereduction of the $r^\prime$ band using stacked frames that reduces the scatter and brings the data into alignment with that from other sources), as well as late-time GTC HiPERCAM observations that detect a bright, slightly extended host galaxy, in agreement with the final detections from \cite{Laskar2015ApJ}. The redshift is given by \cite{Tello2012GCN13118}. The light curve shows several peculiar properties. As has already been remarked upon by \cite{Jelinek2019AN}, at earliest times, the light curve decays in $R_C/r^\prime$, while it rises in $i^\prime$, causing a spectral reversal ($i^\prime>r^\prime$) which is not typically seen in GRB afterglows. We do not discuss this part further. Frm 0.0109 to 0.202 d, the afterglow shows a decay before smoothly turning over into a moderately steep rise to a large, second peak \citep{Urata2014ApJ,Melandri2014AA2}. For his segment, we find $\alpha_{1,1}=0.80\pm0.16$, $\alpha_{rise}=-0.96\pm0.19$, $t_b=0.041\pm0.011$ d, $n=-1$ fixed ($\chi^2/d.o.f.=0.95$). While unable to let $n$ vary freely, we find a smooth rollover fits the light curve better than a sharp break. From 0.285 d onward, the afterglow goes back over into a standard decay, breaking to an extremely steep one. Data in $r^\prime$ from \cite{Laskar2015ApJ} already point to a transition to a host galaxy, which we fully confirm with our late-time GTC HiPERCAM imaging. We find: $\alpha_{1,2}=0.891\pm0.001$, $\alpha_2=3.686\pm0.037$, $t_b=3.417\pm0.014$ d, $n=2$ fixed ($\chi^2/d.o.f.=17.8$). The very bad quality of fit stems from a combination of small errors and scatter, mostly from the data set of \cite{Urata2014ApJ}, even after replacing their $r^\prime$ data with our own rereduction. We note that $\alpha_{1,1}=\alpha_{1,2}$ within errors, typical behavior for an energy injection. The steepness of $\alpha_2$ is dependent on the break sharpness value $n$, we found that fixing it to 2 improves $\chi^2/d.o.f.$ by 5. Allowing it to vary freely leads to a degenerate solution where $n$ continues to become softer while $\alpha_2$ increases even further.

The SED, based on the second fit, is very broad ($uvw2 uvm2 uvw1 u b g^\prime v r^\prime R_C i^\prime z^\prime Y J H K_S$), has small errors and shows some scatter. The UVOT UV and $u$ filters are affected by Lyman damping and are not included. The $b$ band is also anomalously bright and is excluded. Finally, $JHK_S$ data is brighter than a redward extrapolation of the optical data, however, the errors are significantly larger so they are not that strong outliers. Fits with all dust models return negative extinction, and we prefer the fit with no dust: $\beta=0.950\pm0.016$ ($\chi^2/d.o.f.=17.2$ resulting from small errors and the above-mentioned offset of NIR data).

Results from the literature for this GRB vary widely, in part because these publications only cover partial data sets. \cite{Urata2014ApJ} study the late rebrightening. They find $\alpha_{1,2}=0.88...1.01$ in different bands, in full agreement with our value, and $\alpha_2=2.48...2.84$, which is less steep than our result, likely as they do not take the bright host galaxy into account. They derive three different SEDs and find a range of $\beta=1.11...1.44$, somewhat steeper than our value. They estimate $A_V=0.3...0.5$ mag via a joint X-ray-to-optical fit.
\cite{Melandri2014AA2} cover the early light curve and the first days of the rebrightening, but have little data showing the late, very steep decay. For the first part of the light curve, they find  $\alpha_{1,1}=0.50\pm0.05$, $\alpha_{rise}=1.53\pm0.18$, $\alpha_{1,2}=1.77\pm0.11$. This is shallower, steeper, and also steeper than our result. The steeper late-time decay likely results from them fitting additional data beyond the break \cite{Urata2014ApJ} and we find. Spectrally, they perform a X-ray-to-optical fit and find a spectral slope very similar to ours, $\beta=0.88\pm0.03$, but with a very high SMC dust column, $A_V=1.1\pm0.3$ mag. Finally, \cite{Laskar2015ApJ} take the data sets of the other authors into account. For the rebrightening, they find $\alpha_{rise}=-0.52\pm0.06$, $\alpha_{1,2}=1.10\pm0.10$, the first value being shallower than our result and the second being slightly steeper. The post-break decay slope they find is also shallower than our result, with $\alpha_2=2.05\pm0.13$, it is not made clear if the host-galaxy contribution was taken into account (it clearly is in their figure 4, top right). They derive an SED (their figure 3) that looks similar to our own, with a flatter optical part and overly bright NIR emission. Using a joint X-ray-to-optical fit, they find an optical-only slope of $\beta_{opt}=1.80\pm0.16$ and in combination with the X-ray slope $\beta_{X}=0.85\pm0.04$ an extinction of $A_V=0.40\pm0.01$ mag. For the no-dust fit, we find $dRc=-1.571\pm0.006$ mag. At $z=1$, the light curve is only moderately luminous, peaking at $R_C\approx16.3$ mag, however, this is at $\approx0.2$ d. The early afterglow remains at $R_C>17$ mag.

\subsubsection{GRB 120327A, $z=2.81482$}

Data are taken from \cite{DElia2014AA,Steele2017ApJ,Melandri2017AA}, and from GCN circulars \citep{Sudilovsky2012GCN13129,Klotz2012GCN13132}. The redshift is given by \cite{Heintz2018MNRAS}. The afterglow shows a classic rise and smooth decay, which transitions into a shalower decay and then a plateau phase and a subsequent decay. Fitting data up to $\approx0.064$ d, we find: $\alpha_{rise}=-1.70\pm0.17$, $\alpha_1=1.3660\pm0.0096$, $t_b=0.00480\pm0.00013$ d ($t_b=414\pm11$ s), $n=1$ fixed ($\chi^2/d.o.f.=1.75$). Fitting data between the peak and 0.18 d, we find $\alpha_1=1.2959\pm0.0.0041$, $\alpha_{shallow}=0.907\pm0.049$, $t_b=0.0876\pm0.0061$ d, $n=-17.1\pm7.71$ ($\chi^2/d.o.f.=2.43$). Fitting the shallow(er) and plateau phase, we find: $\alpha_{shallow}=1.087\pm0.015$, $\alpha_{plateau}=0.089\pm0.068$, $t_b=0.174\pm0.005$ d, $n=-10$ fixed ($\chi^2/d.o.f.=2.42$), Finally, in $r^\prime i^\prime$ we can fit the transition from plateau to subsequent decay: $\alpha_{plateau}=-0.11\pm0.14$, $\alpha_2=1.607\pm0.034$, $t_b=0.248\pm0.006$ d, $n=10$ fixed ($\chi^2/d.o.f.=2.61$). \cite{Steele2017ApJ} find $\alpha_1=1.22\pm0.02$. \cite{Melandri2017AA} find a similar value $\alpha\approx1.2$. Both are close to our result, likely slightly shallower as the other authors fit the initial decay together with the shallower following one.

We derive three SEDs from our first three fits (the final one uses only two bands), with the first one being the broadest ($uvw1 u b g^\prime v R_C I_C z_G J_G H_G K_G$). $uvw1 u b g^\prime$ data are affected by Lyman damping and are excluded. The SEDs show slight curvature. We find no significant evidence for color change and fit all three SEDs simultaneously. Without dust, we find: $\beta=1.031\pm0.025$ ($\chi^2/d.o.f.=2.28$). MW and LMC dust yield negative extinction, but we find an excellent fit for SMC dust: $\beta=0.656\pm0.090$, $A_V=0.140\pm0.032$ mag ($\chi^2/d.o.f.=1.07$). \cite{Melandri2017AA} find evidence for slight color change, modelling several SEDs either with constant slope and variable extinction, or vice versa. In the first case, they derive $\beta_{opt}=0.55^{+0.05}_{-0.04}$ and $A_V<0.05$ to $A_V=0.06^{+0.04}_{-0.03}$ mag. In the other case, $A_V=0.05\pm0.02$ mag, and the slope spanning from $\beta=0.72^{+0.11}_{-0.09}$ to $\beta=0.81^{+0.08}_{-0.02}$. These values are in decent agreement with our result. Using our SMC fit, we derive $dRc=-3.078^{+0.139}_{-0.136}$ mag. The early afterglow at $z=1$ is luminous, peaking at $R_C\approx11.2$ mag.

\subsubsection{GRB 120404A, $z=2.8767$}

Data are taken from \cite{Guidorzi2014MNRAS,Ershova2020ARep}, from GCN circulars \citep{Xin2012GCN13221,Tristram2012GCN13228,Volnova2012GCN13235,Volnova2012GCN13236}, as well as from our own \emph{Swift} UVOT analysis (first reported in \citealt{Breeveld2012GCN13226}). The redshift is given by \cite{Guidorzi2014MNRAS}. The afterglow shows an initial shallow decay and a plateau phase which goes over into a long rise, smooth turnover and decay. Fitting data starting at 0.016 d, we find: $\alpha_{rise}=-2.89\pm0.11$, $\alpha_1=1.688\pm0.023$, $t_b=0.02257\pm0.00049$ d, $n=0.5$ fixed and no host galaxy  ($\chi^2/d.o.f.=0.78$). The final detection in \cite[][, from GROND]{Guidorzi2014MNRAS} lies $7\sigma$ under the extrapolation of the earlier decay, indicating another break, likely a jet break, has taken place. \cite{Guidorzi2014MNRAS} state the rising slope has a large uncertainty but do not give a value, they measure $\alpha_1=1.9\pm0.1$, and find a peak time $t_p=0.0278\pm0.0069$ d. \cite{Laskar2015ApJ} find $\alpha_{rise}=-1.74\pm0.58$, $\alpha_1=1.71\pm0.17$ ($B$ band, for the $R_C$ band they find $\alpha_1=1.90\pm0.02$), $t_b=0.028\pm0.004$ d, $n=0.78\pm0.43$. These values are generally in agreement with our own.

The SED is very broad ($uvw2 uvm2 uvw1 u B g_G V R_C i^\prime I_C z_G$ $J_G H_G$), with $uvw2 uvm2$ being (deep) upper limits only, $ uvw1 u$ are also strongly affected by Lyman damping and is not included in the fitting. The rest of the SED is moderately red and nearly straight, $J_GH_G$ lie a bit above the redward extrapolation of the optical data. Without dust, we find $\beta=1.112\pm0.041$ ($\chi^2/d.o.f.=3.73$). An SMC dust fit finds negative extinction, and an LMC dust fit yields a fit result that is worse than the fit without dust. However, we find a slight improvement using MW dust, finding: $\beta=1.040\pm0.053$, $A_V=0.064\pm0.029$ mag ($\chi^2/d.o.f.=3.55$). Using an optical-to-X-ray fit and MW, \cite{Guidorzi2014MNRAS} find their best fit results in: $\beta=1.01\pm0.03$, $A_V=0.24\pm0.07$ mag. That is, a very similar spectral slope and a preference for MW dust, but a higher extinction value. \cite{Laskar2015ApJ} find an unextinguished spectral slope $\beta=1.3\pm0.2$ (identical to our result within errors), and derive $A_V=0.13\pm0.01$ mag from a broadband analysis; no dust model is given (close to our result). Using our MW dust result, we derive $dRc=-3.036^{+0.057}_{-0.056}$ mag. At $z=1$, the afterglow is luminous, peaking at $R_C=13.7$ mag. However, the peak is not until $\approx0.015$ d. The early-time afterglow is less luminous, at $R_C\approx15$ mag.

\subsubsection{GRB 120711A, $z=1.405$}

Data are taken from \cite{Martin-Carrillo2014AA}, from GCN circulars \citep{Levato2012GCN13443,Breeveld2012GCN13448}, from unpublished late-time GROND observations, and from our own rereduction of the 0.6m REM/REMIR data, as well as an upper limit from FRAM/PAO. The redshift is given by \cite{Tanvir2012GCN13441}. The REM reanalysis reveals a peculiarity we have not been able to successfully solve. Aside from adding as-yet-unpublished ``late-time'' ($0.06-0.1$ d) $JHK$ observations, the early observations are not in agreement with those published in \cite{Martin-Carrillo2014AA}. The times are earlier and the afterglow exhibits an extreme flare reaching $H\approx8.2$ mag in the second image, exactly at the time the dense optical observations also peak, though they do not exhibit such a strong flaring behavior. The subsequent decay is flatter than that of the data \cite{Martin-Carrillo2014AA} present. A constant shift in time is able to align most of the data but would then move the flare away from the optical peak. The early data in our rereduction is sampled more finely in time which may explain why the flare is not seen in the reduction \cite{Martin-Carrillo2014AA} present.

\subsubsection{GRB 120714B, $z=0.3984$}

Data are taken from \cite{Klose2019AA} as well as from the automatic UVOT analysis page\footnote{\url{https://swift.gsfc.nasa.gov/uvot_tdrss/526642/index.html}, however, this page is not available at time of writing.}. The redshift is given by \cite{Selsing2019AA}. This was a faint GRB at moderately low redshift, with relatively sparse early follow-up, however, the SN phase is well-covered by GROND \citep{Klose2019AA}. The full analysis is given in \cite{Klose2019AA}. We follow \cite{Klose2019AA} and adopt $\beta=0.7\pm0.4$ and no dust extinction, leading to $dRc=2.331^{+0.155}_{-0.158}$ mag. The $g^\prime$-band light curve of this afterglow is unaffected by the accompanying SN 2012eb, and allows follow-up up to $\approx140$ d in the $z=1$ frame. The afterglow is one of the faintest in the entire sample, being $R_C=20.7$ mag (at $z=1$) already at very early times.

\subsubsection{GRB 120729A, $z=0.80$}

Data are taken from \cite{Cano2014AA,Huang2018ApJ}, and from GCN circulars \citep{Wren2012GCN13545,Wang2015GCN18184}. The redshift is given by \cite{Tanvir2012GCN13532}. This was a bright (initial detection $R_C=12.8$ mag correcting for Galactic extinction) afterglow which decayed steeply and showed late-time evidence for a SN bump \cite{Cano2014AA} (see also \cite{Cano2014AA}). Correcting for a small amount of SMC extinction, we find $dRc=0.269^{+0.083}_{-0.084}$ mag and an early afterglow peak magnitude $R_C=13.1$ mag.

\subsubsection{GRB 120815A, $z=2.35820$}

Data are taken from \cite{Kruhler2013AA}, as well as our own analysis of \emph{Swift} UVOT data (first reported in \citealt{Holland2012GCN13666}), and VLT data (X-shooter finding charts, first reported in \citealt{Malesani2012GCN13649}). The redshift is given by \cite{Heintz2018MNRAS}. We note that we added the recommended systematic errors from \cite{Kruhler2013AA} to the data. The light curve shows an initial, shallow rise, which goes over into a shallow decay which steepens mildly in the final data points. We find $\alpha_{rise}=-0.204\pm0.033$, $\alpha_1=0.544\pm0.009$, $t_{b,1}=0.00595\pm0.00023$ d ($t_{b,1}=514.1\pm19.9$ s) ($\chi^2/d.o.f.=0.33$); and $\alpha_1=0.515\pm0.017$, $\alpha_2=0.871\pm0.035$, $t_{b,2}=0.0528\pm0.0066$ d ($\chi^2/d.o.f.=0.43$). This compares well with the results from \cite{Kruhler2013AA}, who find $\alpha_{rise}=-0.18\pm0.02$, $\alpha_1=0.52\pm0.01$, $t_{b,1}=440\pm30$ s, $\alpha_2=0.86\pm0.03$, $t_{b,2}=4300^{+900}_{-600}$ s ($t_{b,2}=0.050^{+0.010}_{-0.007}$ d).

The SED is broad ($ubg_Gvr_GR_Ci_Gz_GJ_GH_GK_G$), with $u$ being affected by Lyman damping and not included. The SED is moderately red and clearly curved, without dust we find $\beta=1.220\pm0.053$ ($\chi^2/d.o.f.=3.86$). A fit with MW dust yields negative extinction and is ruled out. A fit with LMC dust yields an improved result ($\chi^2/d.o.f.=1.60$) but with a negative intrinsic spectral slope, and the $i_G$ detection lies within the 2175 {\AA} bump region and shows there is no bump, so we rule this dust model out as well. The best fit is found with SMC dust, we derive $\beta=0.196\pm0.209$, $A_V=0.423\pm0.085$ mag. This is in agreement with the result from \cite{Japelj2015A&A}, who derive $\beta=0.38^{+0.07}_{-0.05}$, $A_V=0.32\pm0.02$ mag, also using SMC dust. \cite{Kruhler2013AA} also prefer SMC dust and state they find no evidence for a dust bump, especially not from their X-shooter spectrum. \cite{Kruhler2013AA} note they do not fit the $g_G$ data as the filter extends into the Lyman-$\alpha$ region, but we note the fit results are identical whether $bg_G$ are included or not (except for the errors which become larger if the two filters are left out of the fit). \cite{Kruhler2013AA} perform an XRT to optical fit and find no significant evidence for a cooling break between the two regimes, and $\beta=0.78\pm0.01$, $A_V=0.15\pm0.02$ mag. Using our SED, leaving out $bg_G$ and fixing $\beta=0.78$, we derive $A_V=0.13\pm0.02$ mag, confirming the result from \cite{Kruhler2013AA}. Finally, \cite{Zafar2018MNRAS}, using a Fitzpatrick-Massa parametrization, derive and even redder intrinsic afterglow but with somewhat higher extinction compared to \cite{Kruhler2013AA}: $\beta=0.92\pm0.10$, $A_V=0.19\pm0.04$ mag, and $R_V=2.38\pm0.09$. Using our SMC result, we find $dRc=-3.260^{+0.312}_{-0.301}$ mag. The afterglow is moderately luminous, peaking at $R_C\approx14.2$ mag.

\subsubsection{GRB 121024A, $z=2.30244$}

Data are taken from \cite{Wiersema2014Nature,Friis2015MNRAS,Varela2016AA,Jelinek2016AdAst}, from GCN circulars \citep{Klotz2012GCN13887,Zhao2012GCN13898,Pandey2012GCN13904,Cobb2012GCN13928}, and from our own \emph{Swift} UVOT data analysis (first reported in \citealt{Holland2012GCN13901}). The redshift is given by \cite{Heintz2018MNRAS}. This GRB afterglow showed evidence for circular polarization \citep{Wiersema2014Nature}. The afterglow potentially shows a bright, rapidly decaying flash at very early times \citep{Klotz2012GCN13887}. Thereafter, two rebrightenings/plateau phases are seen in the data presented by \cite{Friis2015MNRAS}. We begin our fit after the second plateau, at 0.0148 d. We find the afterglow can be modeled with a smoothly broken power-law, we measure: $\alpha_1=0.939\pm0.014$, $\alpha_2=1.72\pm0.12$, $t_b=0.880\pm0.077$ d, $n=10$ fixed, and the host-galaxy magnitudes given by individual late-time measurements especially from \cite{Friis2015MNRAS} ($\chi^2/d.o.f.=1.41$). \citep{Wiersema2014Nature} find $\alpha_1=0.93\pm0.02$, $\alpha_2=1.25\pm0.04$, $t_b=0.4306\pm0.0081$ d, $n=5.01\pm0.01$. \cite{Varela2016AA} find $\alpha_1=0.71\pm0.03$, $\alpha_2=1.46\pm0.04$, $t_b=0.363\pm0.109$ d, $n=2.7\pm1.1$. Our results are roughly in agreement, however, we find a steeper late-time decay and later break time.

The SED is broad ($u B g_G v r_G R_C i_G I_C z_G J_G H_G K_G$). The $u$ band is definitely Lyman-damped and is excluded. \cite{Friis2015MNRAS,Varela2016AA} comment that $g_G$ is affected by the Lyman$\alpha$ DLA as well and exclude it from their fits. However, we find that an SMC dust fit is essentially identical whether $Bg_G$ are included or not, except that exclusion leads to a significantly larger error. We therefore choose to include these two filters in the SED fit. The SED is moderately red and clearly curved, without dust we find: $\beta=1.446\pm0.060$ ($\chi^2/d.o.f.=2.36$). We find negative extinction for MW dust, LMC dust yields a viable fit ($\chi^2/d.o.f.=1.30$), but with high extinction, a very blue intrinsic SED, and no significant evidence for a 2175 {\AA} bump. We therefore prefer the SMC dust fit, for which we find $\beta=0.52\pm0.46$, $A_V=0.39\pm0.24$ mag ($\chi^2/d.o.f.=0.58$). \citep{Wiersema2014Nature} derive $\beta=0.88\pm0.01$, $A_V=0.22\pm0.02$ mag from a joint X-ray-to-optical fit. \cite{Friis2015MNRAS} find $\beta=0.90\pm0.02$, $A_V=0.088\pm0.059$ mag from a joint X-ray-to-optical fit, using SMC dust. Also using a multi-epoch joint X-ray-to-optical fit (no evidence for color evolution is found), \cite{Varela2016AA} find $\beta=0.86\pm0.02$, $A_V=0.18\pm0.04$ mag. \cite{Zafar2018MNRAS} study the X-shooter spectrum and derive: $\beta=0.85^{+0.09}_{-0.13}$, $A_V=0.26\pm0.07$ mag, $R_V=2.81^{+0.20}_{-0.16}$. We therefore find a bluer intrinsic SED slope, in agreement with a potential cooling break between optical and X-rays. Our extinction value is consequently higher, but only in disagreement with the value \cite{Friis2015MNRAS} derive. Using our SMC fit, we derive $dRc=-3.192^{+0.308}_{-0.312}$ mag, and find a very luminous early afterglow, $R_C\approx9.9$ mag.

\subsubsection{GRB 130427A, $z=0.3399\pm0.0002$}

Data are taken from \cite{Laskar2013ApJ,Xu2013ApJ,Vestrand2014Science,Maselli2014Science,Perley2014ApJ,Melandri2014AA,Coward2017PASA,Becerra2017ApJ,Becerra2017ApJErr}, and from GCN circulars \citep{Morgan2013GCN14453,Yatsu2013GCN14454,Im2013GCN14464,Kuroda2013GCN14465,Wiggins2013GCN14490,Takahashi2013GCN14495,Keel2013GCN14507,Norris2013GCN14511,Kuroda2013GCN14513,vandeStadt2013GCN14521,Flewelling2013GCN14538,Hermansson2013GCN14596}. The redshift is given by \cite{Selsing2019AA}. GRB 130427A was a once-per-several-decades event, the brightest GRB since 1984, however, placing it in contest, it is just a close-by example of the kind of GRBs which are usually detected at $z>1$, earning it the moniker ``a nearby, ordinary monster'' \citep{Maselli2014Science}. It yielded very detailed high-energy observations including the longest-lasting LAT detection to date \citep{Ackermann2014Science,Preece2014Science}. The GRB was associated with SN 2013cq \citep{deUgartePostigo2013GCN14646,Xu2013ApJ,Melandri2014AA}. Its X-ray afterglow was detected for years and shown to not break, implying a wide jet-opening angle and extreme energetics \citep{DePasquale2016MNRAS2}. For this work, it is relevant that we derive $dRc=+2.73$ mag. The extremely bright early prompt flash \citep{Vestrand2014Science} peaks at $R_C\approx9.6$ mag.

\subsubsection{GRB 130831A, $z=0.4791$}

Data are taken from \cite{Cano2014AA,DePasquale2016MNRAS,Klose2019AA}, and from GCN Circulars \citep{Xu2013GCN15142,Xin2013GCN15146,Leonini2013GCN15150,Sonbas2013GCN15161,Khorunzhev2013GCN15244,Gorbovskoy2015GCN18673}. The redshift is given by \cite{Cucchiara2013GCN15144}. This was a low-redshift GRB with extensive follow-up and a late-time SN with spectroscopic confirmation, SN 2013fu \citep{Cano2014AA,Klose2019AA}. The observed afterglow is bright, peaking at $R_C=13.3$ mag. No dust is found, and we find $dRc=1.968^{+0.016}_{-0.017}$ mag, leading to a peak magnitude at $z=1$ of $R_C=15.3$ mag.

\subsubsection{GRB 131030A, $z=1.296$}

Data are taken from \cite{King2014MNRAS,Huang2017PASJ}, from GCN circulars \citep{Xu2013GCN15407,Terron2013GCN15411,Moskvitin2013GCN15412,Hentunen2013GCN15418,Perley2013GCN15423,Tanigawa2013GCN15481,Pandey2013GCN15501}, as well as our own extensive data set, consisting of \emph{Swift} UVOT data (first reported in \citealt{Breeveld2013GCN15414}), very early Ond\v rejov D50 and BART (NF+WF) data (first reported in \citealt{Strobl2013GCN15410}), Watcher data (first reported in \citealt{Kubanek2013GCN15403}), LT data (first reported in \citealt{Virgili2013GCN15406}), LOAO telescope data (first reported by \citealt{ImnChoi2013} and presented but not tabulated in \citealt{Huang2017PASJ}), and late-time GTC HiPERCAM observations which detect a very faint host galaxy (in full agreement with the Subaru detection reported by \citealt{Huang2017PASJ}). The redshift is given by \cite{Selsing2019AA}. This was a very extensively observed afterglow, over 1400 data points have been published so far, and it is the largest data set we present here, with 260 data points, including very early observations from multiple telescopes all the way to very deep host-galaxy observations. It was the first GRB afterglow to be observed by ALMA in open-use \citep{Huang2017PASJ}, and the early afterglow was polarised \citep{King2014MNRAS}.

The afterglow shows classic behavior, with an initial rise which rolls over into a smooth decay, likely the forward-shock rise (initially noted by \citealt{Strobl2013GCN15410}). We find $\alpha_{rise}=-0.54\pm0.11$, $\alpha_1=0.9745\pm0.0036$, $t_{b,peak}=0.00131\pm0.00005$ d ($t_{b,peak}=113.18\pm4.32$ s). The high data density allowed us to leave the break smoothness as a free parameter, $n=3.29\pm0.56$ ($\chi^2/d.o.f.=4.46$ stemming from some scatter and small error bars). At $\approx0.04$ d, the light curve flattens, before completely flattening at $\approx0.095$ d. This is followed by a data gap, the afterglow is decaying again normally when observations resume at $\approx0.2$ d. We fit the afterglow from this point on with a broken power-law and find $\alpha_{1,2}=0.9315\pm0.0067$, $\alpha_2=1.474\pm0.08$, $t_b=1.15\pm0.04$ d, $n=10$ fixed, and using our HiPERCAM host-galaxy magnitudes (we neglected the host galaxy in the NIR as it is very blue). The similarity of $\alpha_1$ and $\alpha_1,2$ supports the rebrightening being an energy injection. \cite{King2014MNRAS} find an early decay slope $\alpha_1=0.78\pm0.02$, flatter than what we find. \cite{Huang2017PASJ}, using later data, find $\alpha_1\approx0.84\pm0.04$, $\alpha_2\approx2.05\pm0.16$, $t_b=2.91\pm0.56$ d. This is a later break and steeper post-break decay than our result. Nonetheless, all evidence points to this being a jet break.

We find no conclusive evidence for color change and fit the SED derived from the early fit ($uvm2uvw1ubvR_C$) and the late fit ($uvw2uvm2uvw1ubg^\prime vr^\prime R_Ci^\prime I_Cz^\prime YJH$) together. UV data was affected by Lyman damping and was not included. A fit without dust finds $\beta=0.838\pm0.035$, indicating dust extinction is likely low. MW dust find negative extinction and is ruled out. A viable solution is found for LMC dust ($\beta=0.24\pm0.16$, $A_V=0.38\pm0.10$ mag, $\chi^2/d.o.f.=3.54$), there is no clear evidence for a 2175 {\AA} bump, however. The best solution is found for SMC dust: $\beta=0.36\pm0.11$, $A_V=0.26\pm0.05$ mag, $\chi^2/d.o.f.=2.83$. \cite{Huang2017PASJ} find no evidence for host-galaxy extinction and $2\sigma$ evidence for a slope change, an early SED showing $\beta=0.57\pm0.09$, while a later one results in $\beta=1.08\pm0.23$. Using our SMC solution, we find $dRc=-1.147^{+0.113}_{-0.115}$ mag. The early afterglow is bright, peaking at $R_C\approx12.9$ mag.

\subsubsection{GRB 140311A, $z=4.954$}

Data are taken from \cite{Littlejohns2015MNRAS,Laskar2018ApJ2}, and from GCN Circulars \citep{Klotz2014GCN15952,D'Avanzo2014GCN15953,Xu2014GCN15956,Yoshida2014GCN15957,Malesani2014GCN15980}. The redshift is given by \cite{Selsing2019AA}. Observations of this event at moderately high redshift are sparse. The afterglows shows an early rise ($\alpha_{rise}=-1.05\pm0.32$, $\chi^2/d.o.f.=0.34$, in full agreement with \citealt{Laskar2018ApJ2}) later followed by a decay with no clear evidence for a further break ($\alpha_1=0.771\pm0.023$, $\chi^2/d.o.f.=1.35$). \cite{Laskar2018ApJ2} find a flatter decay and a break, however, they start their fit from the earliest detection, excluding the following points as a superposed flare.

The SED ($R_Ci^\prime z^\prime Y$) is extremely red, with $R_C$ showing additional evidence for being Lyman-damped. We find $\beta=4.40\pm0.21$ ($\chi^2/d.o.f.=0.02$), fully in agreement with \cite{Laskar2018ApJ2}. It is, however, in strong contrast to \cite{Littlejohns2015MNRAS}, who report $\beta_{opt}\approx0.68$ -- only to later report $A_{V,SMC}=0.45^{+0.08}_{-0.06}$ mag. We fix the spectral slope to an intrinsic value derived from X-rays ($\beta_X=0.584^{+0.200}_{-0.099}$ given on the \emph{Swift} XRT repository) and find that MW dust strongly underestimates the $R_C$ detection. LMC dust ($A_V=1.01\pm0.06$ mag, $\chi^2/d.o.f.=1.40$) and SMC dust ($A_V=0.680\pm0.038$ mag, $\chi^2/d.o.f.=0.30$) yield viable solutions, we prefer the latter, as it needs lower extinction and SMC dust is more typical for GRB host galaxies (the 2175 {\AA} bump would need redder NIR data to be confirmed or rejected). This result is the same order of magnitude as what \cite{Littlejohns2015MNRAS} report. This represents one of the strongest cases for extinction along the line of sight to a GRB at high redshift, and it is therefore an outlier to the findings of \cite{Bolmer2018AA}. The large extinction combined with the high redshift leads to a very large correction $dRc=-8.237^{+0.250}_{-0.251}$ mag. The early flare, after this correction, is very luminous, peaking at $R_C\approx10$ mag.

\subsubsection{GRB 140419A, $z=3.956$}

Data are taken from \cite{Littlejohns2015MNRAS}, from GCN Circulars \citep{Guver2014GCN16120,Cenko2014GCN16129,Kuroda2014GCN16131,Kuroda2014GCN16132,Pandey2014GCN16133,Zheng2014GCN16137,Xu2014GCN16140,Volnova2014GCN16141,Volnova2014GCN16168,Volnova2014GCN16281}, and from our own \emph{Swift} UVOT observations (first reported in \citealt{Kuin2014GCN16130}), as well as our LOAO observations (first reported in \citealt{Choi2014GCN16149}). The redshift is given by \cite{Cucchiara2015ApJ}. This was an intense GRB at moderately high redshift with a very bright early afterglow peaking at $R_C=12.5$ mag. The initial afterglow decay is densely covered. It decays with a straight power-law, $\alpha_1=1.185\pm0.001$ ($\chi^2/d.o.f.=5.19$ stemming from some scatter and very small errors). At $\approx0.07$ d, the decay flattens. Following a data gap, from 0.36 d, it is seen to have resumed its former decay slope, indicating an energy injection. There is no significant evidence for a beak until the last observation at $\approx6.5$ d.

The SED ($ubg^\prime VR_Ci^\prime I_Cz^\prime YJH$) is cleary affected by the redshift of the GRB, showing Lyman-damping all the way into the $R_C$ band. The redder bands show no evidence for curvature, a fit without dust yields $\beta=0.76\pm0.08$. MW and LMC dust lead to negative extinction. SMC dust yields an intrinsically very blue slope whereas the small amount of extinction derived is still 0 within errors, while $\chi^2/d.o.f.$ is higher than for the fit without dust. \cite{Littlejohns2015MNRAS} find only a small amount of dust for the SMC model as well, $A_{V,SMC}=0.11^{+0.06}_{-0.05}$ mag. For the dustless model, we find a correction $dRc=-3.436\pm0.079$ mag. At $z=1$, this is among the most luminous early afterglows every detected, peaking at $R_C\approx9$ mag.

\subsubsection{GRB 140423A, $z=3.26$}

Data are taken from \cite{Littlejohns2015MNRAS,Li2020ApJ}, from GCN Circulars \citep{Ferrante2014GCN16145,Cenko2014GCN16153,Kuroda2014GCN16160,Akitaya2014GCN16163,Pandey2014GCN16164,Harbeck2014GCN16165,Harbeck2014GCN16175,D'Avanzo2014GCN16166,Takahashi2014GCN16167,Cano2014GCN16169,Fujiwara2014GCN16173,Bikmaev2014GCN16185,Sahu2014GCN16272,Sonbas2014GCNR470}, and from our own \emph{Swift} UVOT observations (first reported in \citealt{Chester2014GCN16147}). The redshift is given by \cite{Cucchiara2015ApJ}. This was another bright GRB afterglow at moderately high redshift. The earliest detection \citep{Ferrante2014GCN16145} indicates a steeply decaying flare, which then goes over into a shallower rise and rollover into a typical decay. The rollover shows some substructure, including a short plateau phase from 0.0063 d to 0.0079 d. The beginning of the rise is quite steep ($\alpha_{rise,spl}=-1.71\pm0.10$, $\chi^2/d.o.f.=1.00$), steeper than what is found by fitting the rollover with a smoothly broken power-law: $\alpha_{rise,bpl}=-1.074\pm0.027$, $\alpha_{steep}=1.720\pm0.010$, $t_{b,peak}=0.00352\pm0.00004$ d, $n=1$ fixed ($t_{b,peak}=304.13\pm3.46$ s). This fit, using data until $\approx0.06$ d, has a high $\chi^2/d.o.f.=11.9$ as it does not take the short plateau mentioned above into account. The afterglow decay flattens after this point in time, fitting from after the short plateau to $\approx2$ d yields $\alpha_{steep}=1.849\pm0.017$, $\alpha_1=1.017\pm0.008$, $t_{b,flattening}=0.0385\pm0.0012$ d, $n=-10$ fixed ($\chi^2/d.o.f.=1.53$). This indicates the true decay post-peak is even steeper than the rollover fit indicated. The results indicate this is potentially a reverse- to forward-shock transition -- \cite{Li2020ApJ} interpret it as a transition from a wind to a constant density circumburst medium. The final points indicate a renewed steepening, we find $\alpha_2=2.09\pm0.87$, $n=10$ fixed at $t_b=2.70\pm0.76$ d ($\chi^2/d.o.f.=1.75$). This is probably a jet break, but the lack of further observations prevents us from drawing firmer conclusions. \cite{Li2020ApJ} find a flatter initial rise ($\alpha_{O,I}=-0.59\pm0.04$) but similar values for the steep and shallow decays ($\alpha_{O,II}=1.78\pm0.03$, $\alpha_{O,III}=1.13\pm0.03$), respectively).

The SED shows strong influence from Lyman damping. The afterglow is not detected to deep limits in the UVOT UV and $u$ filters \citep{Sonbas2014GCNR470}, and the $b$ band is also suppressed. The rest of the SED ($g^\prime Vr^\prime R_Ci^\prime I_Cz^\prime YJHK_S$) shows some scatter and is generally flat, without dust we find $\beta=0.75\pm0.06$. \cite{Li2020ApJ}, using only two optical filters, find an X-ray-to-optical spectral slope $\Gamma-1=0.95\pm0.05$, and quite large extinction (using SMC dust) of $A_V=0.54\pm0.16$ mag. Our broader SED indicates the optical SED is flatter than the X-ray SED, indicating the existence of a cooling break between the bands, we therefore fix the intrinsic slope to $\beta=0.55\pm0.08$. We find the best solution for a small amount of LMC dust, with $A_V=0.13\pm0.03$ mag ($\chi^2/d.o.f.=2.56$). A similar low amount of SMC dust cannot be ruled out either ($A_V=0.08\pm0.02$ mag, $\chi^2/d.o.f.=2.66$). However, a small depression in the $z^\prime$ flux is in agreement with a 2175 {\AA} bump at this redshift, making us prefer the LMC fit. Using it, we derive $dRc=-3.249^{+0.139}_{-0.135}$ mag. The very early flash is found to have $R_C=10.1$ mag, with the later peak lying at $R_C=10.9$ mag, making this a very luminous afterglow.

\subsubsection{GRB 140430A, $z=1.601$}

Data are taken from \cite{Kopac2015ApJ}, from GCN Circulars \citep{Malesani2014GCN16193,Kennedy2014GCN16196,Kennedy2014GCN16201}, and from our own \emph{Swift} UVOT observations (first reported in \citealt{Breeveld2014GCN16198}), as well as a single datum from the Ond\v rejov D50. The redshift is given by \cite{Selsing2019AA}. The initial optical light curve (two points) shows a shallow decay, but then the afterglow must fade very rapidly, as it is steeply rising when denser monitoring sets in shortly afterwards. From the rise, the afterglow breaks sharply to a standard decay, we find $\alpha_{rise}=-5.05\pm1.01$, $\alpha_1=0.812\pm0.015$, $t_b=0.00182\pm0.00003$ d ($t_b=157.25\pm2.59$ s), $n=10$ fixed  ($\chi^2/d.o.f.=1.22$). The afterglow then goes over into a plateau phase before resuming decaying again, we find $\alpha_{plateau}=0.237\pm0.035$, $\alpha_2=1.03\pm0.06$, $t_b=0.094\pm0.0015$ d, $n=3$ fixed ($\chi^2/d.o.f.=1.74$). There is no evidence for a further break, however, observations extend only to 1.3 d. \cite{Kopac2015ApJ} differentiate two optical flares in the early data, finding an initial rise in perfect agreement with our results. They find later decay slopes of $\alpha_1\approx1$ and $\alpha_1\approx0.8$, respectively, a reverse of our result, however, no error bars are given so their result generally agrees with ours. 

The SED ($uvw2uvm2uvw1ubVR_CI_C$) is strongly affected by Lyman damping, with the UV filters clearly suppressed. The slope is straight and moderately red, $\beta=1.05\pm0.08$ ($\chi^2/d.o.f.=0.69$). All dust models yield viable results; however, the measured extinction is zero within errors for each, and therefore we use the fit without dust. \cite{Kopac2015ApJ} find a similar slope without extinction, $\beta=0.97\pm0.08$, and a joint X-ray-to-optical fit with SMC dust also only yields an upper limit $A_V<0.14$ mag. For the fit without extinction, we derive $dRc=-1.288\pm0.022$ mag. At $z=1$, the early afterglow is moderately luminous with $R_C=14.5$ mag.

\subsubsection{GRB 140506A, $z=0.88911$}

Data are taken from \cite{Fynbo2014AA,Heintz2017AA,Kann2021AA}. The redshift is given by \cite{Fynbo2014AA}. The afterglow of this otherwise ordinary GRB distinguished itself by showing strong extinction and a peculiar spectrum \citep{Fynbo2014AA,Heintz2017AA}, as well as a late blue bump that can be interpreted as a highly luminous and slow SN component \citep{Kann2021AA}. The full analysis is presented in \cite{Kann2021AA}. Correcting for the very large extinction $A_V\approx1.1$ mag leads to $dRc=-1.537^{+0.285}_{-0.286}$ mag. The afterglow is moderately luminous, peaking at $R_C\approx14.3$ mag at early times.

\subsubsection{GRB 140629A, $z=2.276\pm0.001$}

Data are taken from \cite{Xin2018ApJ,Hu2019AA}, from GCN Circulars \citep{Bikmaev2014GCN16482,Maehara2014GCN16484,Takaki2014GCN16487,Kuroda2014GCN16488,Perley2014GCN16491,Perley2014GCN16498,Garnavich2014GCN16492,D'Avanzo2014GCN16493,Honda2014GCN16496,Moskvitin2014GCN16499,Moskvitin2014GCN16518,Yano2014GCN16501,Pandey2014GCN16517}, as well as our own \emph{Swift} UVOT data set, which expands that presented in \cite{Hu2019AA} and also fixes errors in the upper limits presented in that work. The redshift is given by \cite{Hu2019AA}. The light curve evolution is complex. The very first detections show a decay, before the light curve rises again to a peak, and breaks sharply into a typical decay. Using data up to 0.16 d, we find: $\alpha_{rise}=-0.848\pm0.097$, $\alpha_1=1.138\pm0.005$, $t_b=0.00238\pm0.00007$ d ($t_b=205.6\pm6.0$ s), $n=4.34\pm0.86$ ($\chi^2/d.o.f.=1.45$), with host magnitudes fixed to the values from \cite{Hu2019AA} or estimated (they have little influence as the afterglow is far brighter than the host). After 0.16 d, the afterglow shows a bumpy structure superposed on a steepening decay, likely an achromatic jet break \citep{Hu2019AA}, making the light curve similar to that of GRB 021004 \citep[\eg,][]{deUgarte2005AA} and GRB 060526 \citep{Thoene2010AA}. \cite{Xin2018ApJ} find $\alpha_{rise}=-0.92\pm0.24$, $\alpha_1=1.12\pm0.02$, $t_b=179\pm16$ s, in excellent agreement with our values. \cite{Hu2019AA} find a fit with an additional break in the decaying part of the light curve: $\alpha_{rise}=-0.72^{+0.15}_{-0.33}$, $\alpha_{1a}=0.91^{+0.03}_{-0.04}$, $t_b=176.85^{+3.48}_{-3.22}$ s, $\alpha_{1b}=1.17\pm0.01$, $t_b=638.69^{+126.31}_{-105.89}$ s. As expected their $\alpha_{1a,b}$ results straddle the single values from \cite{Xin2018ApJ} and our work.

The SED spans the full UV-to-NIR width ($uvw2\,uvm2\,uvw1\,$ $ubg^\prime vR_CI_Cz^\prime JHK_S$), with $uvw2\,uvm2\,uvw1\,u$ being affected by Lyman dropout and excluded. We note this is one of the highest redshifts at which an afterglow is still (albeit marginally) detected in $uvw2\,uvm2$. It is moderately red and visibly curved, a fit without dust finds $\beta=1.026\pm0.035$ ($\chi^2/d.o.f.=8.22$). A fit with MW dust yields negative extinction and is ruled out. The LMC dust fit shows a marked improvement but also a negative intrinsic spectral slope. The best fit is found with SMC dust: $\beta=0.197\pm0.136$, $A_V=0.309\pm0.048$ mag ($\chi^2/d.o.f.=2.24$). \cite{Xin2018ApJ} fit the X-ray-to-optical SED with a single power-law with $\beta=0.90\pm0.05$, and negligible extinction. This is based on a $BVR_CI_C$ SED only. \cite{Hu2019AA} tested two SEDs, at 775 s and at 9350 s. In the first there is a significant difference neither in the presence or absence of a cooling break between X-rays and optical, nor in which dust model is preferred. In the second SED, SMC dust is favored, but again the presence of a cooling break cannot be significantly discerned. For SMC dust, they derive $A_V=0.29\pm0.01$ mag and $A_V=0.25\pm0.03$ mag, respectively, in good agreement with our result. The spectral slopes are $\beta=0.489\pm0.01$ and $\beta=0.540\pm0.02$, respectively\footnote{Table 5 of \cite{Hu2019AA} gives $\Gamma=\beta+1=2.040$, but this is likely missing $\Delta\Gamma=0.5$ as in the first SED results of the table for the ``BKP'' models.}, steeper than what we find. Using our SMC dust fit, we derive $dRc=-2.772^{+0.167}_{-0.170}$ mag. The early afterglow peak is luminous, at $R_C=10.5$ mag.

\subsubsection{GRB 140801A, $z=1.3202\pm0.0001$}

Data are taken from \cite{Lipunov2016MNRAS,Pozanenko2017AAC} as well as our own \emph{Swift} UVOT analysis (first reported in \citealt{Hagen2014GCN16662}). Data from \cite{Pozanenko2017AAC} were made brighter by 0.15 mag to bring them into accordance with contemporaneous values from \cite{Lipunov2016MNRAS}.The redshift is given by \cite{Lipunov2016MNRAS}. This GRB represents a rare case of an afterglow which was discovered very early on by wide-field observations of a \emph{Fermi} GBM error box, \emph{without} the benefit of a LAT localization \citep{Lipunov2016MNRAS}. The afterglow shows a classic steep-shallow-steep decay. The initial decay slope is $\alpha_{steep}=1.425\pm0.076$, in full agreement with \cite{Lipunov2016MNRAS}, who find $\alpha_{steep}=1.42\pm0.12$. Following a data gap, the subsequent data decay following a broken power law, with the late steep decay partially based on a deep upper limit, and therefore the late-time decay slope is formally an upper limit, and also the break time may be somewhat different in reality. We find $\alpha_1=0.828\pm0.021$, $\alpha_2=2.37\pm0.16$, $t_b=1.44\pm0.10$ d, $n=5$ fixed. The shallow decay slope is also fully in agreement with \cite{Lipunov2016MNRAS}. The last detection by \cite{Pozanenko2017AAC} shows a rebrightening must have taken place, as those authors already point out.

The SED is relatively broad ($uvw1\,ug^\prime Vr^\prime R_Ci^\prime I_Cz^\prime JH$), with $uvw1$ being affected by Lyman damping and not included. The SED shows some scatter but is generally straight, we find a perfectly acceptable fit with no dust, $\beta=0.67\pm0.16$ ($\chi^2/d.o.f.=0.90$). SMC dust yields negative extinction. For MW and LMC dust, we also find acceptable fits with blue intrinsic spectral slopes and some extinction: $\beta=0.17\pm0.46$, $A_V=0.24\pm0.21$ mag ($\chi^2/d.o.f.=0.86$) for MW dust and $\beta=0.19\pm0.71$, $A_V=0.25\pm0.36$ mag ($\chi^2/d.o.f.=0.97$). These fits do not improve the extinction-less fit significantly and have large errors. Furthermore, \cite{Lipunov2016MNRAS} show the X-ray-to-optical SED is fit perfectly well with no dust and a simple power-law spanning the entire frequency range, with $\beta=0.81^{+0.04}_{-0.03}$, in agreement within errors with our result. We therefore continue with the fit without extinction, for which we derive $dRc=-0.701^{+0.027}_{-0.026}$ mag. The early afterglow is moderately luminous, at $R_C\approx13.3$ mag.

\subsubsection{GRB 141221A, $z=1.452$}

Data are taken from \cite{Bardho2016MNRAS}, from GCN circulars \citep{Guidorzi2014GCN17209,Schweyer2014GCN17212}, and from our own \emph{Swift} UVOT analysis (first reported in \citealt{Marshall2014GCN17219}) and REM seven-band observations (first reported in \citealt{Covino2014GCN17208}). The redshift is given by \cite{Perley2014GCN17228}. The afterglow shows a steep rise, which goes over into a shallow decay/plateau phase \citep{Bardho2016MNRAS} before breaking to a typical decay. We fit the data in two epochs. For the first epoch, we use data until the end of the plateau phase (0.0031 d), we find: $\alpha_{rise}=-4.69\pm0.86$, $\alpha_{plateau}=0.50\pm0.11$, $t_b=0.00117\pm0.0005$ d ($t_b=101.1\pm4.3$ s), $n=1$ fixed, and no host galaxy has been detected ($\chi^2/d.o.f.=2.10$). The second begins at 0.0014 d and yields: $\alpha_{plateau}=0.160\pm0.032$, $\alpha_1=1.221\pm0.015$, $t_b=0.00383\pm0.0012$ d ($t_b=331\pm10$ s), $n=5$ fixed, no host galaxy ($\chi^2/d.o.f.=3.51$, resulting from some scatter in the data). The difference in decay slopes for the plateau is likely a result from the break to the regular decay not being sharp, making it hard to evaluate when to end the first fit. \cite{Bardho2016MNRAS} obtain fits in several single bands, finding \eg, $\alpha_{rise}=-1.6\pm0.9$, $\alpha_{plateau}=0.5\pm0.2$, $t_b=110\pm13$ s. They also find the late decay is better fit by another broken power-law, with data starting at 337 s (almost exactly where we find the second break) they find $\alpha_1=1.0\pm0.2$, $\alpha_2=1.6\pm0.4$. Their rise index is shallower than what we find, they have probably used a sharper break. Their plateau decay is identical to our result, and our late-time decay index represents a mean value of what they find. The afterglow is only significantly detected until 0.11 d.

The SED is broad ($u b g^\prime V r^\prime R_C i^\prime I_C z^\prime J H K$), red and clearly curved. Without dust, we find: $\beta=1.629\pm0.021$ ($\chi^2/d.o.f.=39.6$). MW dust yields negative extinction. LMC dust find a very blue intrinsic slope and high extinction, there is no evidence for a 2175 {\AA} bump and the steepness of the far-UV slope is underestimated. SMC dust yields the only plausible solution, with $\beta=0.26\pm0.09$, $A_V=0.736\pm0.048$ mag ($\chi^2/d.o.f.=2.67$). This $15\sigma$ detection of extinction is one of the most secure measured so far. For a broken power-law X-ray-to-optical fit, \cite{Bardho2016MNRAS} find $\beta=0.3^{+0.3}_{-0.1}$, $A_V=0.569^{+0.063}_{-0.190}$ mag for LMC dust and 
$\beta=0.37^{+0.03}_{-0.09}$, $A_V=0.498^{+0.088}_{-0.147}$ mag for SMC dust. Their spectral slope is slightly redder than our result, and accordingly their extinction result is slightly lower, but generally, they confirm the quite high extinction toward this GRB. For our SMC result, we derive $dRc=-2.536^{+0.112}_{-0.115}$ mag. At $z=1$, the afterglow is luminous, peaking at $R_C\approx12.7$ mag.

\subsubsection{GRB 150301B, $z=1.5169$}

Data are taken from \cite{Gorbovskoy2016MNRAS}, from GCN circulars \citep{Kann2015GCN17522,Guidorzi2015GCN17530}, and from our own \emph{Swift} UVOT analysis (first reported in \citealt{Chester2015GCN17531}), and VLT X-shooter acquisition camera analysis (first reported in \citealt{deUgartePostigo2015GCN17523}). The redshift is given by \cite{Selsing2019AA}. The afterglow evolution is simple, we find it can be fit with a simple power-law decay from the earliest to the last detection (0.00097 to 0.82 d). We find $\alpha_1=0.9781\pm0.0094$ ($\chi^2/d.o.f.=1.70$). Based only on their MASTER data set, \cite{Gorbovskoy2016MNRAS} find $\alpha_1=1.16$. The SED is broad ($uvw1 u b g_G v r_G R_C i_G z_G J_G H_G$) and perfectly well fit by a simple power-law without dust, we find $\beta=0.806\pm0.089$ ($\chi^2/d.o.f.=0.13$). a fit with MW dust yields negative extinction. Fits with LMC and SMC dust are viable but only find small amounts of extinction, 0 within errors. We therefore prefer the fit without dust. \cite{Gorbovskoy2016MNRAS} use only unfiltered data but do find that the X-ray column density toward the GRB is equal to the (low) Galactic value, generally an indication of a low-$A_V$ line of sight (at this redshift). Using this fit, we find $dRc=-1.081\pm0.022$ mag. The early afterglow, which has one of the earliest detections in our sample, is luminous, peaking at $R_C\approx14$ mag.

\subsubsection{GRB 150910A, $z=1.3585$}

Data are taken from \cite{Xie2020ApJ}, from GCN circulars \citep{Kuroda2015GCN18267,Kuroda2015GCN18276,Kuroda2015GCN18301,Xu2015GCN18269,deUgartePostigo2015GCN18274,Moskvitin2015GCN18275,Schmidl2015GCN18277,Yanagisawa2015GCN18278,Cano2015GCN18279,Mazaeva2015GCN18281,Mazaeva2015GCN18289,Mazaeva2015GCN18327,Perley2015GCN18295,Butler2015GCN18302,Butler2015GCN18312,Andreev2015GCN18306,Sonbas2015GCN18314,Volnova2015GCN18319,Volnova2015GCN18320,Rumyantsev2015GCN18556}, and from our own \emph{Swift} UVOT analysis (first reported in \citealt{McCauley2015GCN18270}), as well as our FTN observations (first reported in \citealt{Dichiara2015GCN18266}). The redshift is given by \cite{Xie2020ApJ}. The afterglow shows a short decay at very early times which goes over into a very steep rise to a peak, covered only by our UVOT data. After a short data gap, the afterglow is found to be decaying in our FTN data as well as the observations presented by \cite{Xie2020ApJ}, the dense coverage shows some substructure in this decay. The late evolution is anomalous, the decay seems to steepen after 1.5 d to 2 d, but then slowly rises until at least 4 d, our final low-significance UVOT $white$ detections at 5.9 d shows the decay must have commenced again. We fit the steep rise and following decay, and find: $\alpha_{rise}=-5.60\pm0.40$, $\alpha_1=1.240\pm0.005$, $t_b=0.01061\pm0.0017$ d ($t_b=917\pm15$ s), $n=1$ fixed, no host galaxy ($\chi^2/d.o.f.=2.38$). \cite{Xie2020ApJ}, using a significantly smaller data set, especially at early times, find $\alpha_{rise}=-2.24\pm0.16$, $\alpha_1=1.36\pm0.03$, $t_{peak}=1451\pm51$ s.

The SED is very broad ($uvw2 uvm2 uvw1 u u^\prime b g^\prime v r^\prime R_C i^\prime I_C z^\prime$ $K_S$), the UV filters are affected by Lyman damping and are excluded, the rest of the SED is moderately red and slightly curved. We find $\beta=0.965\pm0.034$, which is similar to $\beta\approx0.9$ that \cite{Xie2020ApJ} find using an X-ray-to-optical fit but only two optical filters. With MW dust, we find negative extinction. We find viable fits for both LMC and SMC dust, for LMC dust: $\beta=0.50\pm0.17$, $A_V=0.245\pm0.085$ mag ($\chi^2/d.o.f.=0.90$), and for SMC dust: $\beta=0.53\pm0.14$, $A_V=0.169\pm0.050$ mag ($\chi^2/d.o.f.=0.42$). There is no clear evidence for a 2175 {\AA} bump so we prefer the SMC fit. Using this fit, we derive $dRc=-1.115^{+0.110}_{-0.112}$ mag, at $z=1$ the afterglow peak is moderately luminous, $R_C\approx14.3$ mag.

\subsubsection{GRB 151027B, $z=4.06463$}

Data are taken from \cite{Greiner2018AA}, from GCN circulars \citep{Watson2015GCN18512,Buckley2015GCN18511,Dichiara2015GCN18520}, from the automatic \emph{Swift} UVOT analysis page\footnote{\url{https://swift.gsfc.nasa.gov/uvot_tdrss/661869/index.html}, offline at the time of writing.}, and from our Ond\v rejov D50 observations (partially already presented in \citealt{Jelinek2019AN}). The redshift is given by \cite{Heintz2018MNRAS}. This is \emph{Swift's} 1000$^{\textnormal{th}}$ detected GRB, which showed strong radio variability \citep{Greiner2018AA}. The afterglow shows an initial rise (based on an upper limit and a single detection), which goes over into a shallow decay, before breaking into a steeper decay. The final detections at $\approx9$ d (which were not included in the fitting) lie $2\sigma$ under the extrapolation of the earlier decay, and even later deep upper limit adds further evidence that another break to an even steeper decay (likely a jet break) must have taken place, however, the lack of data prevents a more quantitative analysis. We find $\alpha_{plateau}=0.42\pm0.10$, $\alpha_1=1.088\pm0.021$, $0.23\pm0.07$ d, $n=10$ fixed, no host galaxy ($\chi^2/d.o.f.=1.10$). \cite{Greiner2018AA} find $\alpha_{plateau}\approx0.4$, $\alpha_1\approx1.4$, the steeper late decay likely stems from their inclusion of the data at $\approx9$ d.

The SED is clearly influenced by the moderately high redshift. The UVOT data yields relevant upper limits in $uvw1$ and $u$. For the detections ($g_G r^\prime r_G R_C i_G i^\prime I_C z_G z^\prime J_G H_G K_G$), bands blueward of $i_G$ are also influenced by Lyman damping and are not included. We find the SED is best fit by a simple power-law with no dust extinction: $\beta=0.526\pm0.087$ ($\chi^2/d.o.f.=1.18$). MW dust yields negative extinction, whereas LMC and SMC dust yield high extinction and negative slopes. From an X-ray-to-optical fit, \cite{Greiner2018AA} find a steeper spectral slope ($\beta=0.81\pm0.01$) but also no evidence for extinction. Using this fit, we find $dRc=-3.265^{+0.086}_{-0.088}$ mag. The early afterglow is moderately luminous, peaking at $R_C\approx14.6$ mag. However, even in the $z=1$ system, the first detection is only 370 s after trigger, at higher redshift, this moves out of the initial detection window of Gamow. We therefore do not include this afterglow in the statistics.

\subsubsection{GRB 161023A, $z=2.71067$}

Data are taken from \cite{deUgartePostigo2018AA}, where a full analysis is presented. The redshift is given by \cite{Heintz2018MNRAS}. The afterglow light curve shows an initial sharp flare likely linked to the prompt emission, followed by a large rise and a rollover into a typical decay, followed by a jet break at several days, all in all a very classical afterglow. The SED is fit by a small amount of SMC dust, and the very luminous afterglow peaks at $R_C=9.75$ mag.

\subsubsection{GRB 161219B, $z=0.1475$}

Data are taken from \cite{Cano2017AA}, \cite{Ashall2017Nature}, \cite{Laskar2018ApJ}, from GCN circulars \citep{Martin-Carrillo2016GCN20305,Fujiwara2016GCN20314,Buckley2016GCN20322,Buckley2016GCN20330}. The redshift is given by \cite{Selsing2019AA}. This is a low-redshift low-luminosity GRB with extensive optical follow-up. The very broad SED yields no evidence for dust and a spectral slope $\beta=0.497\pm0.011$, leading to $dRc=4.603^{+0.006}_{-0.007}$ mag. Based on mostly the \emph{Swift} UV data, which is unaffected by SN 2016jca, the afterglow is detected up to 100 d in the observer frame. In the $z=1$ frame, it is one of the least luminous detected afterglows, being $R_C\approx20.5$ mag even at very early times.

\subsubsection{GRB 180325A, $z=2.2486$}

Data are taken from \cite{Zafar2018ApJ,Becerra2021ApJ}, from GCN Circulars \citep{Littlefield2018GCN22538,Lipunov2018GCN22543,Malesani2018GCN22551}, as well as our own \emph{Swift} UVOT analysis (first reported in \citealt{Marshall2018GCN22549}), REM data (first reported in \citealt{DAvanzo2018GCN22536}), LT data (first reported in \citealt{Guidorzi2018GCN22534}), as well as data from the Ond\v rejov D50, first reported in \cite{Strobl2018GCN22541}, which will not be published in this work. The redshift is given by \cite{Zafar2018ApJ}. After initial rapid variability, the afterglow exhibits a bright rapidly rising and decaying flare (see also \citealt{Becerra2021ApJ}), for which we find $\alpha_{flare,rise}=-8.75\pm1.80$, $\alpha_{flare,decay}=2.12\pm0.14$, $t_{b,flare}=0.00112\pm0.00005$ d ($t_b=96.77\pm4.32$ s), $n=1.23\pm0.61$, and no host ($\chi^2/d.o.f.=0.92$). Beginning at 0.005 d, the afterglow can be described by a second smoothly broken power-law with $\alpha_1=0.442\pm0.008$, $\alpha_2=1.521\pm0.030$, $t_b=0.086\pm0.0015$ d ($t_b=7456\pm128$ s), a very sharp break $n=25.6\pm14.3$, and no host ($\chi^2/d.o.f.=1.59$). \cite{Zafar2018ApJ}, using a significantly sparser data set, find a similar break time ($t_b=6480\pm1200$ s) but more extreme slopes ($\alpha_1=0.02\pm0.03$, $\alpha_2=2.00\pm0.13$). They do not report a break sharpness, but a smaller value combined with the sparser data may explain the shallower, then steeper decay they report. \cite{Becerra2021ApJ} report $\alpha_1=0.46\pm0.01$, $\alpha_2=1.48\pm0.18$, in good agreement with our values (they also note the predicted decay slope from the $\alpha-\beta$ relations should be $\alpha=1.52$, in perfect agreement with our measurement).

The broad SED ($ubg_Gvr^\prime i^\prime z^\prime J_GHK_G$) has been extensively discussed in \cite{Zafar2018ApJ}, being very red and showing clear evidence for a 2175 {\AA} bump. For a fit without dust, we find $\beta=2.537\pm0.017$ ($\chi^2/d.o.f.=65.9$). Generally, $\chi^2/d.o.f.$ values are very high in this case as the well-defined optical light curve leads to very small errors in the normalization which form the SED, and even small deviations lead to large values. The existence of the bump also leads to SMC dust not yielding a viable result, with $\beta=2.598\pm0.052$, $A_V=0.027\pm0.022$ mag ($\chi^2/d.o.f.=76.4$). MW dust, while yielding a much better fit, is not able to account for the steepness of the SED, we find $\beta=1.969\pm0.040$, $A_V=0.431\pm0.028$ mag ($\chi^2/d.o.f.=35.2$). LMC dust, finally, yields a somewhat worse fit but values in agreement with GRB theories, namely $\beta=1.137\pm0.104$, $A_V=0.819\pm0.060$ mag ($\chi^2/d.o.f.=47.1$). To directly compare with the results of \cite{Zafar2018ApJ}, we fix the spectral slope to the value derived from the X-rays (\citealt{Zafar2018ApJ} find $\beta_X=0.84\pm0.10$, in full agreement with the XRT repository value of $\beta_X=0.79\pm0.09$), and find $A_V=1.011\pm0.010$ mag. Using a multiparameter Fitzpatrick-Massa parameterization, \cite{Zafar2018ApJ}, using four different SEDs, derive a higher $A_V\approx1.6$ mag, albeit also with a much higher value $R_V\approx4.5$. For the values we derive, we calculate a very large $dRc=-4.957\pm0.056$ mag. Correcting for this, we find the early flare of GRB 180325A was very luminous, peaking at $R_C\approx11$ mag.

\subsubsection{GRB 180720B, $z=0.654$}

This GRB does not yet have any optical observations published beyond GCN Circulars. We here present our own observations from telescopes of the LCOGT (the FTS, with later observations obtained by the FTN and the Haleakala Observatory; first reported by \citealt{Martone2018GCN22976}), the OSN T90 and T150 telescopes, first reported in \cite{Kann2018GCN22985}, the JAST/T80 telescope, first reported in \cite{Izzo2018GCN23040}, the REM telescope, first reported in \cite{Covino2018GCN23021}, as well as data from the Ond\v rejov D50, first reported in \cite{Jelinek2018GCN23024}, which will not be published in this work. \emph{Swift} UVOT did not observe the GRB due to a very bright star in the FoV. We also add data from GCN Circulars \citep{Sasada2018GCN22977,Crouzet2018GCN22988,Schmalz2018GCN23020,Lipunov2018GCN23023,Zheng2018GCN23033}. We furthermore obtained a late-time host-galaxy observation in $u^\prime g^\prime r^\prime i^\prime z^\prime$ with the GTC HiPERCAM, discovering the host and detecting it in all bands. The redshift is given by \cite{Vreeswijk2018GCN22996}. This was a very bright GRB at moderately low redshift which was also the first GRB for which evidence of ultrahigh-energy emission was discovered \citep[][, however, the early high-GeV/TeV emission of GRB 190114C was reported first, \citealt{MAGIC2019Nature1}]{Abdalla2019Nature}. This discovery resulted in extensive discussion in the literature \citep{Duan2019ApJ,Wang2019ApJ,Fraija2019ApJ,Ronchi2020AA,Sahu2020ApJ,Huang2020ApJ,Chen2021ApJ}. The GRB was also detected prominently in the radio range \citep{Rhodes2020MNRAS}.

The early afterglow is among the brightest ever discovered, at $R_C\approx9.4$ mag \citep{Sasada2018GCN22977}. We present an optical light curve of this event here for the first time based on our extensive observations spanning from early to late times. It initially shows a steep decay going over into an ill-defined plateau phase, we find $\alpha_{steep}=1.702\pm0.051$, $\alpha_{plateau}=-0.12\pm1.69$, $t_{b,early}=0.0126\pm0.0016$ d ($t_{b,early}=1089\pm134$ s), $n=-10$ fixed ($\chi^2/d.o.f.=0.04$). The initial steep decay is in accordance with an interpretation of a reverse-shock flash. While it may be assumed the plateau could be the onset of the forward-shock afterglow caught at peak, a back-extrapolation of the later data slightly underestimates this data, indicating another steeper decay must have taken place that then transits into the shallower decay we find next. The rest of our data is described by a smoothly broken power-law with $\alpha_1=0.856\pm0.021$, $\alpha_2=1.574\pm0.020$, $t_b=0.434\pm0.023$ d, $n=10.10\pm3.95$, and host galaxy values based on our GTC HiPERCAM detections. No SN component in the light curve has been reported so far, our data do not cover the time frame when a SN is expected. The broad SED ($Bg^\prime Vr^\prime R_Ci^\prime I_Cz^\prime JHK$) shows a small amount of scatter, but is otherwise straight. All three dust models yield slightly negative extinction, we prefer the fit with no dust, which yields $\beta=0.851\pm0.078$ ($\chi^2/d.o.f.=0.23$). Using this spectral slope, we derive $dRc=1.114\pm0.016$ mag.

\subsubsection{GRB 181201A, $z=0.450$}

Data are taken from \cite{Laskar2019ApJ,Belkin2020AstL}, and from GCN and ATel Circulars \citep{Kong2018GCN23475,Heintz2018GCN23478,Watson2018GCN23481,Ramsay2018GCN23503,Srivastava2018GCN23510,Kumar2019ATel12383}. The redshift is given by \cite{Izzo2018GCN23488}. This was a bright INTEGRAL GRB at moderately low redshift. It shows evidence for a rising SN component at late times \citep{Belkin2020AstL}. The SED shows no evidence for dust, we find $\beta=0.578\pm0.030$ and derive $dRc=1.984\pm0.010$ mag. The early peak reaches $R_C=14.1$ mag in the $z=1$ frame, but early coverage is sparse and the true peak mag is likely brighter.

\subsubsection{GRB 190114C, $z=0.4250$}

Data are taken from \cite{MAGIC2019Nature2,Jordana-Mitjans2020ApJ,Misra2021MNRAS,Melandri2022AA}, Th\"one et al., in prep., and GCN Circulars \citep{Kim2019GCN23732,Kim2019GCN23734,Mazaeva2019GCN23741,Watson2019GCN23751,Im2019GCN23757}, as well as some automatically reduced UVOT data (\url{https://swift.gsfc.nasa.gov/uvot_tdrss/883832/index.html}). The SN component (SN 2019jrj) was first reported by \cite{Melandri2019GCN23983} and is discussed in more detail in \cite{Melandri2022AA}. The host galaxy was first reported by \cite{deUgartePostigo2019GCN23692} and has been discussed in detail in \cite{deUgartePostigo2020AA}. The redshift is given by \cite{Kann2019GCN23710}. GRB 190114C was an exceptionally bright GRB, similar to GRB 130427A, it was a ''full-fledged cosmological'' GRB that occurred at a relatively low redshift. It is especially exceptional as a rapid reaction by the Major Atmospheric Gamma Imaging Cherenkov Telescopes (MAGIC Florian Goebel Telescopes) led to an early detection of radiation up to TeV energies \citep{MAGIC2019Nature1,MAGIC2019Nature2}. The bright prompt emission was detected by many satellites \citep{Ajello2020ApJ,Ursi2020ApJ,Minaev2019GCN23714,Xiao2019GCN23716}. The afterglow was detected right after the GRB at 10$^{\textnormal{th}}$ magnitude \citep{Jordana-Mitjans2020ApJ}, yielding a very rich data set, despite the high line-of-sight extinction \citep{Kann2019GCN23710,MAGIC2019Nature2,Misra2021MNRAS}. \cite{Campana2021AA} detect a decrease by a factor of two in the late X-ray column density, which they explain as a compact absorbing cloud that initially covers the entire radiating surface of the jet.

\subsubsection{GRB 210312B, $z=1.0690\pm0.0005$}

Data are taken from Jelinek et al., in prep. The redshift is given by \citet[][see also Jelinek et al., in prep.]{Kann2021GCN29655}. This was a relatively short ($T_{90}\approx5$ s) and faint GRB localized by INTEGRAL/IBAS \citep{Mereghetti2021GCN29650}. It was followed-up rapidly by the Ond\v rejov D50 telescope, which discovered the afterglow \citep{Jelinek2021GCN}, which decayed rapidly \citep{Kann2021GCN29653,Kann2021GCN29655}. The full analysis is presented in Jelinek et al., in prep.; we here focus on the late-time afterglow. We fit a small flaring episode and the subsequent decay with a smoothly broken power-law, finding $\alpha_1=-0.16\pm0.75$, $\alpha_2=1.18\pm0.09$, $t_b=0.046\pm0.014$ d. We fixed $n=10$, and all data have been host-galaxy subtracted. The afterglow is very faint, being $R_C\approx24.7$ mag at one day already. The SED ($Bg^\prime Vr^\prime R_Ci^\prime I_Cz^\prime$) is red but shows no significant sign of curvature. All three dust fits yield negative extinction, and therefore we proceed with a simple power-law fit with no dust which is perfectly acceptable: $\beta=1.186\pm0.077$, $\chi^2/d.o.f.=0.77$. Using this slope, we find $dRc=-0.188\pm0.003$ mag. Both observationally and at $z=1$, the afterglow of GRB 210312B represents one of the faintest ever followed-up in such detail.

\clearpage
\onecolumn

\begin{longtable}{llcll}
\caption{\label{obslog} GRB afterglow/supernova/host observations.}\\
\hline
\hline
GRB & $\Delta$t (days) & mag & filter & Telescope/Instrument\\
\hline
\endfirsthead
\caption{continued.}\\
\hline\hline 
GRB & $\Delta$t (days) & mag & filter & Telescope/Instrument\\
\hline  
\endhead
\hline
\endfoot
\vspace{1mm}
080413A	&	0.061861	& $	19.249	^{+	0.263	}_{-	0.212	}$ &	$b$	&	\emph{Swift} UVOT	\\\vspace{1mm}
	&	1.587929	& $ >	22.091					$ &	$b$ UL	&	\emph{Swift} UVOT	\\\vspace{1mm}
	&	1.718968	& $ >	21.616					$ &	$b$ UL	&	\emph{Swift} UVOT	\\\vspace{1mm}
	&	0.003005	& $	14.605	^{+	0.099	}_{-	0.091	}$ &	$v$	&	\emph{Swift} UVOT	\\\vspace{1mm}
	&	0.003177	& $	14.701	^{+	0.072	}_{-	0.068	}$ &	$v$	&	\emph{Swift} UVOT	\\\vspace{1mm}

\end{longtable}
\tablefoot{This table is available in its entirety at the CDS. All data are in AB magnitudes and not corrected for Galactic foreground extinction. Midtimes are derived logarithmically. $t=10^{([log(t_1-t_0)+log(t_2-t_0)]/2)}$, hereby $t_{1.2}$ are the absolute start and stop times, and $t_0$ is the GRB trigger time.}\\

\end{appendix}

\end{document}